\documentclass[%
reprint,
superscriptaddress,
showpacs,
 amsmath,amssymb,
 aps,
prb,
floatfix
]{revtex4-1}

\usepackage[utf8]{inputenc}
\usepackage[T1]{fontenc}
\usepackage{lmodern}

\usepackage[usenames]{color}
\usepackage{graphicx,subfigure}

\graphicspath{{figs/}}

\usepackage{booktabs}
\usepackage{dcolumn}
\usepackage{bm}
\usepackage[colorlinks=false]{hyperref}

\AtBeginDocument{
\heavyrulewidth=.08em
\lightrulewidth=.05em
\cmidrulewidth=.05em
\belowrulesep=.65ex
\belowbottomsep=0pt
\aboverulesep=.4ex
\abovetopsep=0pt
\cmidrulesep=\doublerulesep
\cmidrulekern=.5em
\defaultaddspace=.5em
}

\newcommand{\baslik}{Structural, Vibrational and Electronic Properties of \\Single Layer Hexagonal Crystals of Groups IV and V}

\newcommand{\wmax}{\omega_\mathrm{max}}

\begin{document}

\title{\baslik}

\author{B. \"{O}zdamar}
\affiliation{Izmir Institute of Technology, Department of Materials Science and Engineering, 35430 Urla, Izmir Turkey}
\author{G. \"{O}zbal}
\affiliation{Izmir Institute of Technology, Department of Physics, 35430 Urla, Izmir Turkey}
\author{M. N. \c{C}{\i}nar}
\affiliation{Izmir Institute of Technology, Department of Materials Science and Engineering, 35430 Urla, Izmir Turkey}
\author{K. Sevim}
\affiliation{Izmir Institute of Technology, Department of Physics, 35430 Urla, Izmir Turkey}
\author{G. Kurt}
\affiliation{Izmir Institute of Technology, Department of Materials Science and Engineering, 35430 Urla, Izmir Turkey}
\author{B. Kaya}
\affiliation{Izmir Institute of Technology, Department of Materials Science and Engineering, 35430 Urla, Izmir Turkey}
\author{H. Sevin\c{c}li}
\email{haldunsevincli@iyte.edu.tr}
\affiliation{Izmir Institute of Technology, Department of Materials Science and Engineering, 35430 Urla, Izmir Turkey}

\date{\today}

\begin{abstract}
Using first-principles density functional theory calculations, we investigate a family of stable two-dimensional crystals with chemical formula $A_2B_2$, where $A$ and $B$ belong to groups IV and V, respectively ($A$ = C, Si, Ge, Sn, Pb; $B$ = N, P, As, Sb, Bi).
Two structural symmetries of hexagonal lattices $P\bar{6}m2$ and $P\bar{3}m1$ are shown to be dynamically stable, named as $\alpha$- and $\beta$-phases correspondingly.
Both phases have similar cohesive energies, and the $\alpha$-phase is found to be energetically favorable for structures except CP, CAs, CSb and CBi, for which the $\beta$-phase is favored.
The effects of spin-orbit coupling and Hartree-Fock corrections to exchange-correlation are included to elucidate the electronic structures.
All structures are semiconductors except CBi and PbN, which have metallic character.
SiBi, GeBi and SnBi have direct band gaps, whereas the remaining semiconductor structures have indirect band gaps.
All structures have quartic dispersion in their valence bands, some of which make the valence band maximum and resemble a Mexican hat shape.
SnAs and PbAs have purely quartic valence band edges, \textit{i.e.} $E{\sim}{-}\alpha k^4$, a property reported for the first time.
The predicted materials are candidates for a variety of applications.
Owing to their wide band gaps, CP, SiN, SiP, SiAs, GeN, GeP can find their applications in optoelectronics.
The relative band positions qualify a number of the structures as suitable for water splitting, where CN and SiAs are favorable at all pH values.
Structures with quartic band edges are expected to be efficient for thermoelectric applications.

\end{abstract}

\maketitle

\section{\label{sec:intro}Introduction}

Successful exfoliation of graphene in 2004 aroused intensive research interest towards prospective two-dimensional (2D) monolayers possibly having novel electronic, structural, optical and thermoelectric properties \cite{novoselov2004electric}. Subsequently, synthesis of graphene analogs belonging to the same column of the periodic table, silicene, germanene and stanene (monolayers of silicon, germanium and tin, respectively) emphasized the significance of monoelemental single layer materials and their diverse fields of application \cite{balendhran2015elemental,wang2016van,cahangirov2009two,ezawa2015monolayer}. 
There also exist theoretical predictions and experimental realizations of  stable 2D honeycomb lattices of group-V elements (pnictogens), namely nitrogene, phosphorene, arsenene, antimonene and bismuthene.~\cite{lee2015two,kadioglu2017functionalization,
zhang2018recent,liu2014phosphorene,qiao2014high,li2014black,pizzi2016performance,madhushankar2017electronic,ji_two-dimensional_2016,zhang_atomically_2015}

\begin{figure}[b]
	\includegraphics[width=0.499\textwidth]{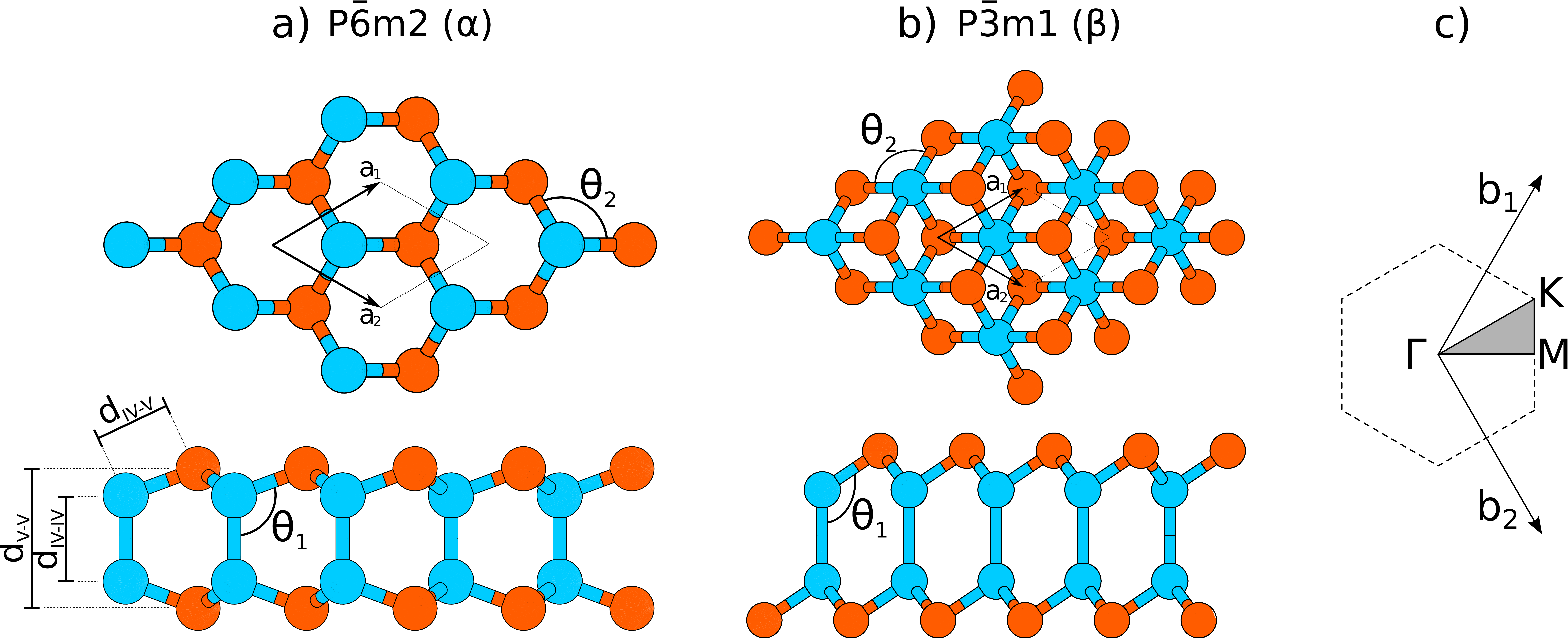}
	\caption{\label{fig:struct} Top and side views of the crystal lattice structure of group IV-V systems in two different space groups. (a) $P\bar{6}m2$ ($\alpha$ phase) and (b) $P\bar{3}m1$ ($\beta$ phase). (c) Reciprocal lattice of the hexagonal symmetry and the high symmetry points. Lattice vectors are represented by $\mathbf{a}_1$ and $\mathbf{a}_2$ which are equal in length $a$. Primitive unit cell comprises of four atoms, each of two are from either species. Group IV and V atoms are shown in blue and orange, respectively. Related structural parameters are detailed in Table \ref{tab:the-table} for $P\bar{6}m2$ and Table \ref{tab:AB-the-table} for $P\bar{3}m1$.}
\end{figure}

Besides group-IV and group-V monolayers, theoretical and experimental studies also concentrated on the other possibly stable compounds belonging to the cross columns of the periodic table, such as single layers of group-III \cite{demirci2017structural,guo2017defects} and group-IV monochalcogenides \cite{hu2016two,li2018recent}, V-IV-III-VI \cite{lin2017single} as well as I-III-VI$_{2}$ \cite{nagatani2015structural} compounds and their diverse fields of application. Among these, molecular electronics\cite{gao2013nanostructured,wang2014role}, energy conversion and storage devices such as photoelectrochemical water splitting cells \cite{chowdhury2017monolayer,lin2017single} and solar cells \cite{zhao2018band}, gas sensors \cite{late2012gas}, photon-counting detectors in ultraviolet-visible region \cite{liu2014high,hu2013highly}, and second harmonic generation (SHG) \cite{hu2017peculiar} are some of the most topical examples.


\begin{table*}[t]
\caption{\label{tab:the-table} Structural and electronic properties of IV-V monolayers (IV = C, Si, Ge, Sn, Pb; V = N, P, As, Sb, Bi) in $P\bar{6}m2$ space group ($\alpha$ phase). Lattice constant ($a$), bond length ($d_\mathrm{IV-IV}$, $d_\mathrm{V-V}$, $d_\mathrm{IV-V}$), bond angle ($\theta_1$, $\theta_2$), band gap calculated with two XC functionals PBE and HSE06 ($E_\mathrm{g}^\mathrm{PBE}$ and $E_\mathrm{g}^{HSE06}$, respectively), band gap calculated in PBE with spin-orbit coupling included ($E_\mathrm{g}^{SO}$), cohesive energy ($E_\mathrm{c}$) and charge transfer ($\Delta\rho$ = $\rho_\mathrm{IV}-\rho_\mathrm{V}$). FIC stands for the fractional ionic character. See Fig. \ref{fig:struct} for length and angle references.}
\begin{ruledtabular}
\begin{tabular}{lcccccdllccdd}

 & $a$ &$d_\mathrm{IV-IV}$ & $d_\mathrm{V-V}$ & $d_\mathrm{IV-V}$ & $\theta_1$ & \multicolumn{1}{c}{$\theta_2$} & \multicolumn{1}{c}{$E_\mathrm{g}^\mathrm{PBE}$} & \multicolumn{1}{c}{$E_\mathrm{g}^\mathrm{HSE06}$} & $E_\mathrm{g}^\mathrm{SO}$ & $E_c$ & \multicolumn{1}{c}{$\Delta\rho$} & \multicolumn{1}{c}{FIC} \\

 & ($\AA$) & (\AA) & (\AA) & (\AA) & (deg) &  \multicolumn{1}{c}{(deg)} & \multicolumn{1}{c}{(eV)} & \multicolumn{1}{c}{(eV)} & (eV) & (eV/atom) & \multicolumn{1}{c}{(e$^{-}$)} & \multicolumn{1}{c}{(\%)} \\

\hline

CN   & 2.38 & 1.63 & 2.65 & 1.46 & 110.35 & 108.58 & 3.71 (K$\Gamma$-M) & 5.14 (K$\Gamma$-M) & 3.71 & 6.16  & -1.85 & 5.8\\
CP\footnotemark[1]  & 2.90 & 1.55 & 3.29 & 1.88 & 117.50 & 100.39 & 1.82 (K) & 2.74 ($\Gamma$-K) & 1.81 & 5.35  &  0.27 &3.2\\
CAs\footnotemark[1]  & 3.11 & 1.53 & 3.46 & 2.04 & 118.27 & 99.41  & 1.21 ($\Gamma$-K) & 1.95 ($\Gamma$-K) & 1.12 & 4.79  &  0.41 &3.4\\
CSb\footnotemark[1]  & 3.41 & 1.53 & 3.70 & 2.25 & 118.97 & 98.52  & 0.28 ($\Gamma$-K) & 0.85 ($\Gamma$-K) & 0.13 & 4.42  &  1.18 &6.1\\
CBi\footnotemark[1]  & 3.60 & 1.49 & 3.80 & 2.38 & 119.09 & 98.37  & Metallic & Metallic & Metallic & 4.13  &  0.57 &6.8\\
\rule{0pt}{3ex}SiN  & 2.90 & 2.43 & 3.54 & 1.76 & 108.40 & 110.52 & 1.74 (K$\Gamma$-M) & 2.73 (K$\Gamma$-M) & 1.74 & 5.59  & -4.04 & 27.7\\
SiP  & 3.53 & 2.37 & 4.41 & 2.28 & 116.50 & 101.62 & 1.52 (K$\Gamma$-M) & 2.22 (K$\Gamma$-M) & 1.51 & 4.19  & -2.84 & 2.1\\
SiAs & 3.70 & 2.36 & 4.57 & 2.40 & 117.38 & 100.54 & 1.63 ($\Gamma$-M) & 2.27 ($\Gamma$-M) & 1.54 & 3.85  & -2.52 & 1.9\\
SiSb & 4.02 & 2.36 & 4.82 & 2.62 & 117.92 & 99.85  & 1.18 ($\Gamma$-M) & 1.76 ($\Gamma$-M) & 0.99 & 3.50  & -0.52 & 0.6\\
SiBi & 4.17 & 2.35 & 4.92 & 2.73 & 118.07 & 99.66  & 0.64 ($\Gamma$) & 1.15 ($\Gamma$) & 0.27 & 3.31  & -0.66& 0.4 \\
\rule{0pt}{3ex}GeN  & 3.10 & 2.57 & 3.90 & 1.91 & 110.40 & 108.53 & 1.17 (K$\Gamma$-$\Gamma$) & 2.25 (K$\Gamma$-$\Gamma$) & 1.17 & 4.28  & -2.77& 23.3 \\
GeP  & 3.66 & 2.51 & 4.65 & 2.37 & 116.91 & 101.12 & 1.35 (K$\Gamma$-M) & 2.05 (K$\Gamma$-M) & 1.34 & 3.60  & -1.52& 0.8 \\
GeAs & 3.82 & 2.50 & 4.80 & 2.49 & 117.54 & 100.33 & 1.20 ($\Gamma$-M) & 1.81 ($\Gamma$-M) & 1.08 & 3.36  & -1.39 &0.7\\
GeSb & 4.12 & 2.50 & 5.01 & 2.69 & 117.84 & 99.95  & 0.65 ($\Gamma$-M) & 1.15 ($\Gamma$-M) & 0.43 & 3.12  & -0.40 &0.0\\
GeBi & 4.26 & 2.49 & 5.09 & 2.78 & 117.85 & 99.94  & 0.22 ($\Gamma$) & 0.67 ($\Gamma$) & Metallic & 2.98  & -0.44 & 0.0\\
\rule{0pt}{3ex}SnN  & 3.42 & 2.97 & 4.44 & 2.11 & 110.30 & 108.63 & 0.12 (K$\Gamma$-$\Gamma$) & 0.89 (K$\Gamma$-$\Gamma$) & 0.13 & 3.78  & -3.52& 25.3 \\
SnP  & 3.95 & 2.89 & 5.22 & 2.56 & 117.05 & 100.94 & 1.29 (K$\Gamma$-M) & 1.91 (K$\Gamma$-M) & 1.28 & 3.28  & -2.57 & 1.3\\
SnAs & 4.09 & 2.88 & 5.37 & 2.67 & 117.74 & 100.04 & 1.14 ($\Gamma$) & 1.72 ($\Gamma$-M) & 1.05 & 3.10  & -2.39& 1.2\\
SnSb & 4.39 & 2.87 & 5.58 & 2.87 & 118.13 & 99.59  & 0.80 ($\Gamma$-M) & 1.28 ($\Gamma$-M) & 0.54 & 2.89  & -1.75 &0.2\\
SnBi & 4.51 & 2.86 & 5.66 & 2.96 & 118.26 & 99.43  & 0.48 ($\Gamma$) & 0.92 ($\Gamma$) & Metallic & 2.78  & -1.48 & 0.1\\
\rule{0pt}{3ex}PbN  & 3.63 & 3.17 & 4.78 & 2.24 & 110.92 & 107.98 & Metallic & Metallic & Metallic & 3.04  & -2.90 & 11.8\\
PbP  & 4.12 & 3.06 & 5.49 & 2.67 & 117.08 & 100.91 & 0.40 (K$\Gamma$-M) & 0.76 (K$\Gamma$-M)  & 0.38 & 2.83  & -1.81 & 0.5\\
PbAs & 4.25 & 3.05 & 5.62 & 2.77 & 117.66 & 100.18 & 0.36 (K$\Gamma$-M) & 0.67 ($\Gamma$-M) & 0.25 & 2.72  & -1.63 &0.6\\
PbSb & 4.53 & 3.03 & 5.82 & 2.96 & 118.01 & 99.75  & 0.28 ($\Gamma$-M) & 0.51 ($\Gamma$-M) & Metallic & 2.58  & -0.93& 1.9\\
PbBi & 4.63 & 3.02 & 5.88 & 3.03 & 118.10 & 99.62  & 0.06 ($\Gamma$) & 0.35 ($\Gamma$-M) & 0.01 & 2.50  & -0.91&2.4 \\

\end{tabular}
\end{ruledtabular}
\footnotetext[1]{Have lower total energies in $P\bar{3}m1$ ($\beta$) symmetry. Also see Table \ref{tab:AB-the-table} for structural properties of the marked compounds in $P\bar{3}m1$ space group.}
\end{table*}

Single-layer group-III monochalcogenides in hexagonal structure \cite{demirci2017structural,zhuang2013single,sundarraj2014recent,sun2017band,ayadi2017ab} gathered attention also for their intriguing thermoelectric properties. Interestingly, theoretical and experimental results demonstrate that they present so-called Mexican-hat shape dispersion at their valence band maximum (VBM), and consequently gives rise to Van-Hove singularity \cite{VanHove:pr:1953} in the density of states (DOS) near VBM \cite{li2014controlled,cao2015tunable,aziza2017tunable,chen2018large}. This phenomena gives rise to a large temperature-independent thermopower along with linear-temperature resistivity \cite{sevinccli2017quartic,newns1994quasiclassical}. Theoretical figure of merit ($ZT$) of these systems are found to increase dramatically upon reducing to monolayers from their corresponding bulk counterparts \cite{wickramaratne2015electronic}. Given the electronic configuration of 2D ``naturally bilayer'' group III-VI materials and their stabilities in hexagonal symmetry with honeycomb structure \cite{demirci2017structural}, group IV-V monolayers generated tremendous research interest with the purpose of disclosing their potential authentic characteristics.~\cite{barreteau2016high,zhang2016two,wadsten1967crystal,cheng2018monolayered,wu2016stability,zhou2018geas,shojaei2016electronic,huang2015highly,wang2016carbon,singhsingle,matta:beilstein:2018,xu:jcms:2018,miao:chemmater:2016,ashton2016computational}

Recent synthesis of two-dimensional SiP, SiAs, GeP, and GeAs binary compounds \cite{barreteau2016high} invoked significant amount of research interest in both experimental and theoretical aspects \cite{zhang2016two}. Bulk GeAs and SiAs are known to crystallize in a layered structure with monoclinic space group of $C2/m$ \cite{wadsten1967crystal,cheng2018monolayered}, having theoretical band gaps of 0.41 eV and 0.93 eV, respectively \cite{wu2016stability}. However, given the relatively low interlayer formation energies, GeAs and SiAs monolayers are manageable to be fabricated by mechanical cleaving from their bulk structures. Upon cleavage, monolayers of GeAs and SiAs in C2/m space group are calculated to have band gaps of 2.06 eV (direct) and 2.50 eV (indirect), respectively, using HSE06 functional \cite{zhou2018geas}. Furthermore, application of in-plane strain converts GeAs to a direct-gap material, while this is not true for the GeP monolayer, therefore rendering the latter is impractical for optoelectronic applications \cite{shojaei2016electronic}.

Various stoichiometries of Si$_x$P$_y$ (y/x $\geqslant$ 1) monolayers are studied by Huang et al. \cite{huang2015highly} in order to explore prospective stable or metastable structures of this promising compound by the use of global structural search algorithm and first-principles calculations. SiP monolayer in $P\bar{6}m2$ space group is found to be more stable than its bulk structure. 
Also, structural, electronic, vibrational, optical and thermoelectric properties of CP monolayers in different crystal structures are studied theoretically.~\cite{wang2016carbon,singhsingle}
It is very recently reported that $AB$ type monolayers, consisting of two sublayers ($A=$~C, Si, Ge, Sn and $B=$~Sb) are thermally and kinetically unstable due to imaginary frequencies in their phonon spectra, which can be stabilized by surface functionalization.
Also 16 of the $A_2B_2$ type monolayer structures of groups IV-V with $P\bar{6}m2$ symmetry were studied theoretically by Ashton \textit{et al.}~\cite{ashton2016computational} 

Despite fruitful outcomes of group IV-V compounds both experimentally and theoretically, the literature lacks a complete and exhaustive database for the mentioned set of compounds, especially the hexagonal lattices. Based on this fact, we systematically studied the structural, electronic and vibrational properties of experimentally or otherwise available 2D group IV-V binary monolayers (IV = C, Si, Ge, Sn, Pb; V = N, P, As, Sb, Bi) with hexagonal crystal structure belonging to the space groups of $P\bar{6}m2$ and $P\bar{3}m1$.
We find that both phases are dynamically stable and have small differences in their cohesive energies. Hence polymorphism is quite likely to take place.
Nevertheless the electronic structures of $\alpha$- and $\beta$-phases are quite similar due to their structural similarity.
We find a wide range of band gap values between 0.35 to 5.14~eV, which point to various possible applications for these structures such as water splitting.
The quartic band dispersions are of particular importance, since they may give rise to interesting magnetic transitions~\cite{sevinccli2017quartic,cao:prl:2015} and thermoelectric performance.~\cite{sevinccli2017quartic,wickramaratne2015electronic}

\section{\label{sec:method}Computational Details}

First-principles calculations were performed using \texttt{VASP} package~\cite{kresse1996efficient} in density functional theory (DFT) framework by employing projector augmented wave  (PAW) method and Perdew-Burke-Ernzerhof (PBE) exchange correlation (XC) functionals.~\cite{perdew1996generalized} Hybrid  Heyd-Scuseria-Ernzerhorf (HSE06) functionals are also employed to predict the energy band gaps correctly.~\cite{heyd2003hybrid} 
The systems were first subjected to various tests regarding the choice of plane-wave cut-off energies, and k-point grids in order to obtain optimum values for efficient calculations. The energy cut-off for plane-wave basis sets were taken to be ranging from 160~eV (PbBi) to 500~eV (CSb).
Brillouin zone (BZ) is sampled by $n{\times}n{\times}1$  ($n$ ranging between 7 to 14) k-point grids in the Monkhorst-Pack scheme~\cite{monkhorstpack} according to the test results.
A sufficiently large vacuum spacing of at least 15~{\AA} in the direction orthogonal to the monolayer was utilized to hinder the interaction between periodically repeated images. The convergence criteria for electronic and ionic relaxations are set to $10^{-6}$~eV and $10^{-3}$~eV/{\AA}, respectively.

Electronic band structures are calculated using both PBE and hybrid HSE06 functionals. HSE06, whose correlation part is only contributed from the PBE, mixes 25\% of the exact Hartree-Fock exchange and 75\% PBE exchange. The influence of the spin-orbit coupling (SOC) is also taken into consideration by employing the fully unconstrained noncollinear magnetic approach \cite{hobbs2000fully}. Force constants are calculated using Density Functional Perturbation Theory (DFPT)~\cite{dfpt} with super cell sizes ranging from  $4{\times}4{\times}1$ to $6{\times}6{\times}1$. k-point grids are chosen to be $10{\times}10{\times}1$. 
\texttt{Phonopy} package~\cite{togo2015first} was used to compute the phonon dispersion relations and the thermal properties such as heat capacity ($C_v$). 
Effective charge analysis is performed by using Bader's method. This method utilizes zero-flux surfaces in order to partition the charge distribution \cite{bader19981997,henkelman2006fast}. Cohesive energies ($E_c$) per atom is defined by the following expression,
\begin{equation}
E_c = |E_\mathrm{tot} - (n_\mathrm{IV} \cdot E_\mathrm{IV} + n_\mathrm{V} \cdot E_\mathrm{V})|/(n_\mathrm{IV}+n_\mathrm{V}),
\end{equation}
where $E_\mathrm{IV}$ and $E_\mathrm{V}$ are energies of neutral atoms belonging to groups IV and V, respectively; and $E_\mathrm{tot}$ is the total energy of the system calculated by the conjugate gradient (CG) geometry optimization method. $n_\mathrm{IV}(n_\mathrm{V})$ is the number of group IV(V) atoms in the unit cell, two atoms from each. Fractional ionic character (FIC) is calculated as
$FIC{=}1{-}\exp\left[{-}(\chi_1{-}\chi_2)^2/4\right]$,
where $\chi_1$ and $\chi_2$ are the electronegativities of the constituent atoms.  

\begin{figure}[b]
\includegraphics[width=0.38\textwidth]{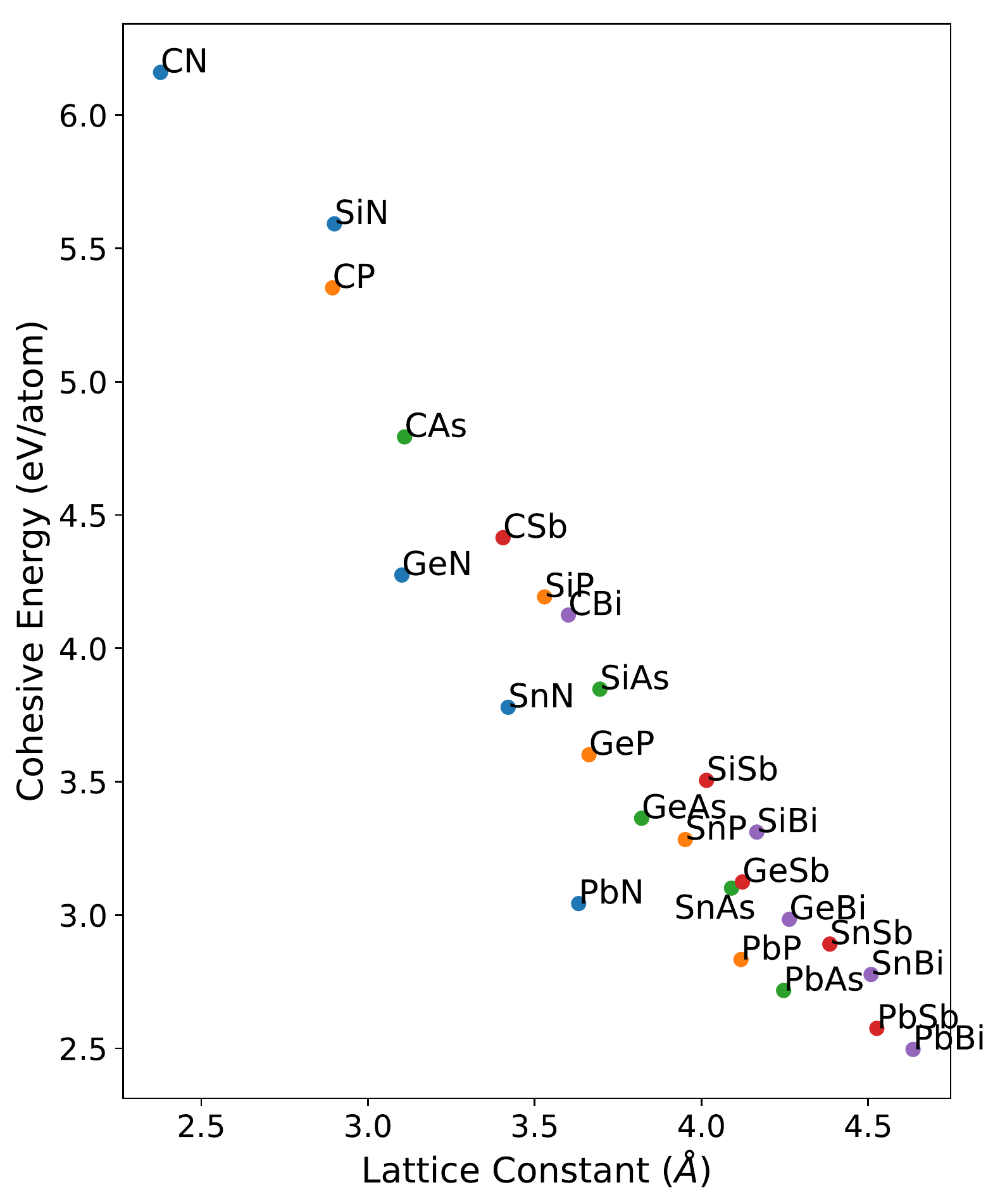}
\caption{\label{fig:a-vs-coh} The change in the cohesive energy with the lattice constant of structures in the $\alpha$-phase. Group V elements are shown in their respective colors.}
\end{figure}

\begin{table*}
\caption{\label{tab:AB-the-table}Structural and electronic properties of IV-V monolayers (IV = C, Si, Ge, Sn, Pb; V = N, P, As, Sb, Bi) in $P\bar{3}m1$ ($\beta$) symmetry. Lattice constant ($a$), bond length ($d_\mathrm{IV-IV}$, $d_\mathrm{V-V}$, $d_\mathrm{IV-V}$), bond angle ($\theta_1$, $\theta_2$), band gap calculated with two XC functionals PBE and HSE06 ($E_\mathrm{g}^\mathrm{PBE}$ and $E_\mathrm{g}^\mathrm{HSE06}$, respectively), band gap calculated in PBE with spin-orbit coupling included ($E_\mathrm{g}^{SO}$), cohesive energy ($E_\mathrm{c}$) and charge transfer ($\Delta\rho$ = $\rho_\mathrm{IV}-\rho_\mathrm{V}$). FIC stands for the fractional ionic character. See Figure~\ref{fig:struct} and Figure~\ref{fig:ab} for structural references and related plots, respectively.}
\begin{ruledtabular}
\begin{tabular}{lcccccdcccccc}

 & $a$ &$d_\mathrm{IV-IV}$ & $d_\mathrm{V-V}$ & $d_\mathrm{IV-V}$ & $\theta_1$ & \multicolumn{1}{c}{$\theta_2$} & $E_\mathrm{g}^\mathrm{PBE}$ & $E_\mathrm{g}^\mathrm{HSE06}$ & $E_\mathrm{g}^\mathrm{SO}$ & $E_\mathrm{c}$ & $\Delta\rho$ & FIC \\


 & ($\AA$) & ($\AA$) & ($\AA$) & ($\AA$) & (deg) &  \multicolumn{1}{c}{(deg)} & (eV) & (eV) & (eV) & (eV/atom) & (e$^{-}$) & \% \\

\hline

$\beta$-CP    & 2.90 & 1.53 & 3.67 & 1.89 & 117.35 & 100.56 & 1.91 ($\Gamma$-M)  & 2.77 ($\Gamma$-M) & 1.89 & 5.37 & 1.86 & 3.2\\
$\beta$-CAs   & 3.12 & 1.50 & 3.86 & 2.04 & 118.02 &  99.73 & 1.14 ($\Gamma$-M)  & 1.84 ($\Gamma$-M) & 1.08 & 4.82 & 0.38 & 3.4\\
$\beta$-CSb   & 3.42 & 1.49 & 4.16 & 2.25 & 118.77 & 98.77  & Metallic           & 0.43 ($\Gamma$-M) & Metallic & 4.45  & 1.11 & 6.1\\
$\beta$-CBi   & 3.60 & 1.44 & 4.33 & 2.39 & 119.63 & 97.67  & Metallic           & Metallic & Metallic & 4.18  & 0.62 & 6.8\\
\end{tabular}
\end{ruledtabular}
\end{table*}

\begin{figure*}[t]
     \begin{center}
        \subfigure
        {%
            \label{fig:first}
            \includegraphics[width=0.225\textwidth]{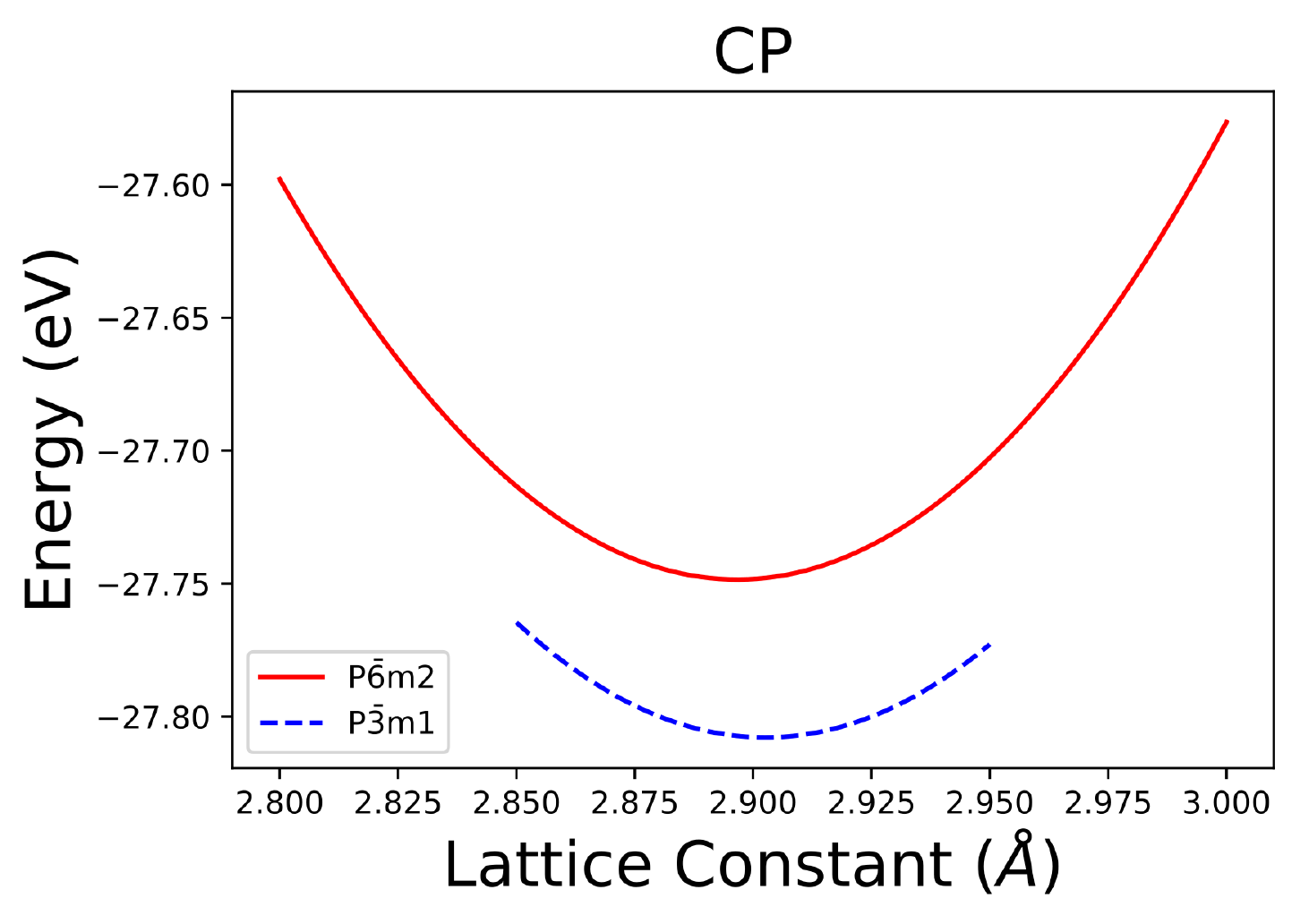}
			\includegraphics[width=0.225\textwidth]{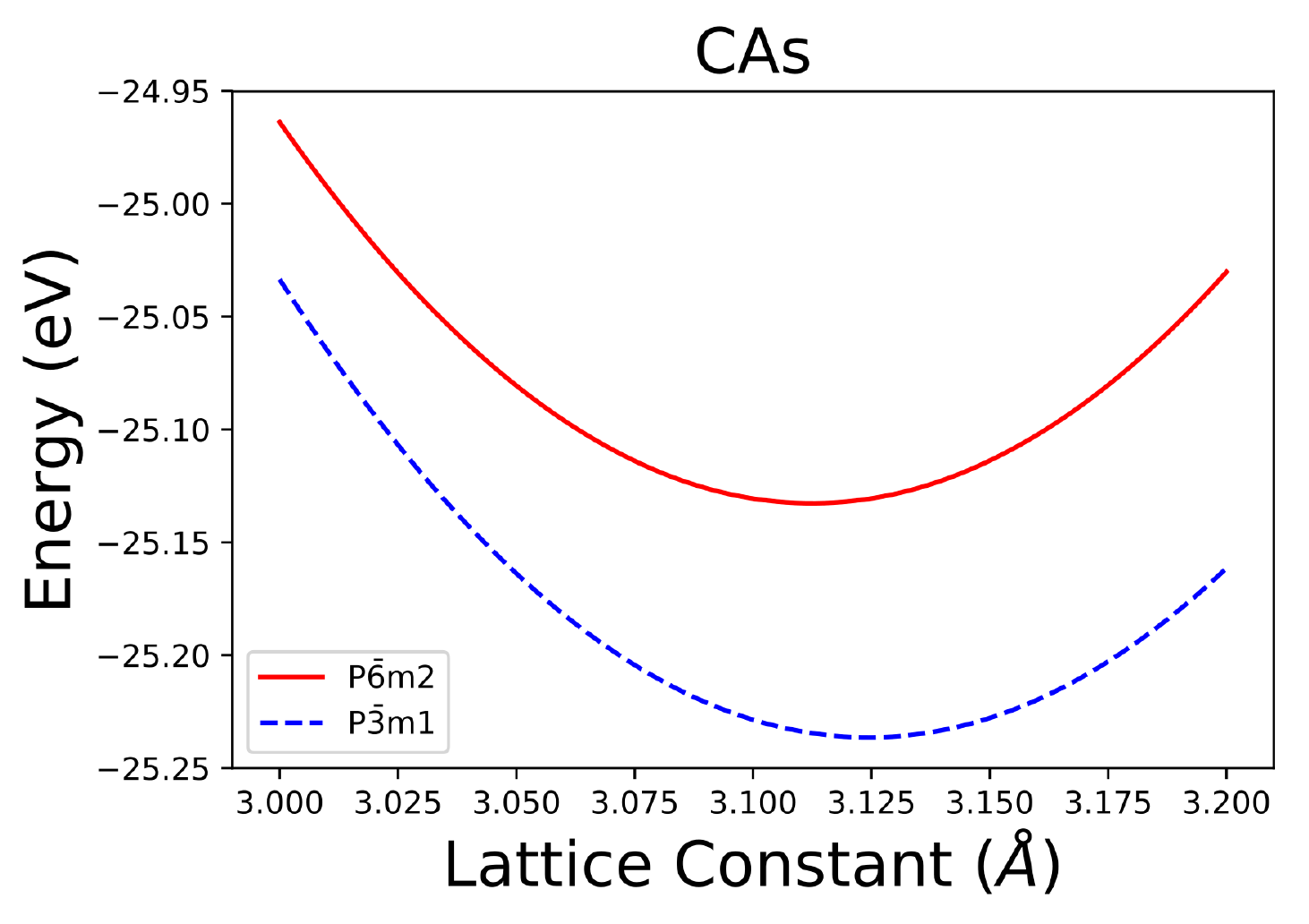}
			\includegraphics[width=0.225\textwidth]{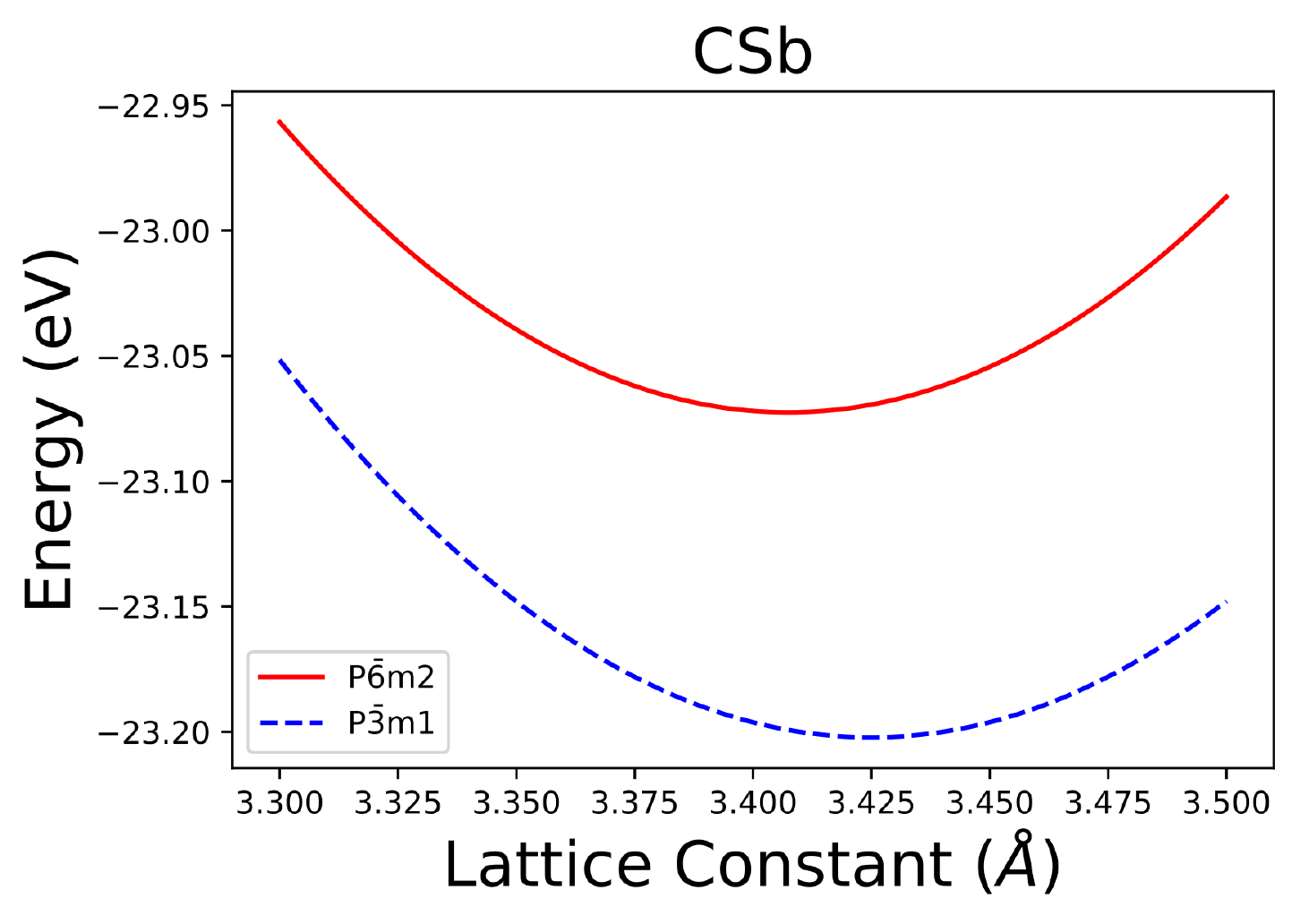}
			\includegraphics[width=0.225\textwidth]{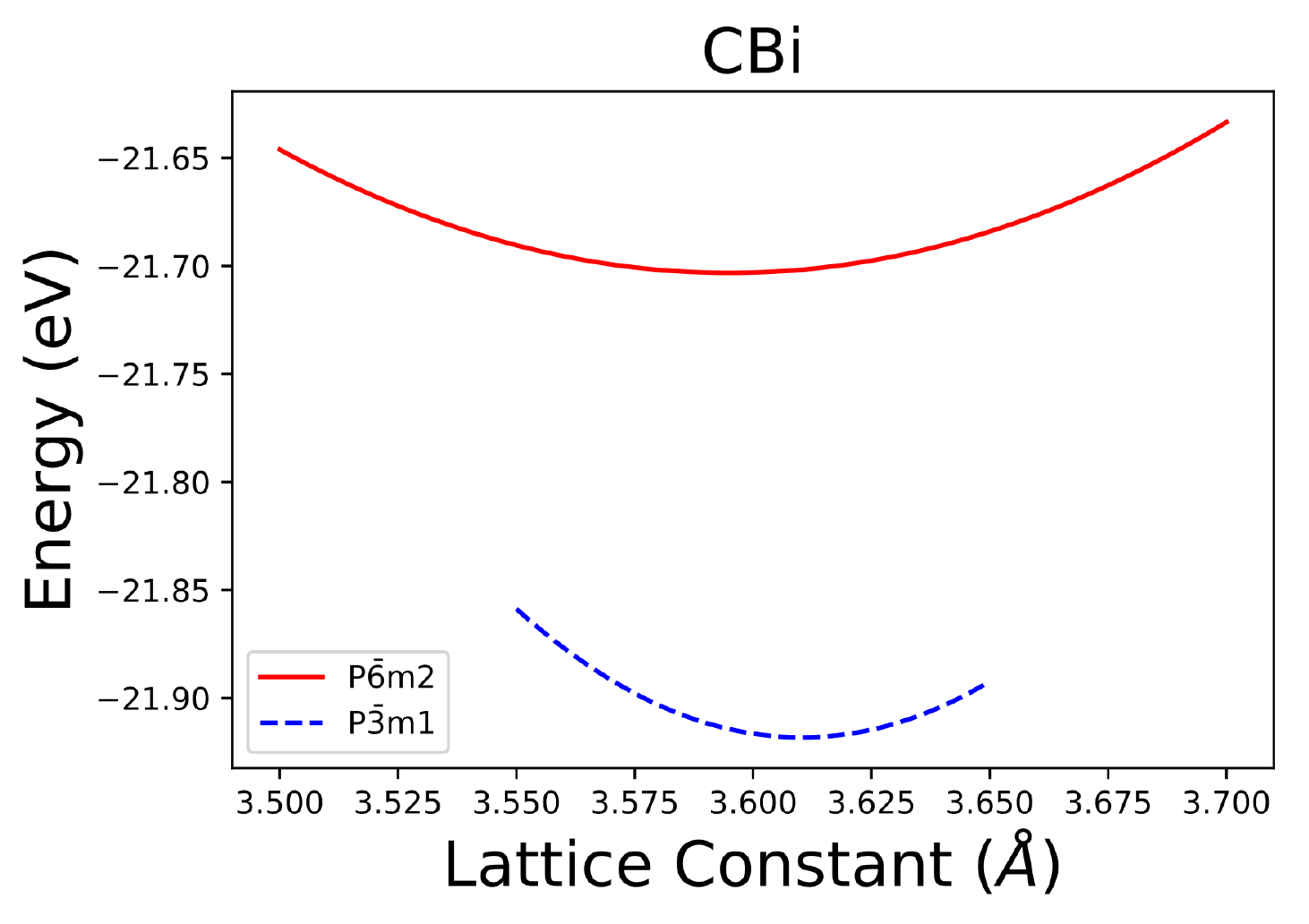}
        }\\
        \subfigure
        {%
            \label{fig:fourth}
			\includegraphics[width=0.225\textwidth]{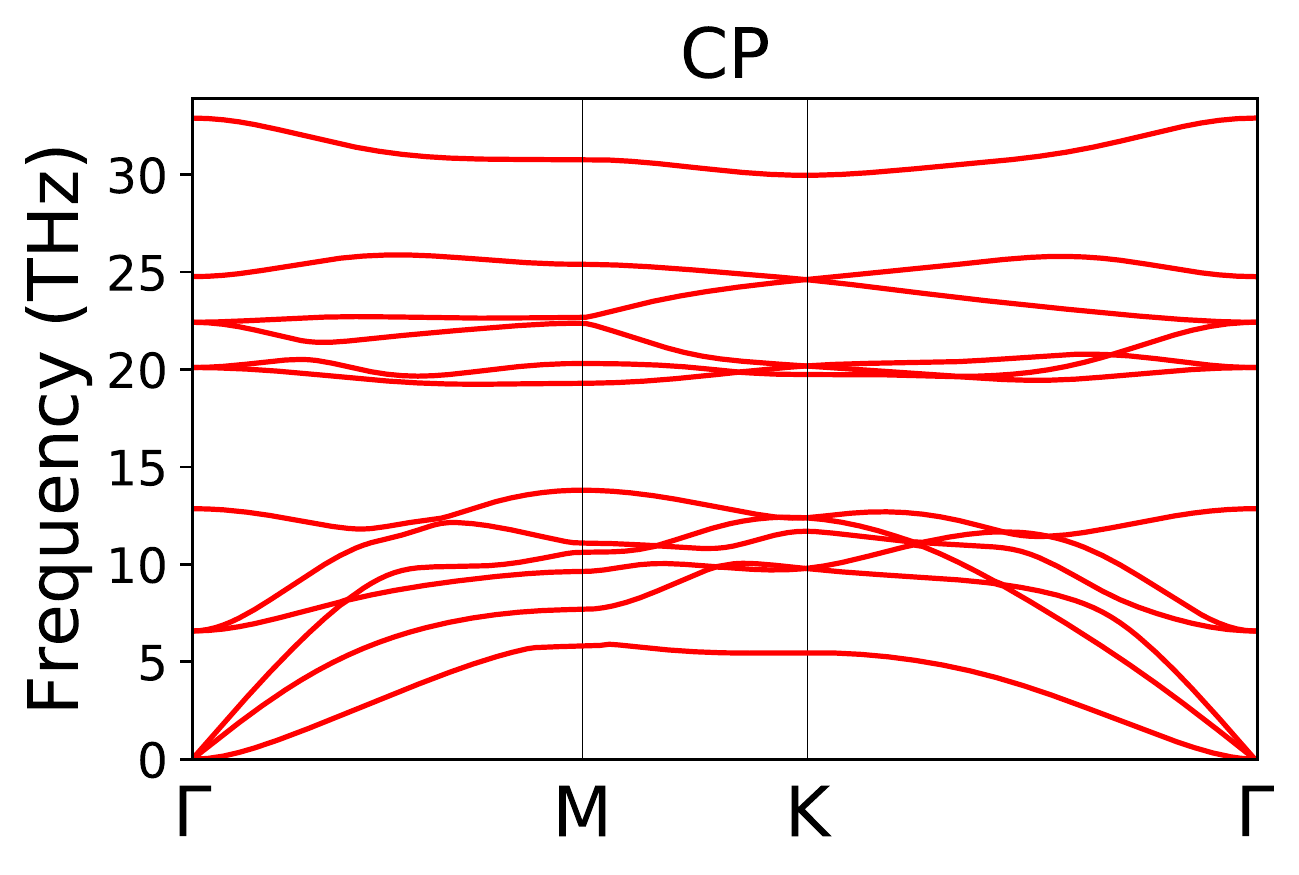}
			\includegraphics[width=0.225\textwidth]{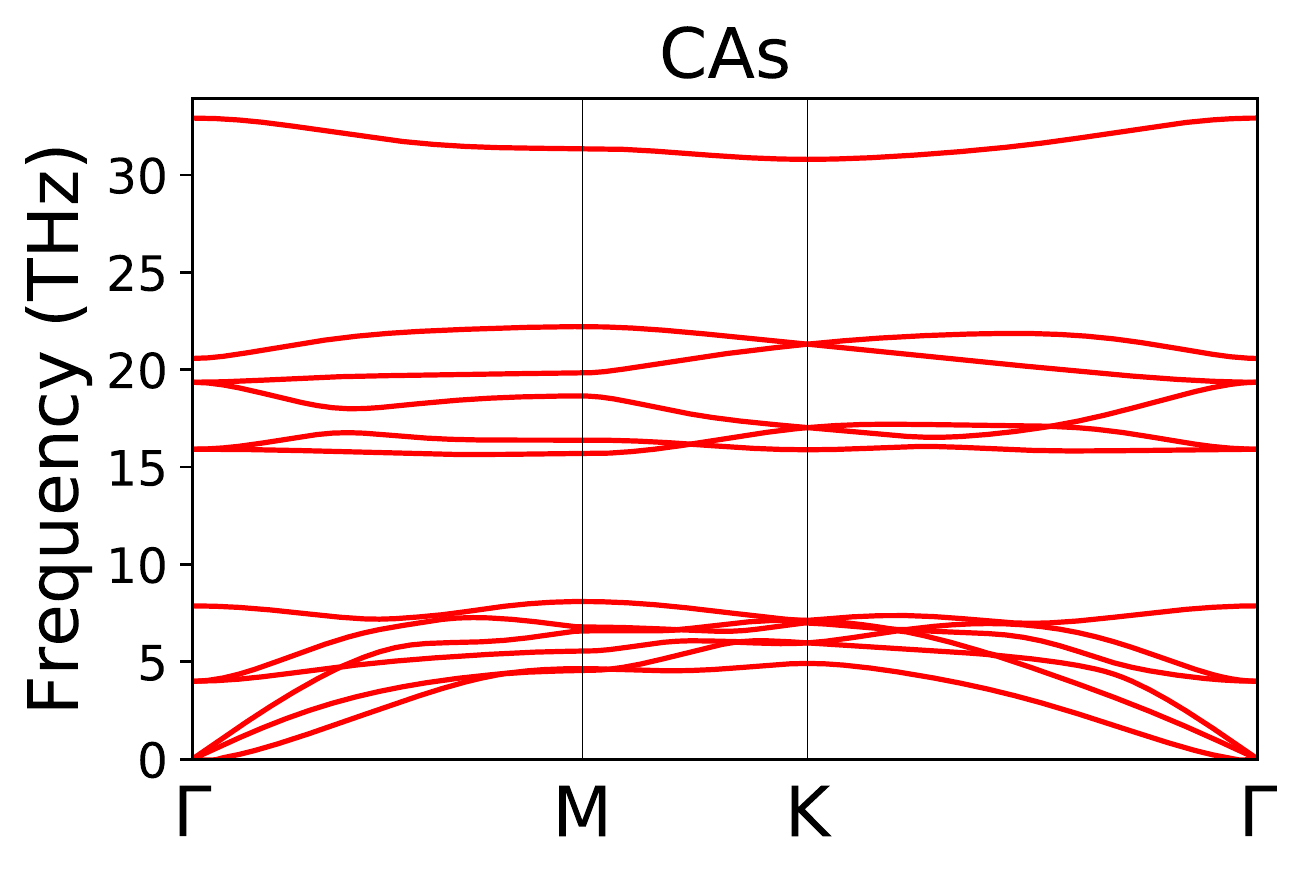}
			\includegraphics[width=0.225\textwidth]{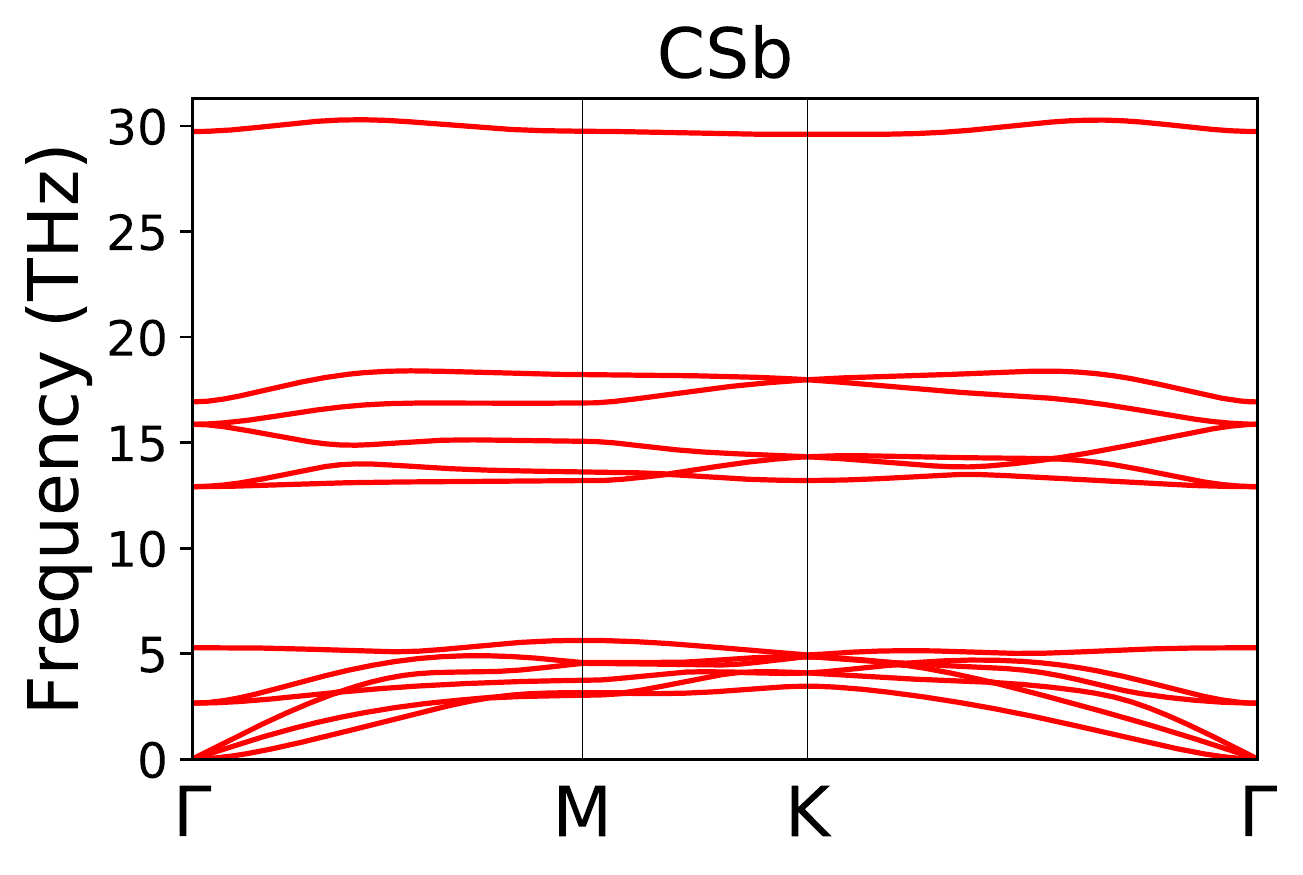}
			\includegraphics[width=0.225\textwidth]{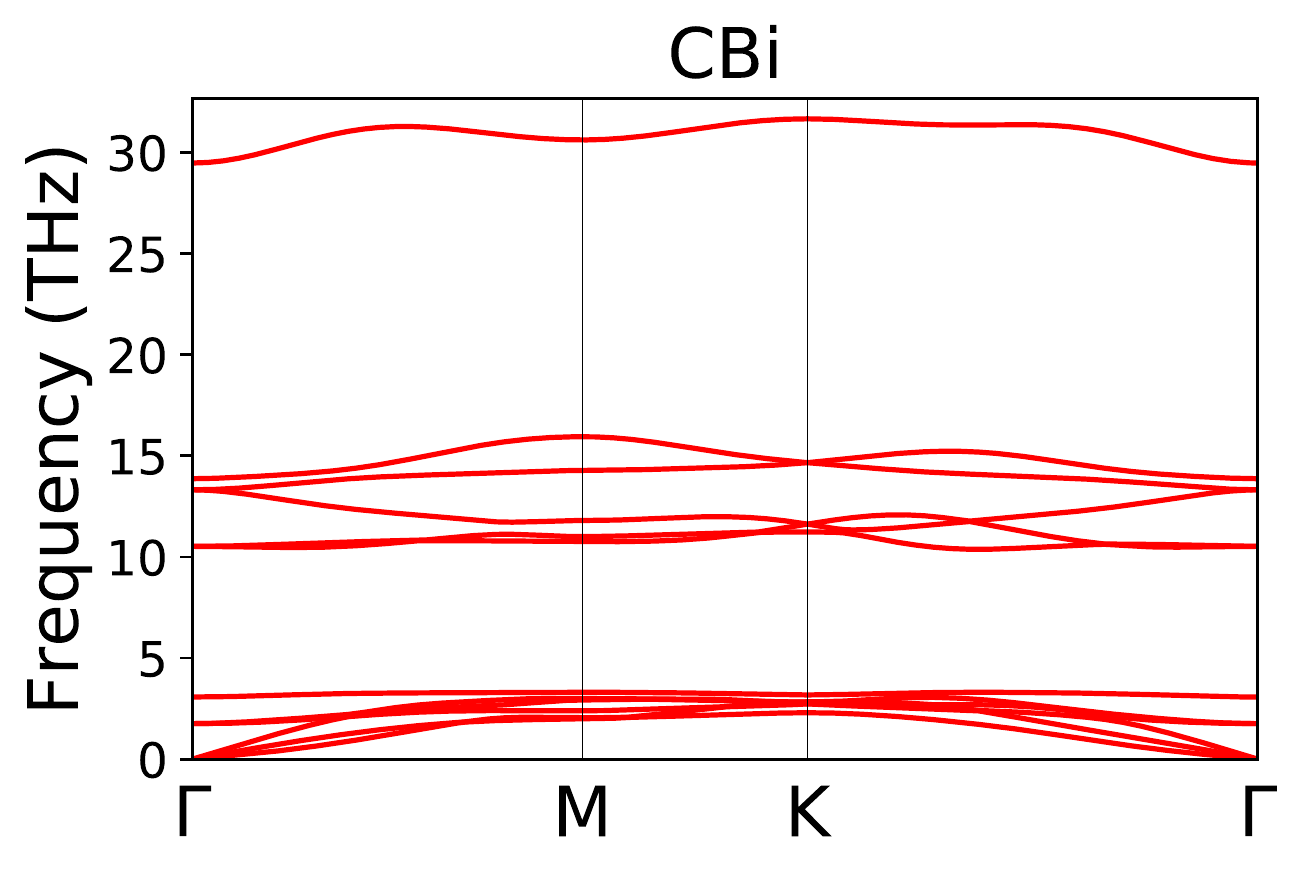}
        }\\
    \end{center}
    \caption{
    	The change in total energies with lattice constants of the two configurations ($P\bar{6}m2$ and $P\bar{3}m1$) shown in solid (red) and dashed (blue), respectively (top panel). Phonon dispersion relations of $\beta$-CP, $\beta$-CAs, $\beta$-CSb, $\beta$-CBi belonging to the $P\bar{3}m1$ symmetry (bottom panel). 
    	See Table \ref{tab:AB-the-table} for structural and electronic properties.}%
   \label{fig:ab-latt-and-phonon}
\end{figure*}

\section{\label{sec:results}Results and Discussion}

\subsection{Structural Properties}

We consider $\alpha$- and $\beta$-phases, which belong to $P\bar{6}m2$ and $P\bar{3}m1$ space groups, respectively.
$\alpha$-SiP was recently shown to be stable.~\cite{huang2015highly}
Also, group IV-V elements are expected to have structurally equivalent compounds with group III-VI monolayers, which have stable $\alpha$-phases.~\cite{demirci2017structural}
Both $\alpha$- and $\beta$-phases have $A_2B_2$ stoichiometry ($A$=C, Si, Ge, Sn, Pb; $B$=N, P, As, Sb, Bi) with $B{-}A{-}A{-}B$ stacking.
Stable geometric structures of the $\alpha$-phase is illustrated in Figure~\ref{fig:struct}(a), which is found to be more stable than the $\beta$-phase for most of the structures.
Structural and electronic properties of the $\alpha$-phase compounds are detailed in Table~\ref{tab:the-table}. Lattice constants, bond lengths ($d_\mathrm{IV-IV}$, $d_\mathrm{V-V}$, $d_\mathrm{IV-V}$), and bond angles were obtained by performing structural optimization. Lattice constant $a$ tends to increase steadily for the compounds within the same group IV. Bond lengths $d_\mathrm{V-V}$ and $d_\mathrm{IV-V}$ also follow the same trend. 
In contrast, $d_\mathrm{IV-IV}$ follows an opposite trend compared to the other distances. 
It tends to decrease within the group IV. Bond angle $\theta_1$, on the other hand, increase in the same group IV compounds, however very slightly. It is also noteworthy that only the nitrides possess a relatively narrower $\theta_1$. This can be explained by the highest electronegativity that nitrogen bears in its group. In comparison, $\theta_2$ decreases within the same group IV, which is also associated with the increasing trend that the bond length $d_\mathrm{V-V}$ demonstrates. Figure~\ref{fig:a-vs-coh} shows the variation of the cohesive energy $E_c$ with the lattice constant. $E_c$ has a tendency to decrease systematically as one goes down in group IV. High cohesive energy is a result of the stability of the material, which implies that nitrides are more stable than its neighbors in group IV.

The $\beta$-phase compounds are energetically more favorable for four structures, CP, CAs, CSb, and CBi.
In Figure~\ref{fig:ab-latt-and-phonon}(a), the total energies of $\alpha$- and $\beta$-phases of these compounds are plotted as functions of lattice constants, where it is observed that the minimum energies for the $\beta$-phases are 25 to 110~meV/atom lower than their $\alpha$-phase counterparts. (see Figure~\ref{fig:ab-latt-and-phonon}(a))
The phonon dispersions in Figure~\ref{fig:ab-latt-and-phonon}(b) show that the $\beta$-phases are dynamically stable.
An interesting point is the charge transfer characteristics of these compounds. 
As summarized in Table \ref{tab:the-table}, the only group of compounds that represent charge transfer from group-IV to group-V elements are the compounds whose energies are lower in $P\bar{3}m1$ symmetry. 
This is probably due to Coulombic repulsion between group-V atoms.
Table \ref{tab:AB-the-table} presents structural and electronic parameters of the mentioned compounds. Lattice constants, bond lengths $d_\mathrm{V-V}$ and $d_\mathrm{IV-V}$, and additionally bond angle $\theta_1$ follow an increasing trend when going down in the same group. Bond length $d_\mathrm{IV-IV}$ and bond angle $\theta_2$ represents an inverse behavior. When compared, cohesive energies are at most 1.21\% larger than that of their $P\bar{6}m2$ counterparts.

\begin{figure*}[t]
   \centering
\begin{tabular}{cccccc}
\\
\includegraphics[width=0.19\textwidth]{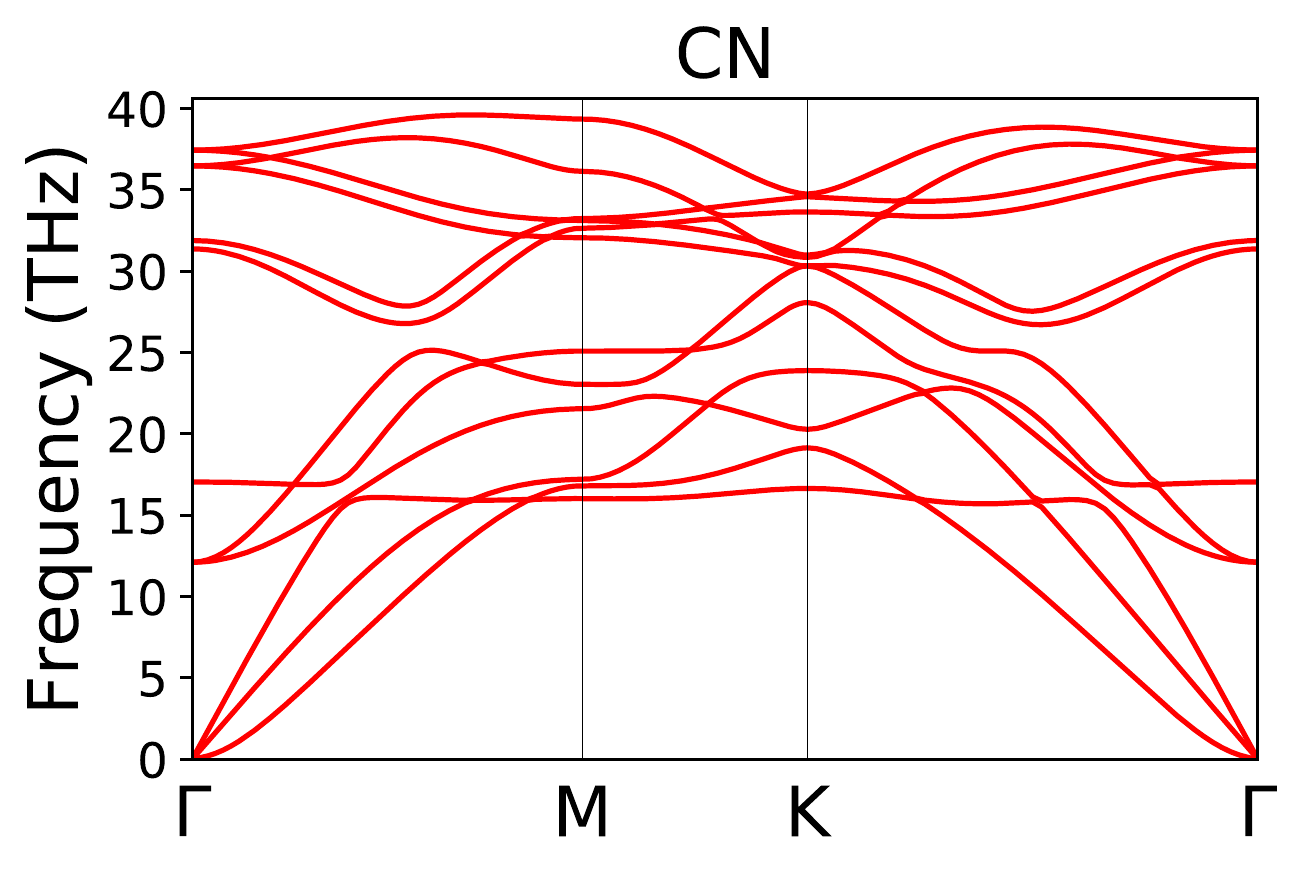}&
\includegraphics[width=0.19\textwidth]{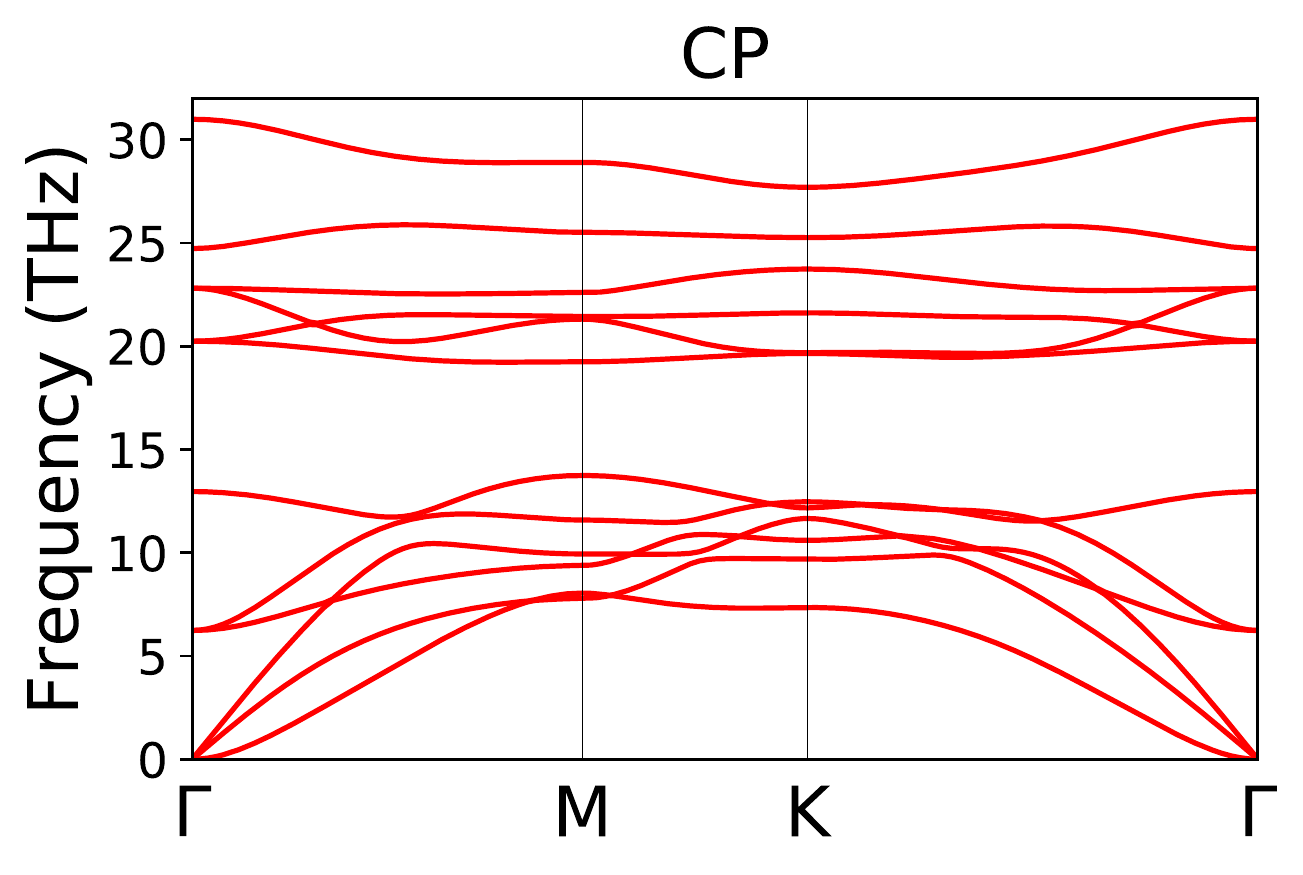}&
\includegraphics[width=0.19\textwidth]{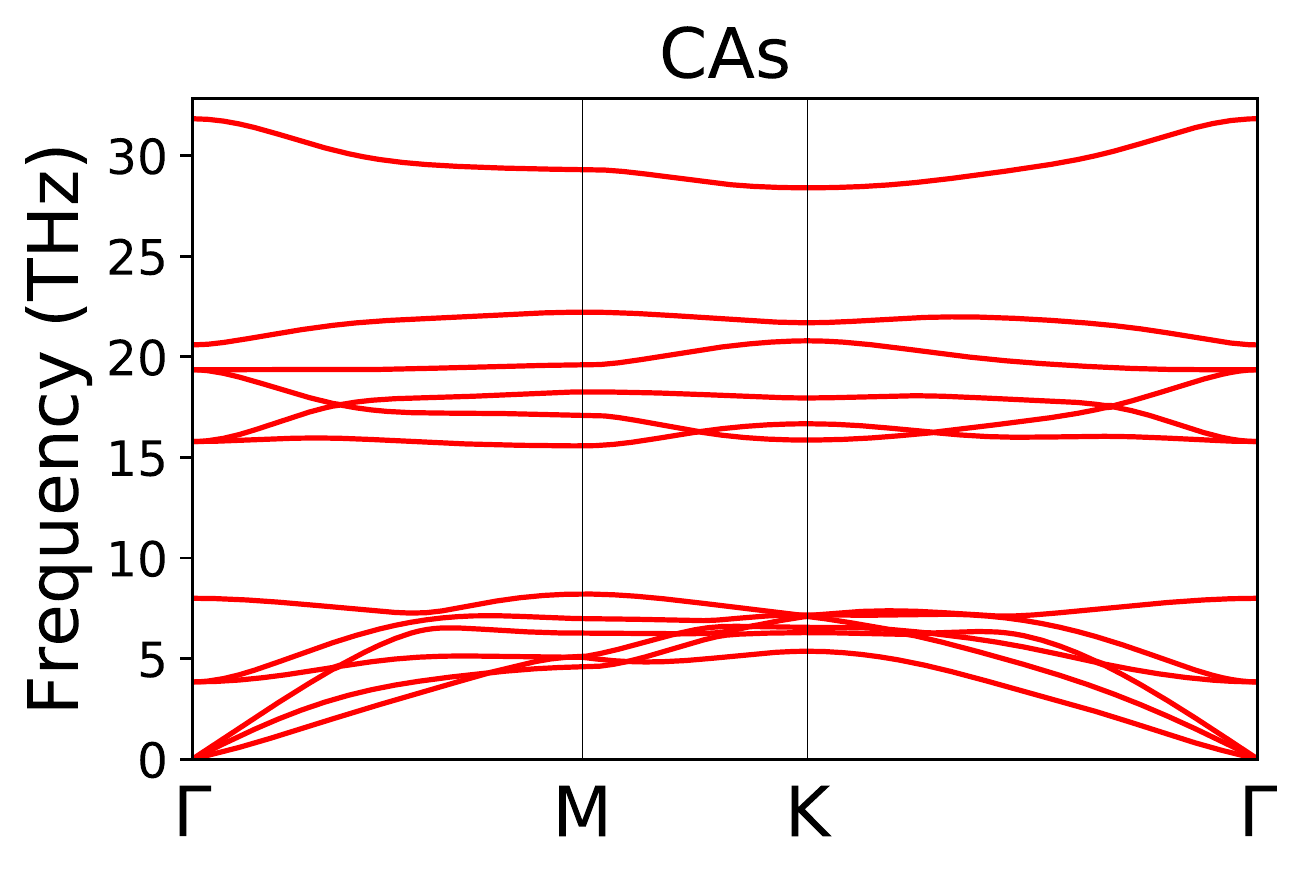}&
\includegraphics[width=0.19\textwidth]{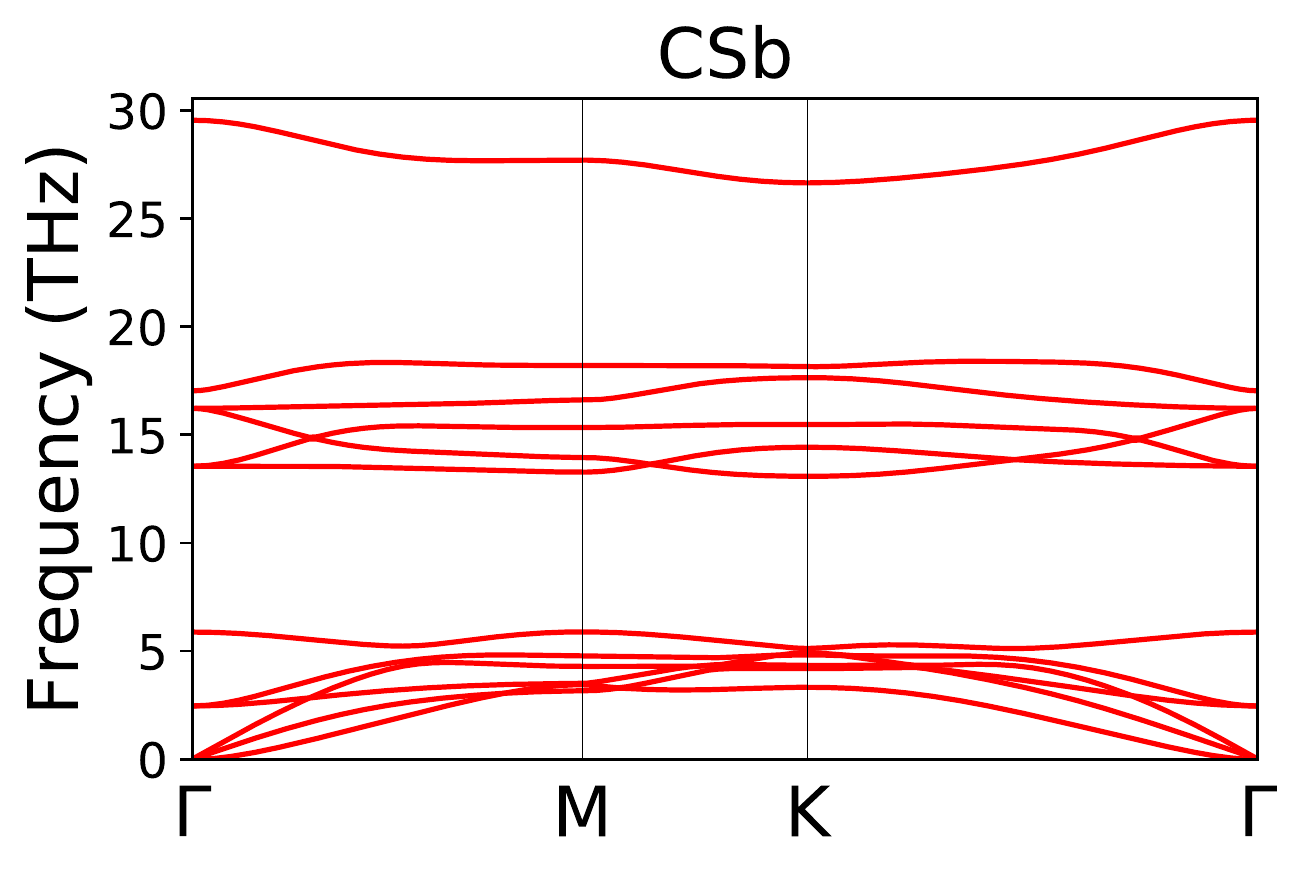}&
\includegraphics[width=0.19\textwidth]{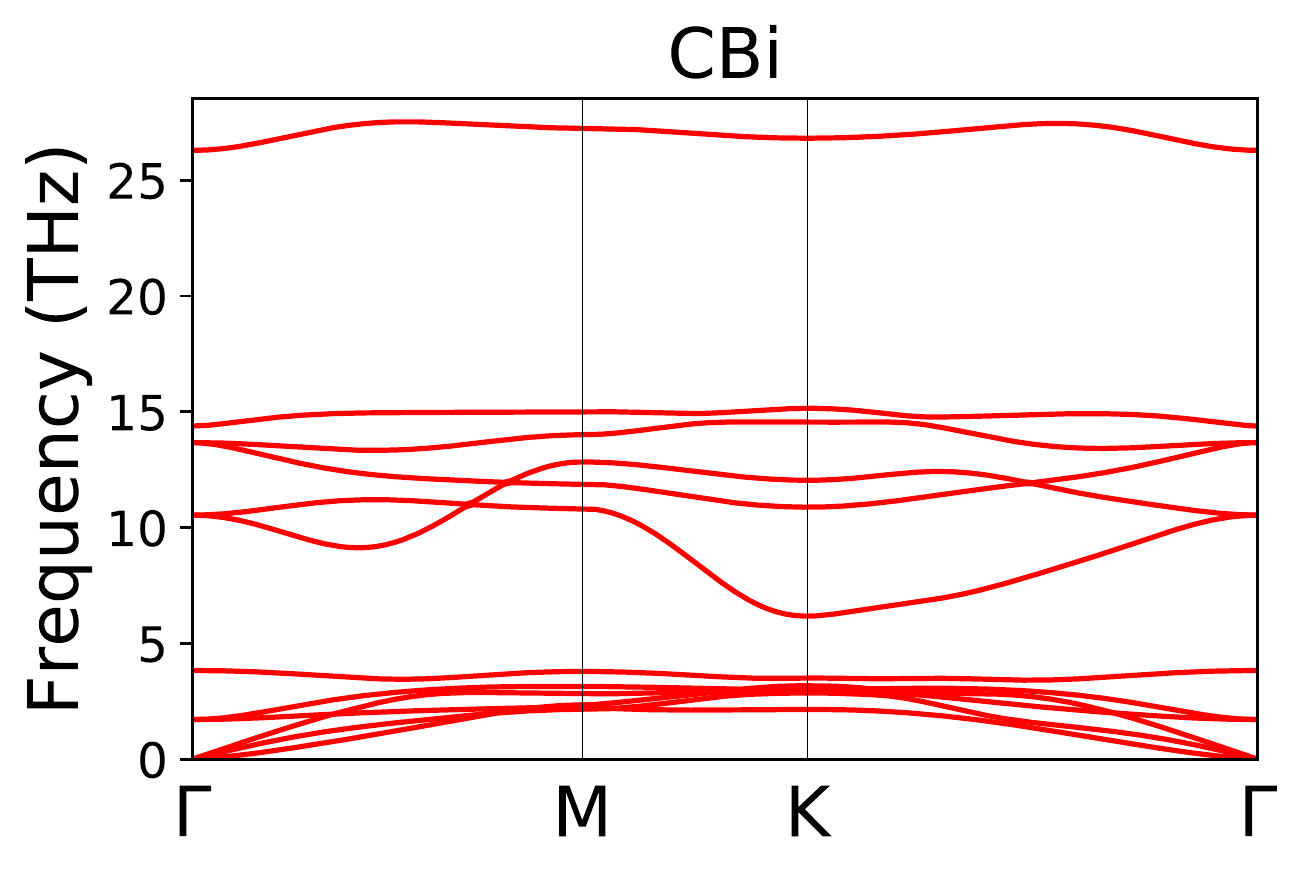}&\\

\includegraphics[width=0.19\textwidth]{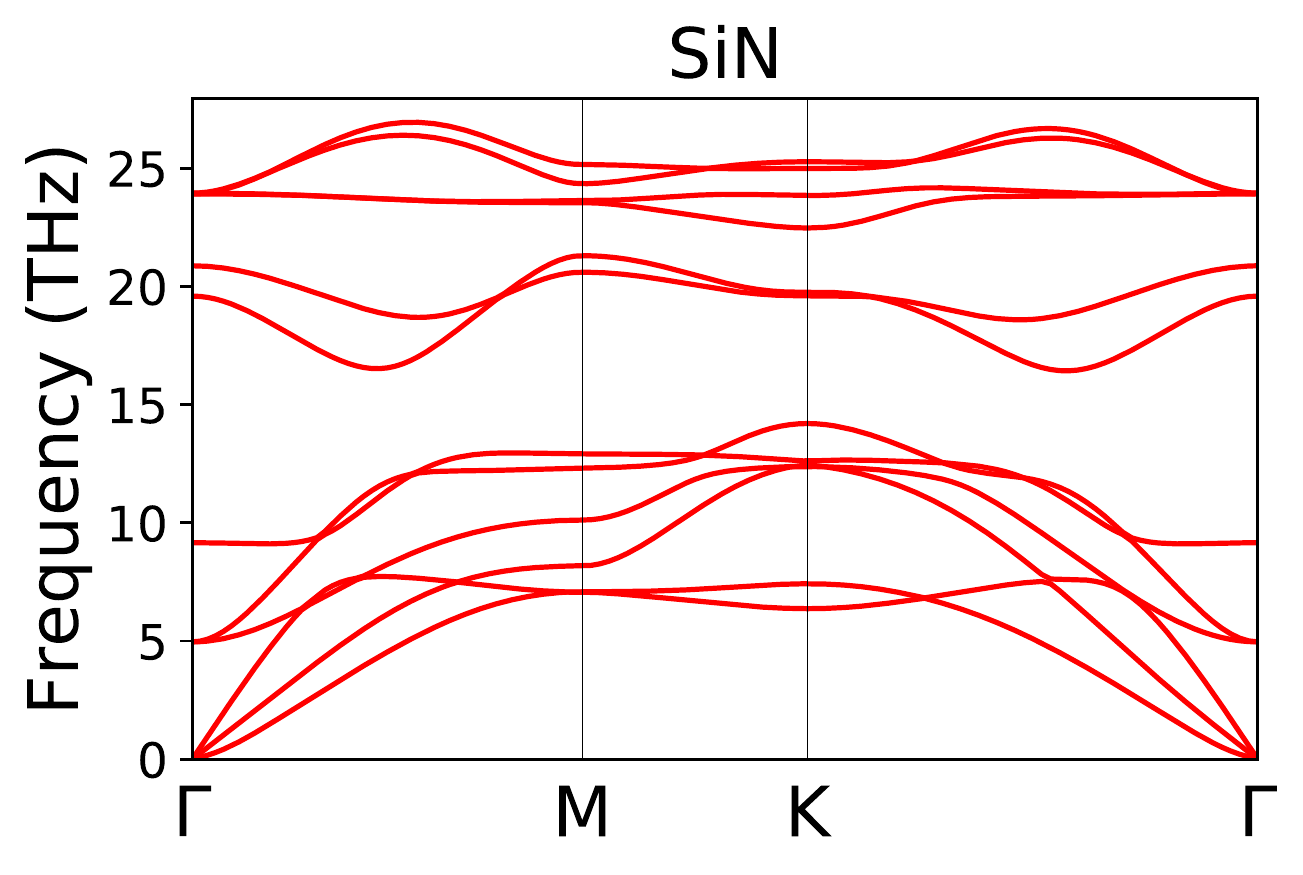}&
\includegraphics[width=0.19\textwidth]{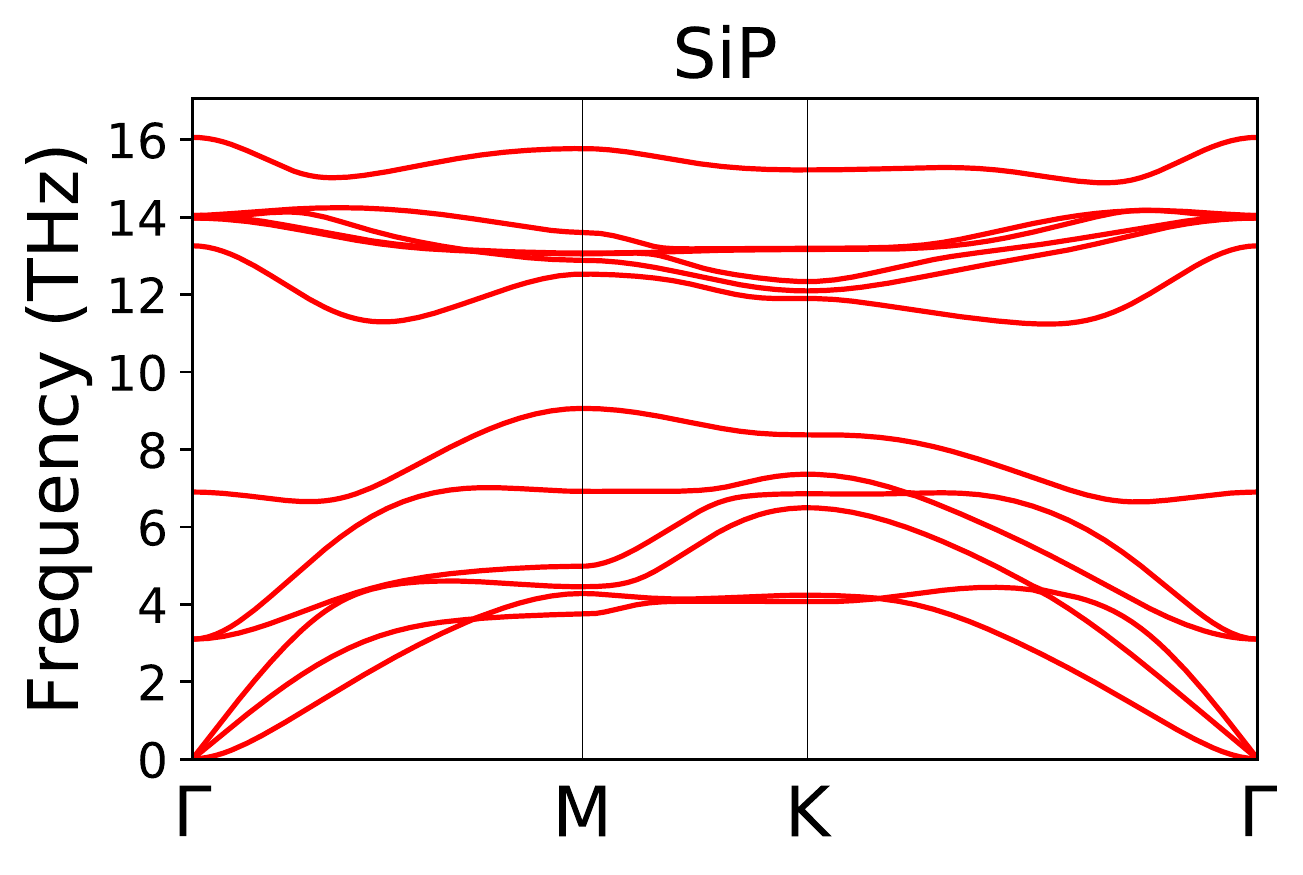}&
\includegraphics[width=0.19\textwidth]{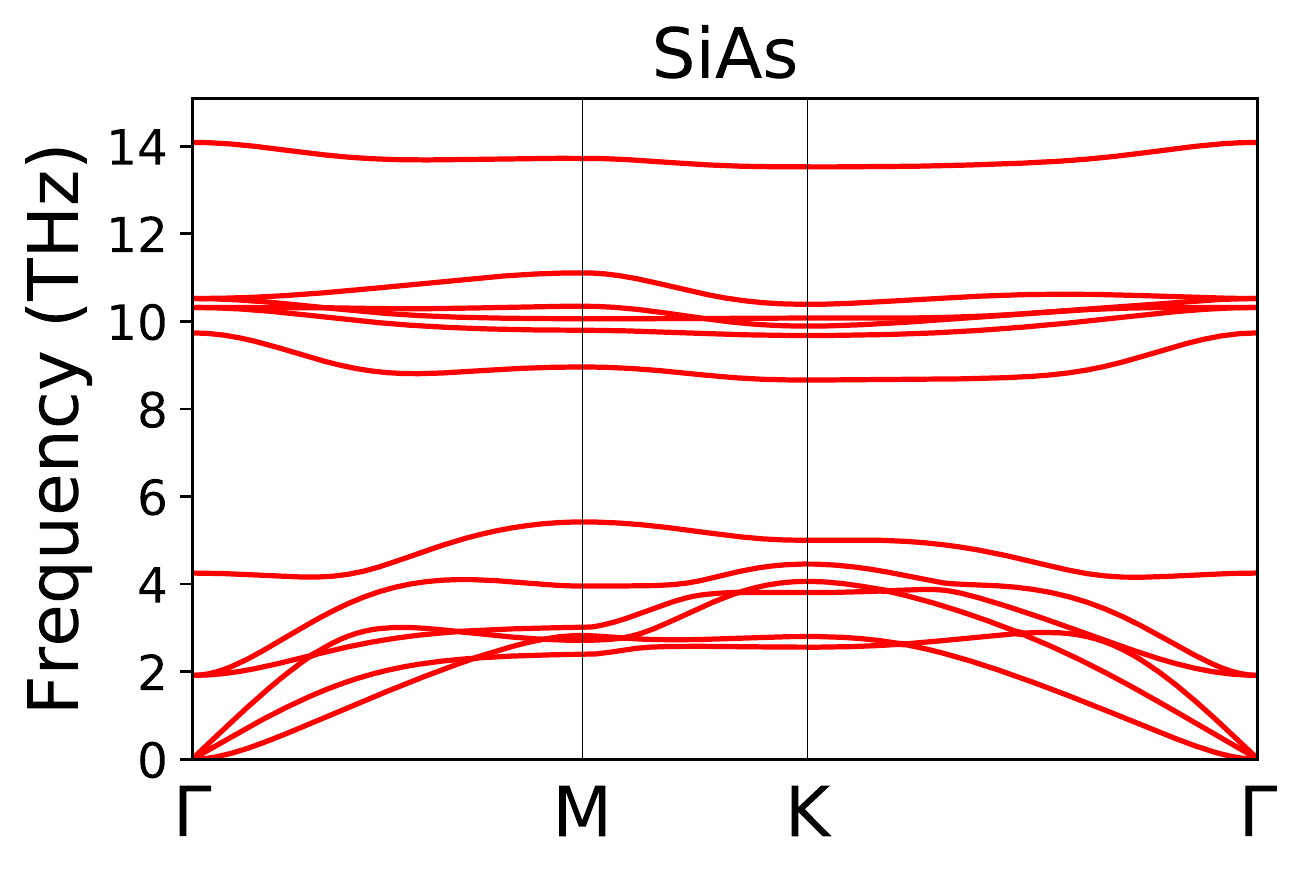}&
\includegraphics[width=0.19\textwidth]{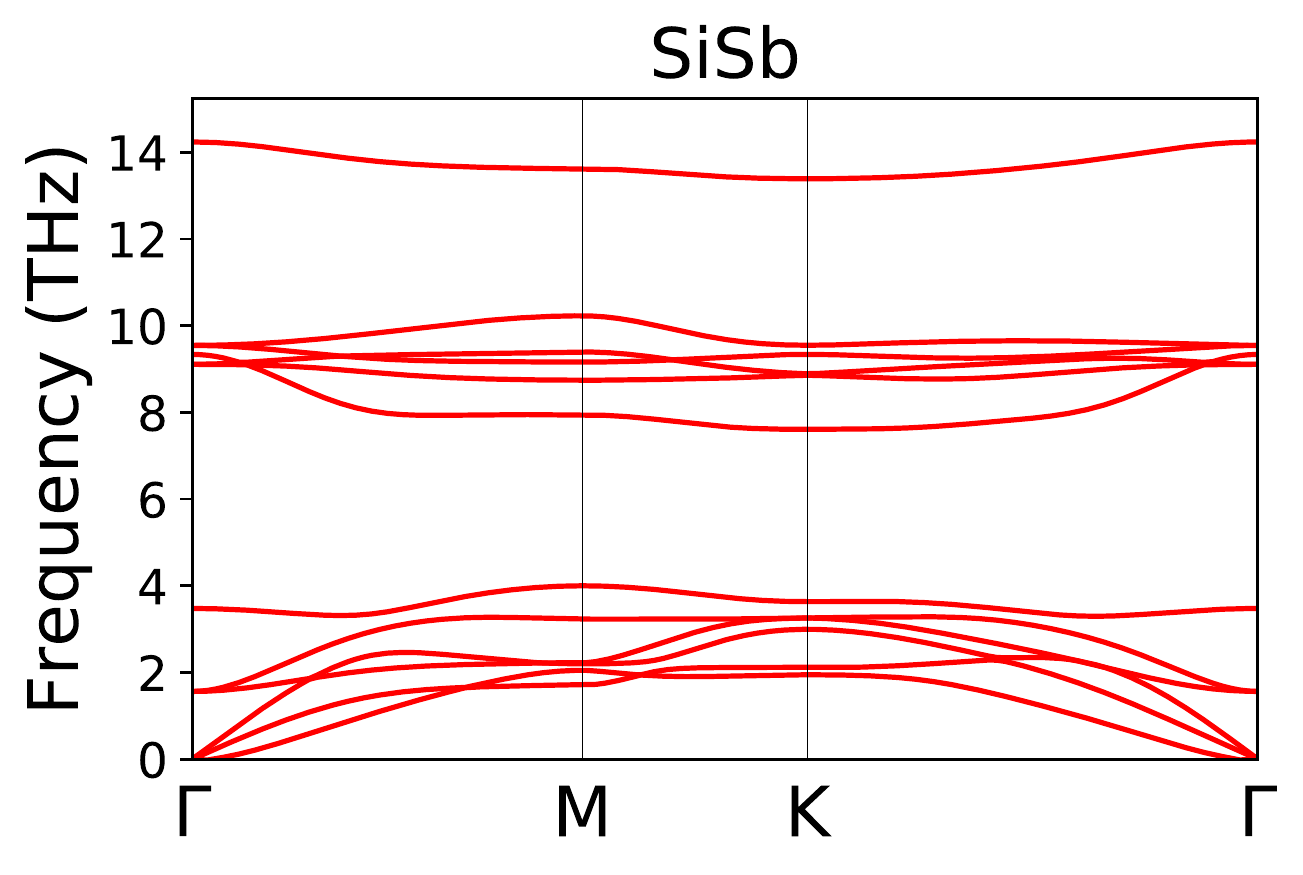}&
\includegraphics[width=0.19\textwidth]{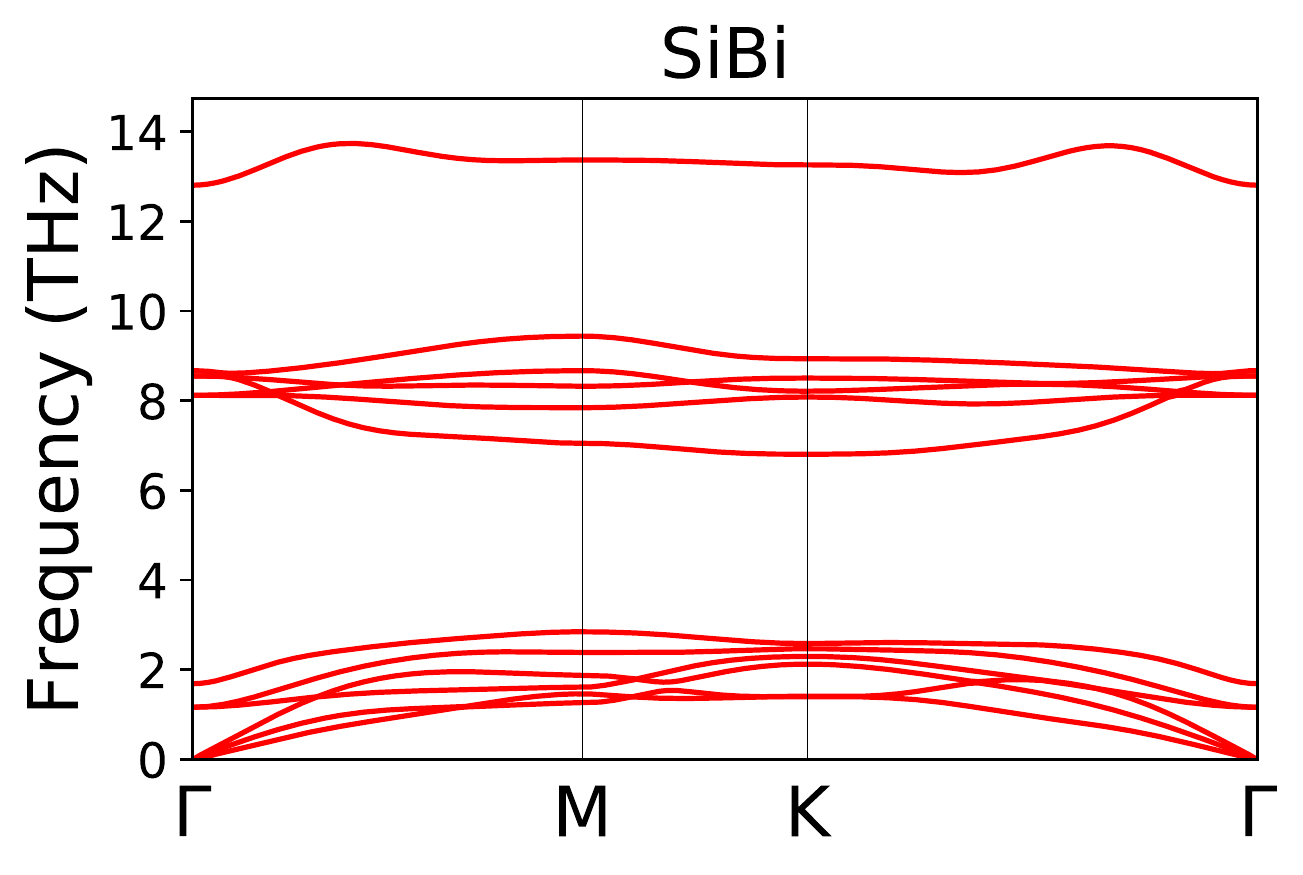}&\\

\includegraphics[width=0.19\textwidth]{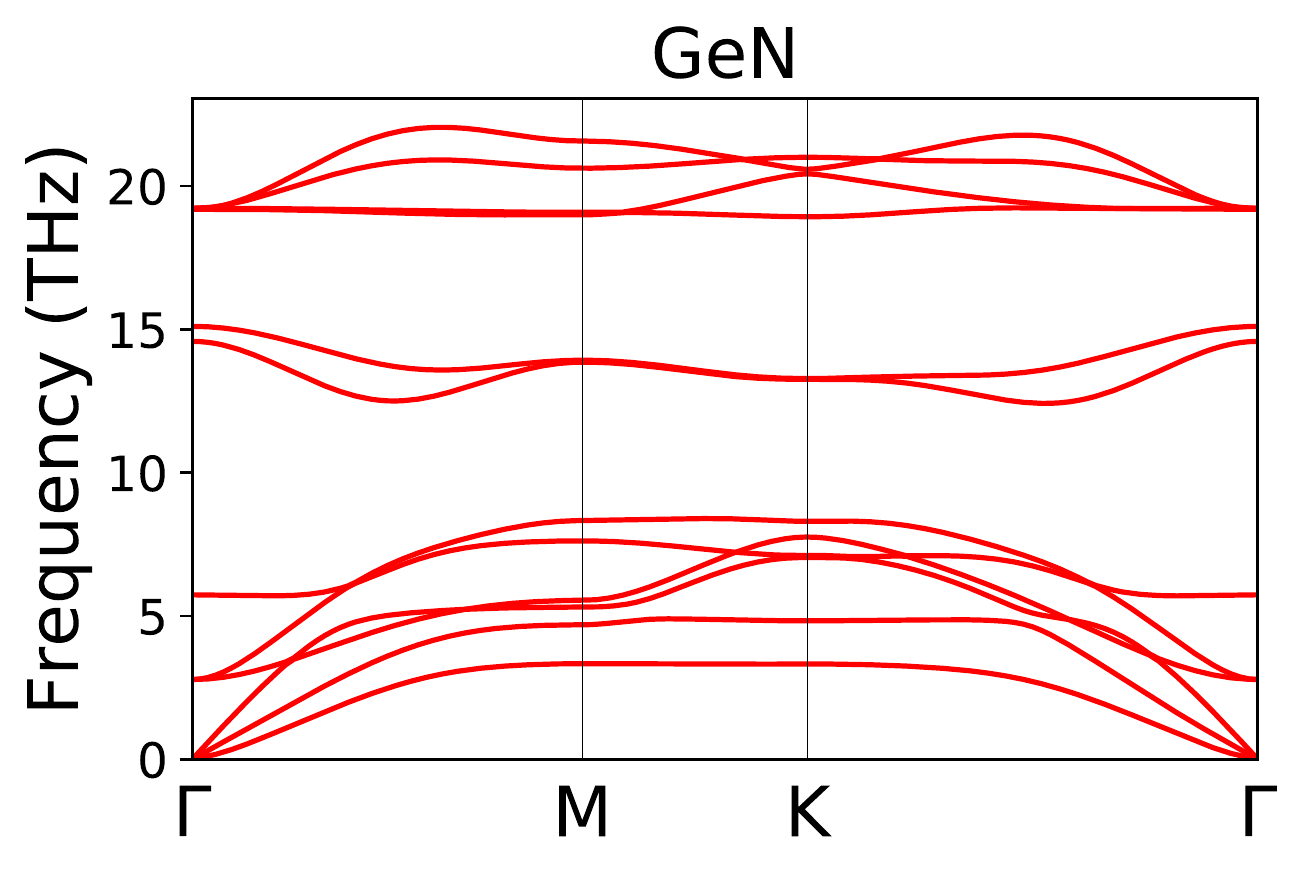}&
\includegraphics[width=0.19\textwidth]{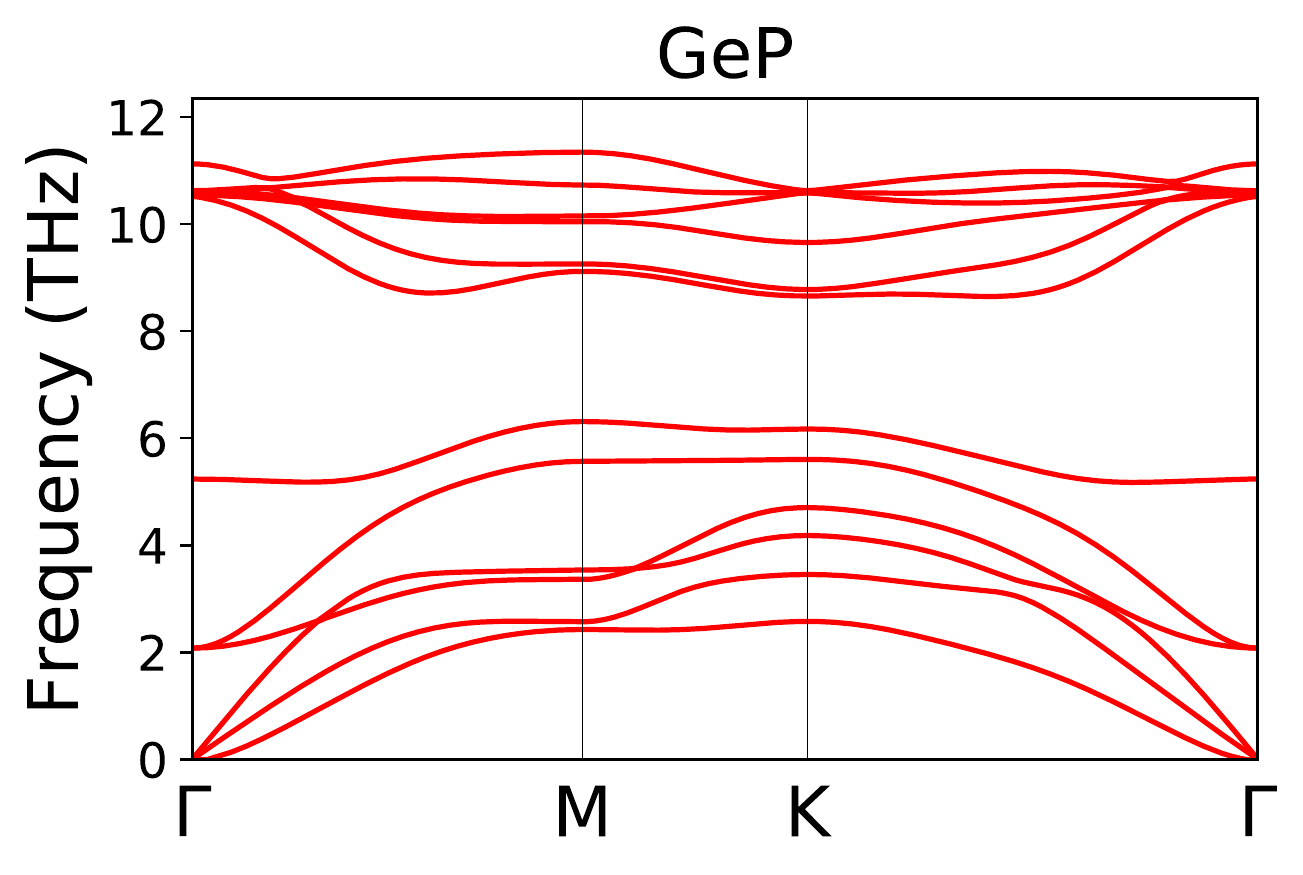}&
\includegraphics[width=0.19\textwidth]{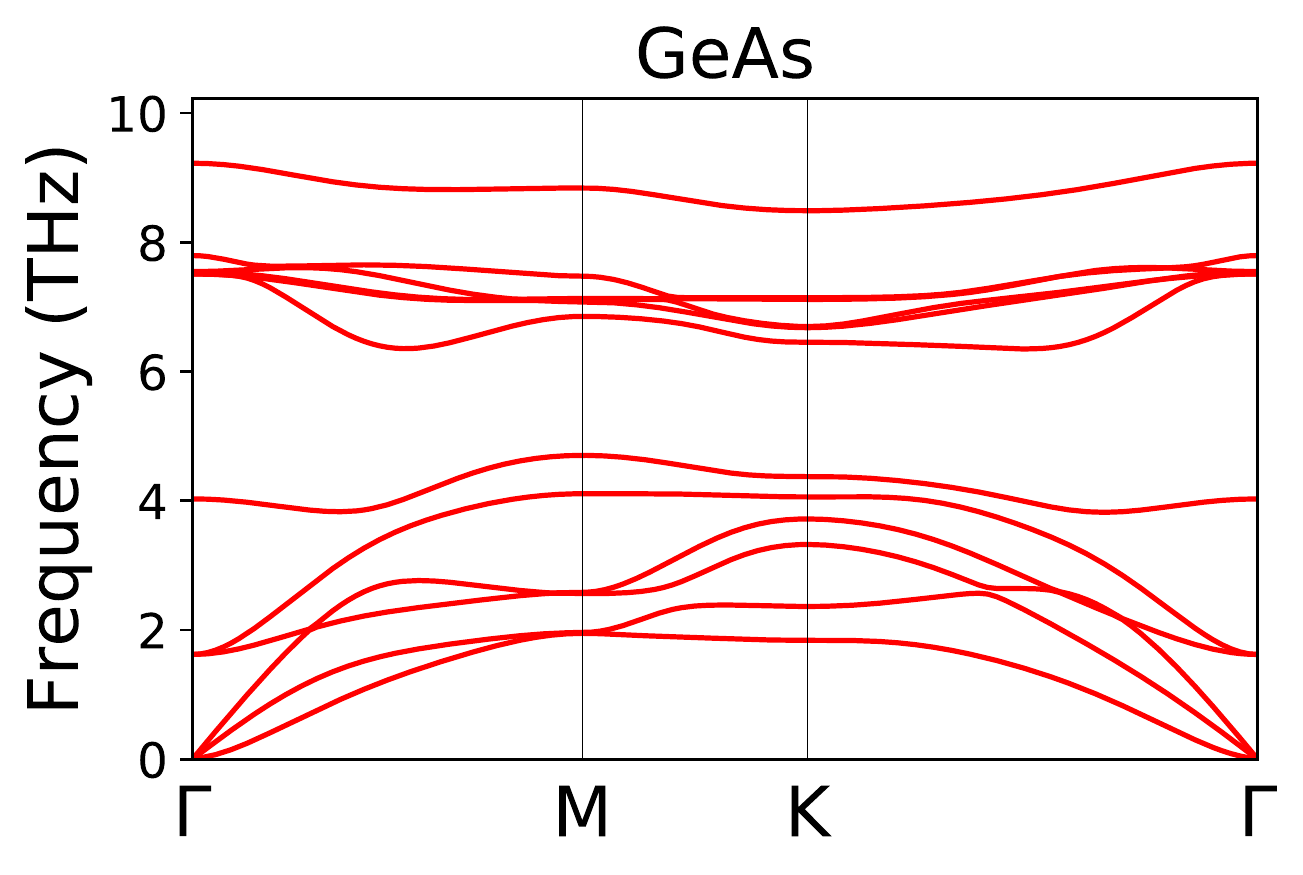}&
\includegraphics[width=0.19\textwidth]{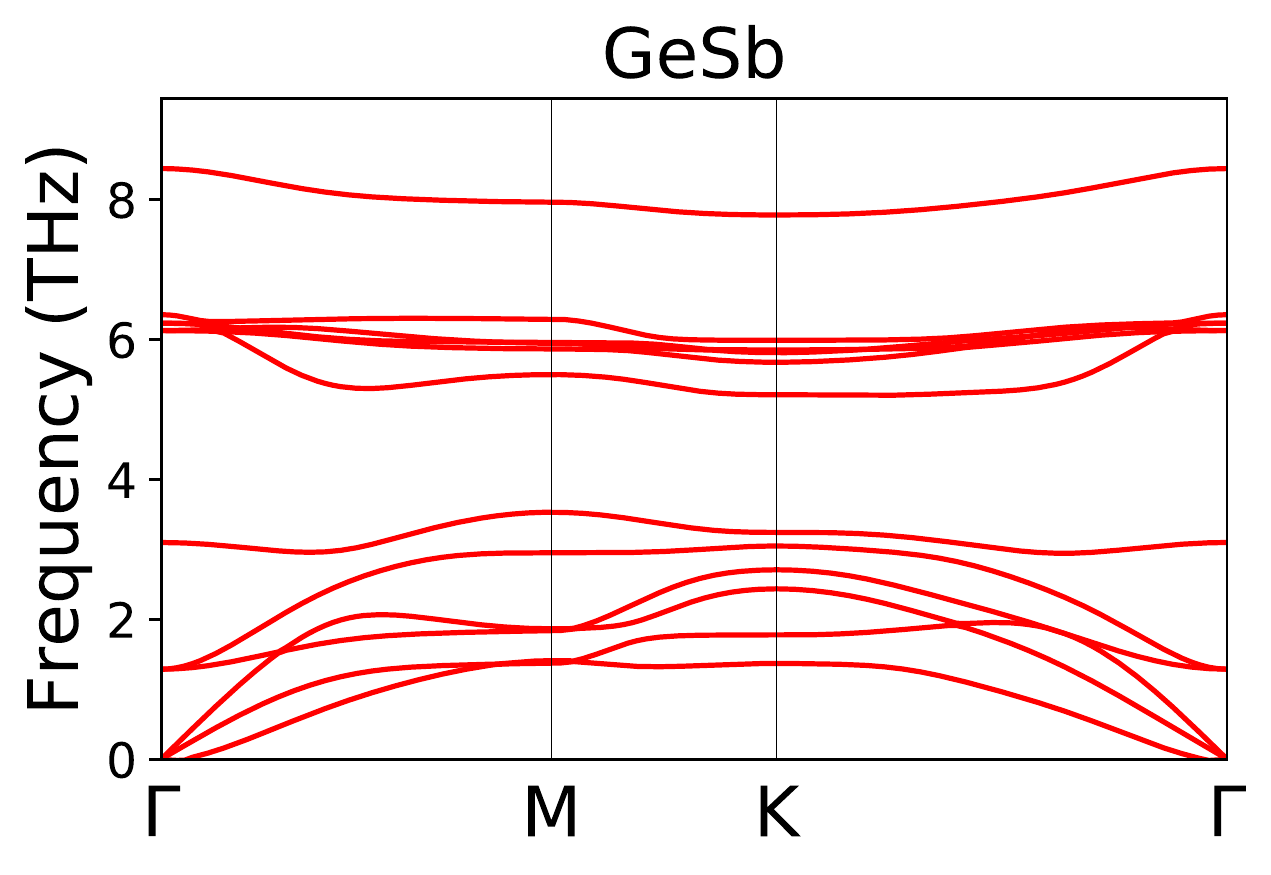}&
\includegraphics[width=0.19\textwidth]{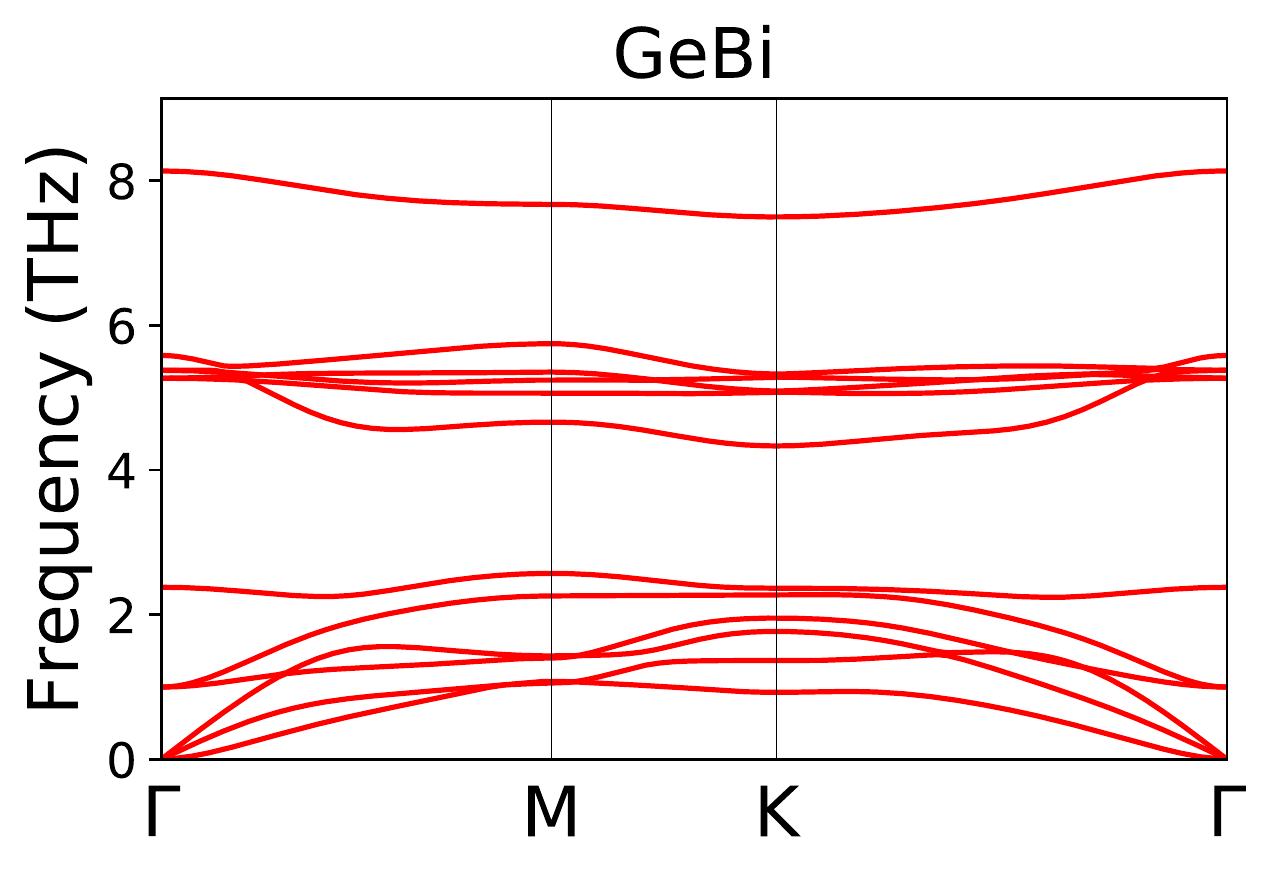}&\\

\includegraphics[width=0.19\textwidth]{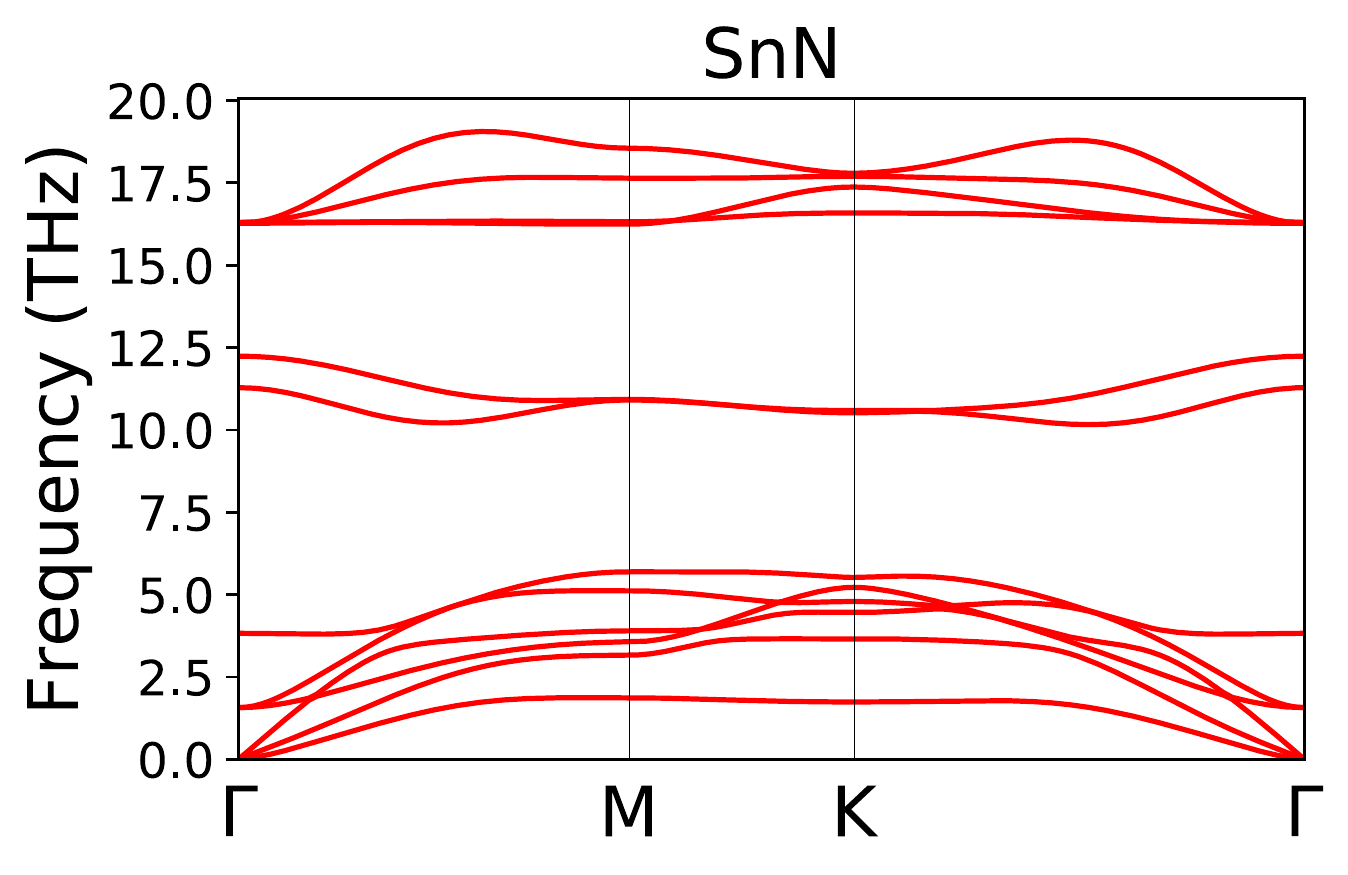}&
\includegraphics[width=0.19\textwidth]{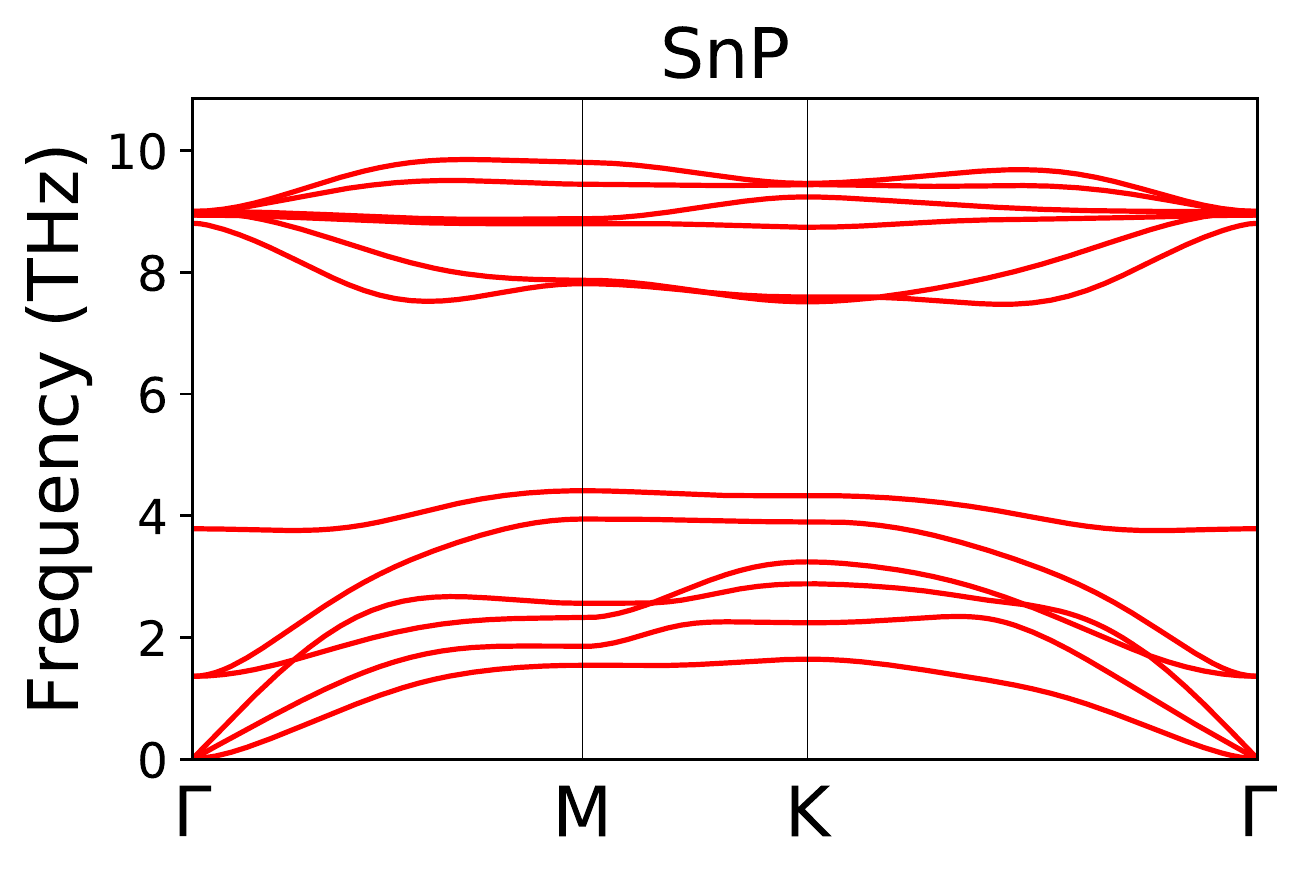}&
\includegraphics[width=0.19\textwidth]{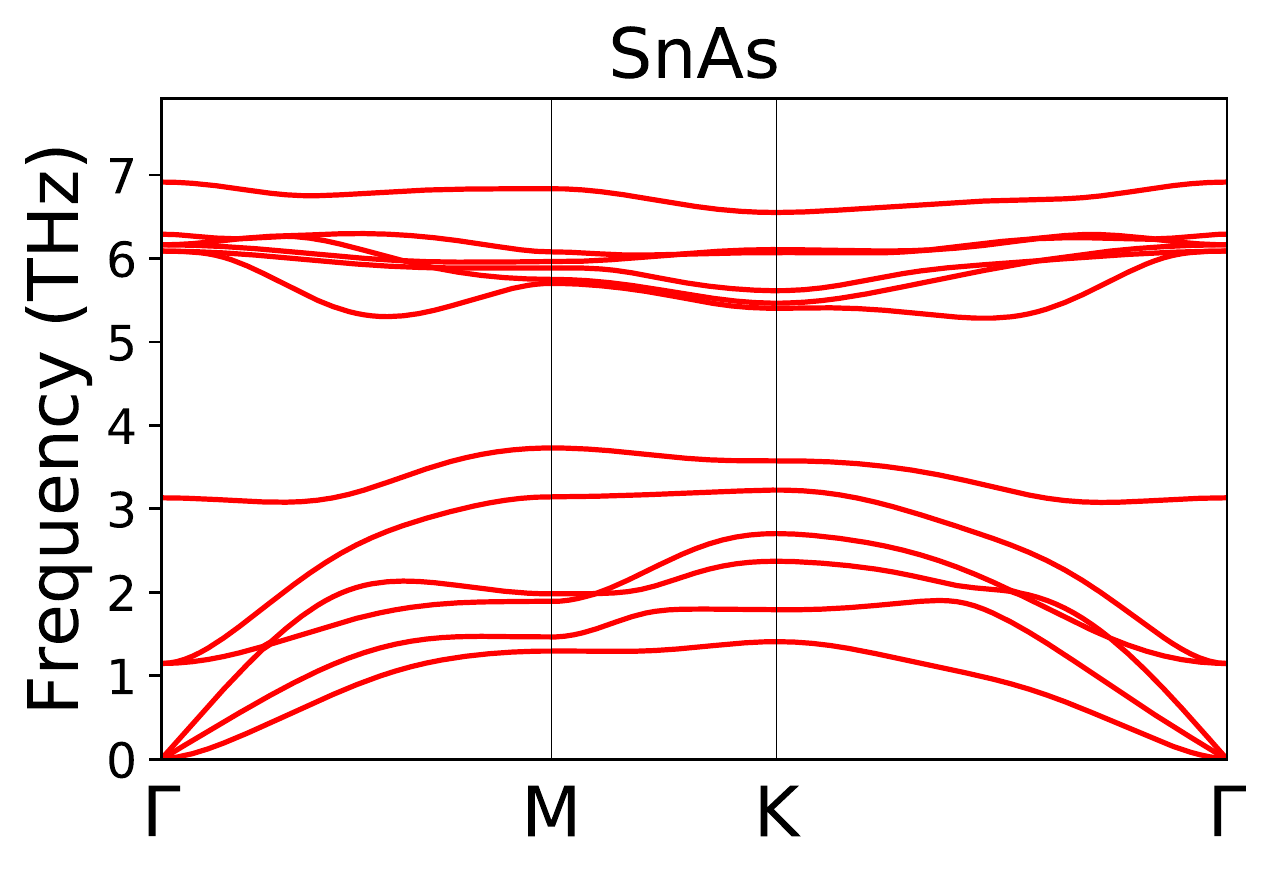}&
\includegraphics[width=0.19\textwidth]{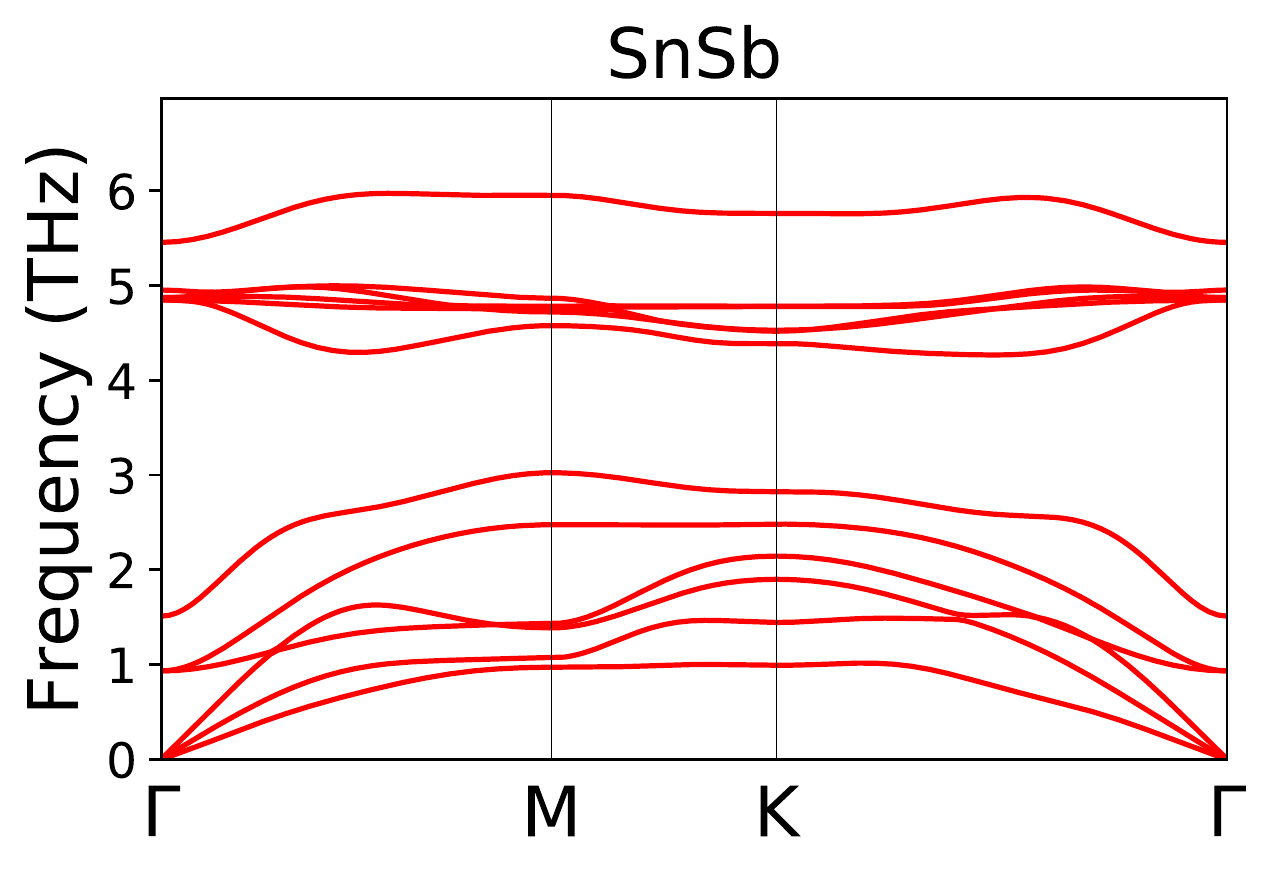}&
\includegraphics[width=0.19\textwidth]{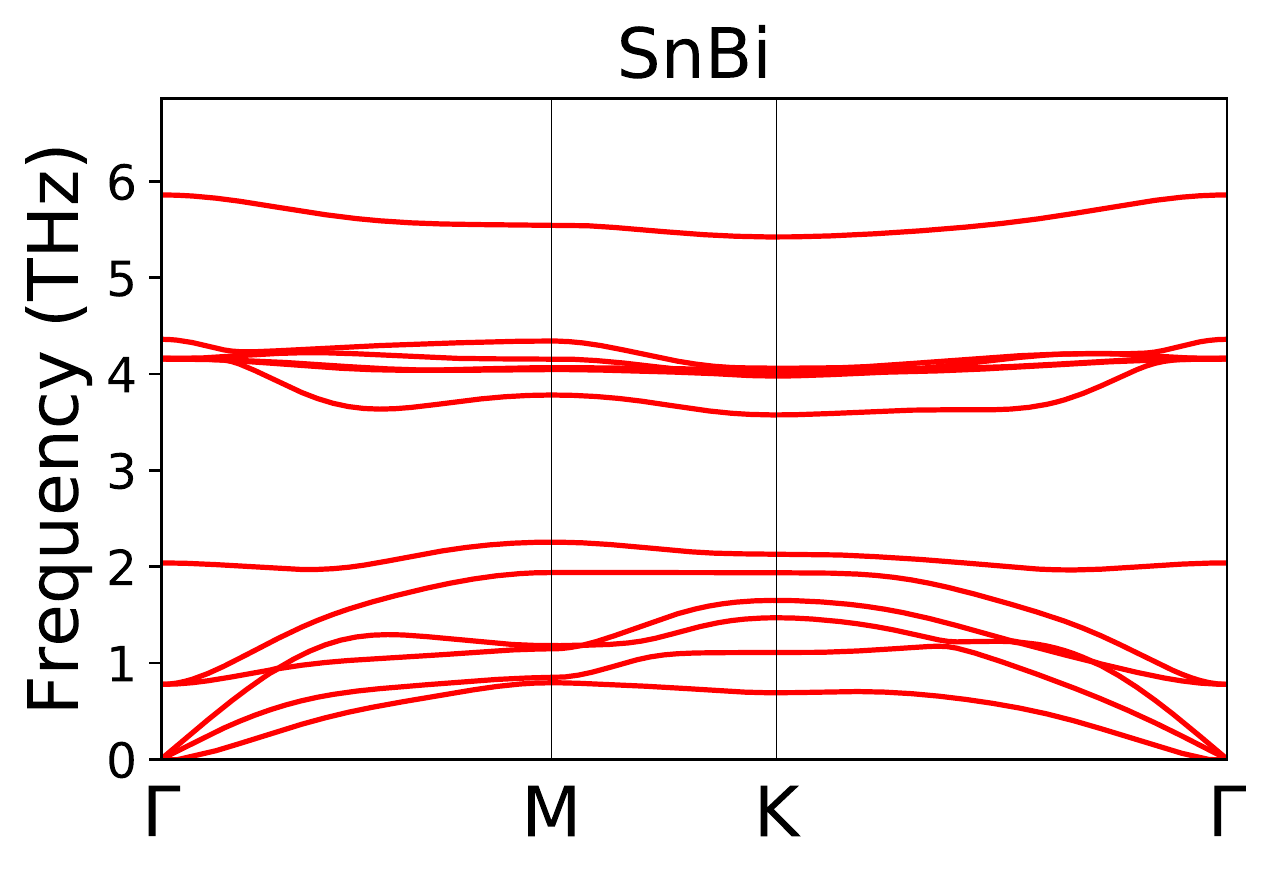}&\\

\includegraphics[width=0.19\textwidth]{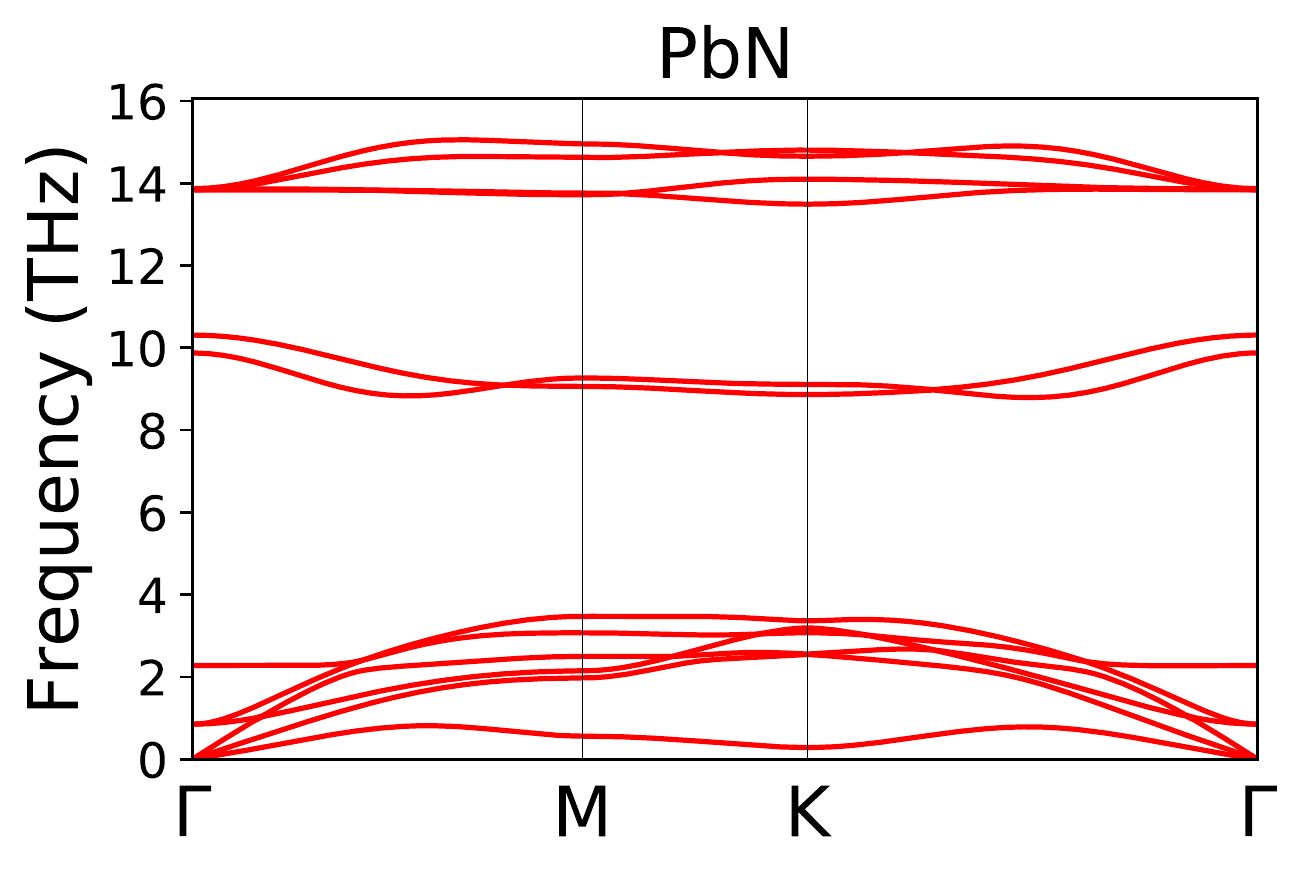}&
\includegraphics[width=0.19\textwidth]{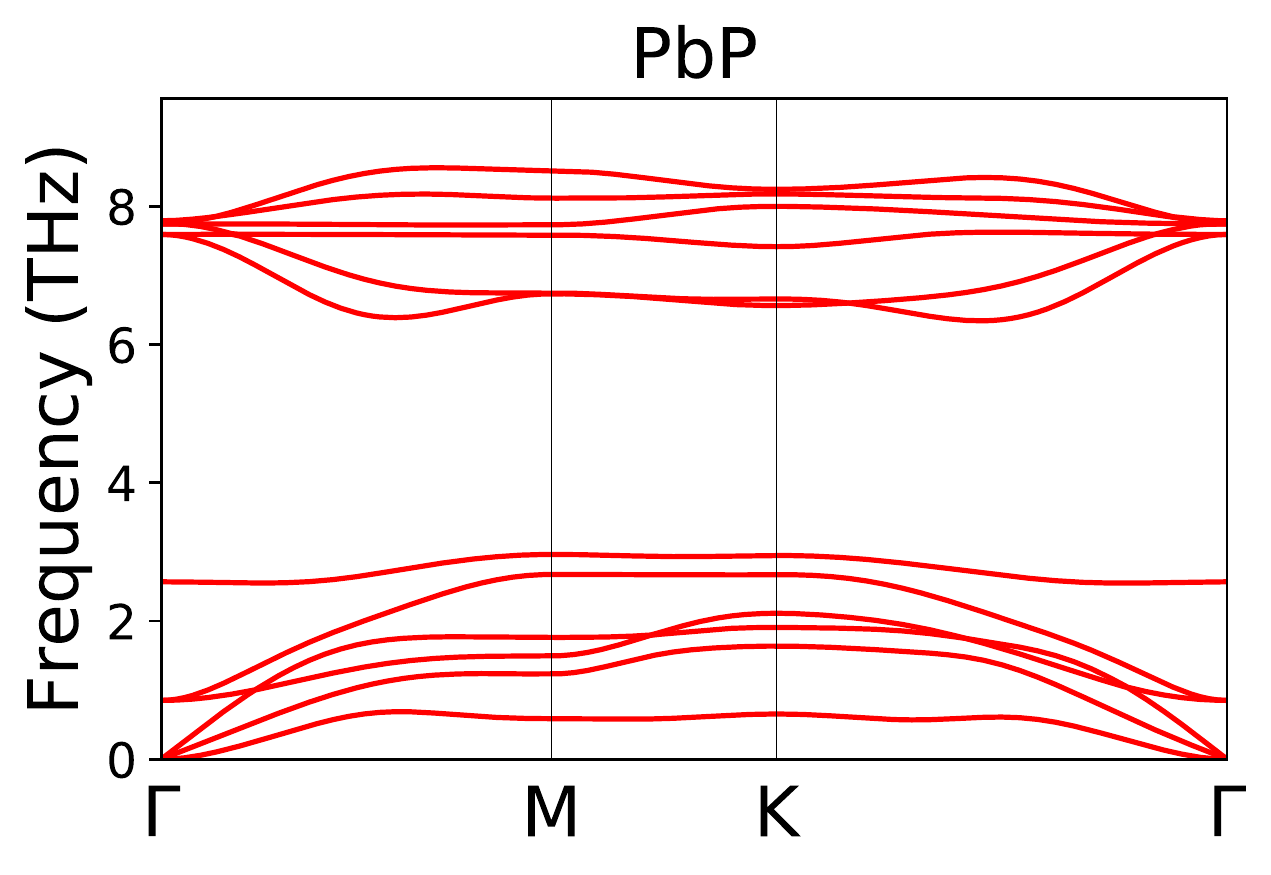}&
\includegraphics[width=0.19\textwidth]{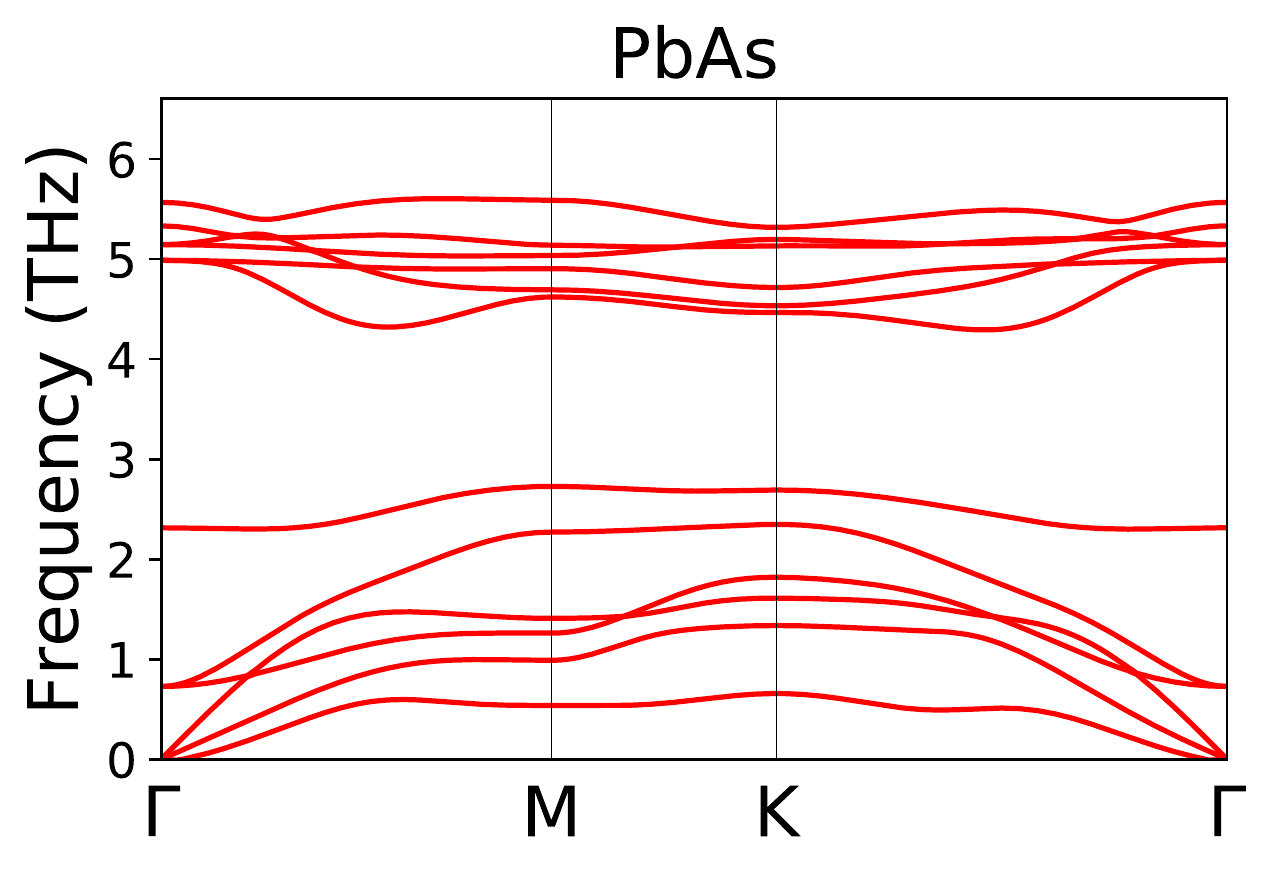}&
\includegraphics[width=0.19\textwidth]{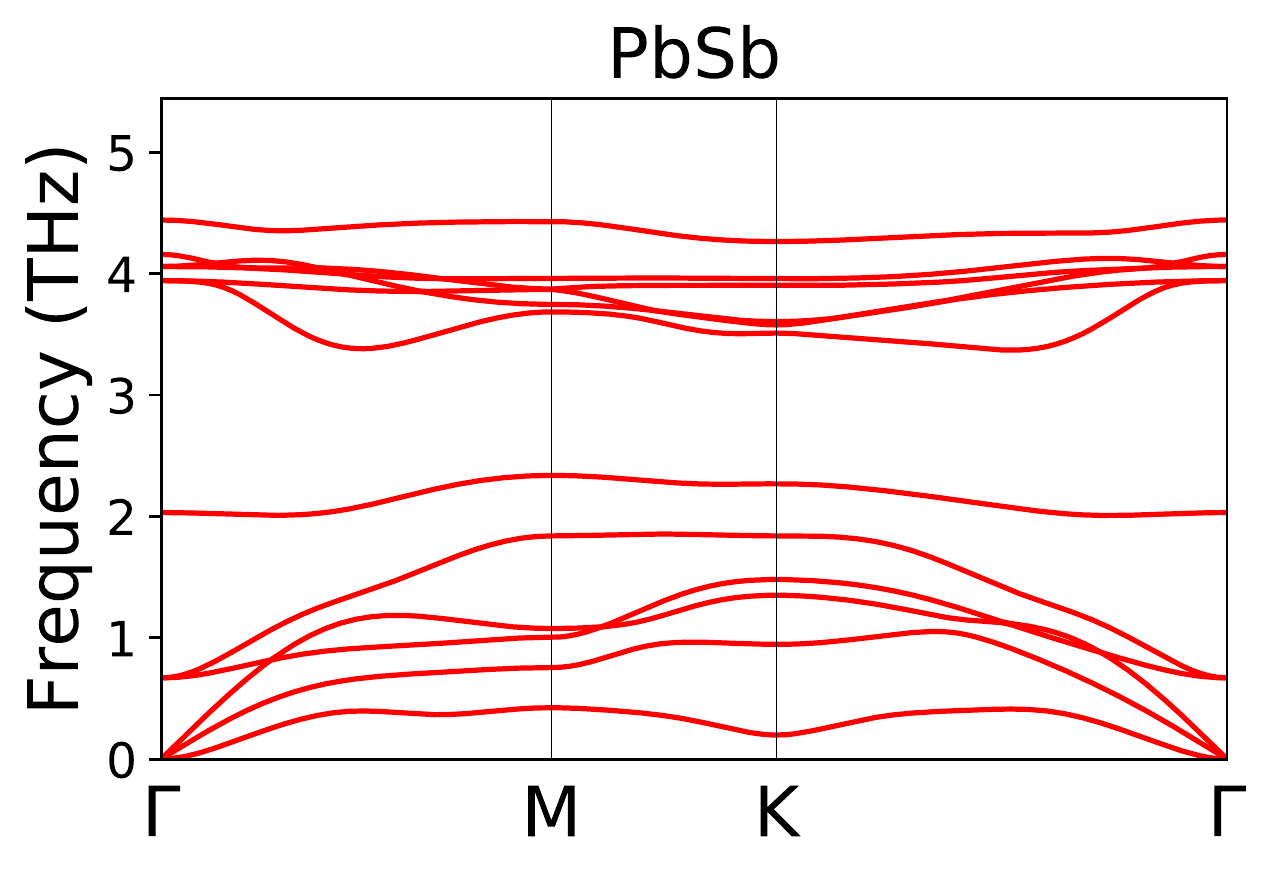}&
\includegraphics[width=0.19\textwidth]{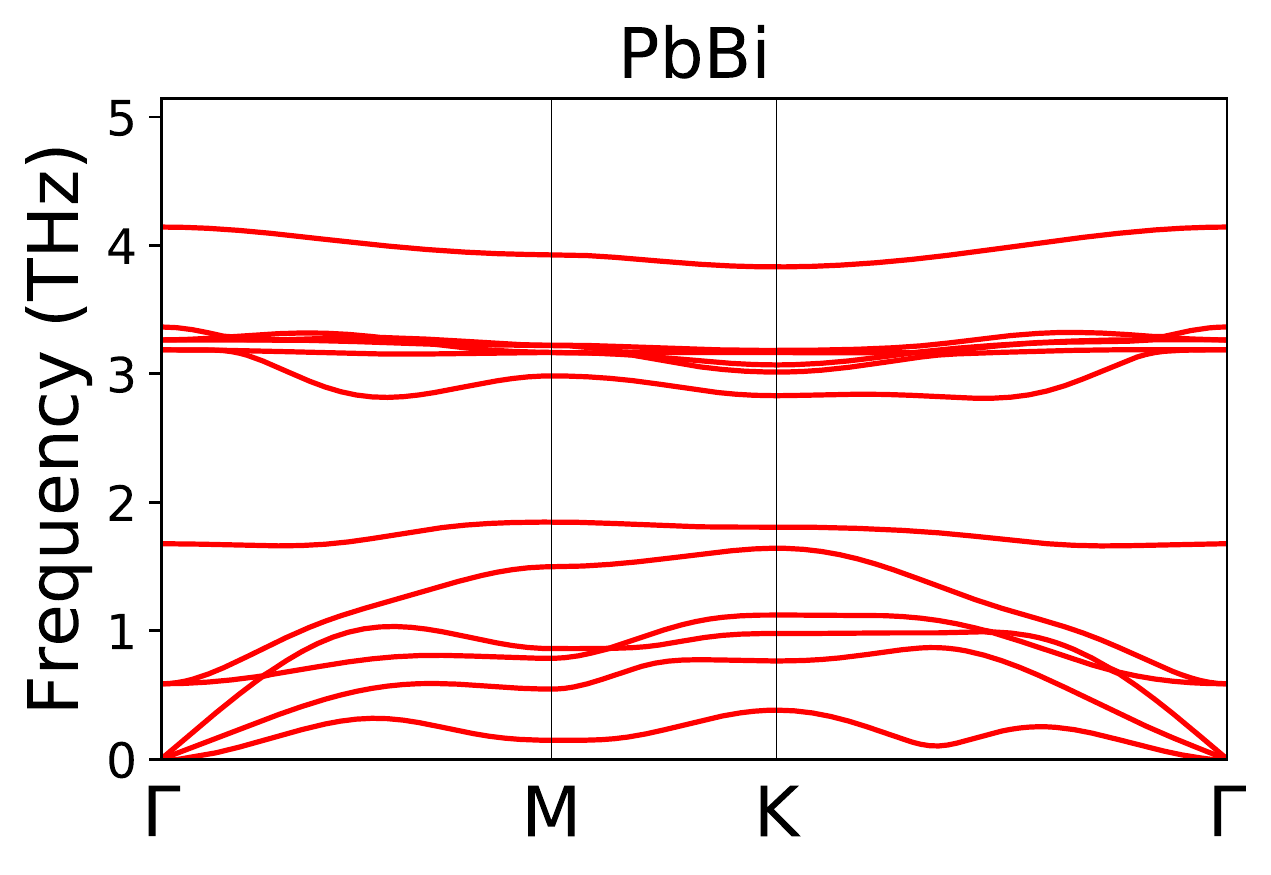}&\\
\end{tabular}

\caption{Phonon dispersion relations of $\alpha$-phases of group IV-V monolayers. Absence of negative frequencies is an indication for the dynamic stability of these compounds.}
    \label{fig:phonon-bands}
\end{figure*}

\subsection{Vibrational Properties}

\begin{figure}[t]
\includegraphics[width=0.45\textwidth]{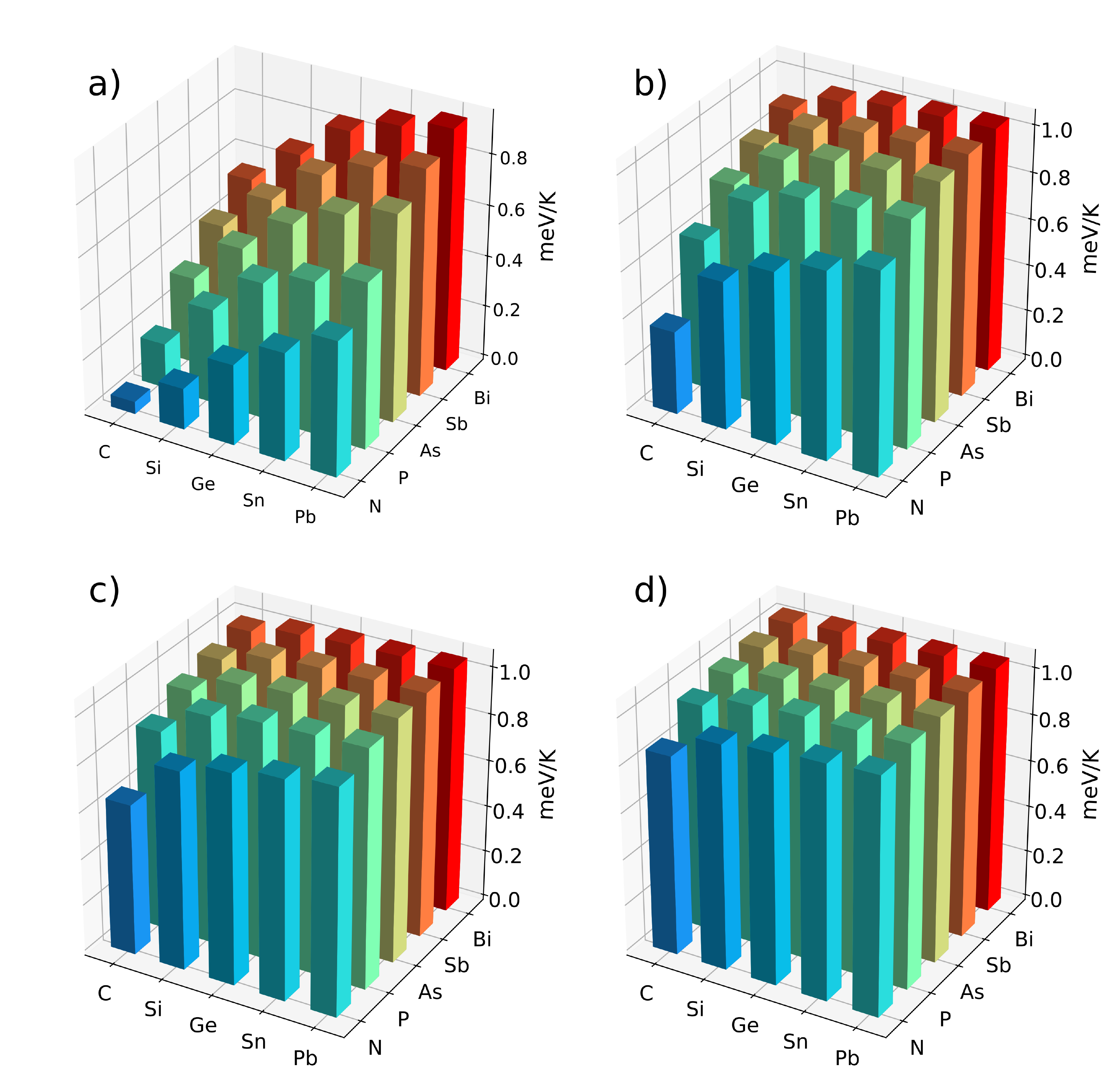}
\caption{Constant volume vibrational heat capacities ($C_v$) at 100 K, 300 K, 500 K and 800 K, from a) to d), respectively.}
    \label{fig:3d-thermal-prop}
\end{figure}
Phonon dispersion relation is an important benchmark to evaluate the stability of the systems at hand. As seen in Figure~\ref{fig:phonon-bands}, all of the $\alpha$-compounds demonstrate positive phonon frequencies around $\Gamma$ point, linear in-plane (longitudinal acoustic and transverse acoustic) and quadratic out-of-plane (ZA) modes. This implies that the monolayers of group IV-V are dynamically stable. 
In a recent study, $\alpha$-phases of compounds including carbon and nitrogen were reported to be highly unstable.~\cite{ashton2016computational}.
However, our computations with increased accuracy and denser k-point grids reveal that all these mentioned structures are dynamically stable, that is
no negative frequency is associated with these materials. 
Having said that, it is a well-known fact that acoustic modes may bear minuscule imaginary frequencies around the $\Gamma$-point, and this may be stemming from numerical inaccuracies rather than the real instability of the system. 
On the other hand, $\alpha$-PbN and $\alpha$-SnP have negative frequencies in their out-of-plane acoustic modes around the K point, when the electronic temperature is low.
One way to remedy this is to employ an enhanced smearing during DFPT computations. 
The Fermi-Dirac smearing function dictates the electronic temperature and occupation probability of the electronic states~\cite{mermin1965thermal}. An increase in the smearing value corresponds to an increase in the effective temperature, therefore stability in higher temperatures can be examined. 
We have performed calculations for different smearing values ($\sigma$) and found that $\sigma=0.5$~eV (0.1~eV) is required for PbN (SnP) for obtaining real and positive vibrational frequencies.
We note that such onset of imaginary frequencies at the boundaries of the Brillouin zone could also be related to charge density wave formation at low temperatures.~\cite{singh:prb:2017}

The maximum phonon frequency of a given structure ($\wmax$) decreases steadily with increasing atomic masses in the unit cell, as expected. However detailed analysis of the force constant matrices and phonon dispersions show that the decrease is not only due to the increased mass, but also because of weaker inter atomic force constants.
Assuming the masses to be those of heavier atoms in the group always overestimates $\wmax$.
Phonon band gaps also show a particular trend.
As a rule of thumb the phonon band gaps increase with increasing mass difference between the constituent elements in the unit cell, and it decreases with increasing atomic masses as the overall spectrum is squeezed.
Both reduced of $\omega_\mathrm{max}$ and wider phonon band gaps decrease phonon thermal transport. 
Therefore those structures can be expected to have better thermoelectric performances.

Vibrational heat capacities at constant volume are calculated using
\begin{equation}
C_v = k_\mathrm{B} \int {d\omega}\,\rho(\omega)\,p(\omega,T),
\label{eq:heat-cap}
\end{equation}
where $\rho$ is the phonon density of states and $p(x){=}{-}x^2\partial{f_\mathrm{BE}}/\partial{x}$ with $f_\mathrm{BE}{=}1/(\mathrm{e}^x{-}1)$ and $x{=}\hbar\omega/k_\mathrm{B}T$.
In Figure~\ref{fig:3d-thermal-prop}, the vibrational heat capacities are plotted at $T=100$~K, 300~K, 500~K, and 800~K.
The structures with slower sound velocities, \textit{i.e.} those with heavier elements, have larger phonon DOS at lower frequencies.
At lower temperatures, the function $p(x)$ in Equation~\ref{eq:heat-cap} filters out higher frequency modes.
Therefore they have considerably higher heat capacities at 100~K, compared to e.g. carbon and nitrogen compounds.
At higher temperatures, $p(x)$ changes slowly with respect to $\omega$ and it is approximately equal to 1 in the entire phonon spectrum. Therefore $C_v$ of all structures reach to $12k_\mathrm{B}$ at high temperatures, 12 being the number of modes per unit cell.
Heat capacities are also calculated for the $\beta$-phases and almost identical numerical results have been obtained with the corresponding $\alpha$-phases, as expected.

\subsection{Electronic Properties}

\begin{figure*}[t]
   \centering
\begin{tabular}{ccccc}
\\
\includegraphics[height=0.19\textwidth]{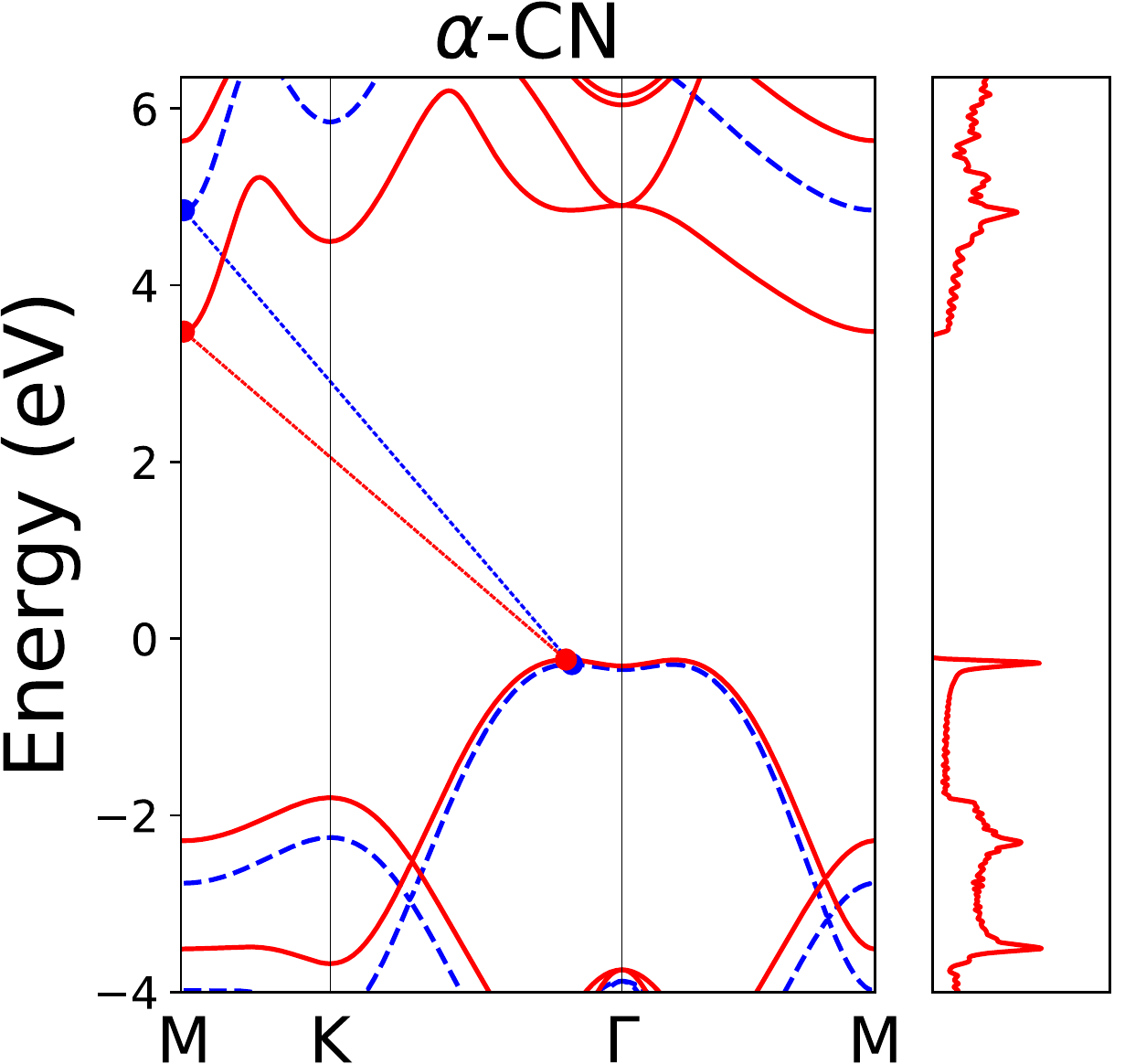}&
\includegraphics[height=0.19\textwidth]{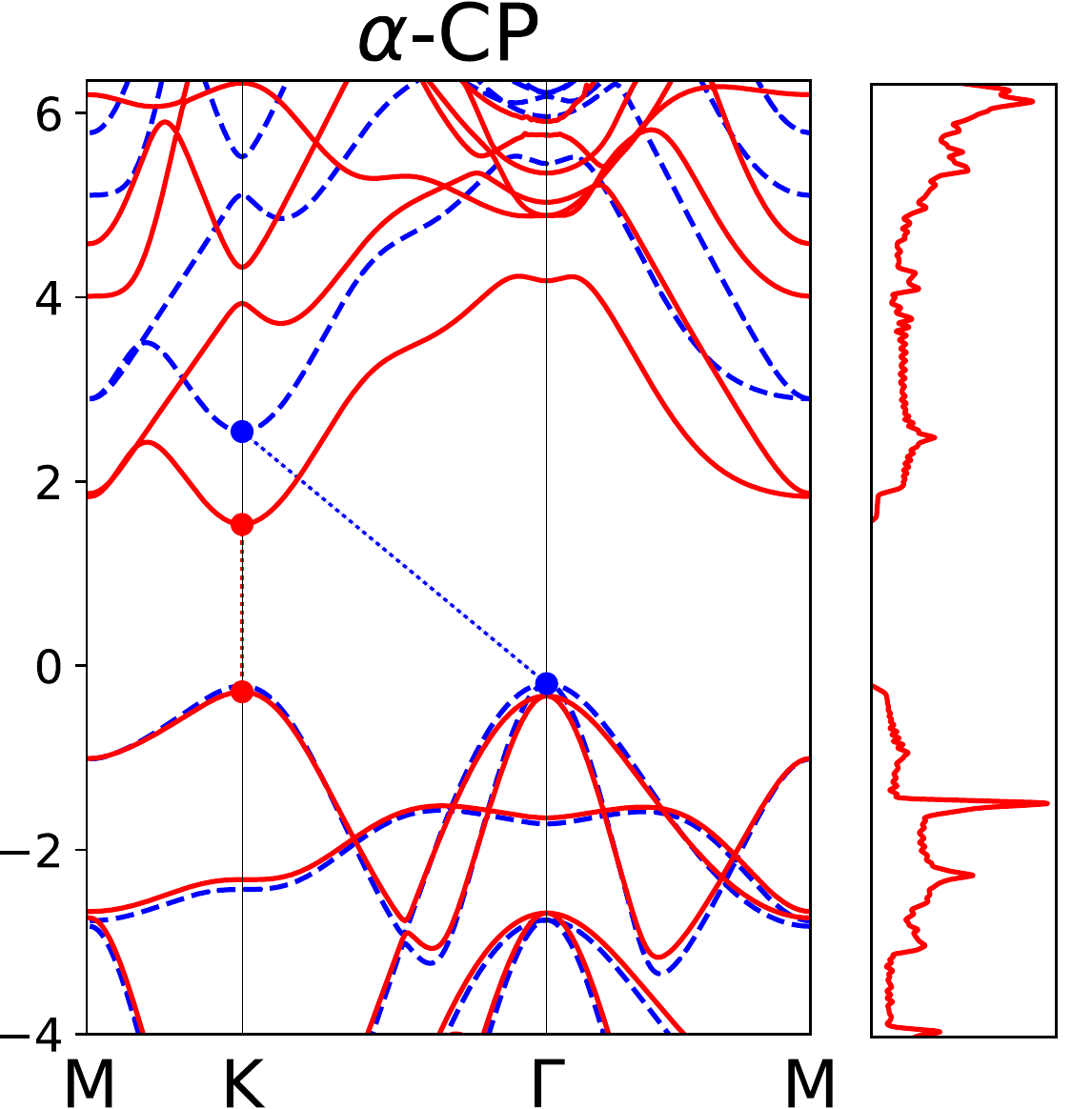}&
\includegraphics[height=0.19\textwidth]{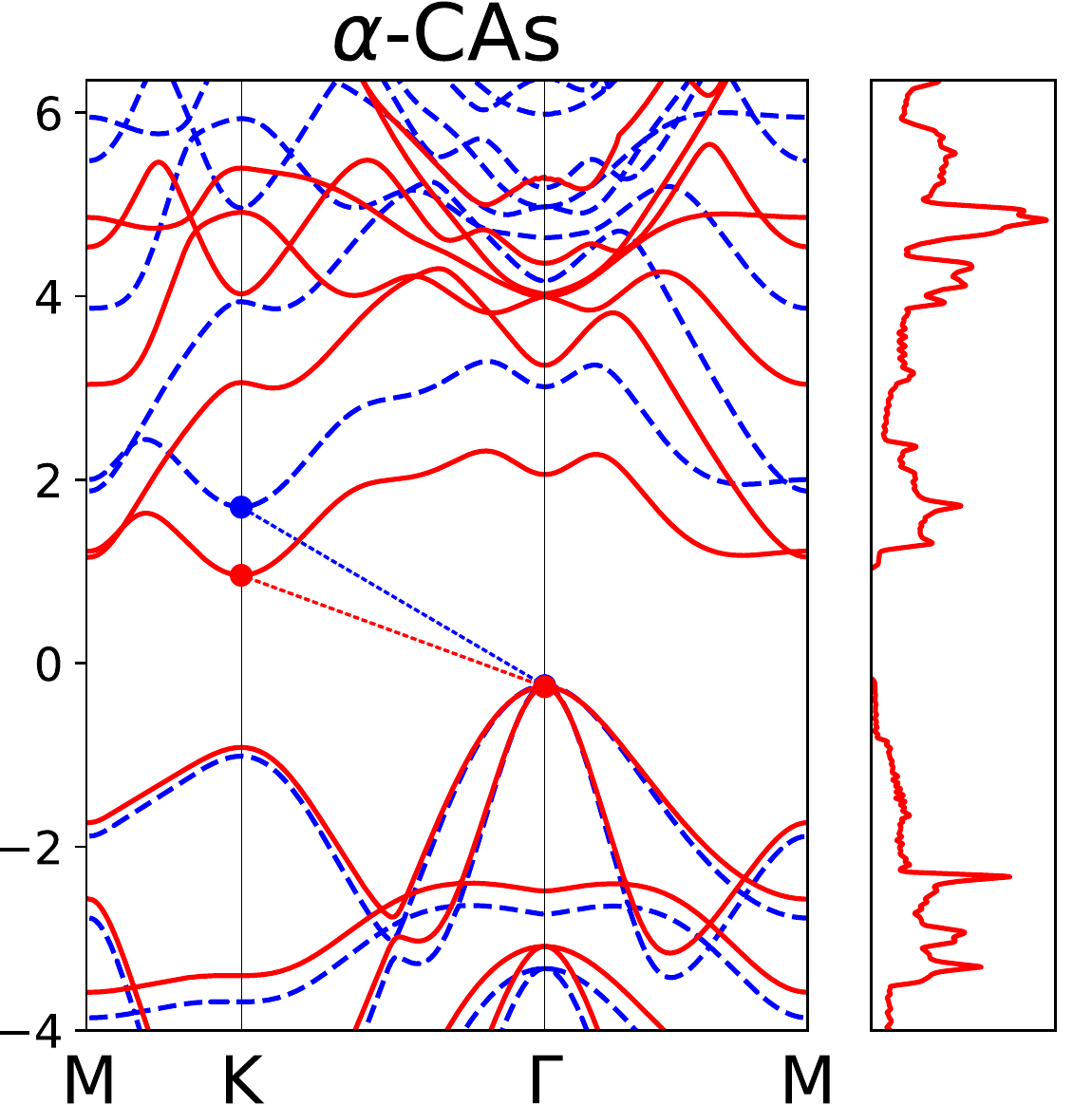}&
\includegraphics[height=0.19\textwidth]{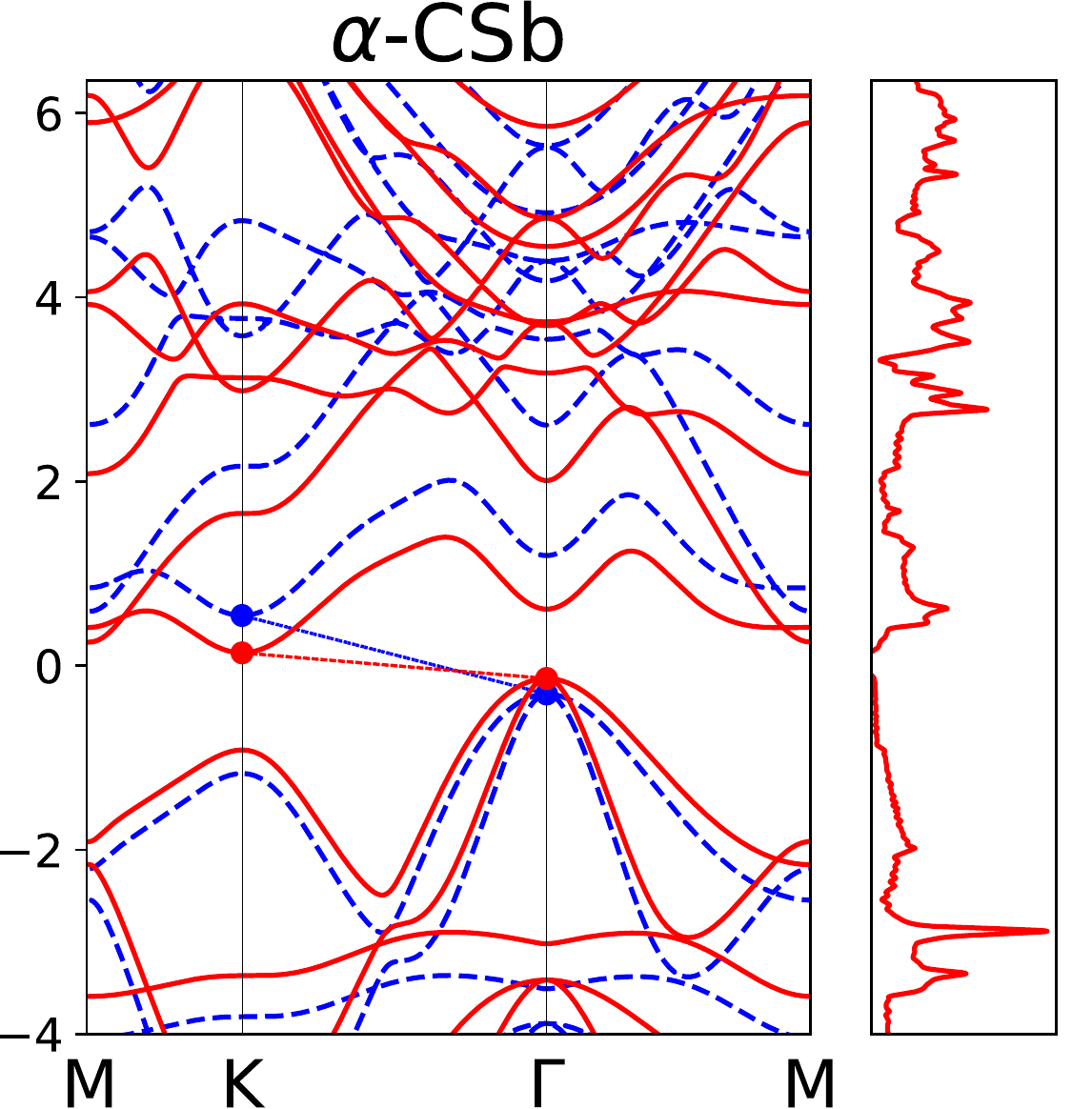}&
\includegraphics[height=0.19\textwidth]{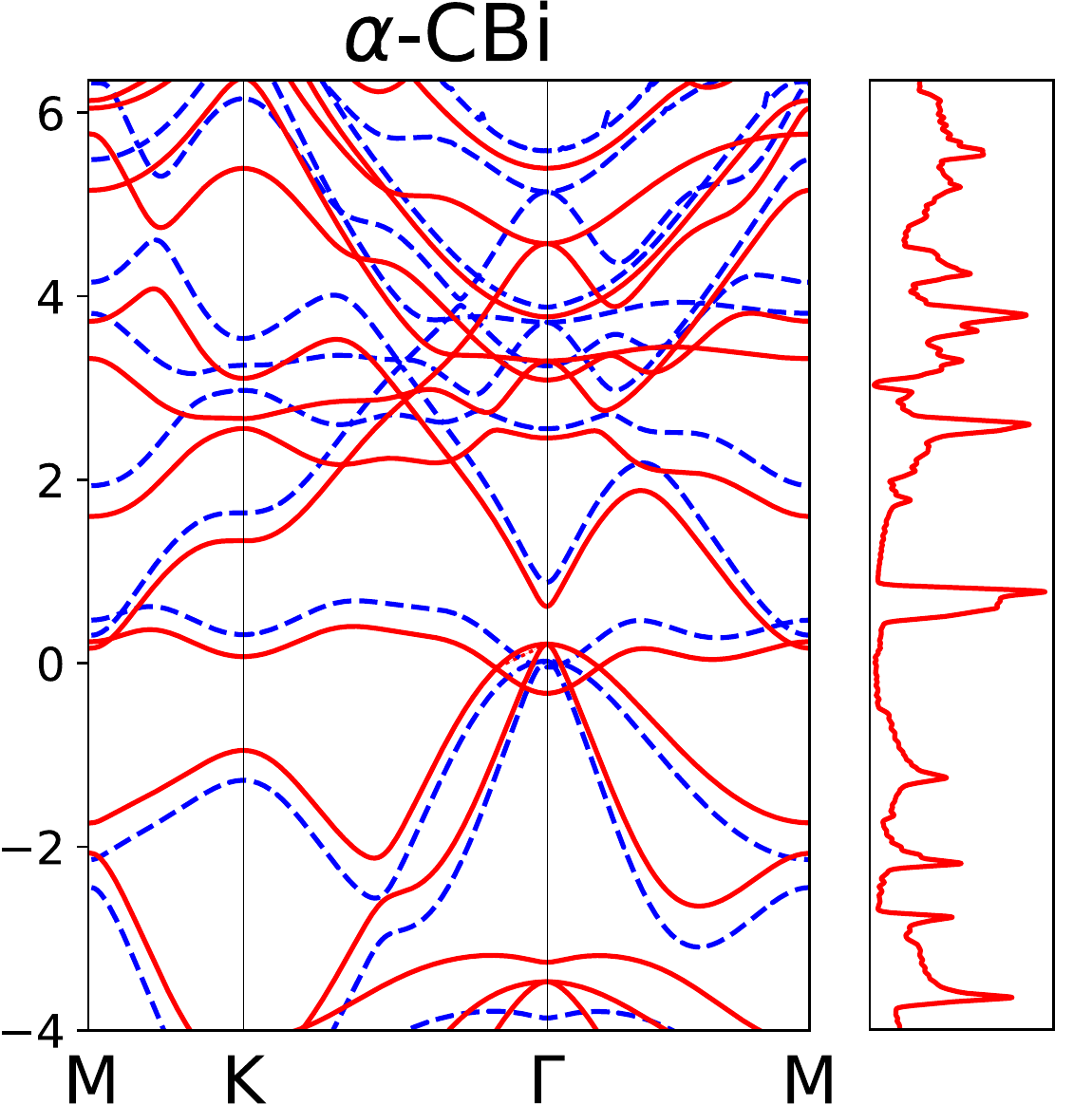}\\

\includegraphics[height=0.19\textwidth]{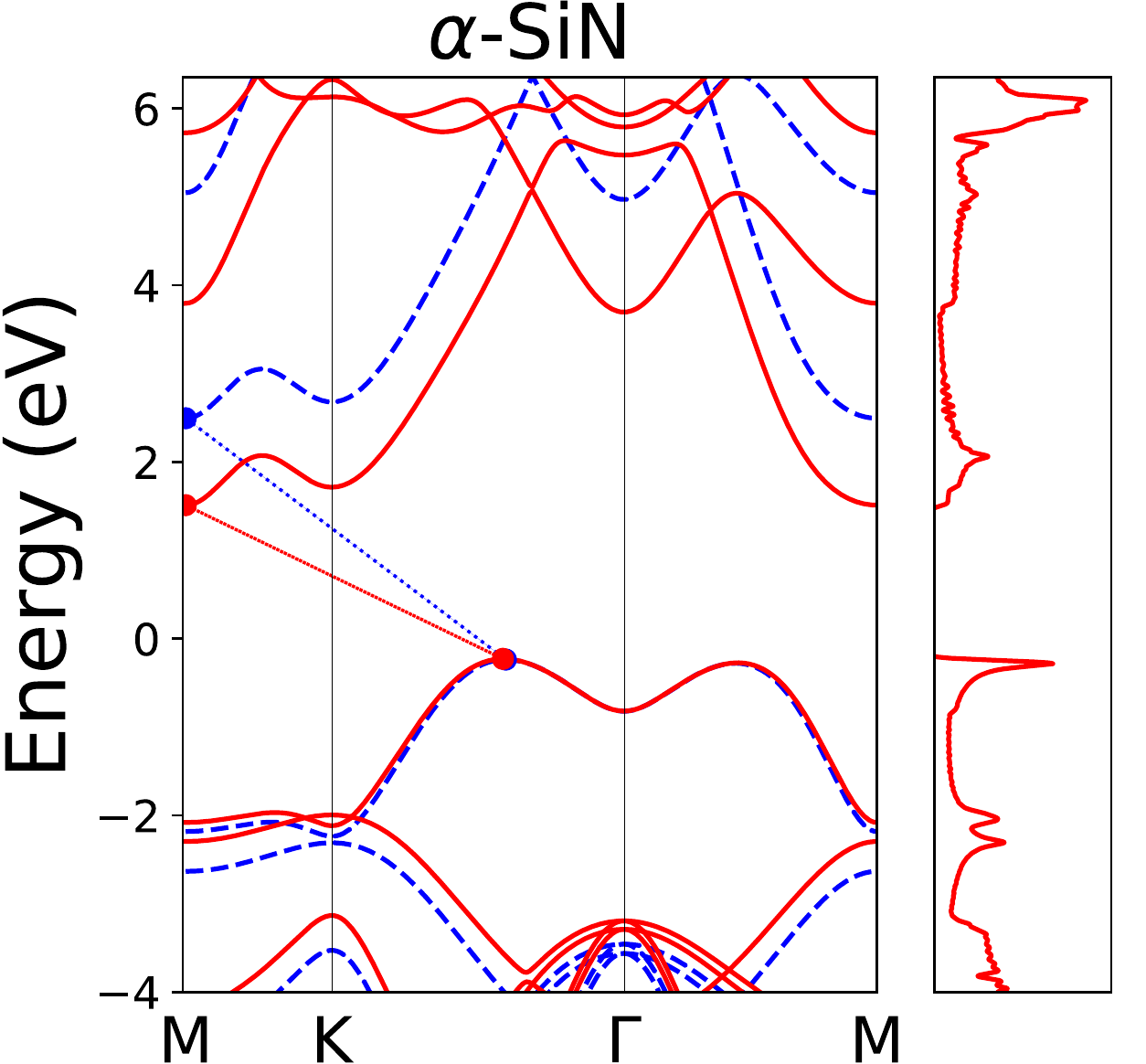}&
\includegraphics[height=0.19\textwidth]{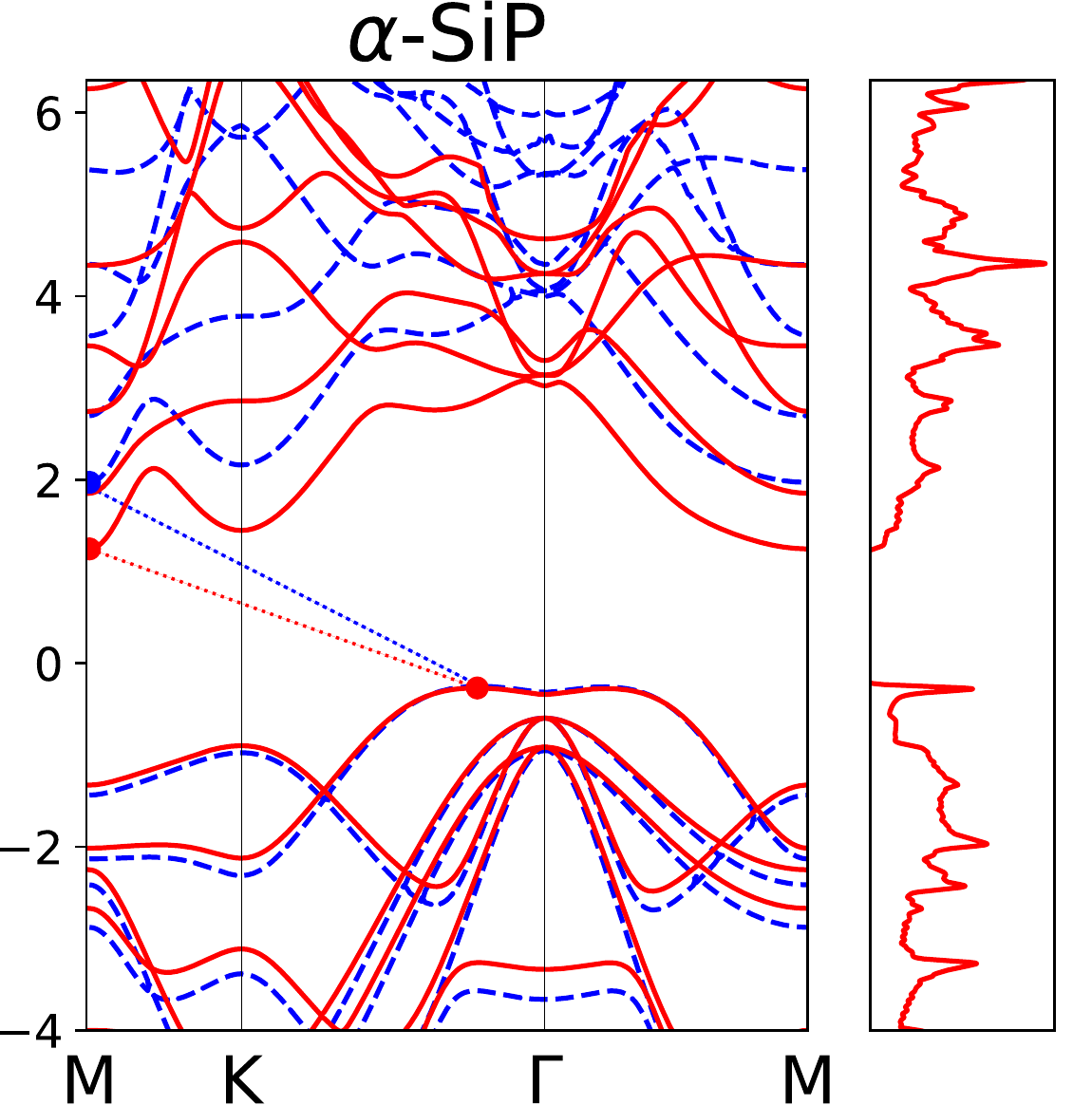}&
\includegraphics[height=0.19\textwidth]{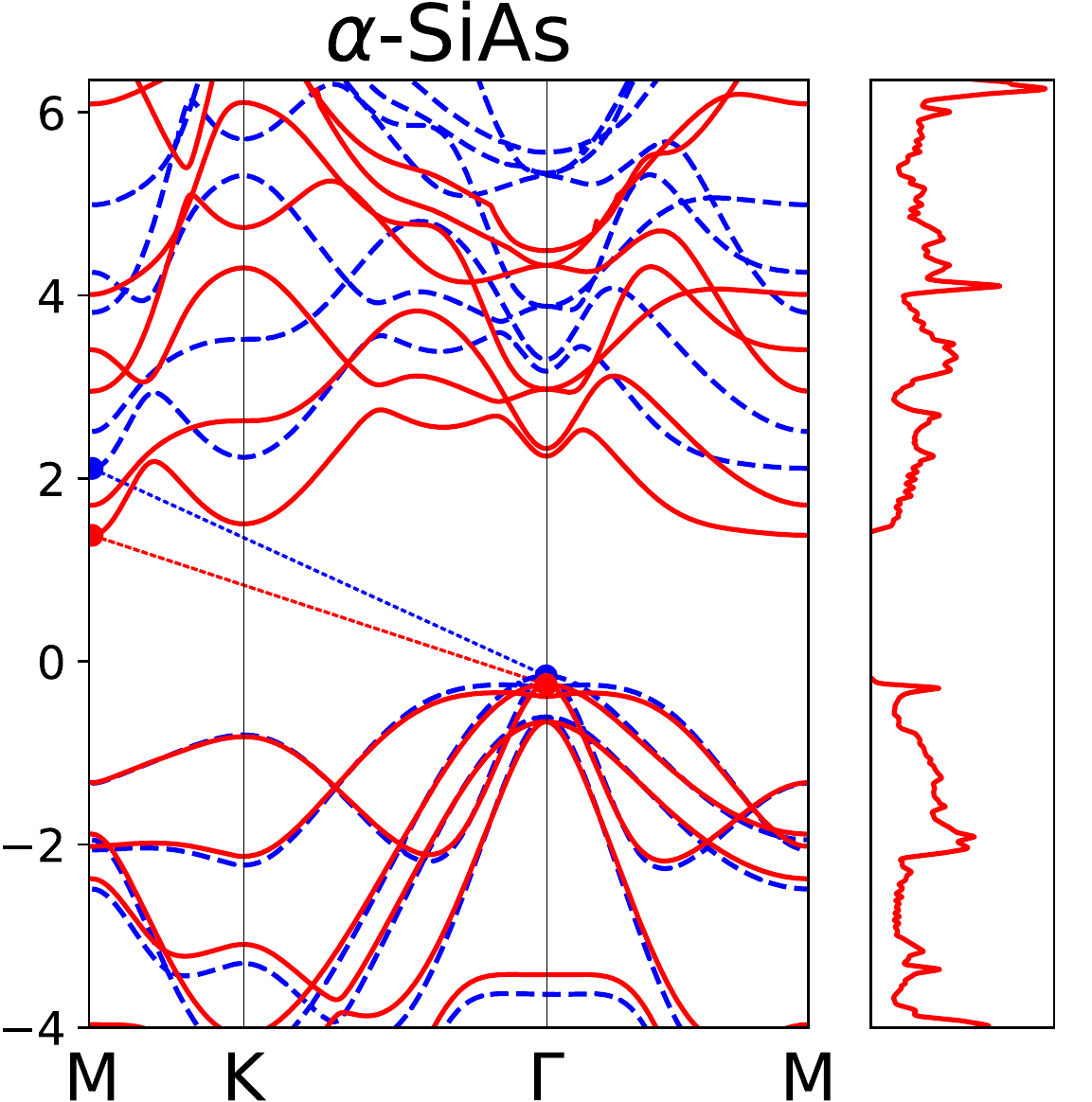}&
\includegraphics[height=0.19\textwidth]{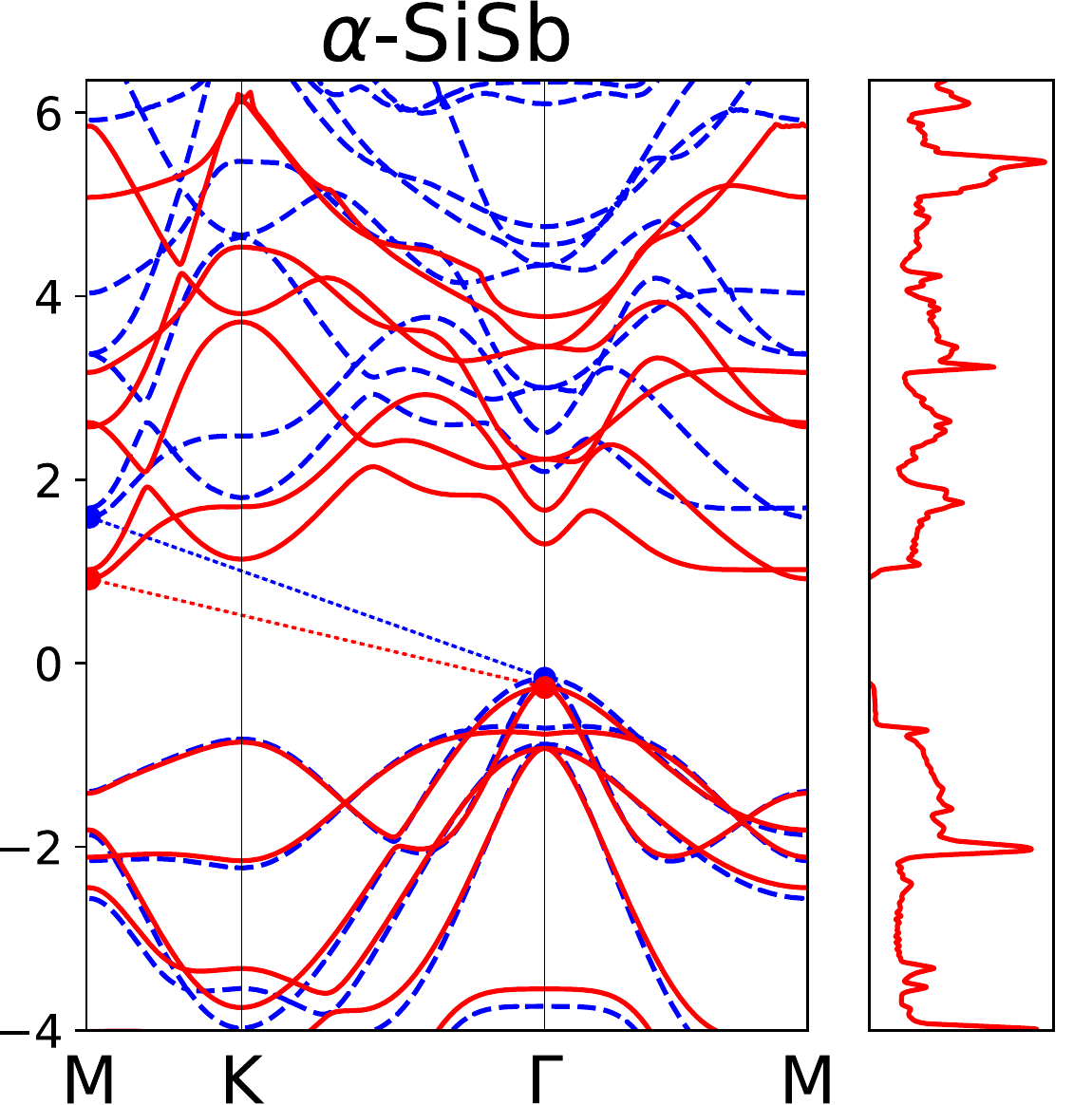}&
\includegraphics[height=0.19\textwidth]{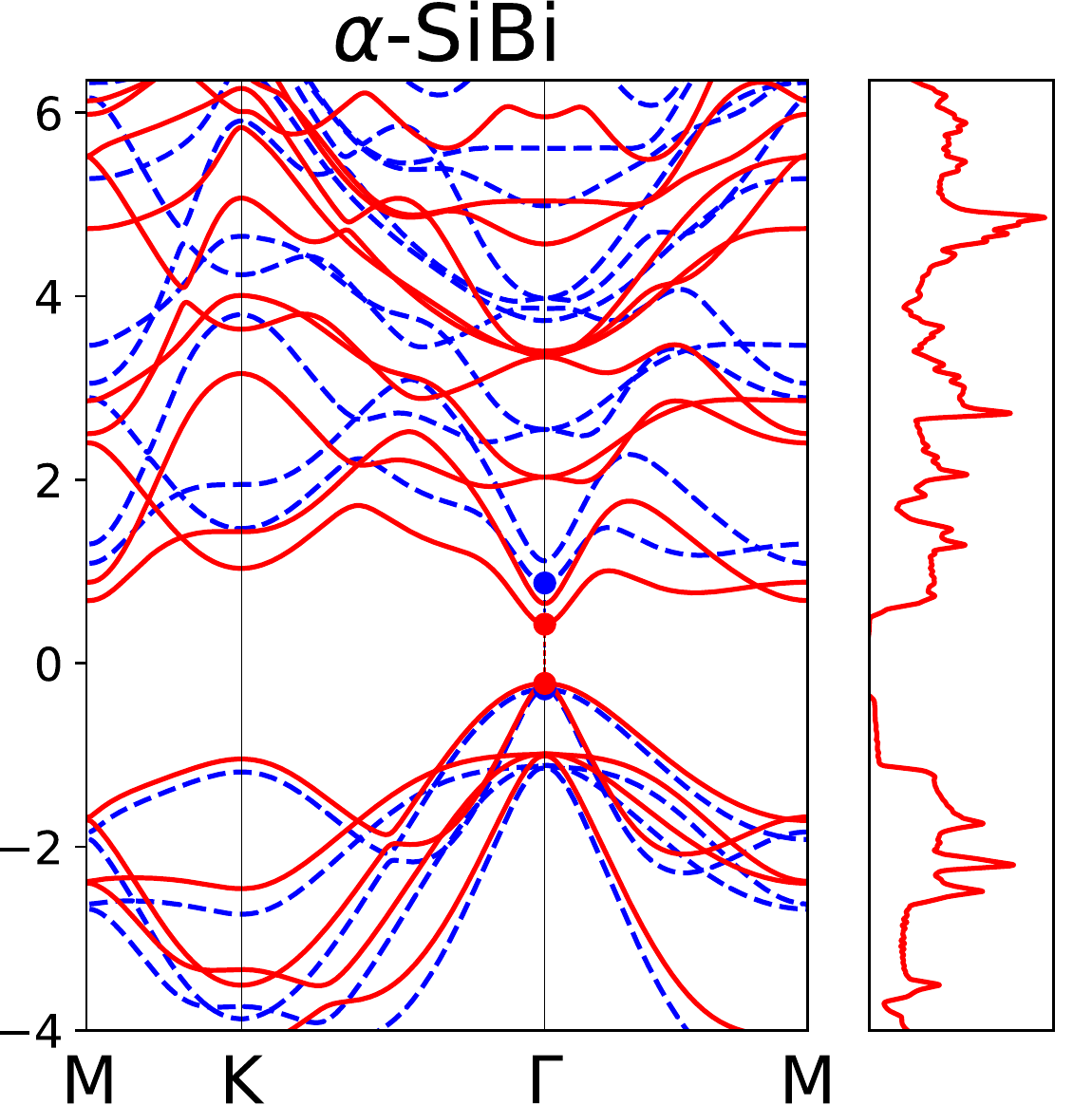}\\

\includegraphics[height=0.19\textwidth]{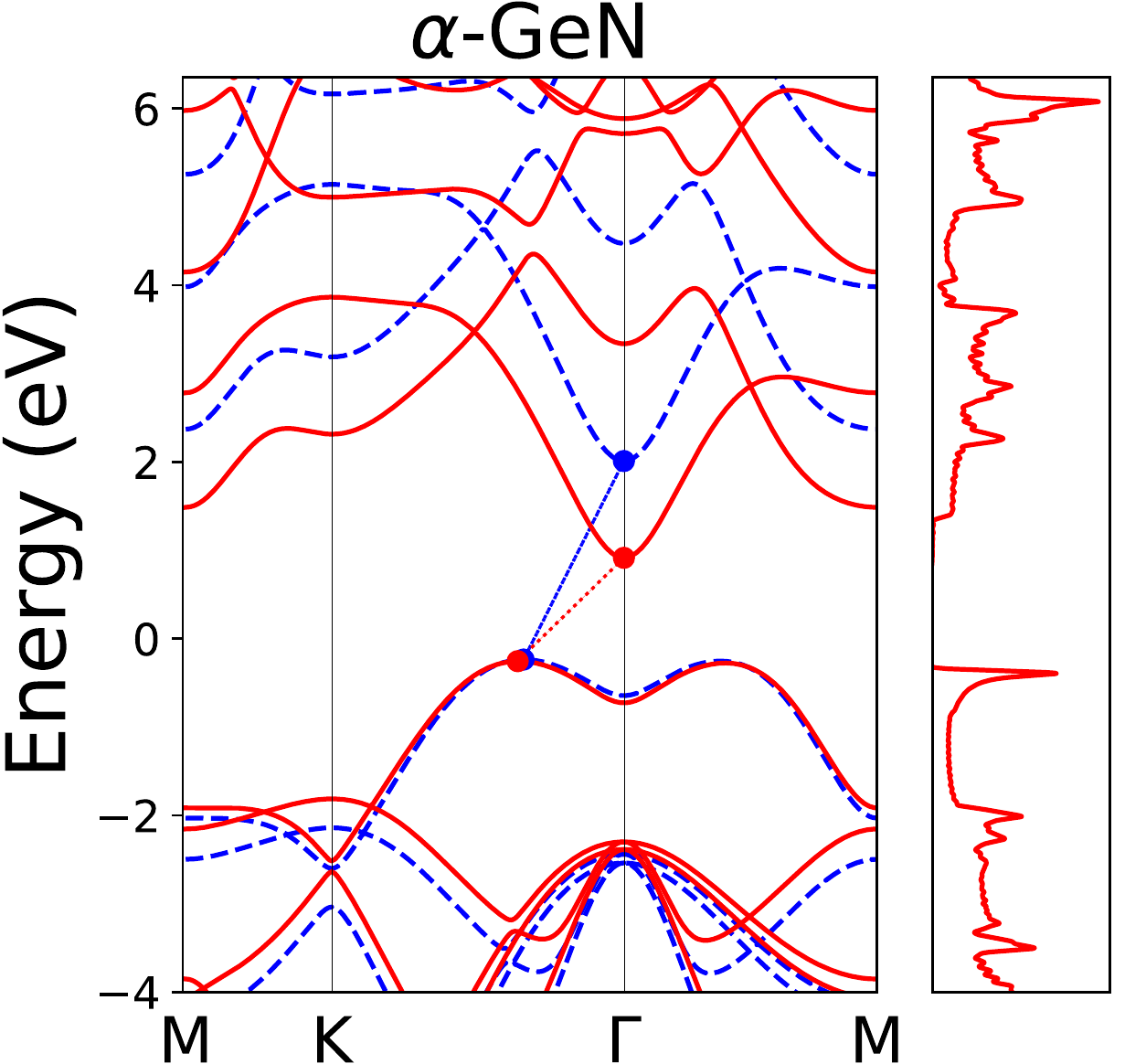}&
\includegraphics[height=0.19\textwidth]{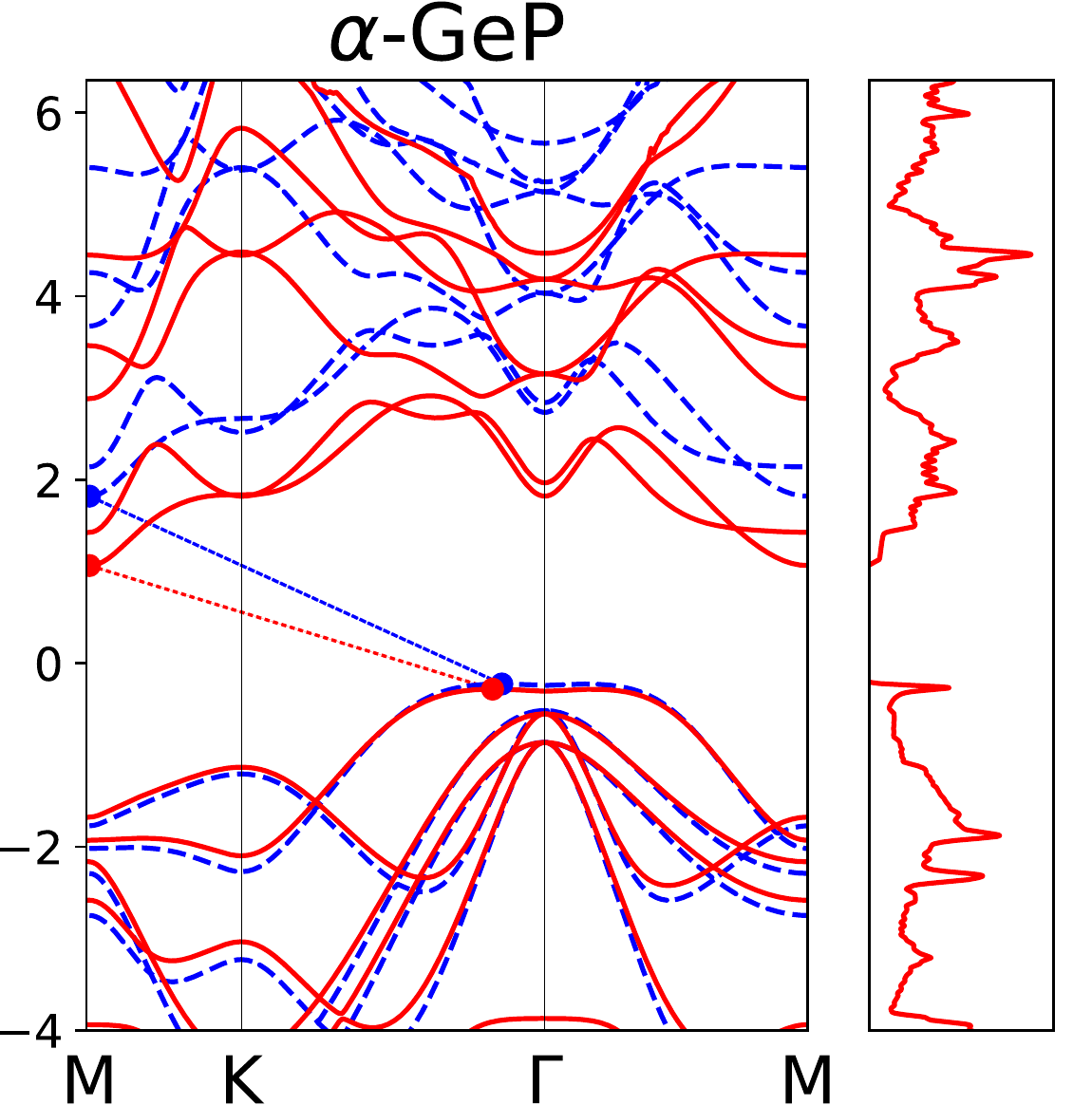}&
\includegraphics[height=0.19\textwidth]{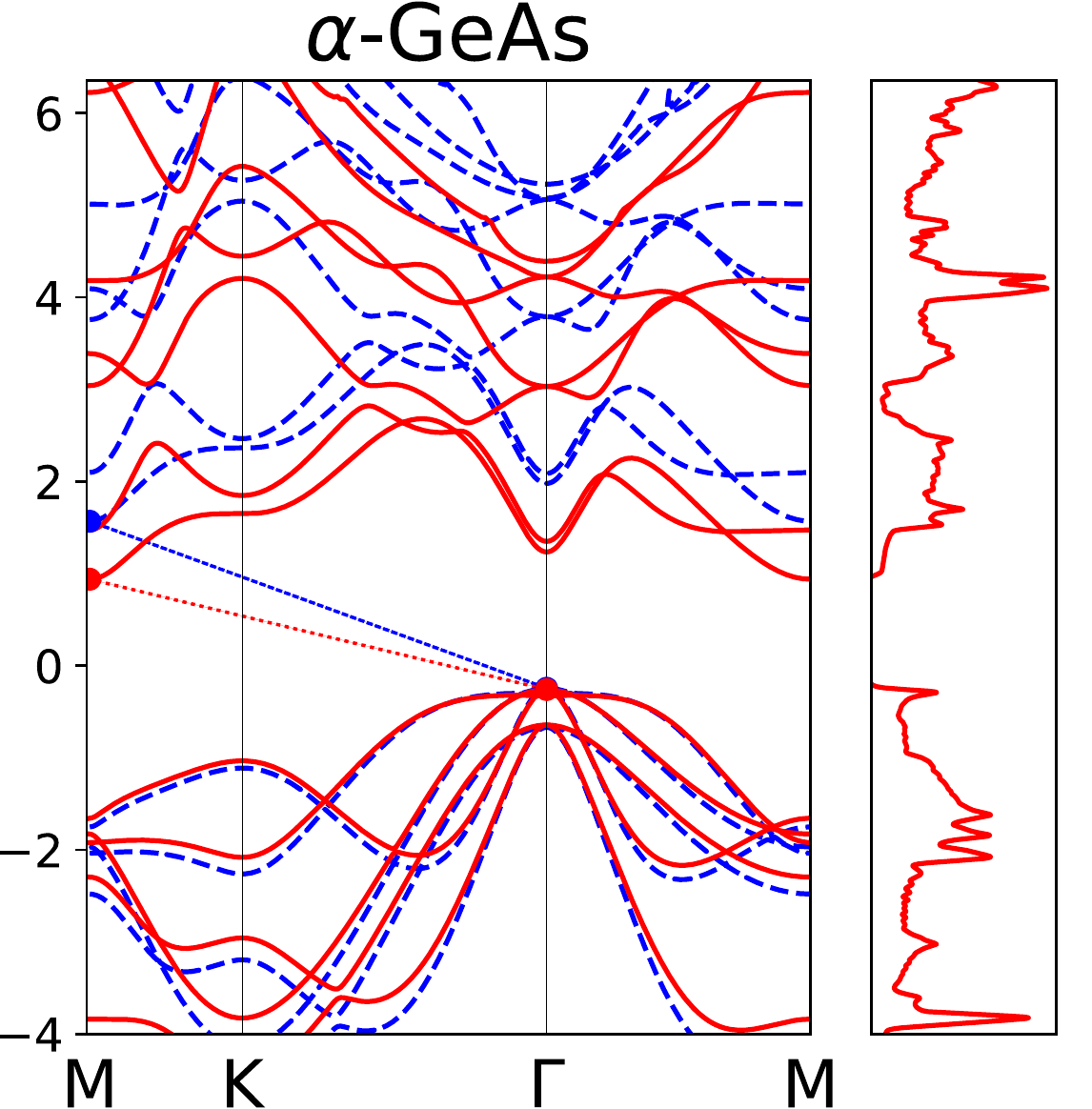}&
\includegraphics[height=0.19\textwidth]{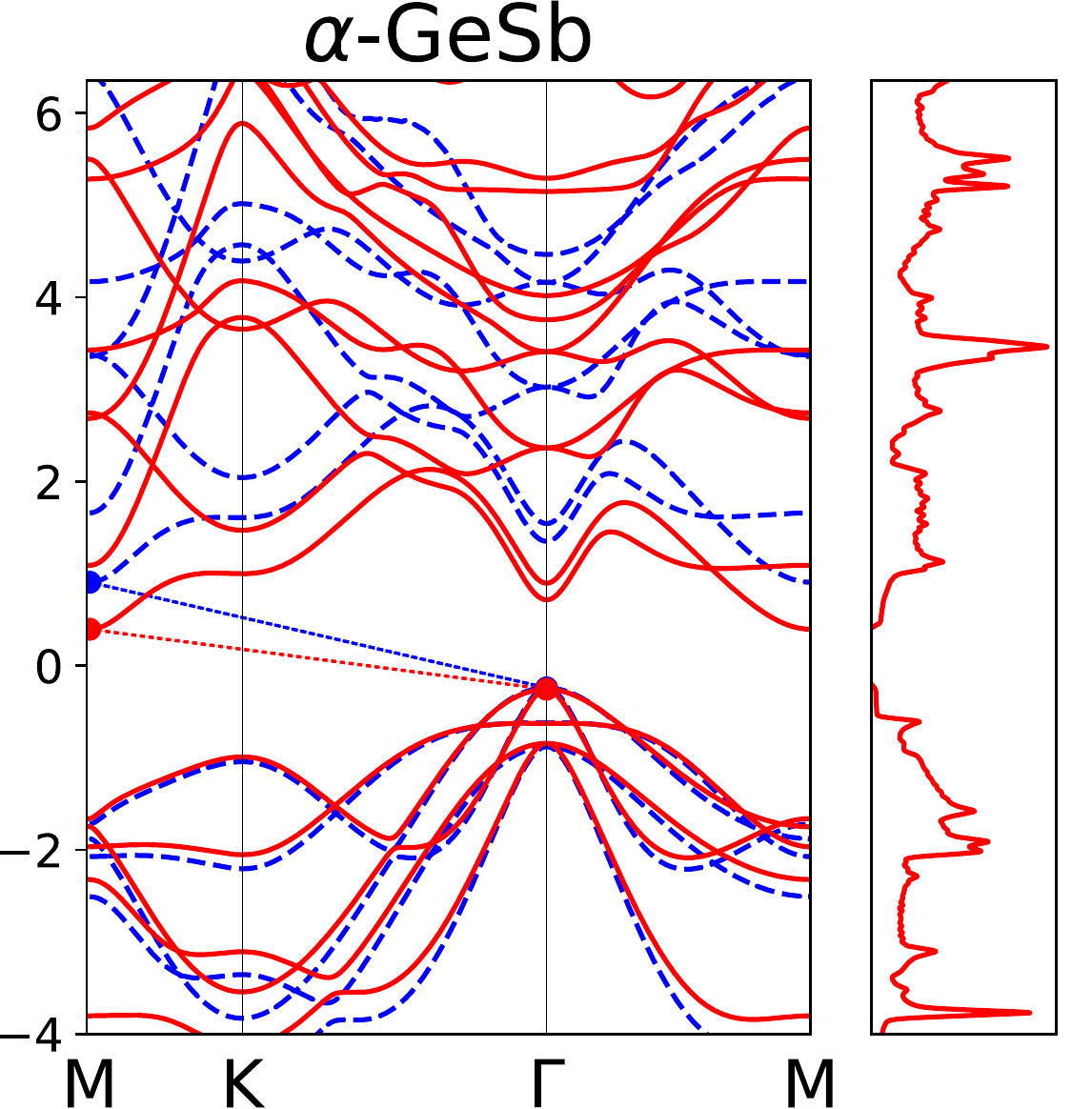}&
\includegraphics[height=0.19\textwidth]{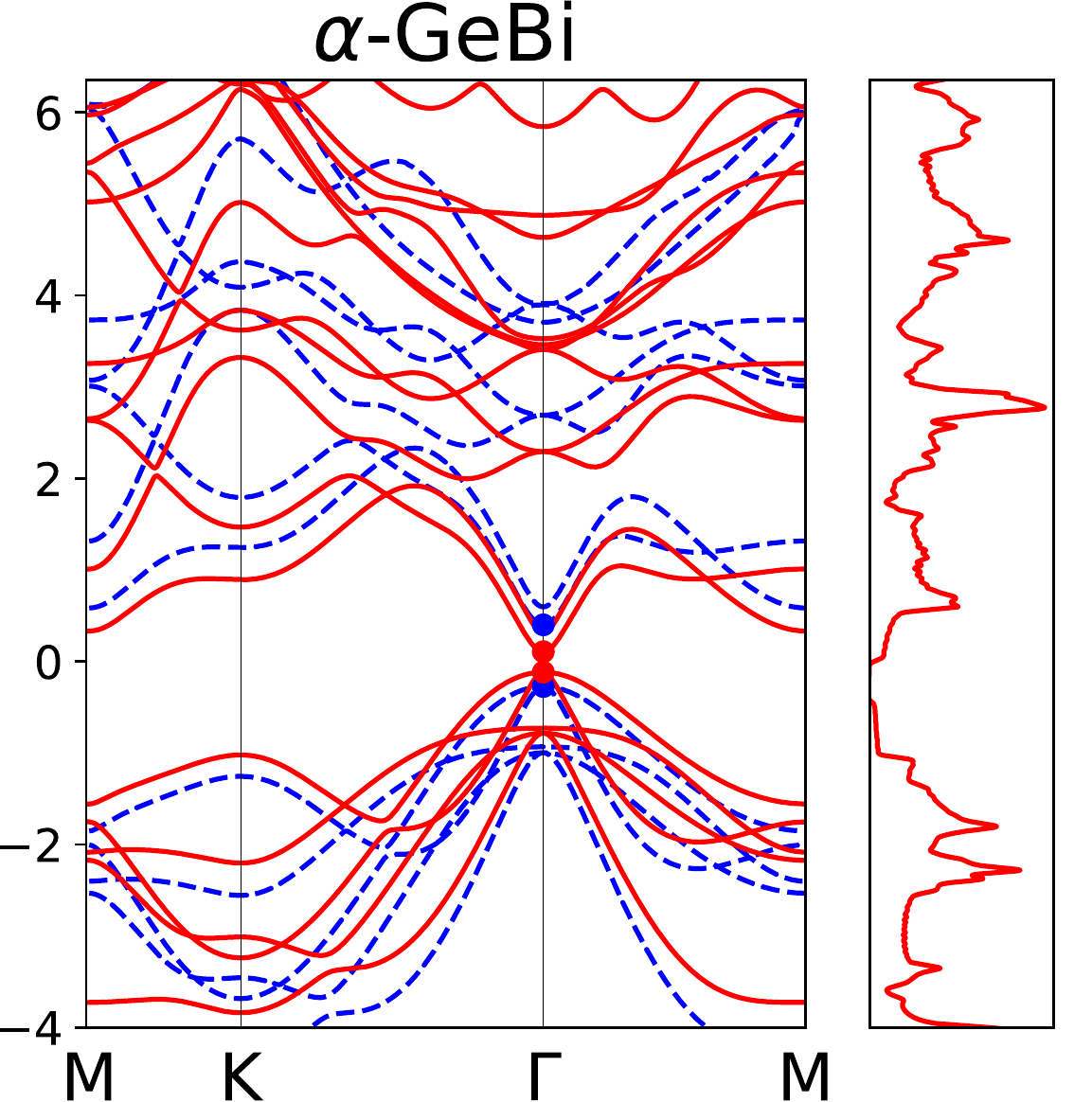}\\

\includegraphics[height=0.19\textwidth]{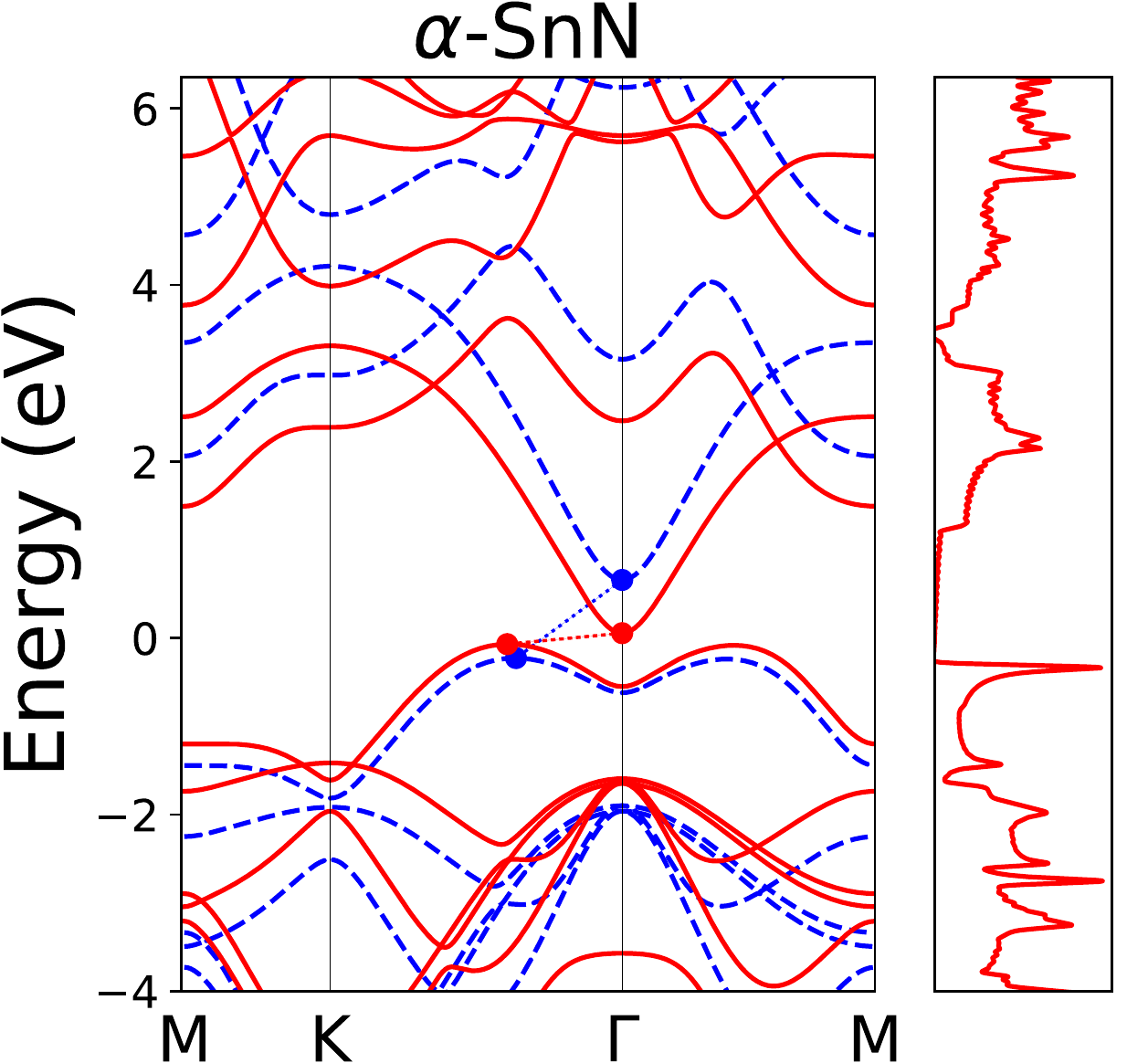}&
\includegraphics[height=0.19\textwidth]{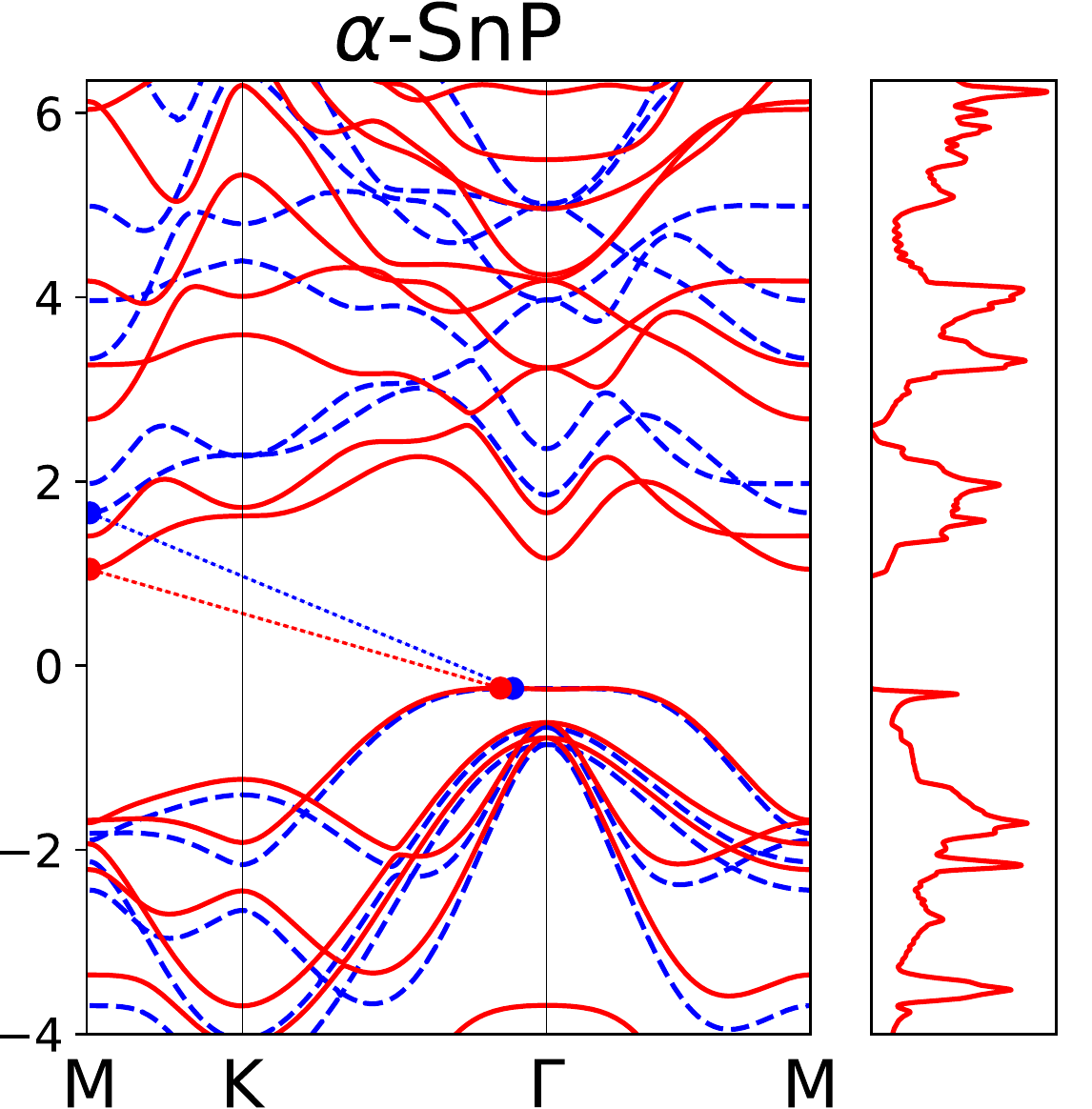}&
\includegraphics[height=0.19\textwidth]{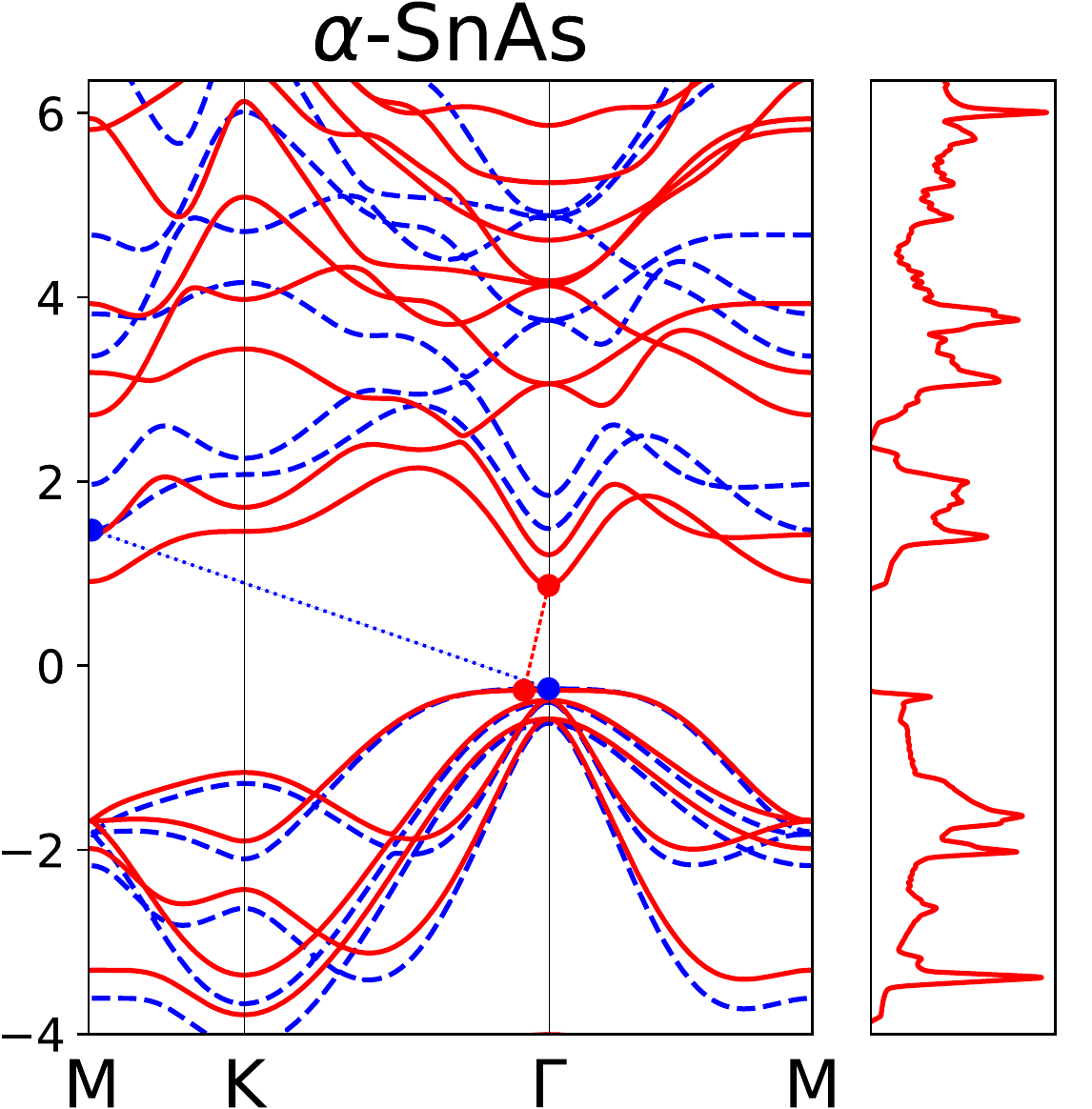}&
\includegraphics[height=0.19\textwidth]{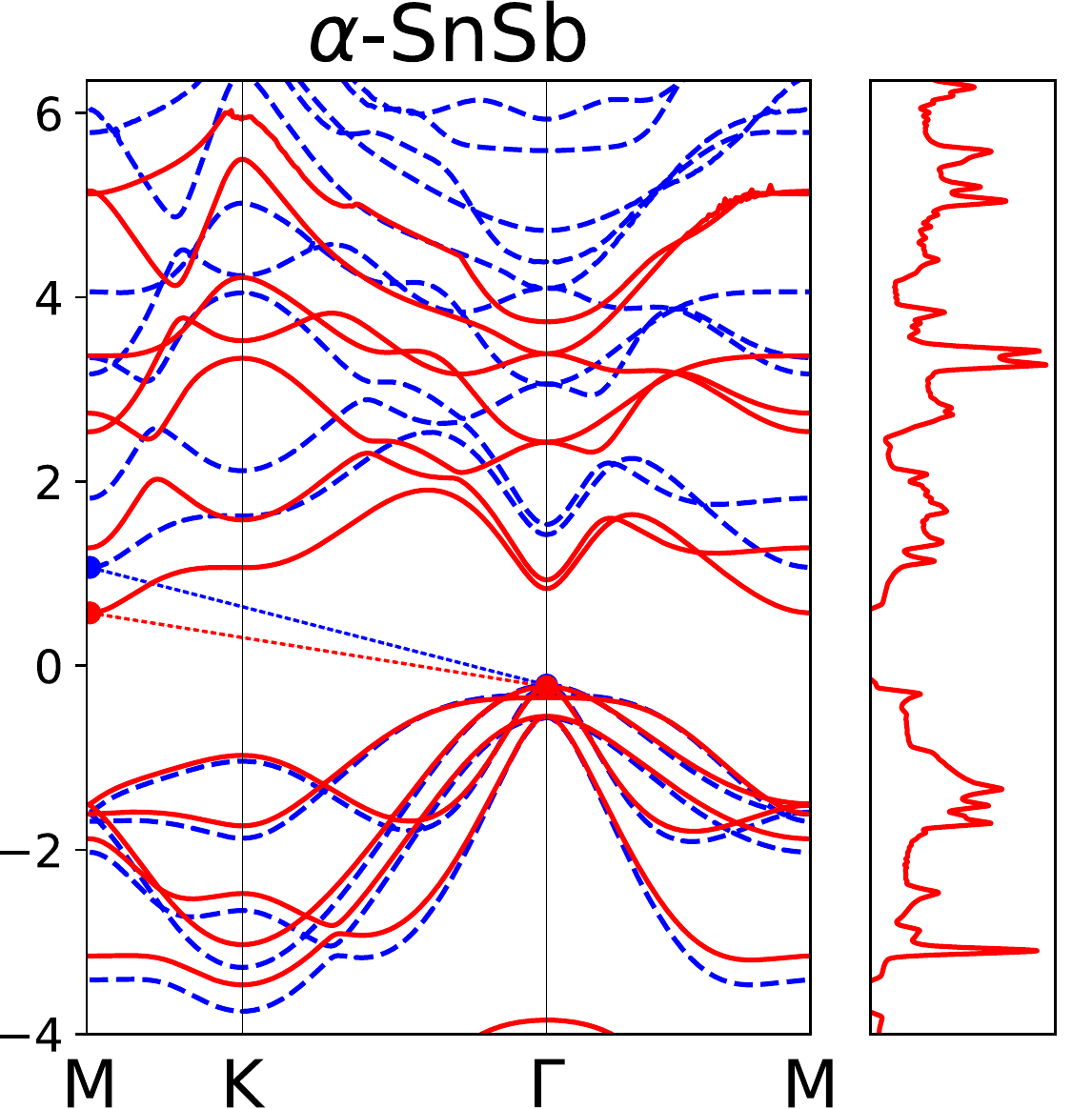}&
\includegraphics[height=0.19\textwidth]{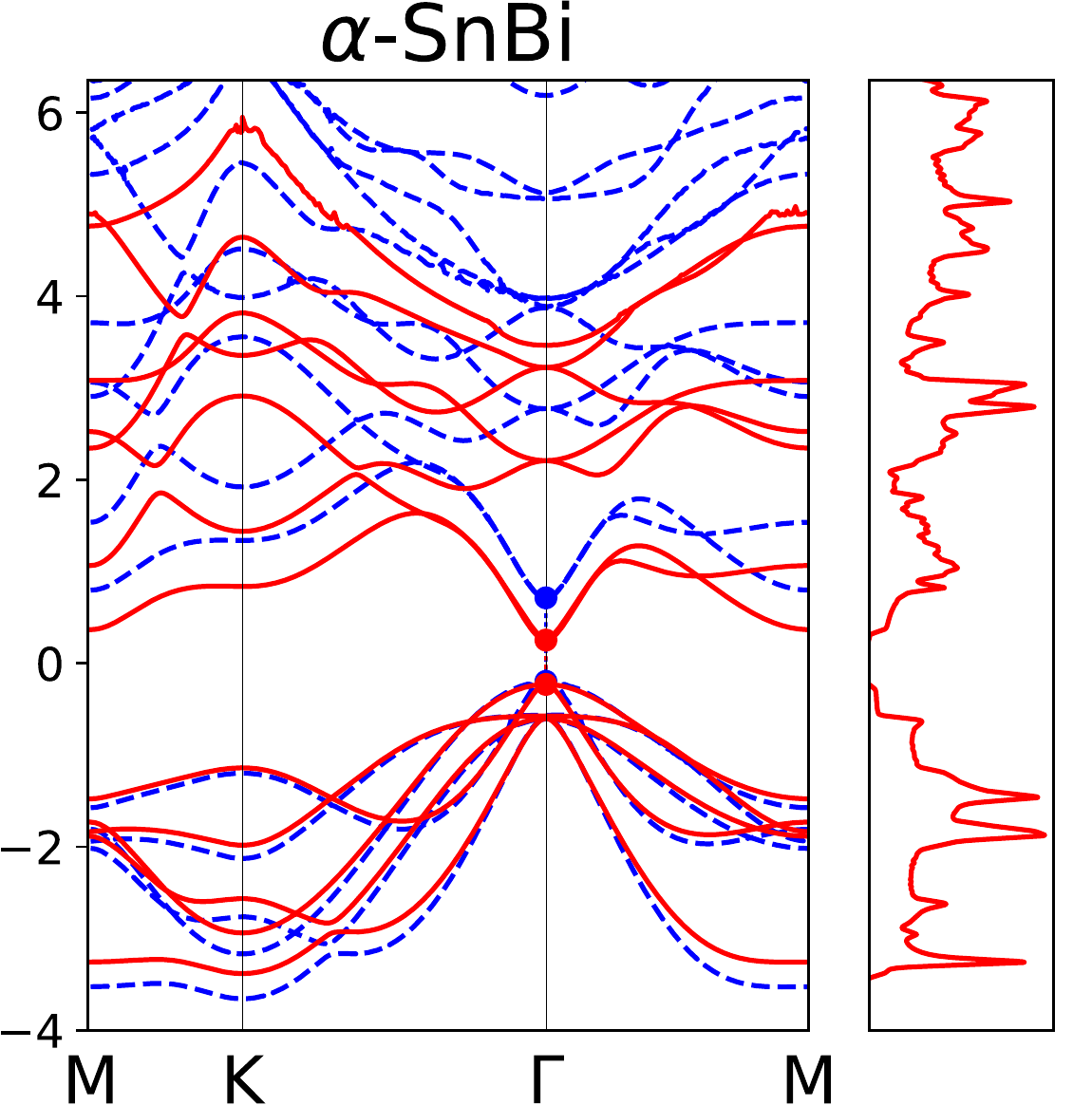}\\

\includegraphics[height=0.19\textwidth]{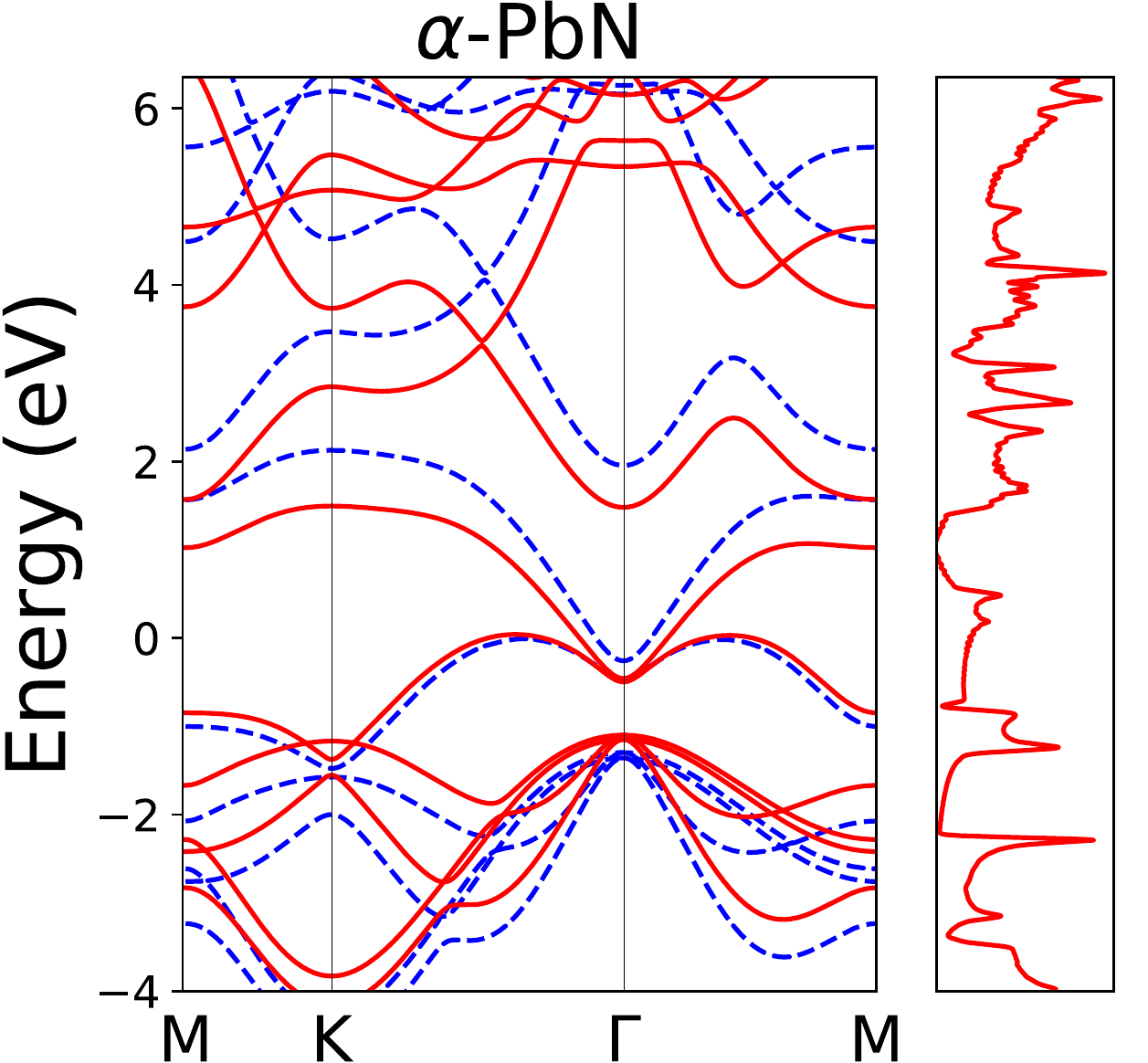}&
\includegraphics[height=0.19\textwidth]{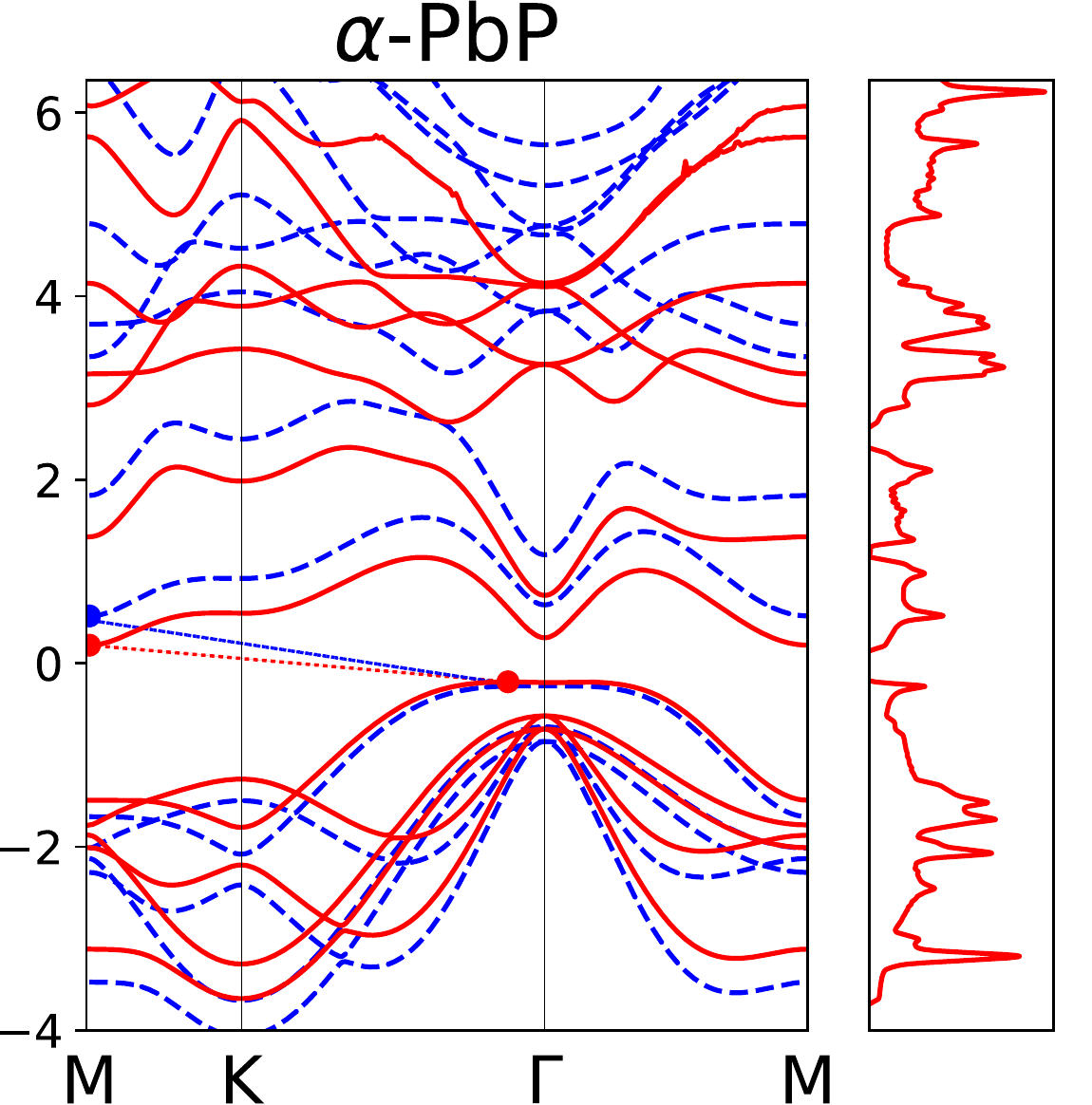}&
\includegraphics[height=0.19\textwidth]{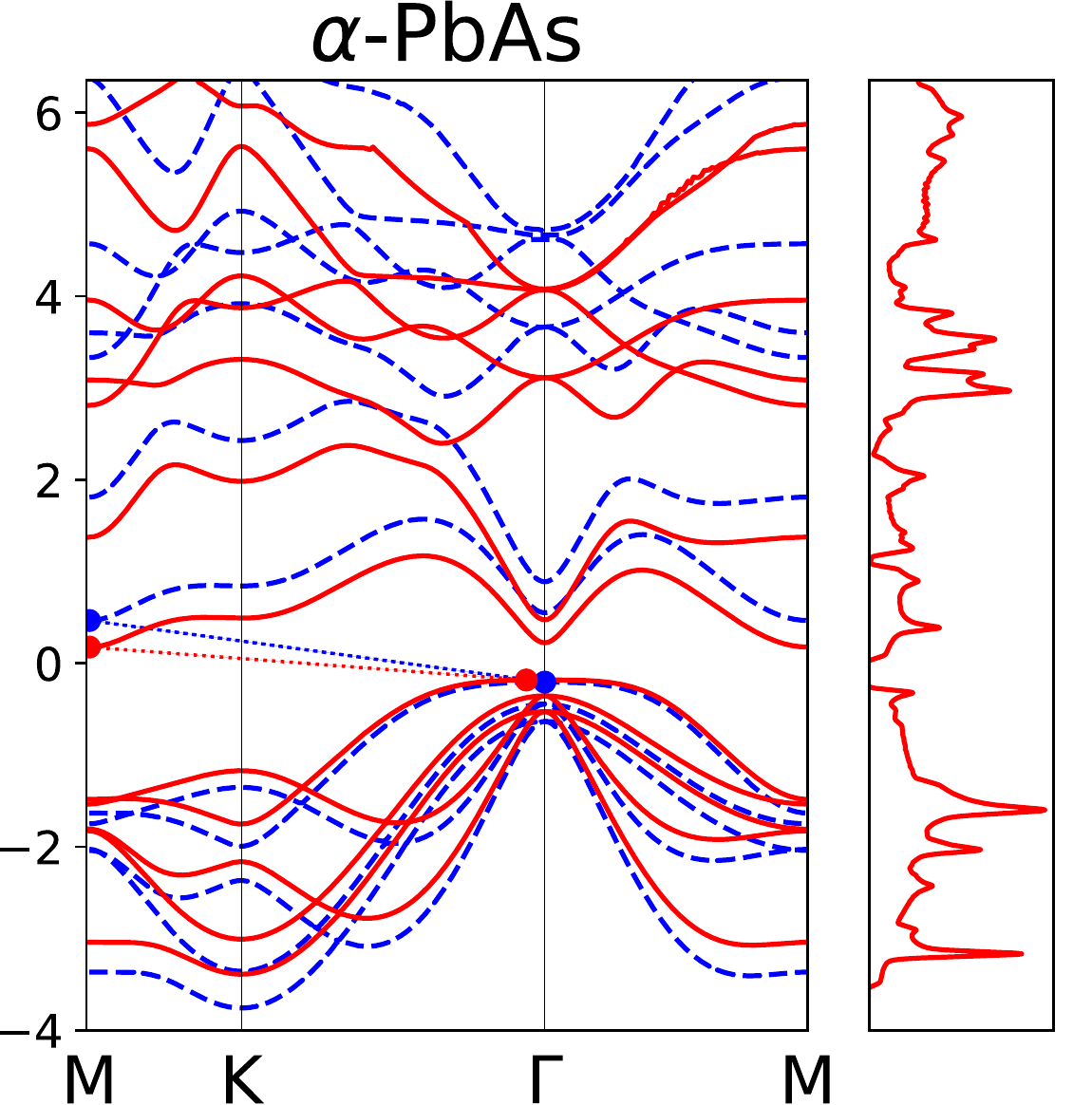}&
\includegraphics[height=0.19\textwidth]{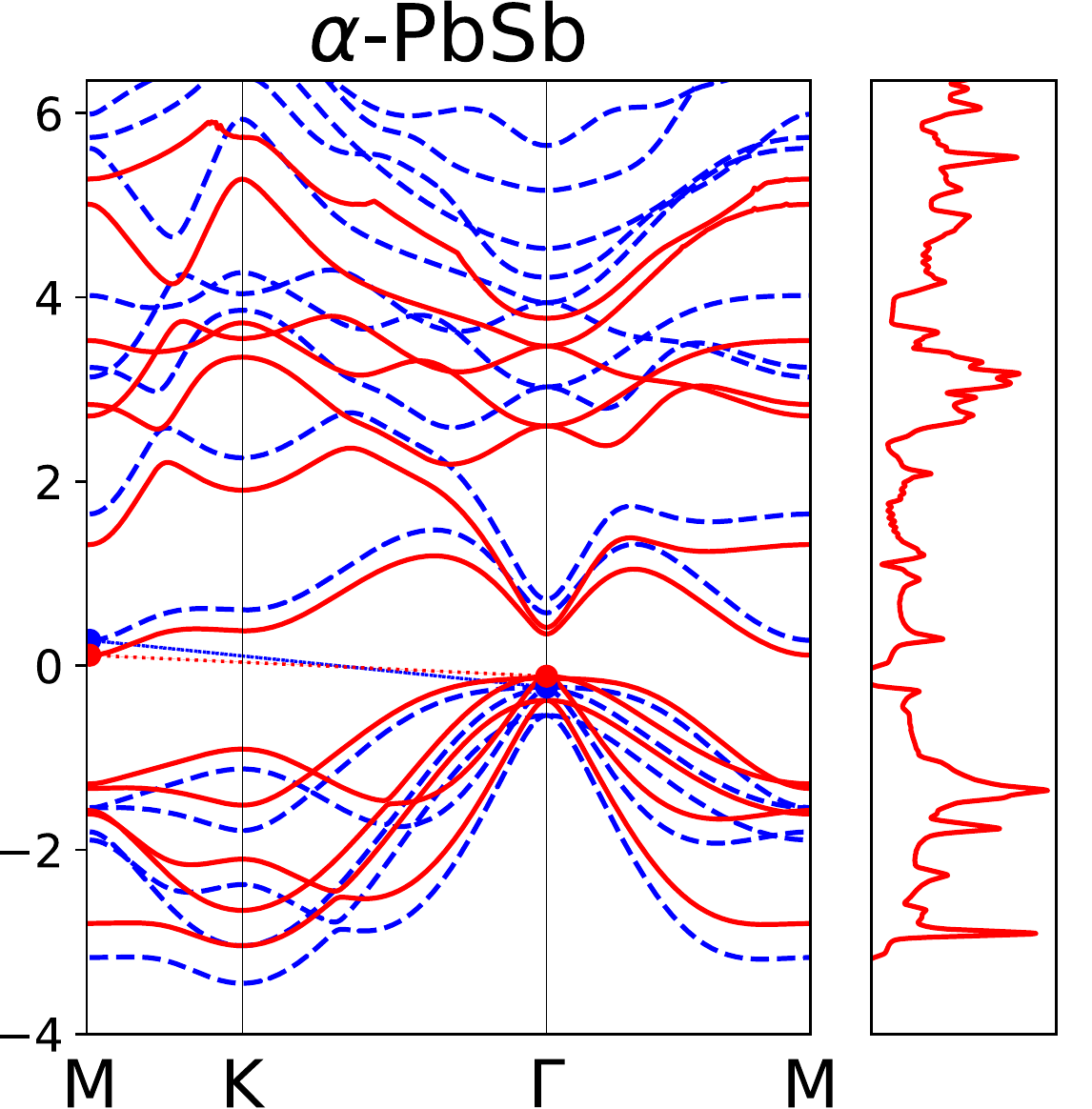}&
\includegraphics[height=0.19\textwidth]{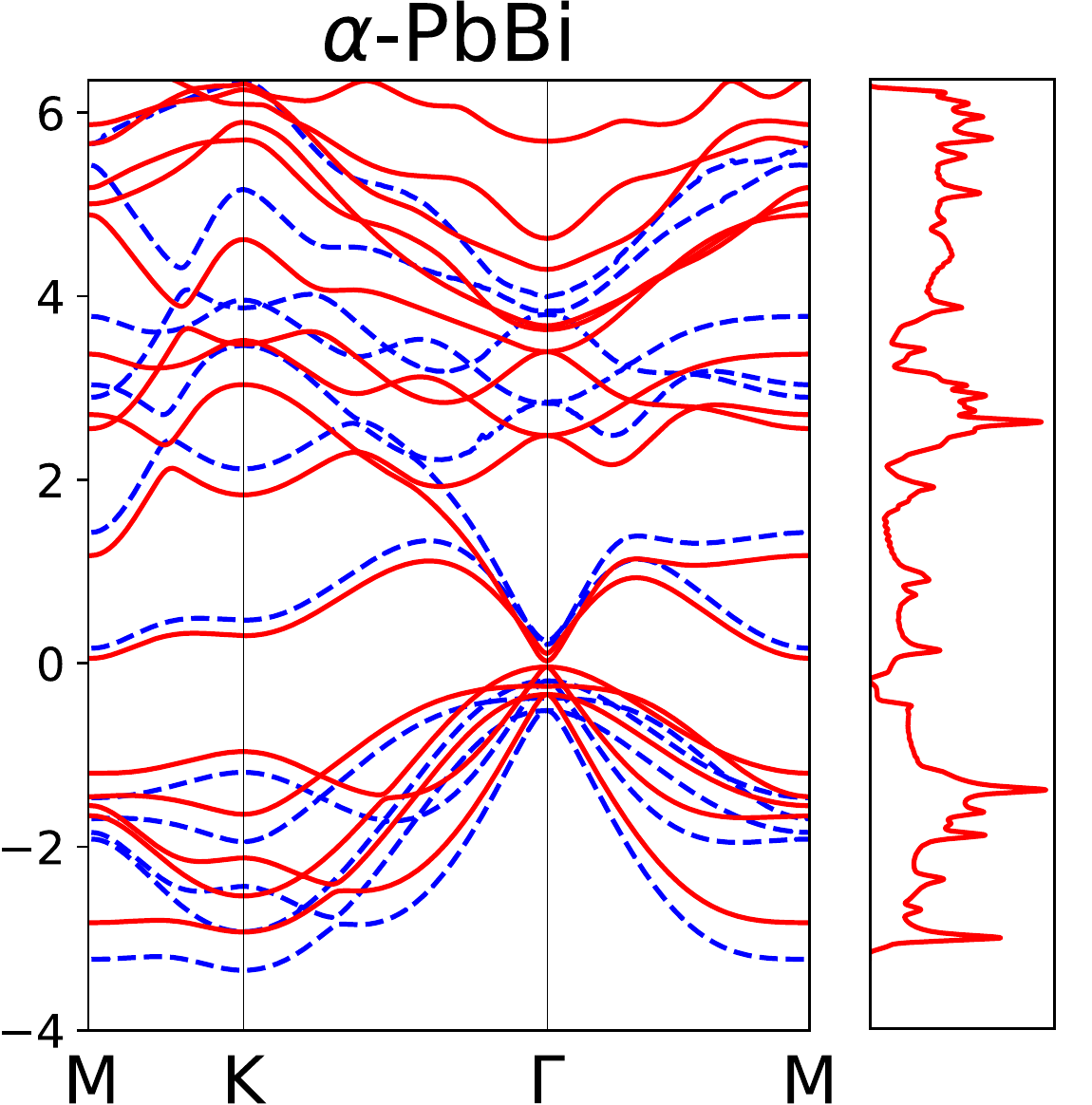}\\
\end{tabular}

\caption{Electronic band diagrams of the $\alpha$-phases as calculated with PBE (solid red) and hybrid HSE06 (dashed blue) functionals.  Energy gaps are given in Table \ref{tab:the-table}. Conduction band minima and the valence band maxima are joined to indicate the band gaps. 
}
    \label{fig:elband-dos}
\end{figure*}

\begin{figure}[h!]
	\begin{tabular}{ccc}
		\\
		\includegraphics[width=0.24\textwidth]{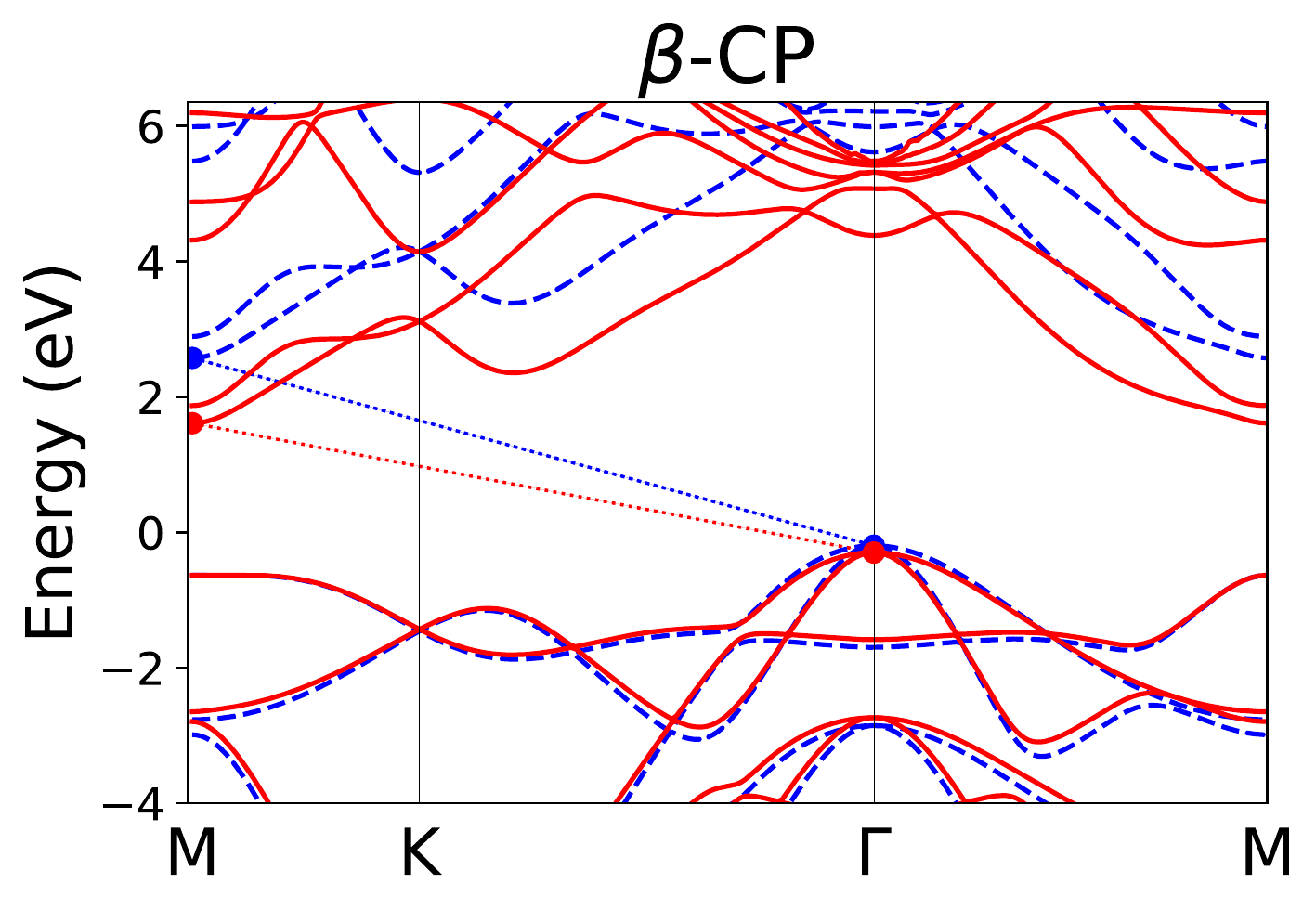}
		\includegraphics[width=0.24\textwidth]{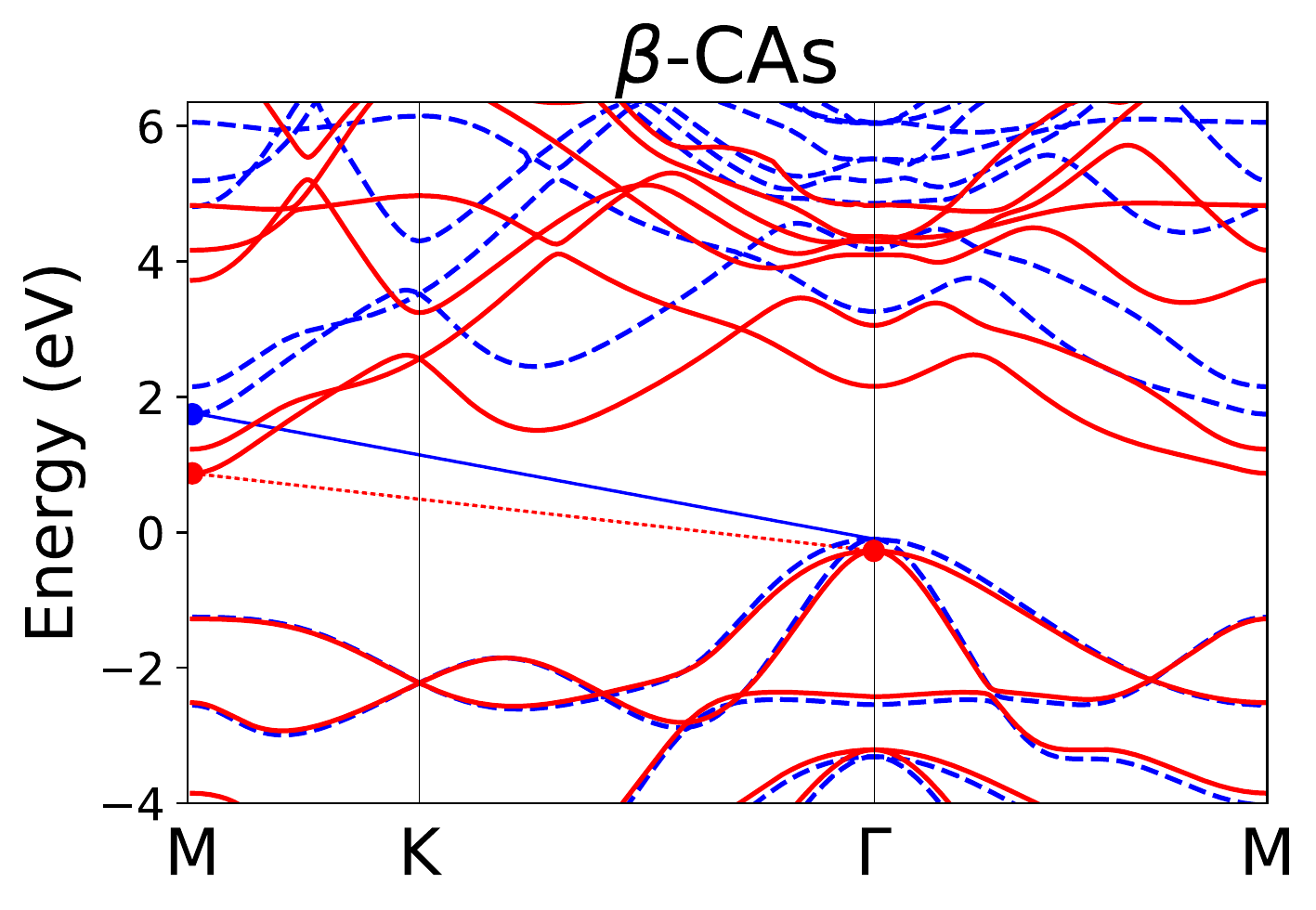}\\
		
		\includegraphics[width=0.24\textwidth]{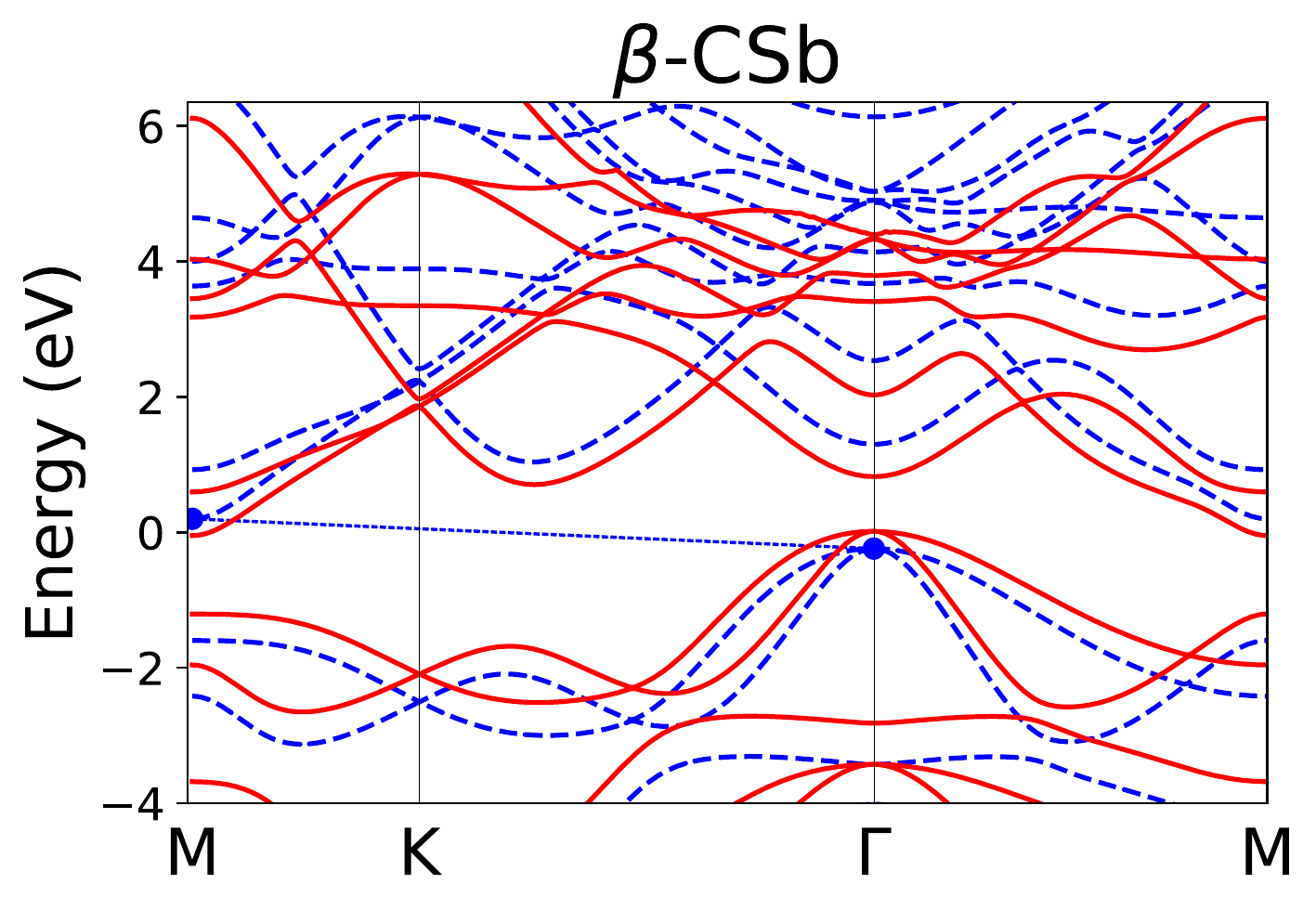}
		\includegraphics[width=0.24\textwidth]{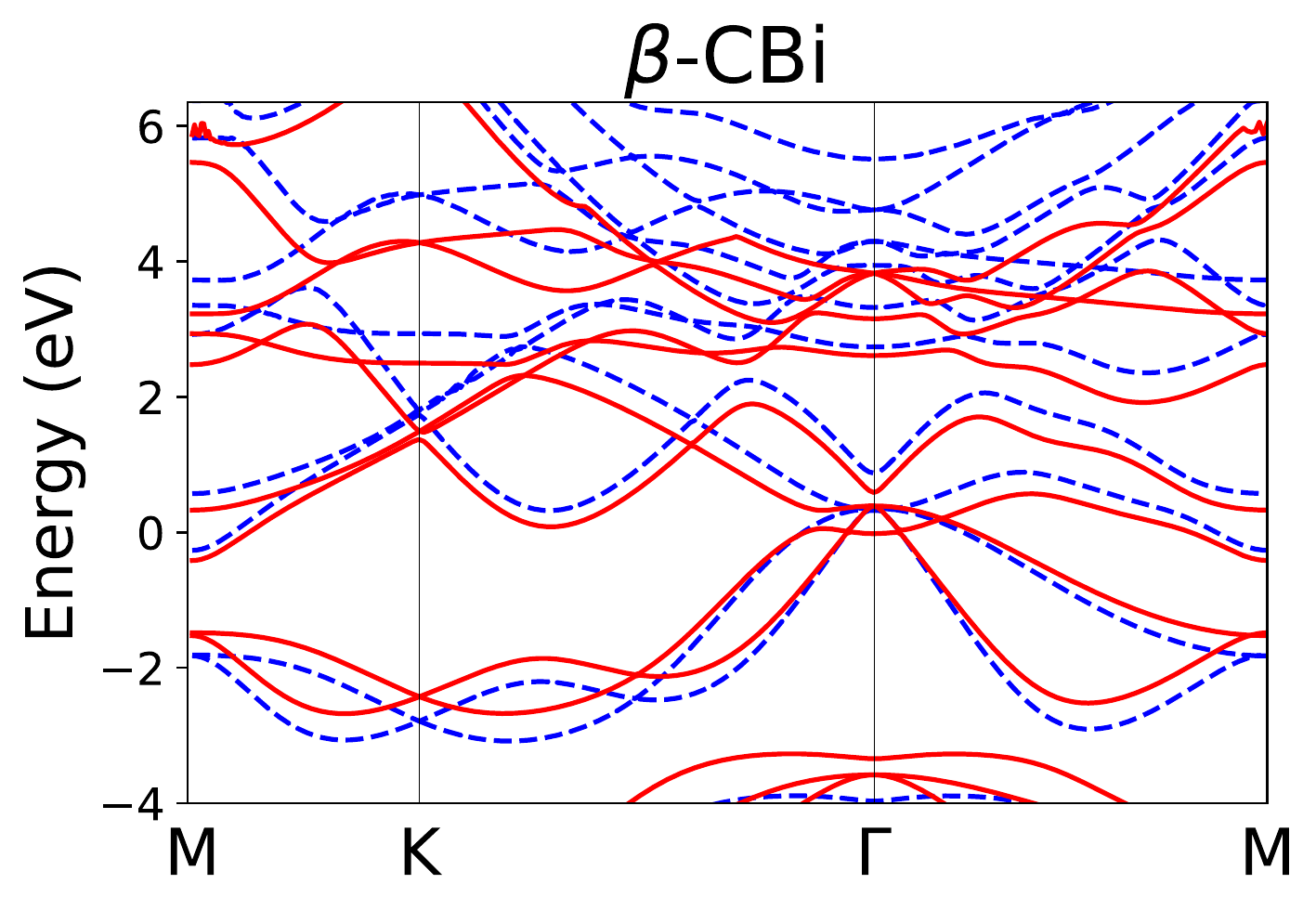}
	\end{tabular}
	
	\caption{Electronic band structures of CP, CAs, CSb, CBi in P$\bar{3}$m1 ($\beta$) symmetry calculated with both PBE and HSE06 functionals, given in solid red and dashed blue lines, respectively. See Table \ref{tab:AB-the-table} for structural and electronic properties.}
	\label{fig:ab}
\end{figure}

The electronic structure is studied using GGA-PBE with and without including spin-orbit interactions and also using hybrid HSE06 functionals.
The electronic band diagrams of $\alpha$-phase structures as obtained from PBE and HSE06 functionals are presented in Figure~\ref{fig:elband-dos}. 
The band diagrams of the $\beta$-phase compounds are plotted in Figure~\ref{fig:ab}.

We first discuss  the $\alpha$-phase. 
Total of six compounds; CP (K), SiBi ($\Gamma$), GeBi($\Gamma$) SnAs ($\Gamma$), SnBi ($\Gamma$), PbBi($\Gamma$) have direct band gaps according to PBE. 
The k-points at which the direct transitions takes place are indicated in parenthesis. 
The band gap characteristics of the mentioned compounds, except CP, are identical with HSE06. 
Ten of the $\alpha$-phase structures (CN, SiN, SiP, GeN, GeP, SnN, SnP, SnAs, PbP, PbAs) have their valence band edges between $\Gamma$ and K.
In fact, the valence band maxima (VBM) for these structures resemble a Mexican hat.
In other words, the VBM occurs not on a single point and it is highly degenerate.
The Mexican hat shaped quartic dispersions are discussed separately below.

HSE06 calculations yield band structures, which are of the same character with the PBE results.
The band gap values are increased by up to 1.43~eV ($\alpha$-CN) with HSE06. (see Figure~\ref{fig:elband-dos})
Some of the studied materials are wide band gap semiconductors, having band gaps greater than 2 eV. 
These are CP (2.737 eV), SiN (2.732 eV), SiP (2.223 eV), SiAs (2.267 eV), GeN (2.247 eV), GeP (2.047 eV). 
These materials can be used for UV-light applications such as, UV-light detection and photodetectors \cite{monroy2003wide}. 
A recent study reports nitrides and phosphides of Si, Ge and Sn in hexagonal symmetry \cite{lin2017single}. Electronic structures of these compounds are quite consistent with our results, except those of antimony. While the difference in band gaps of Si and Ge nitrides and phosphides are at most 12\%, this is almost 50\% for SnN. Given that the structural parameters such as the lattice constants and layer heights of the mentioned compounds are very comparable with our findings, the difference in electronic structure may arise from the parameters of the HSE06 functional.
According to Figure~\ref{fig:elband-dos}, compounds containing bismuth generally tend to have direct ($\Gamma$) band gaps ranging between 0.06 and 0.64 eV for PBE; and 0.35 to 1.15 eV for HSE06 calculations. The exception is PbBi when it comes to HSE06. In this case, the band gap points in the $\Gamma$-M direction. 
Our band structure results are in agreement with those available in the literature.~\cite{ashton2016computational}
Additionally, the effect of spin-orbit coupling (SOC) on the electronic states is also studied. Electronic band structures calculated with PBE and SOC included can be found in the Supporting Information. (see Figure~\ref{fig:alpha-soc})
It is observed that direct/indirect character of interband transitions are not effected by the SOC.
The predicted band gaps with SOC included ($E_\mathrm{g}^\mathrm{SOC}$) are close to those obtained from PBE without SOC. (see Table~\ref{tab:the-table})

The electronic structures of the $\beta$-phase compounds are also studied in detail.
The $AB$ layers, which make up the $\alpha$-$A_2B_2$ and $\beta$-$A_2B_2$ structures are identical for both phases.
The neighboring atom types are identical and the distances are almost the same  up to the third nearest neighbors for $A$ type and the second nearest neighbors for $B$ type atoms.
As a result, the lattice parameters and the band structures change only slightly between $\alpha$- and $\beta$-phases.
Still, there are some quantitative differences.
For example, five more structures in the $\beta$-phase have quartic valence band edges compared to the $\alpha$-phase, which make fifteen such structures in total ($\beta$-CN, $\beta$-SiN, $\beta$-SiP, $\beta$-SiAs, $\beta$-GeN, $\beta$-GeP, $\beta$-GeAs, $\beta$-SnN, $\beta$-SnP, $\beta$-SnAs, $\beta$-PbP, $\beta$-PbAs, $\beta$-PbSb).
The band gap values are approximately the same for $\alpha$- and $\beta$-phases.
(see Figure~\ref{fig:beta-pbe})

\begin{table*}[t]
	\begin{ruledtabular}
		\begin{tabular}{ccccccccccc}
			& 	CN		&	SiN		&	GeN 	& 	SnN		& 	SiP		&	GeP		&	SnP		&	PbP		&	SnAs	&	PbAs	\\
			\midrule
			$k_c$ (\AA$^{-1}$)				&	0.32	& 	0.57	&	0.47	&	0.47	&	0.24	&	0.18	&	0.16	&	0.11	&	0.0		&	0.0\\
			$\alpha$ (eV$\cdot$\AA$^{4}$)	&	6.516	&	5.359	&	9.254	&	9.853	&	20.448	&	21.807	&	19.845	&	54.385	&	41.929	&	126.039
		\end{tabular}
	\end{ruledtabular}
	\caption{Parameters for quartic dispersion formula belonging to the given structures in $\alpha$-phase. (see Equation~\ref{eqn:quartic})}
	\label{table:quartic}
\end{table*}

A distinctive feature of the hexagonal structures of group IV-V elements is the onset of quartic bands in their valence bands.
It is well known that $A_2B_2$ type lattices of groups III-VI with $P\bar{6}m2$ symmetry  have quartic dispersions in their valence bands, which is also referred to as the Mexican hat dispersion.~\cite{ma:pccp:2013,zolyomi:prb:2013,zolyomi:prb:2014,rybkovskiy:prb:2014}
Of those structures, the quartic dispersions form the valence band edge for $\alpha-$phases of BO, BS and $A_2B_2$ with $A{=}$Ga, In, Al and $B{=}$S, Se, Te. \cite{demirci2017structural}
Quartic dispersion gives rise to a strong singularity ($1/\sqrt{E}$) in the DOS, which gives rise to novel exciting properties~\cite{stauber:prb:2007} such as tunable magnetism~\cite{cao2015tunable} and multiferroic phase~\cite{seixas:prl:2016}, namely simultaneous presence of ferromagnetism and ferroelasticity.
Quartic dispersion also gives rise to a step-like change in the transmission spectrum, which is the reason for temperature independent thermopower and efficient thermoelectric transport.~\cite{wickramaratne2015electronic,sevinccli2017quartic}

It was shown for elemental lattices of group-V elements also display quartic dispersion~\cite{ozcelik2015prediction,zhu:prl:2014,kamal2015arsenene,akturk2015single,akturk2016single}, and that the appereance of quartic bands is because of the hexagonal symmetry and that the dispersion relation can be expressed as~\cite{sevinccli2017quartic}
\begin{equation}
	\label{eqn:quartic}
	E=E_v-\alpha(k^2-k_c^2)^2,
\end{equation}
where $E_v$ is the band edge, and $k_c$ is the radius of the circular band maximum.
This expression is obtained from a series expansion around the center of the Brillouin zone. 
Higher order terms ($k^n$, with $n\geq6$), which break the circular symmetry and establish a hexagonal one, can also be included~\cite{zolyomi:prb:2013,zolyomi:prb:2014} but we limit our attention to the quartic case.

Layered hexagonal lattices of group IV-V elements also exhibit quartic dispersion as already shown in Figure~\ref{fig:elband-dos}.
The valence band edges are formed by the quartic bands in 10 out of 25 structures, which are CN, SiN, GeN, SnN, SiP, GeP, SnP, PbP, SnAs and PbAs.
Different from the PBE bands, the critical wave-vector $k_c$  shifts towards the $\Gamma$ point in SnAs and PbAs such that $k_c=0$.
That is a purely quartic dispersion, $E{-}E_v{=}{-}\alpha k^4$, is obtained.
We obtained the values of $\alpha$ for these structures by using the $k_c$, $E_v$ and $E_0{=}E_{k{=}0}{-}E_v$ values as obtained from DFT calculations. (see Table~\ref{table:quartic})

\begin{figure}[t]
\includegraphics[width=0.48\textwidth]{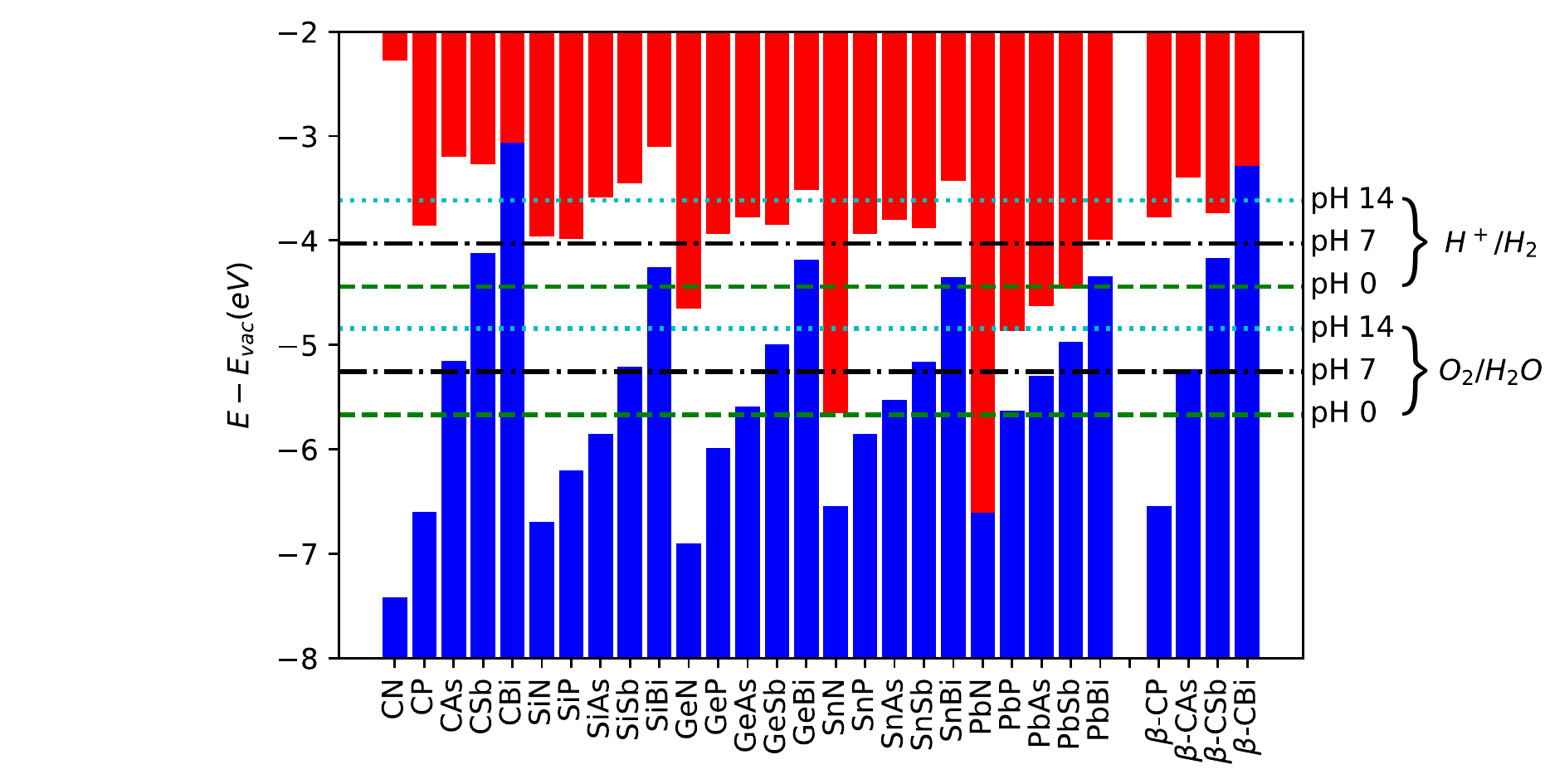}
\caption{\label{fig:redox-pot} Band edge positions of all the compounds studied vs the redox potential of water. Dashed green, dashdotted black and dotted cyan horizontal lines depict the corresponding potentials at pH 0, pH 7 and pH 14, respectively. The compounds starting with $\beta$- are in $P\bar{3}m1$ space group.}
\end{figure}

Photocatalytic water splitting~\cite{ni2007review,maeda2010photocatalytic} is a promising field aiming to dissociate water to its constituents, hydrogen and oxygen solely using light. The main purpose is to use hydrogen for fuel. The absolute band edges of the material is important for hydrogen generation. Therefore, band edge positions of the studied systems are calculated by using HSE06 functional, and compared with the redox potentials of water in Figure~\ref{fig:redox-pot}. Dashed green, dashdotted black
and dotted cyan lines corresponds to the absolute electrode potentials  \cite{trasatti1986absolute} in three different pH levels (0, 7 and 14, respectively). Given the relative band edge positions and the mentioned redox potentials of water in different pH environments; CN, CP, SiN, SiP, SiAs, GeP, SnP, and $\beta$-CP are favorable in pH 0; CN, CP, SiN, SiP, SiAs, GeP, GeAs, SnP, SnAs and $\beta$-CP are favorable in pH 7, lastly CN, CAs, SiAs, SiSb and $\beta$-CAs are favorable in pH 14. The common compounds in all three pH conditions are CN and SiAs. However, as the band gap of CN is more than four times the required 1.23 eV of water splitting gap, which limits its efficiency.
Provided its availability and abundance on earth, SnP may be an excellent candidate for this application in both acidic and neutral conditions.

\subsection{Conclusion}
We have presented a detailed study of group IV-V monolayers.
A total of 50 structural configurations are investigated and tabulated, most of which are predicted for the first time.
The small energy difference between the $\alpha$- and $\beta$-phases ($P\bar{6}m2$ and $P\bar{3}m1$ space groups, respectively) suggest that polymorpic structures should be expected.
Two of the materials  are metallic, while the rest span a wide range of energy band gap values between 0.35 to 5.14~eV.
Quartic energy dispersion with a Mexican hat shape  is a common feature of all structures in their valence band, which make the valence band maximum in some of the structures. $\alpha$-SnAs and $\alpha$-PbAs have purely quartic valence band edges.
CN and SiAs are predicted to be useful for water splitting in terms of their relative band positions.
Nonetheless SnP is an outstanding candidate regarding efficiency and environmental effects.

\begin{acknowledgments}
We acknowledge support from Scientific and Technological Research Council of Turkey (TÜBİTAK) Grant No. 117F131.
\end{acknowledgments}


%

\clearpage

\onecolumngrid
\vspace{200mm}
\begin{center}
\huge{Supplemental Material}\\
\vspace{10mm}
\large{\textbf{\baslik}}
\end{center}

\renewcommand\thefigure{S\arabic{figure}}
\renewcommand\thetable{S-\Roman{table}}

\setcounter{figure}{0}
\setcounter{page}{1}

\begin{figure}[b]
   \centering
\begin{tabular}{ccccc}
\\
\includegraphics[width=0.2\textwidth]{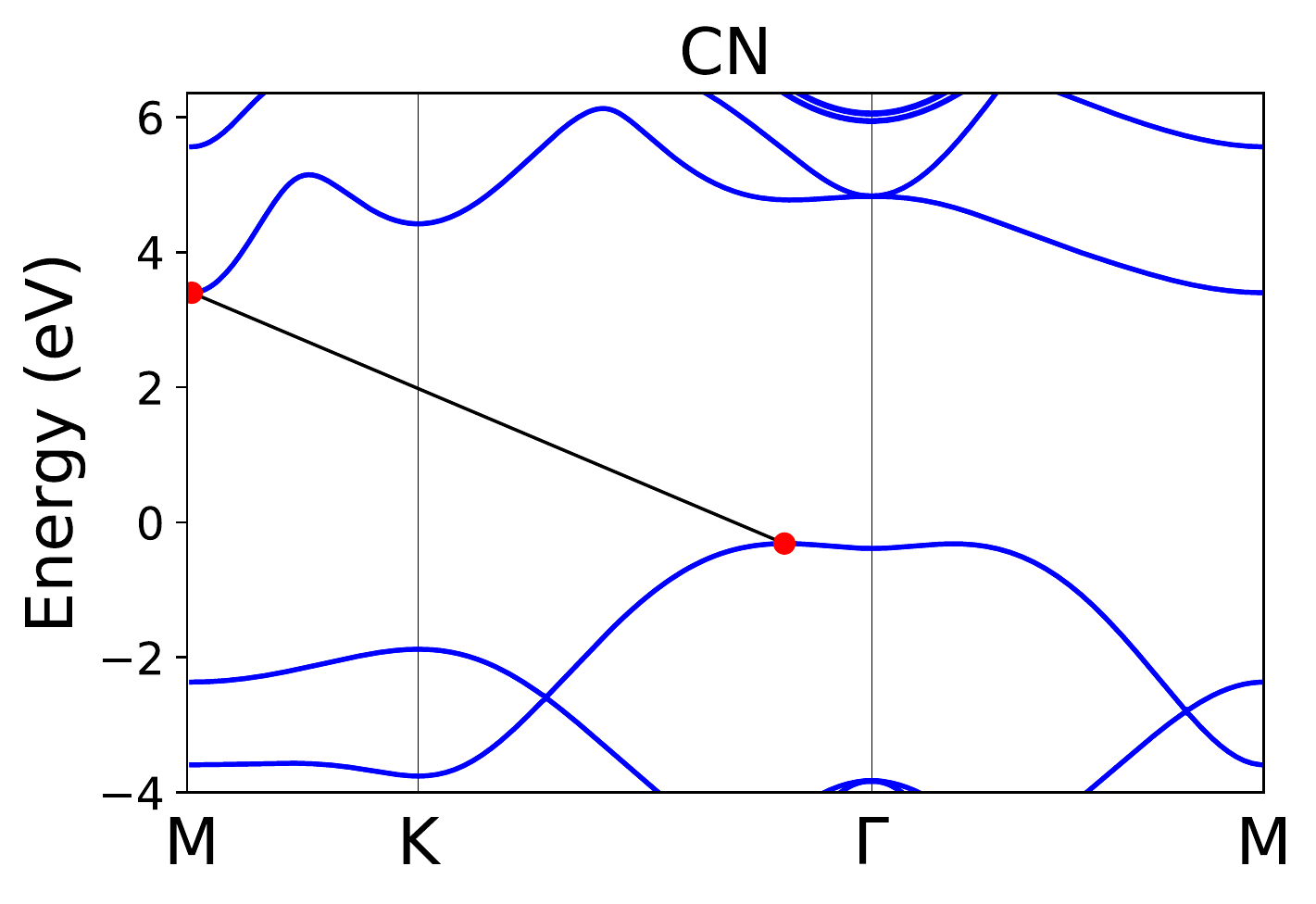}
\includegraphics[width=0.2\textwidth]{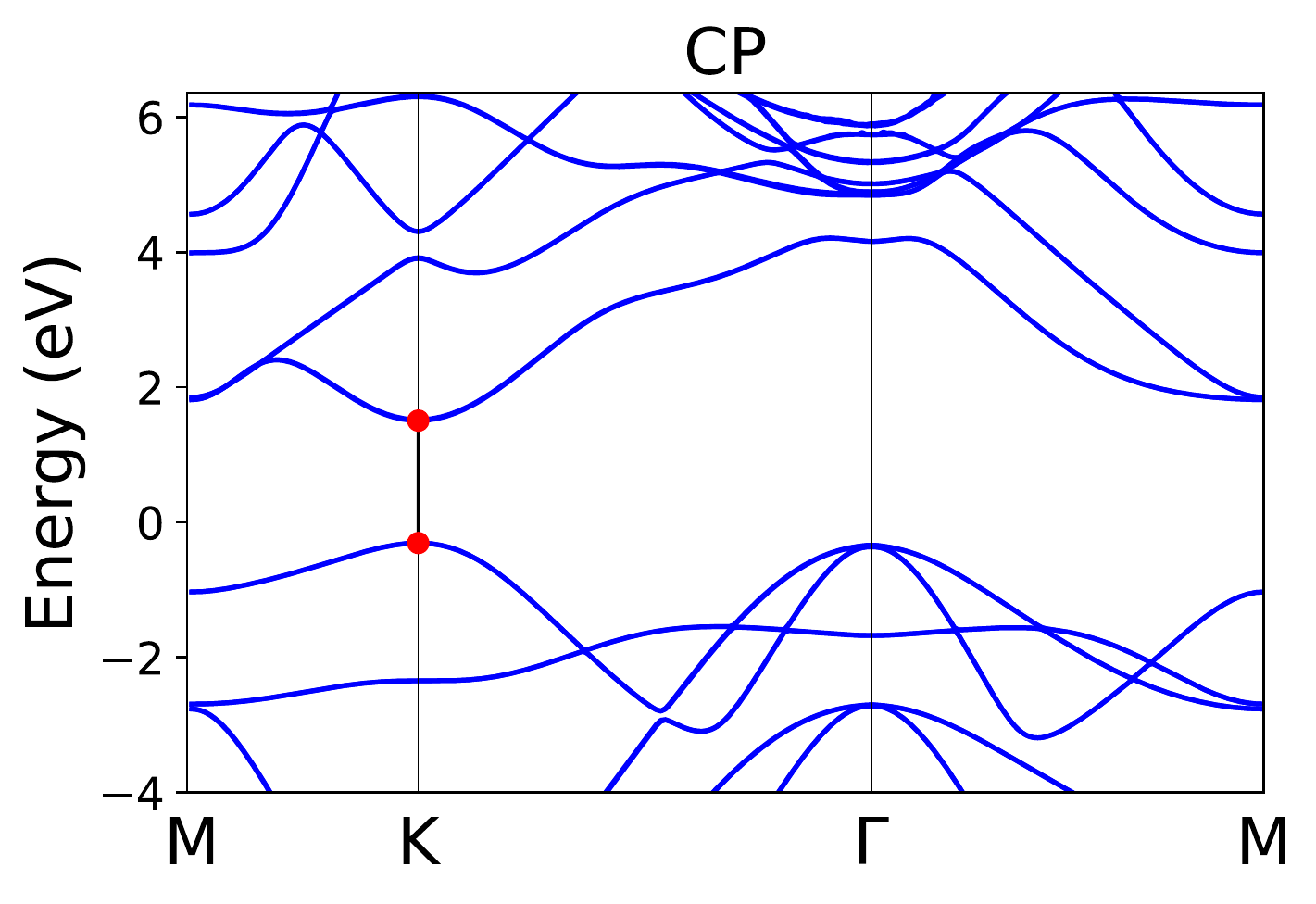}
\includegraphics[width=0.2\textwidth]{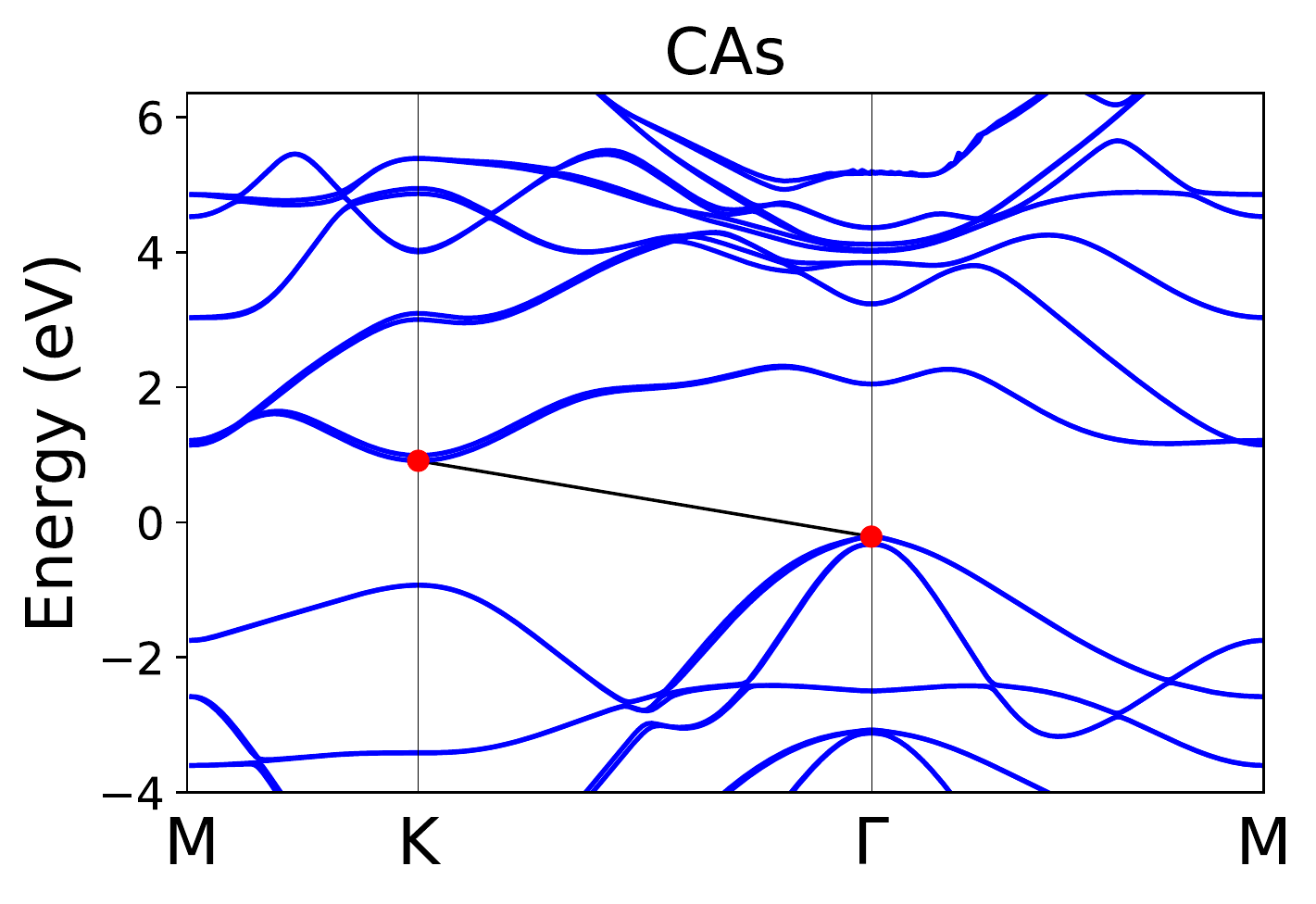}
\includegraphics[width=0.2\textwidth]{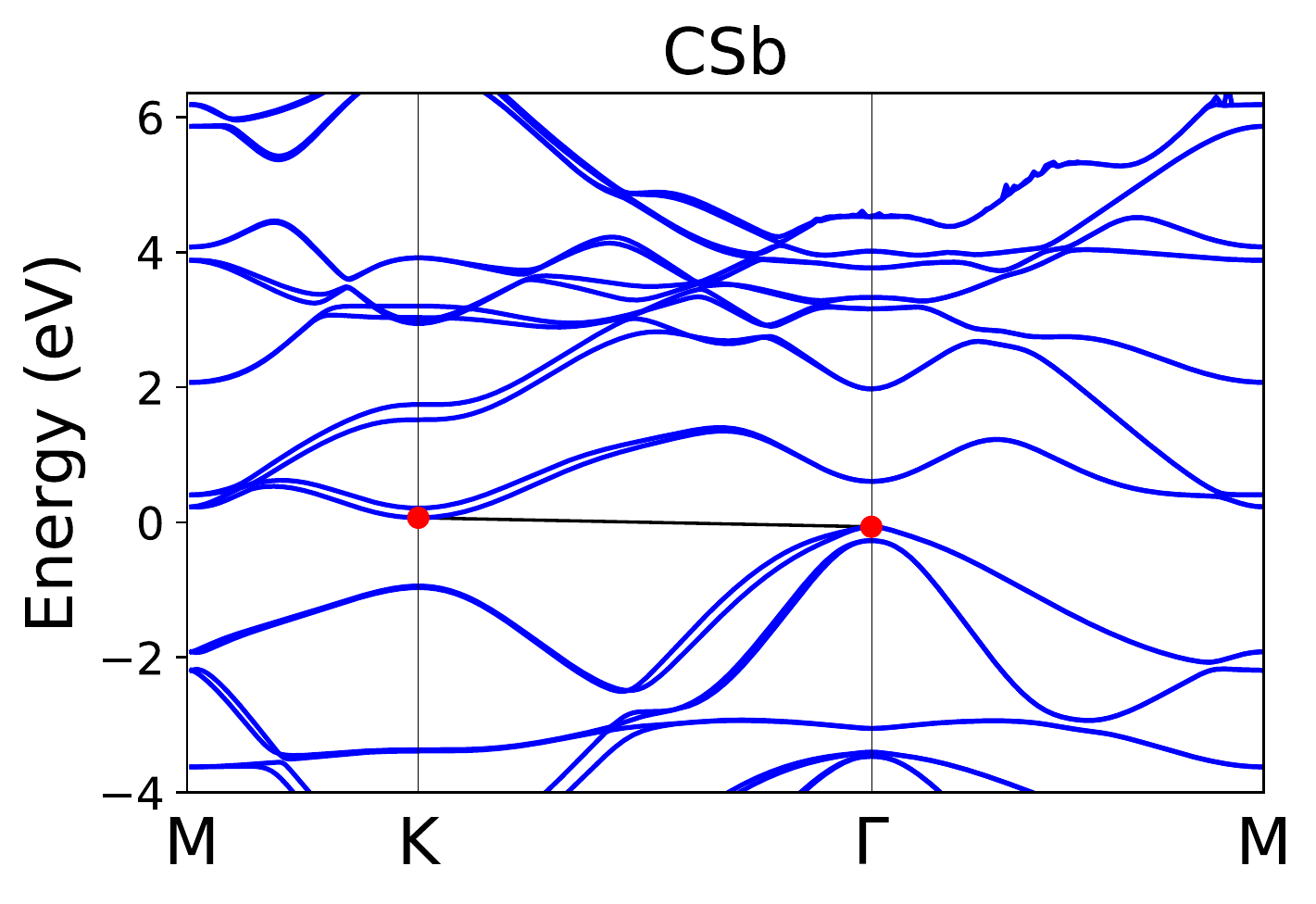}
\includegraphics[width=0.2\textwidth]{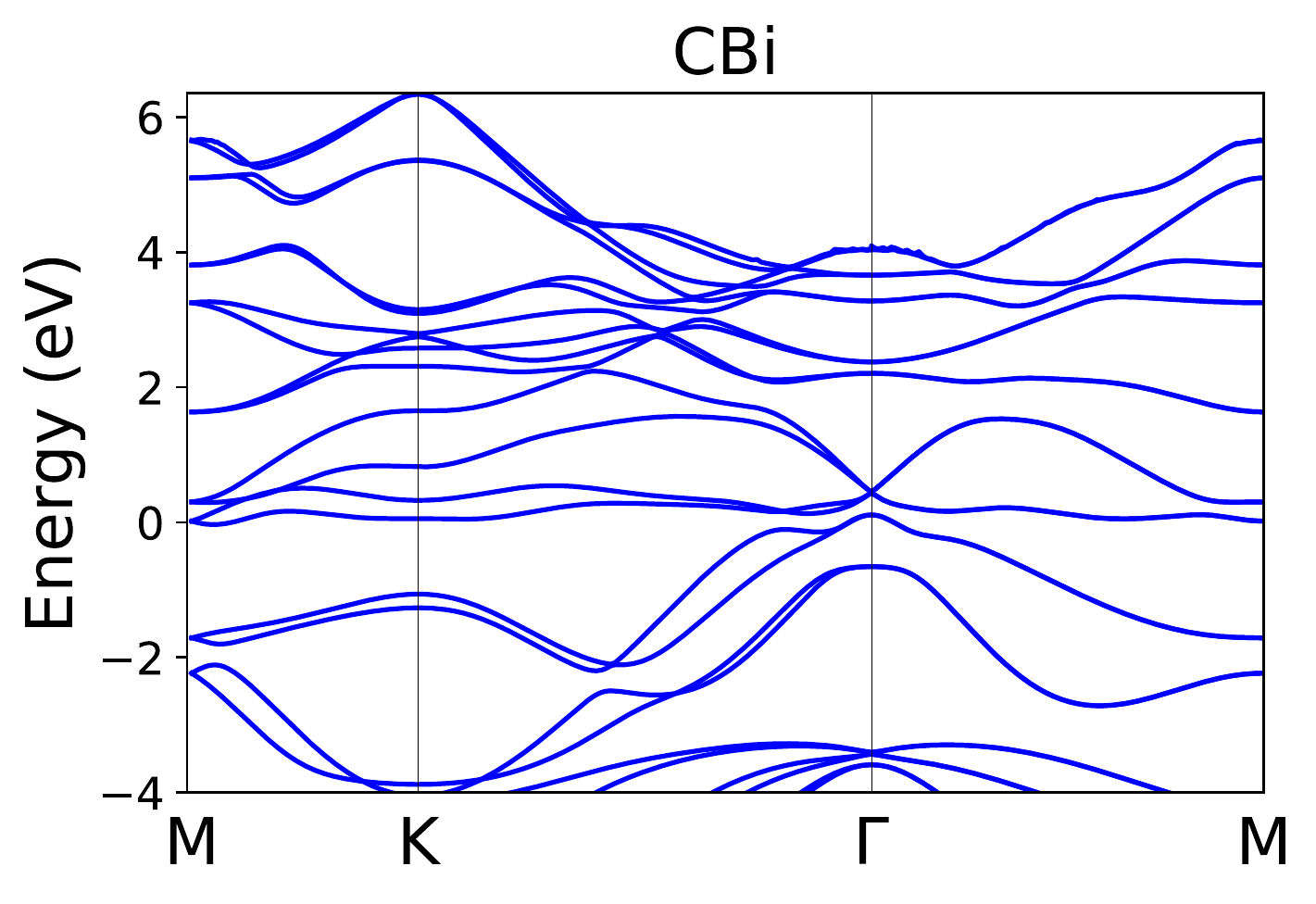}\\

\includegraphics[width=0.2\textwidth]{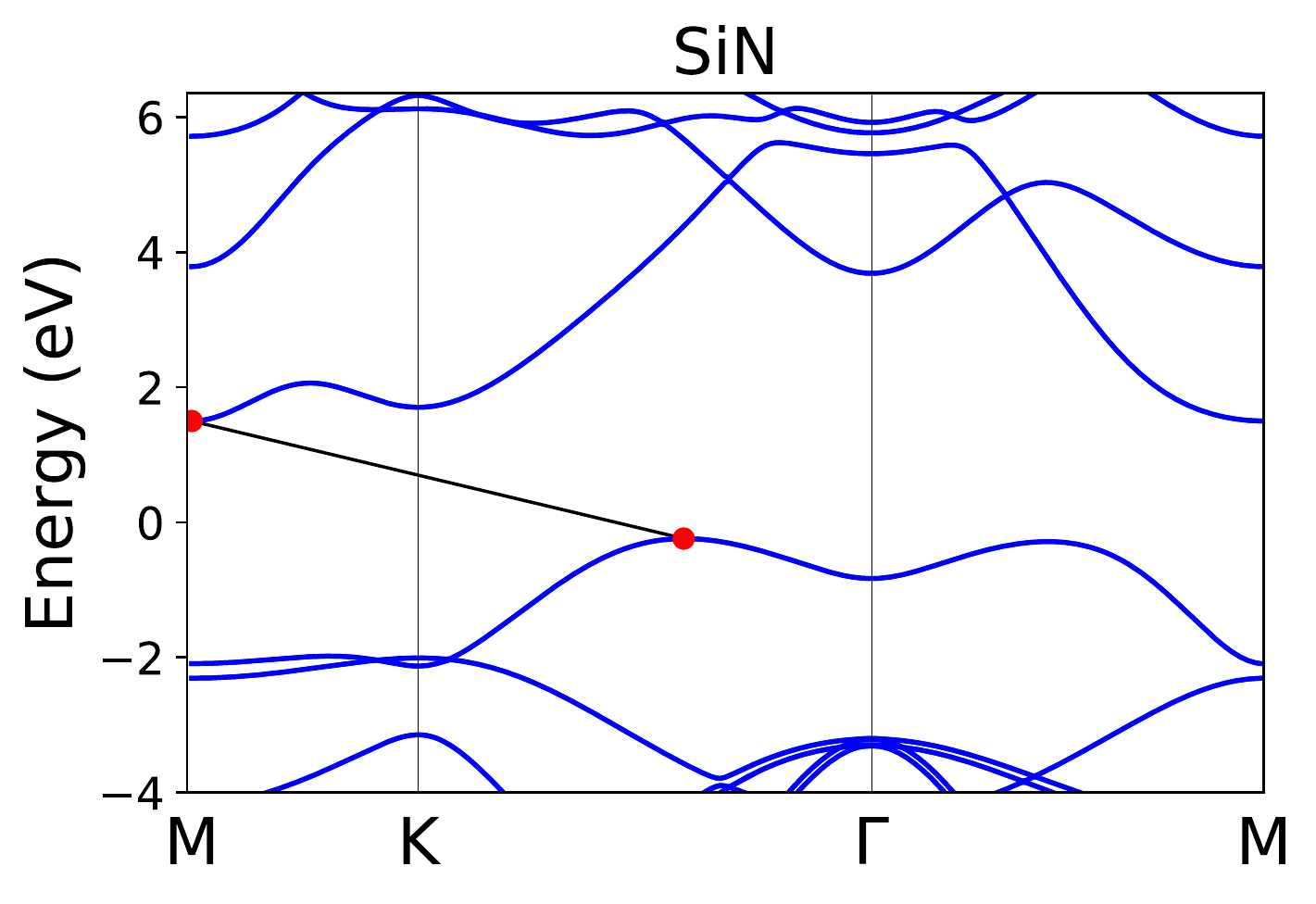}
\includegraphics[width=0.2\textwidth]{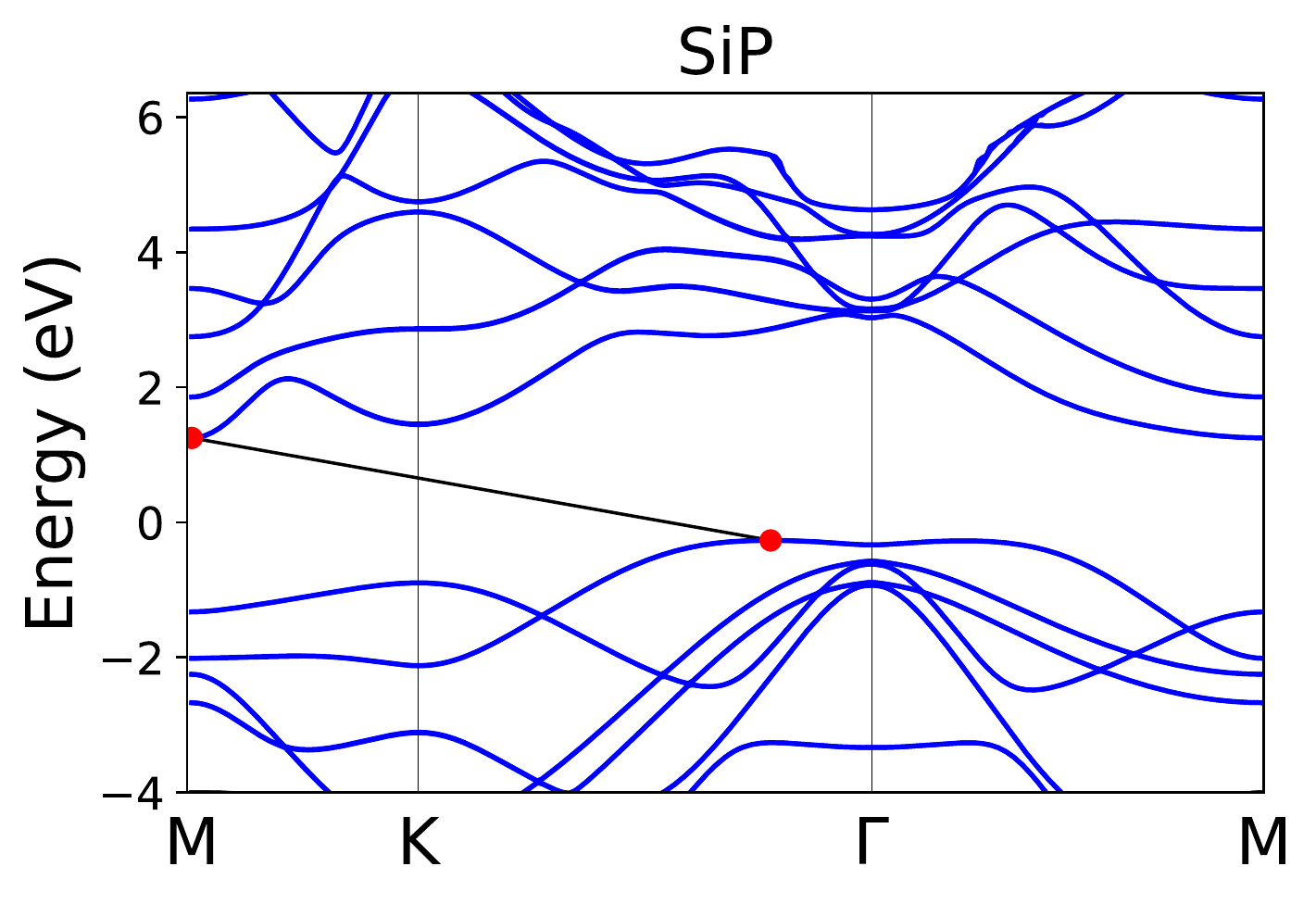}
\includegraphics[width=0.2\textwidth]{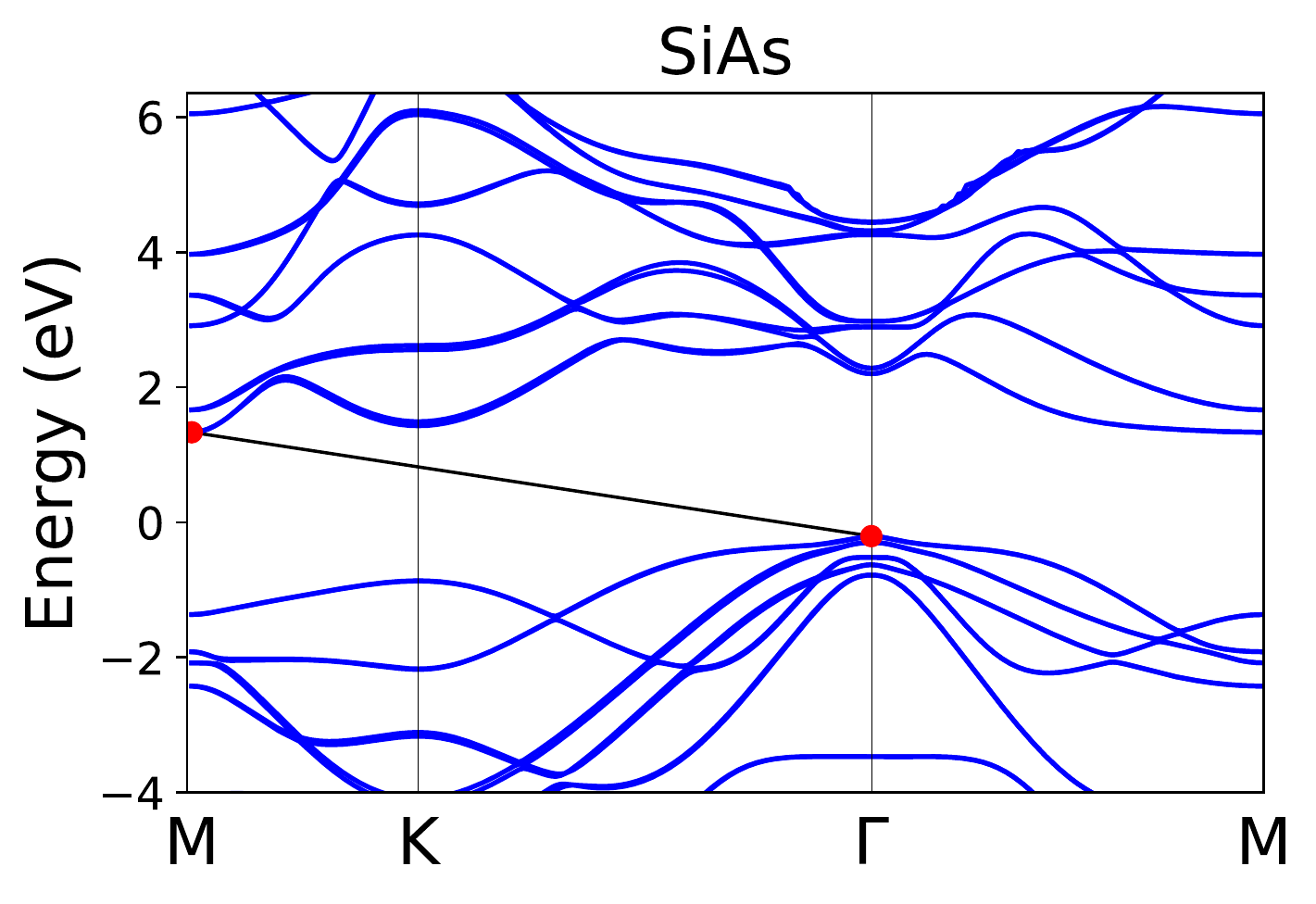}
\includegraphics[width=0.2\textwidth]{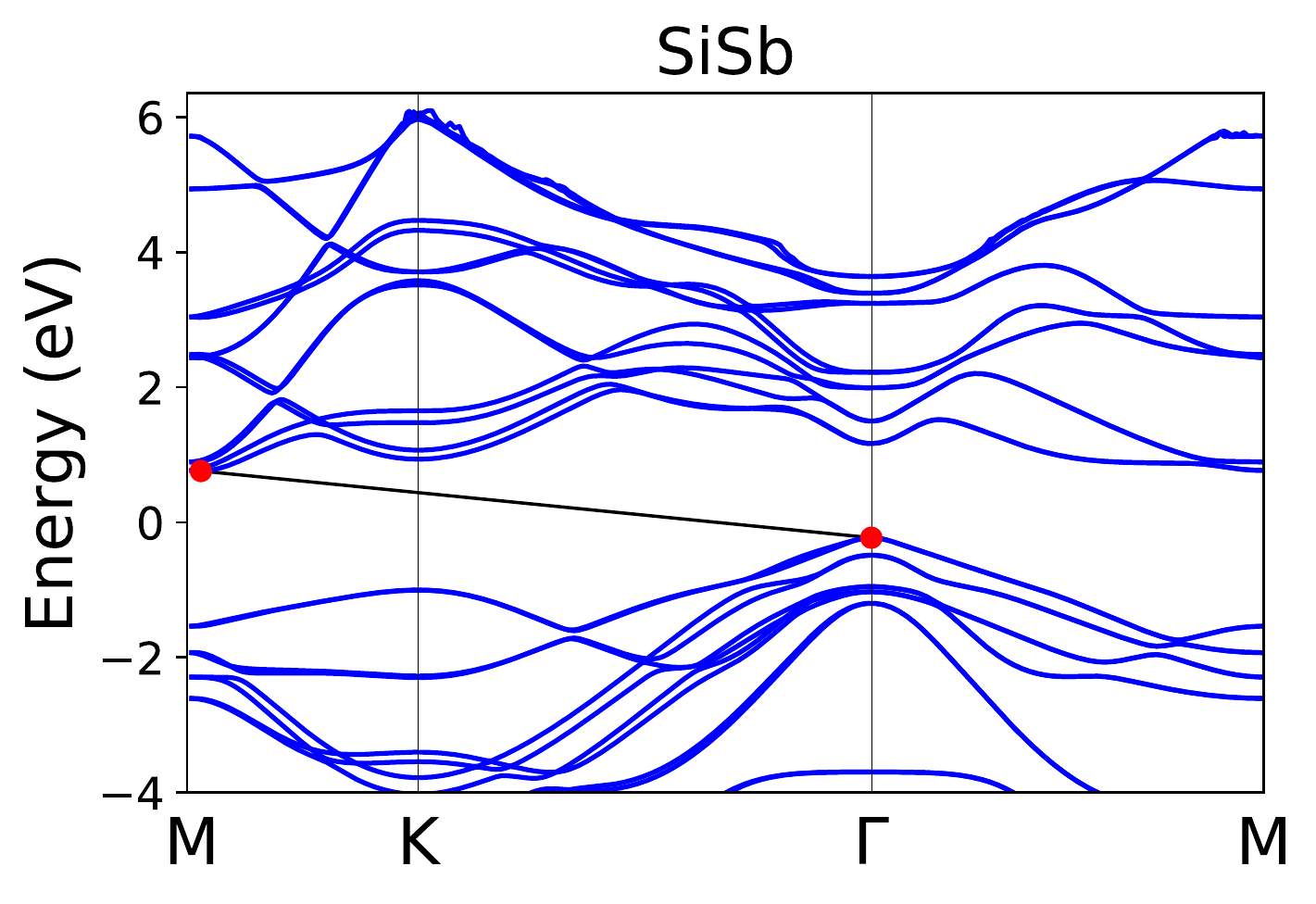}
\includegraphics[width=0.2\textwidth]{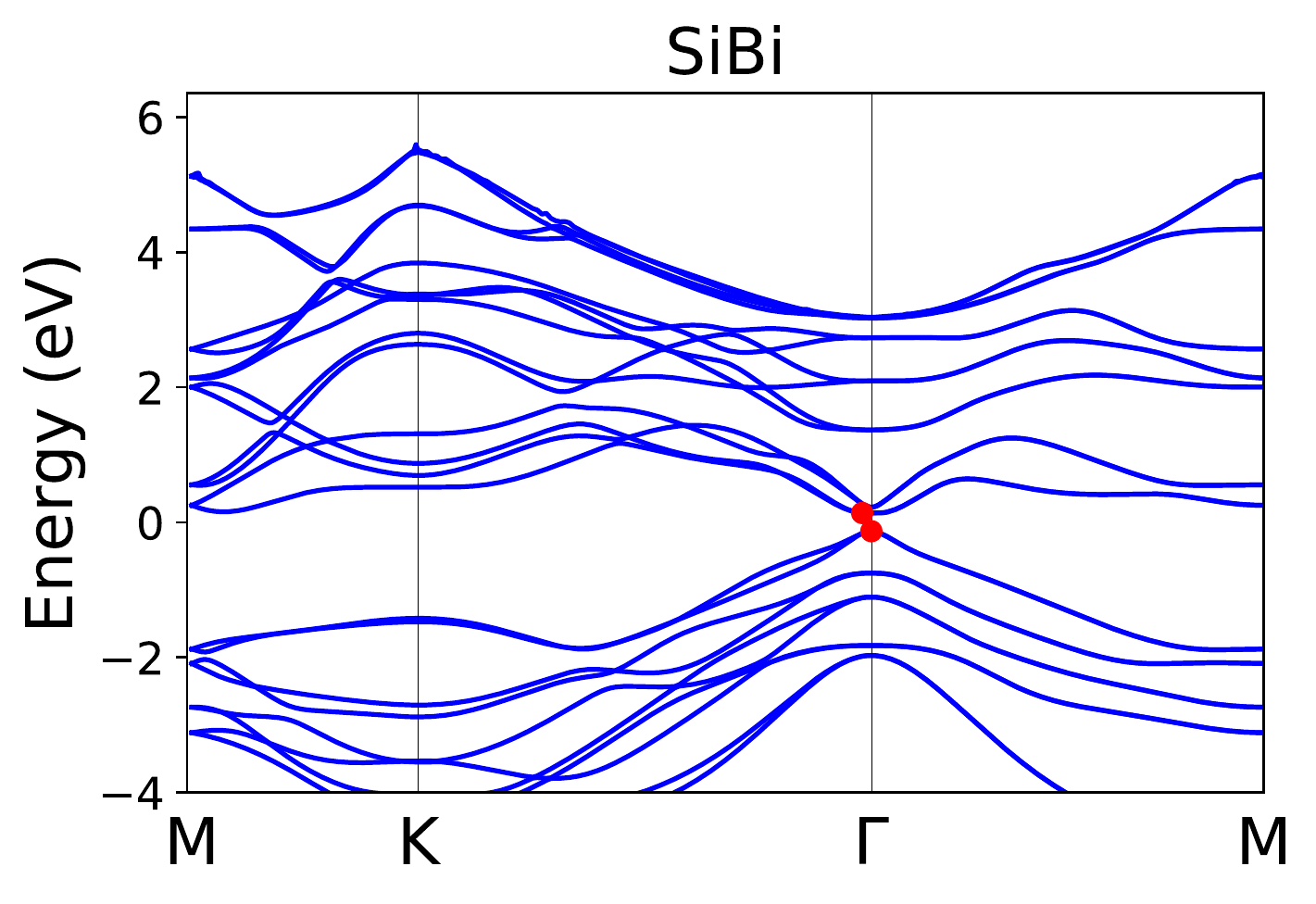}\\

\includegraphics[width=0.2\textwidth]{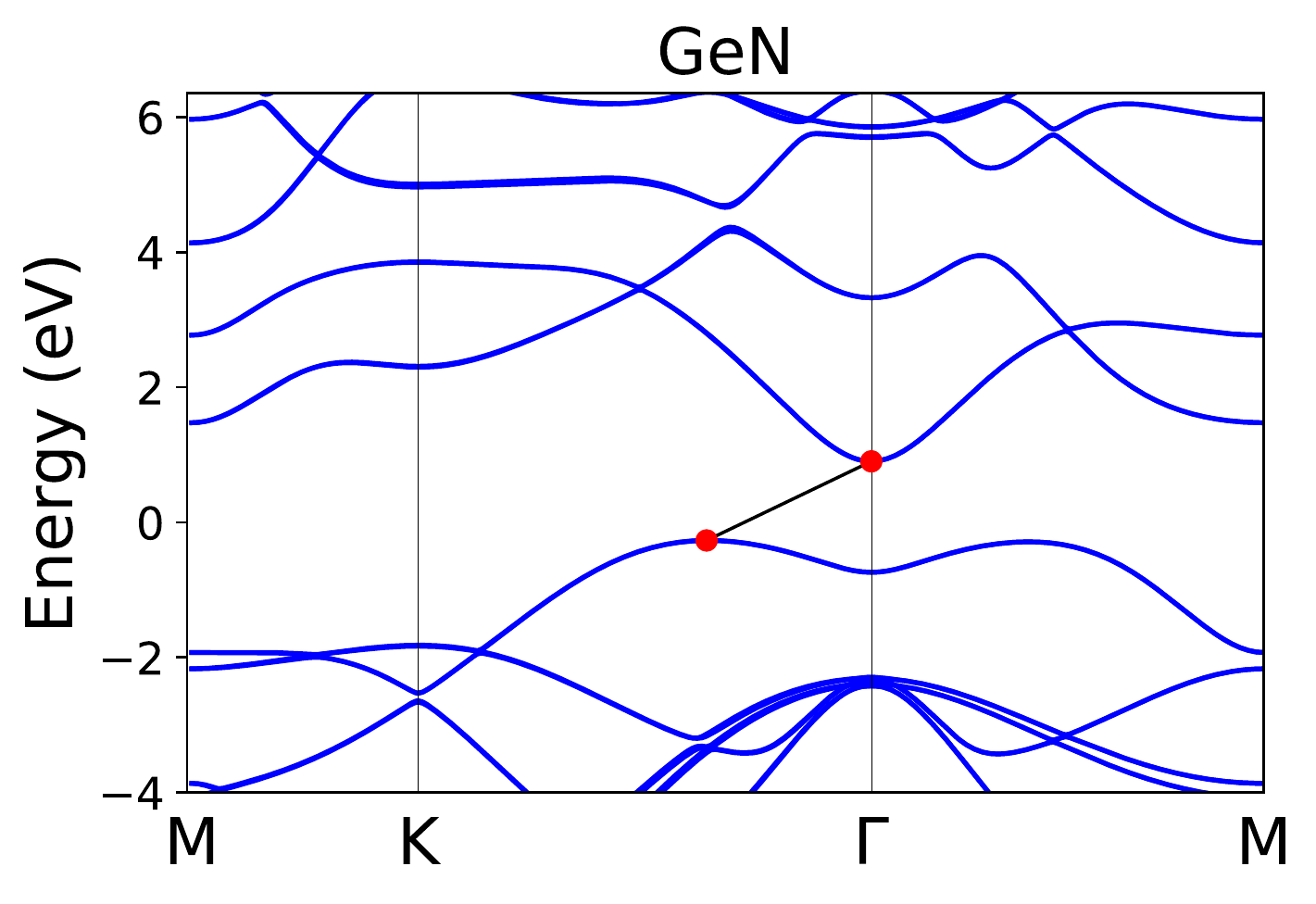}
\includegraphics[width=0.2\textwidth]{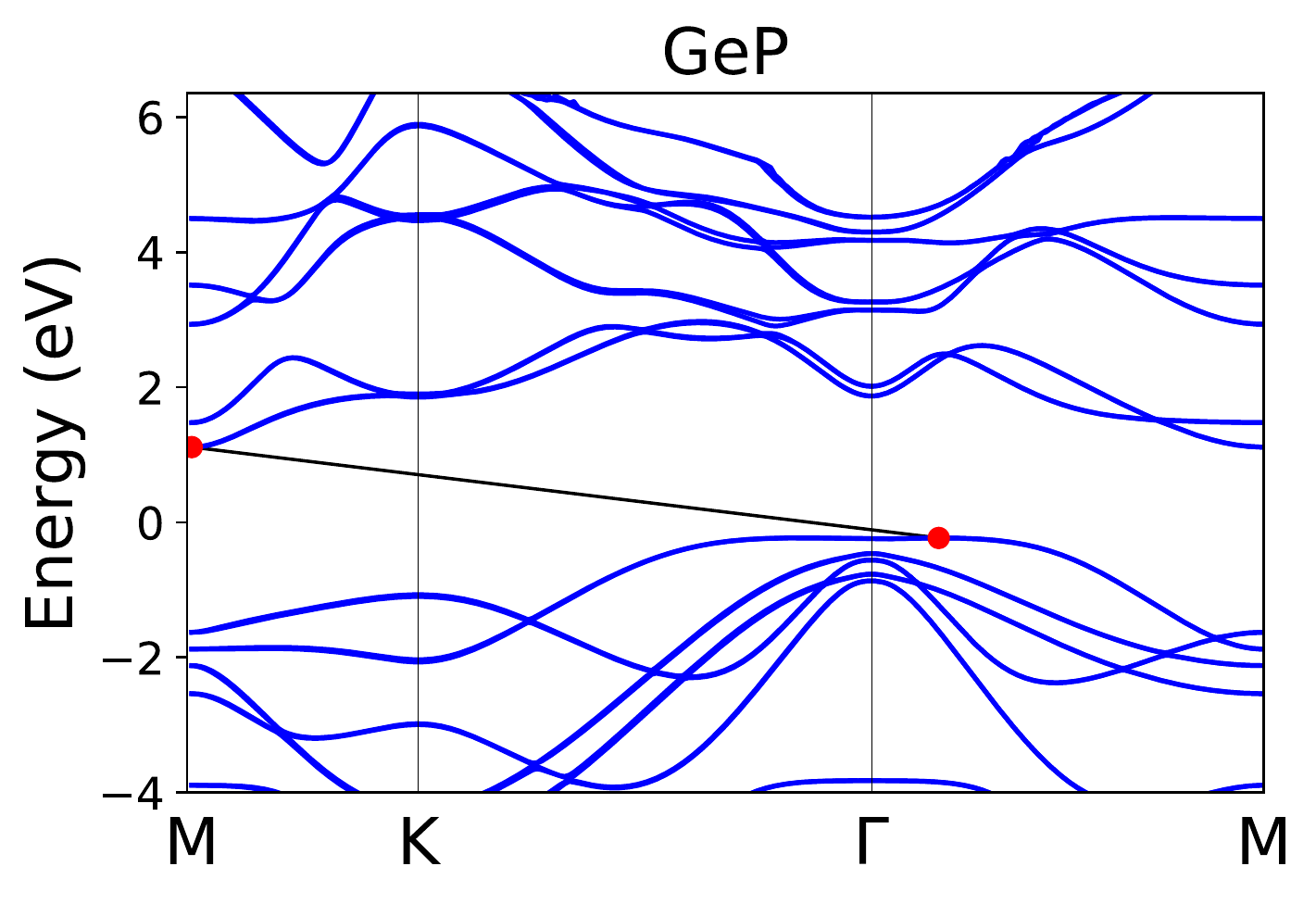}
\includegraphics[width=0.2\textwidth]{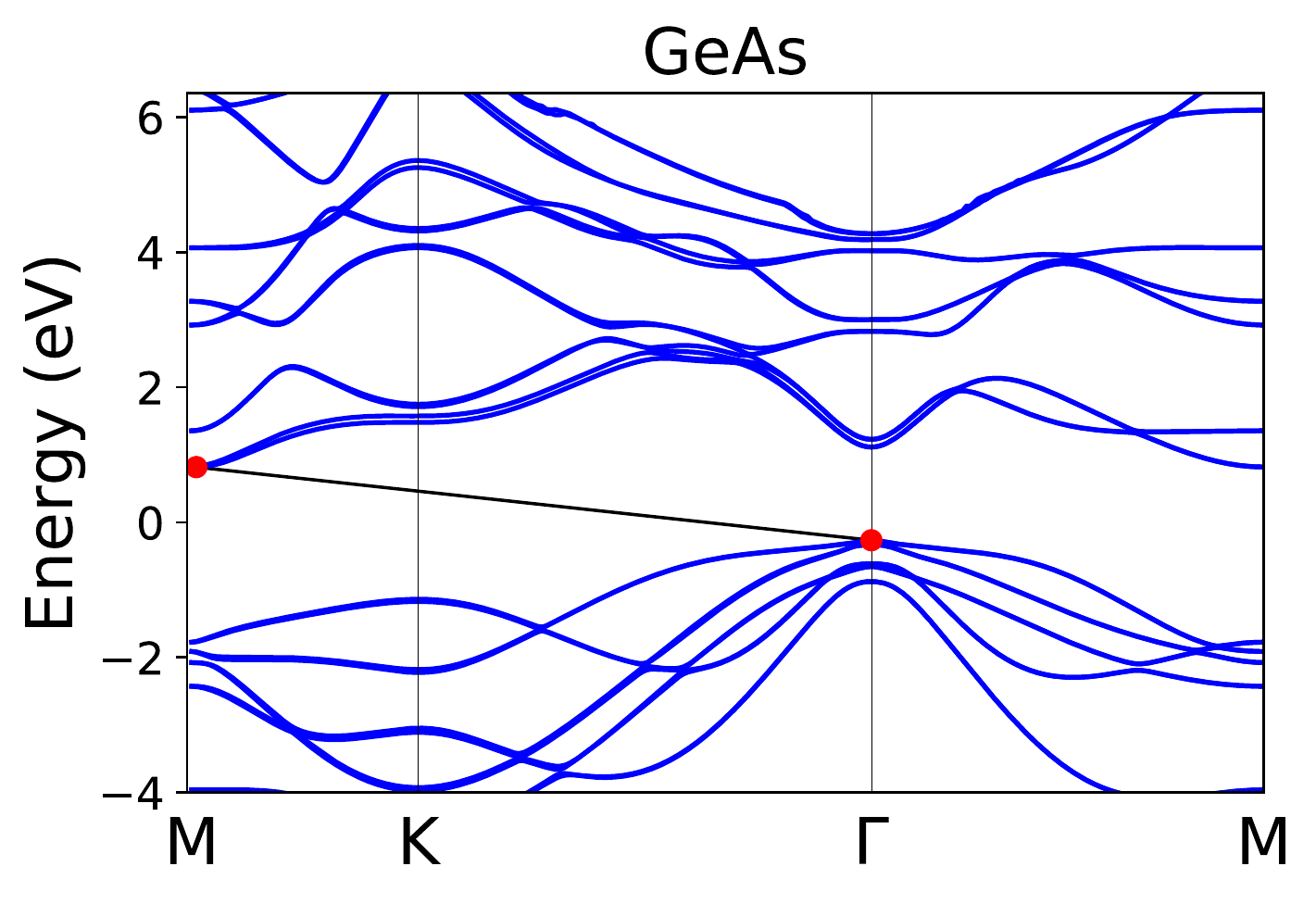}
\includegraphics[width=0.2\textwidth]{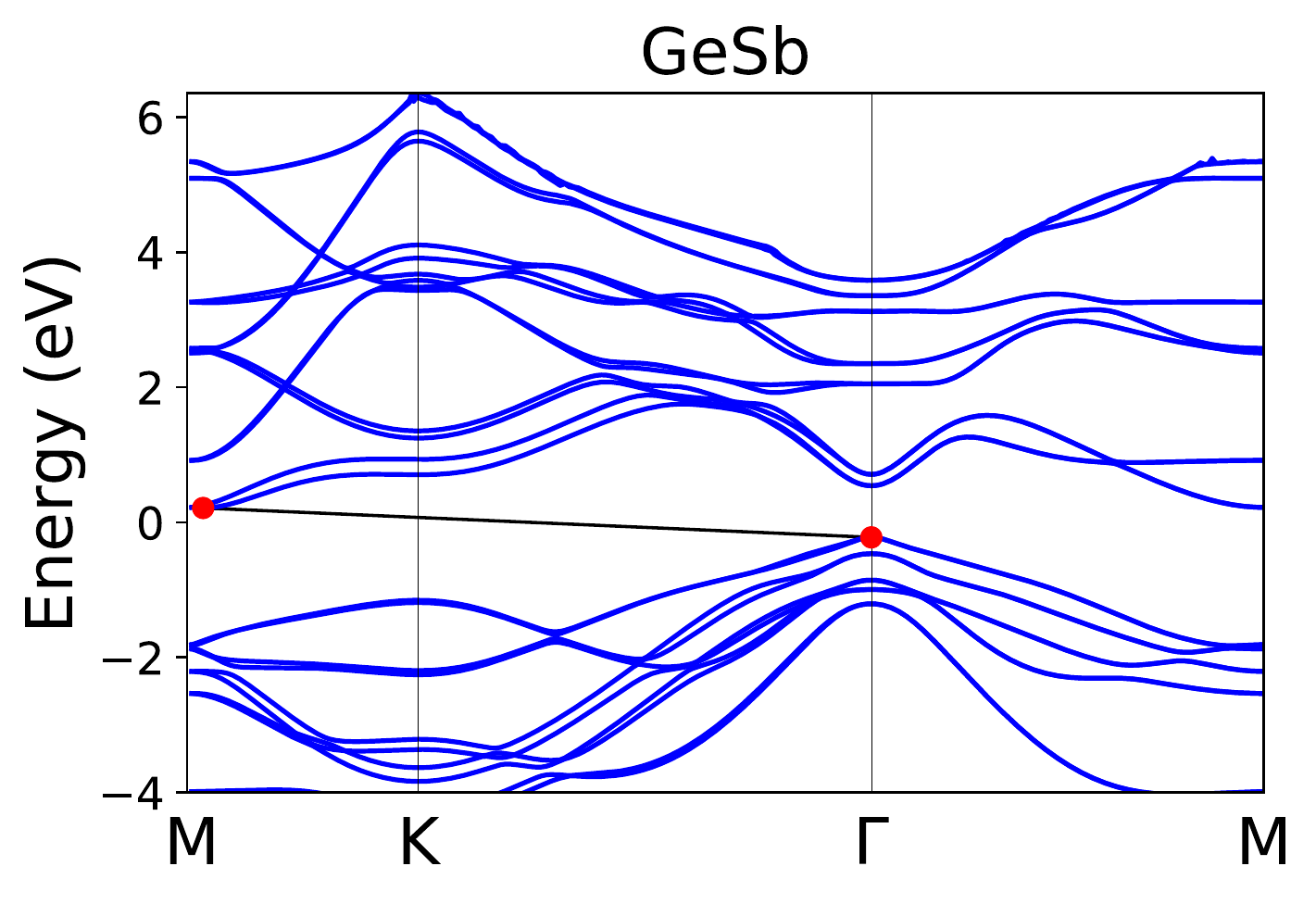}
\includegraphics[width=0.2\textwidth]{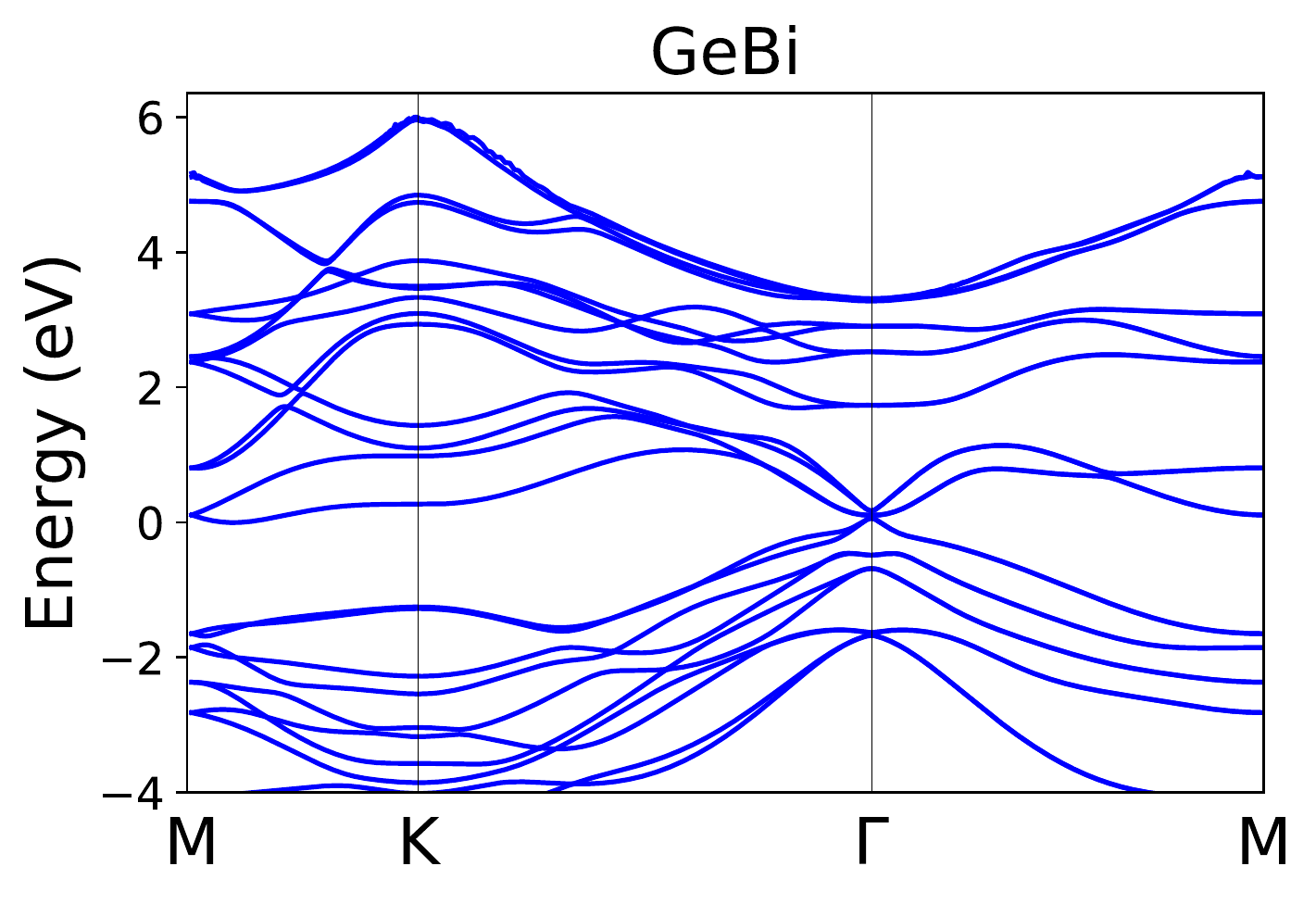}\\

\includegraphics[width=0.2\textwidth]{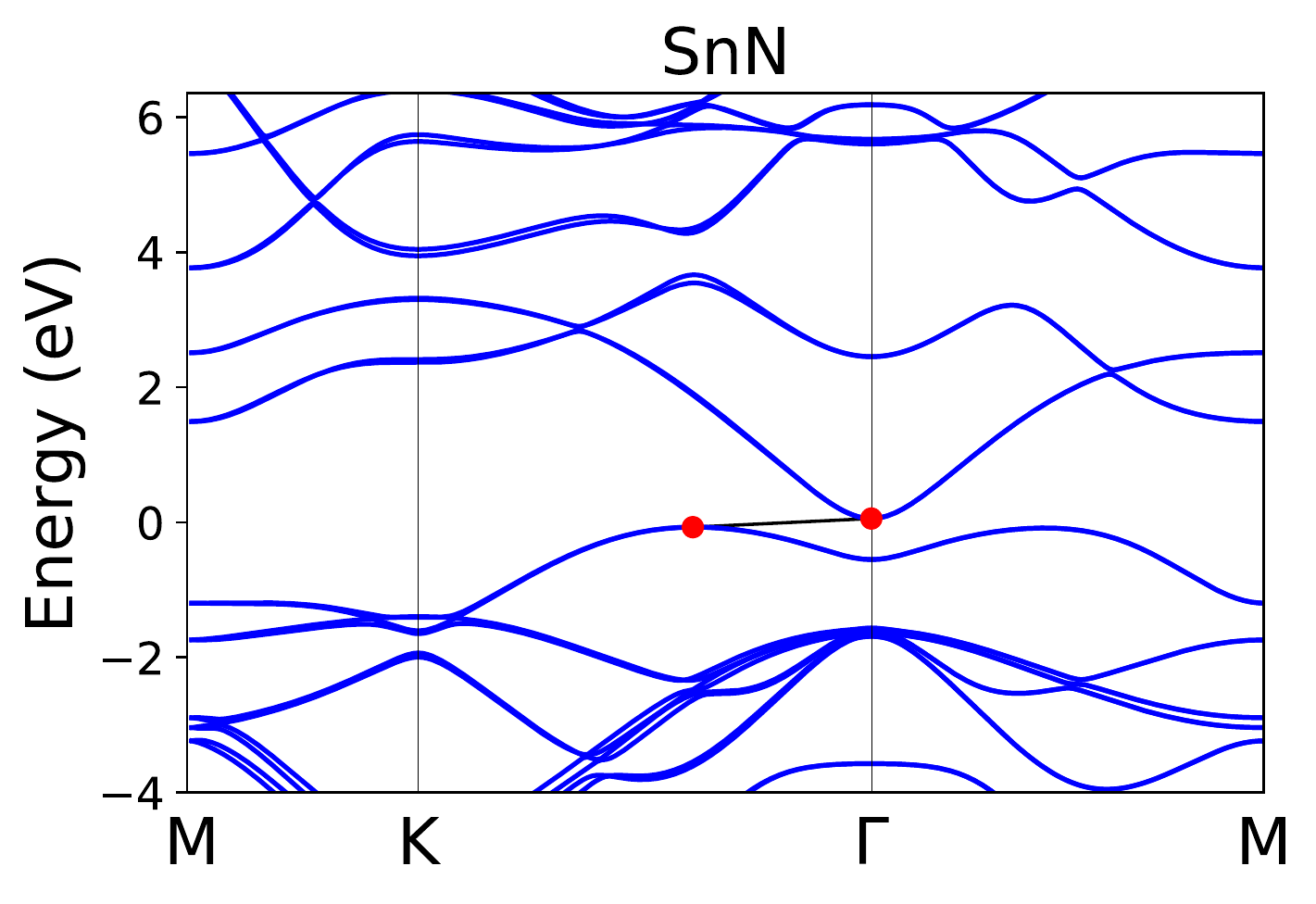}
\includegraphics[width=0.2\textwidth]{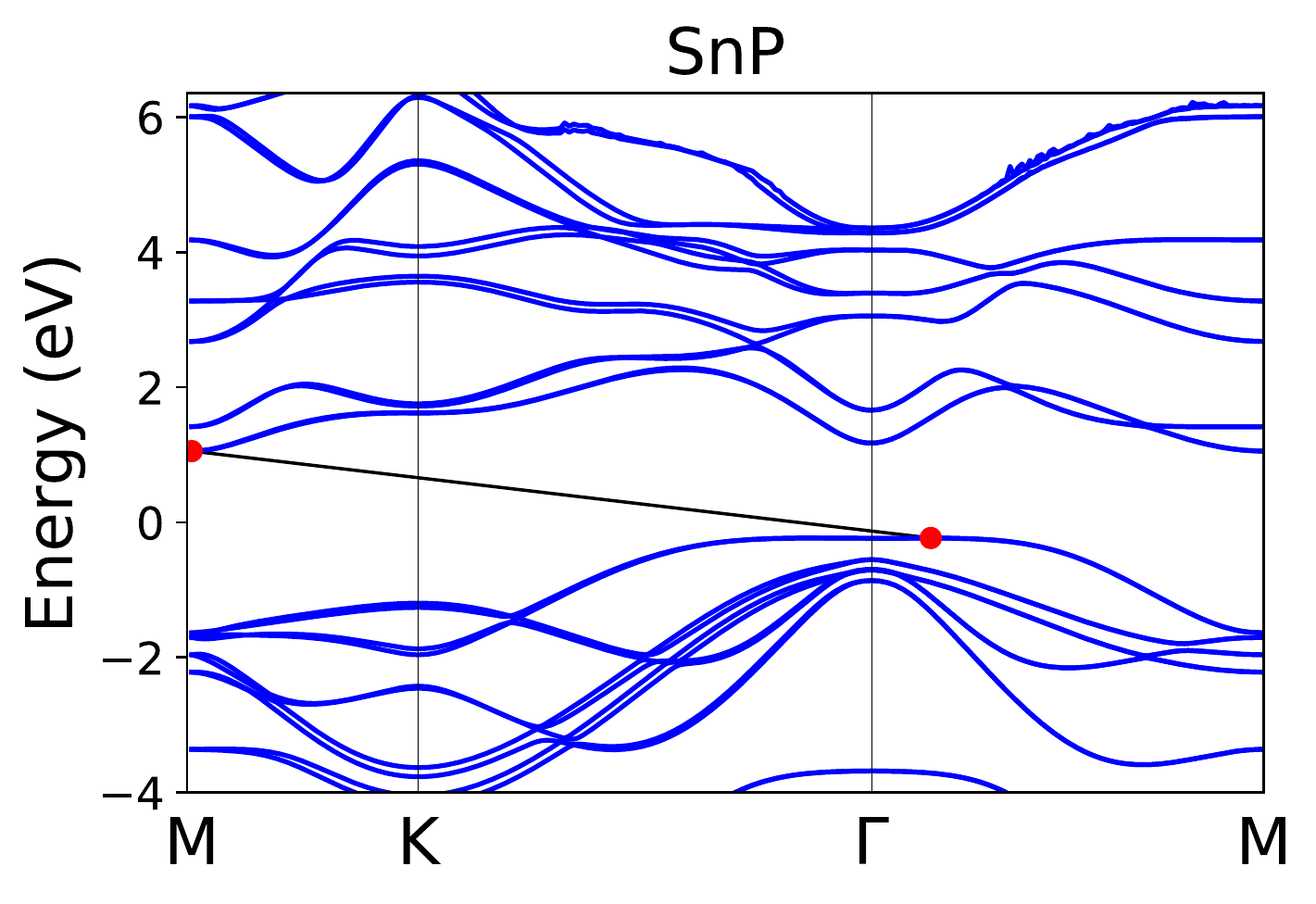}
\includegraphics[width=0.2\textwidth]{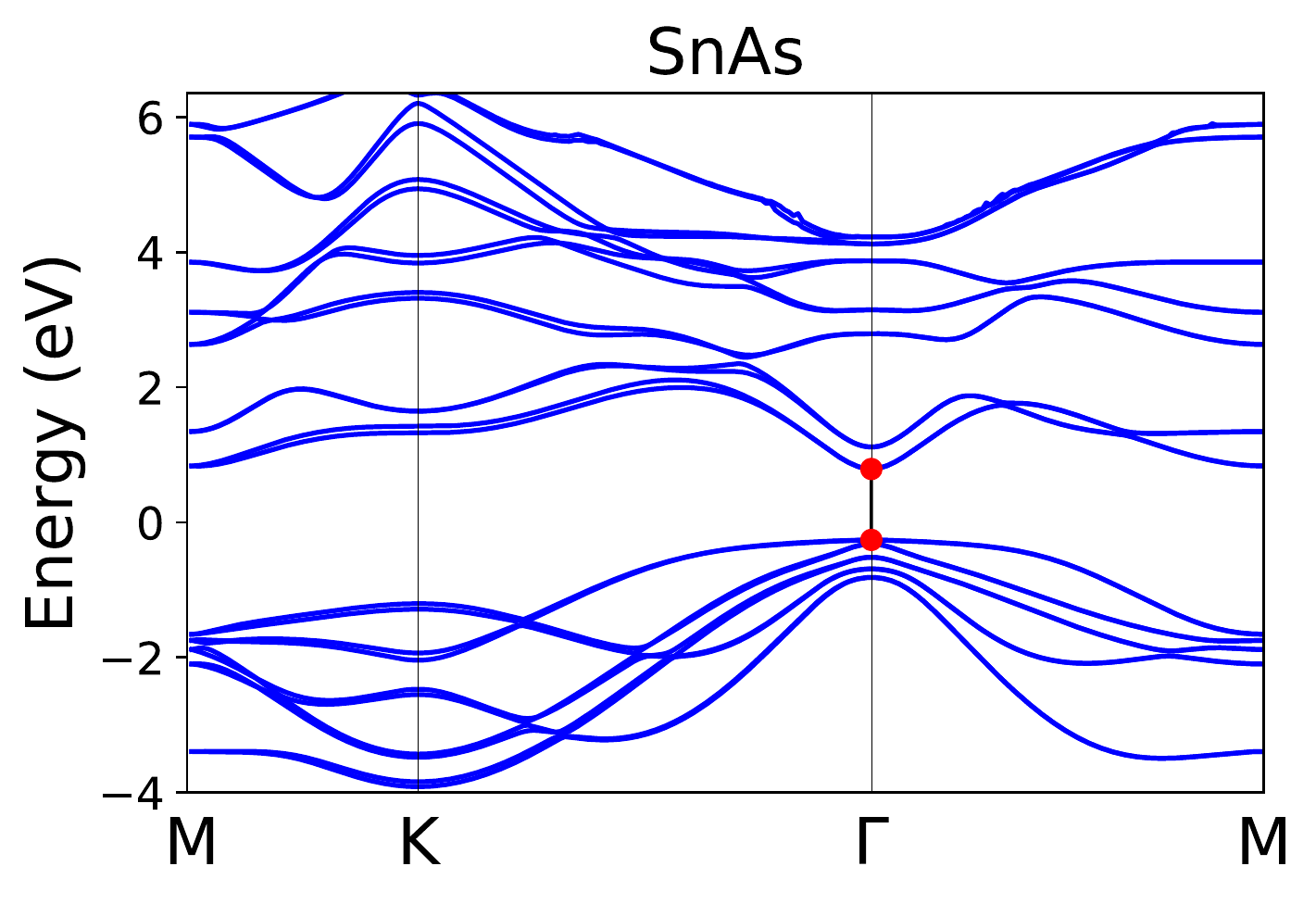}
\includegraphics[width=0.2\textwidth]{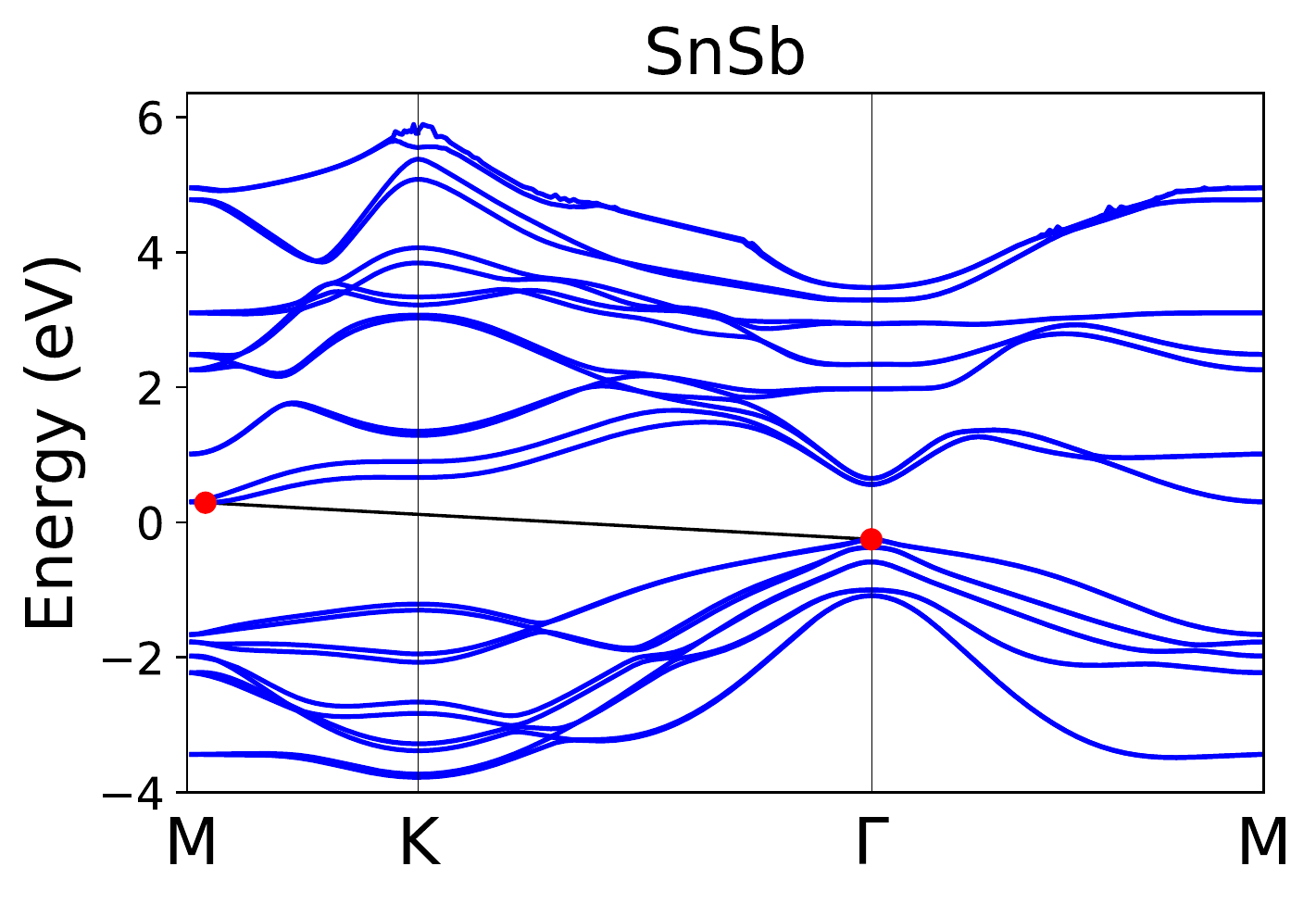}
\includegraphics[width=0.2\textwidth]{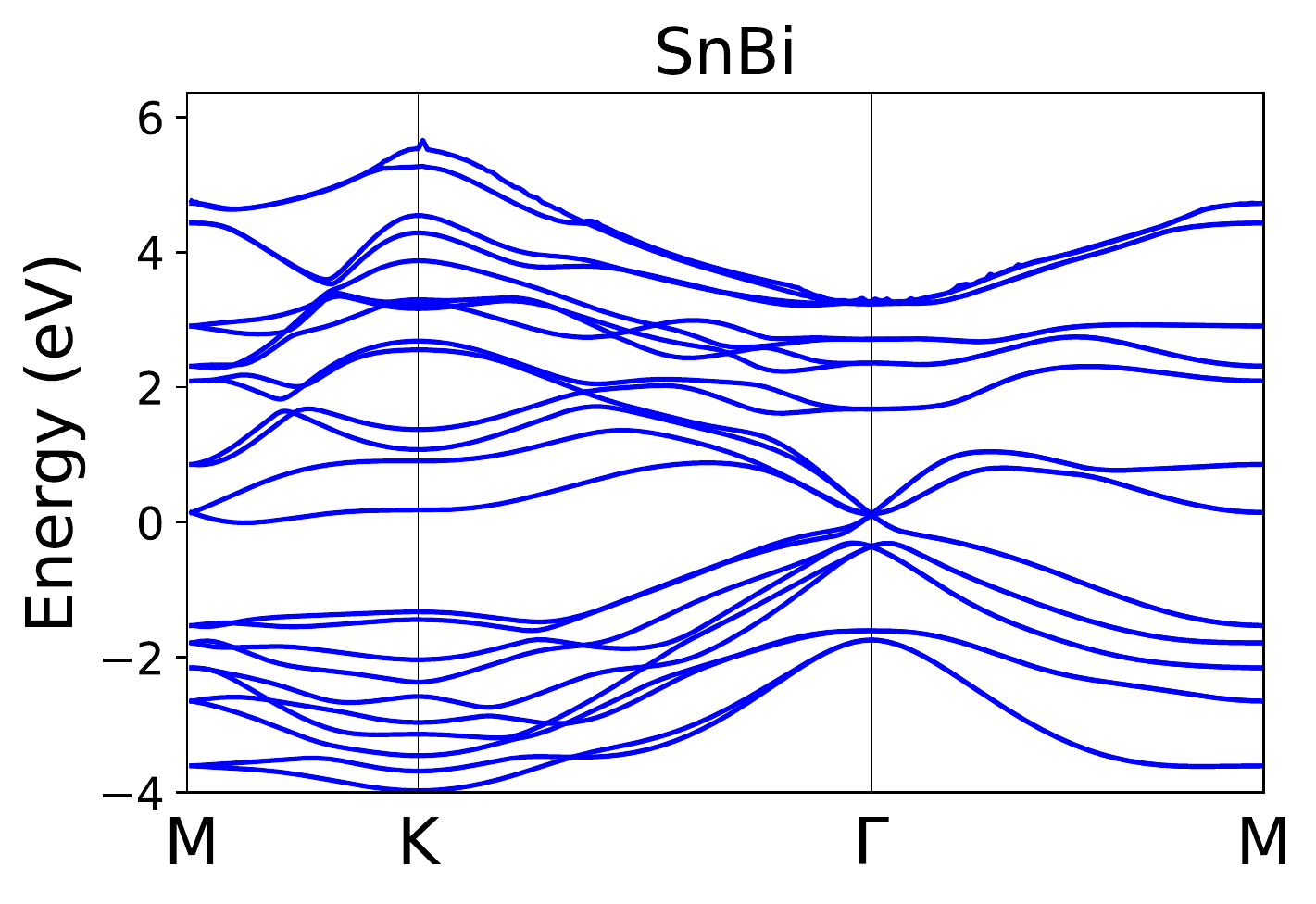}\\

\includegraphics[width=0.2\textwidth]{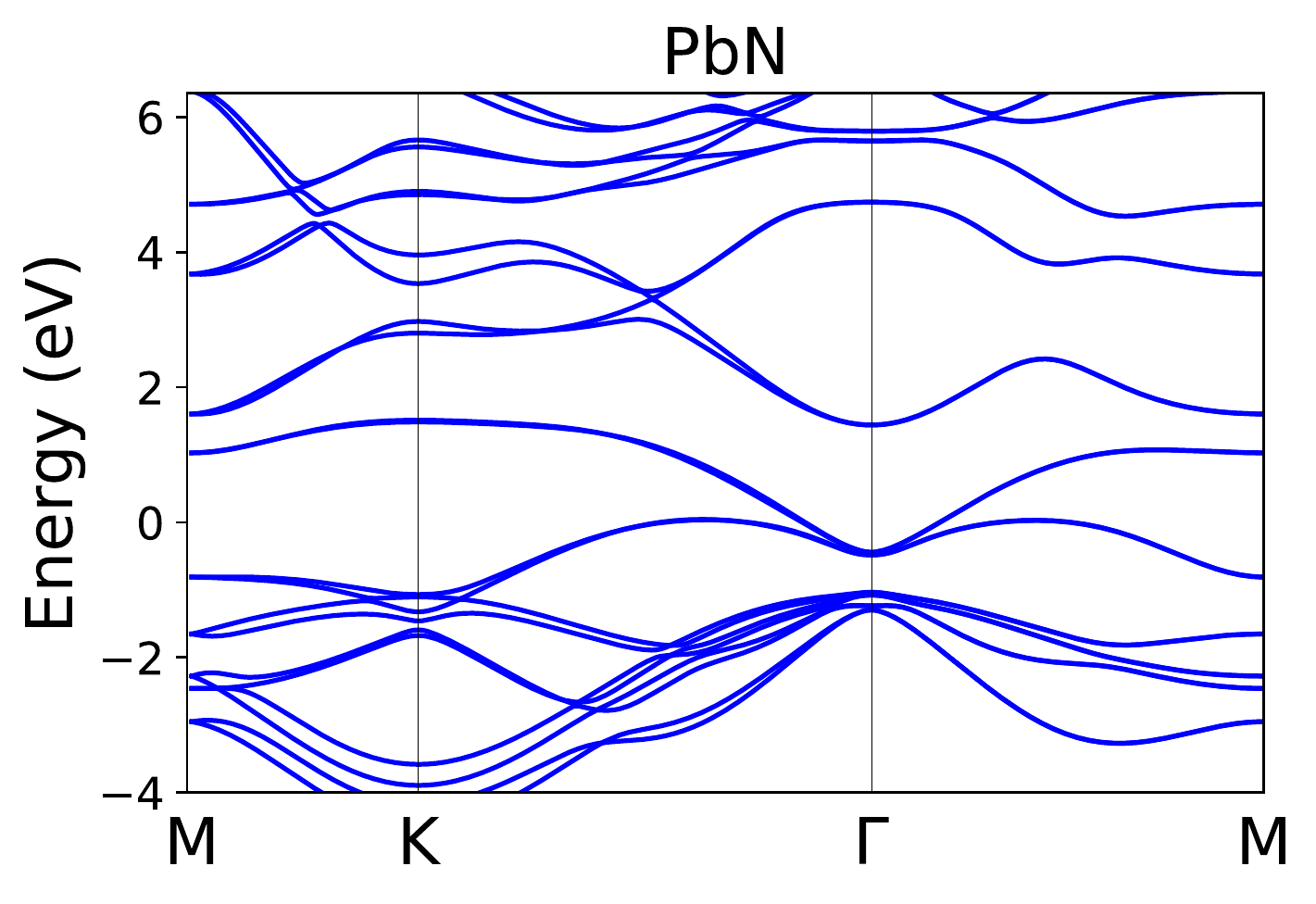}
\includegraphics[width=0.2\textwidth]{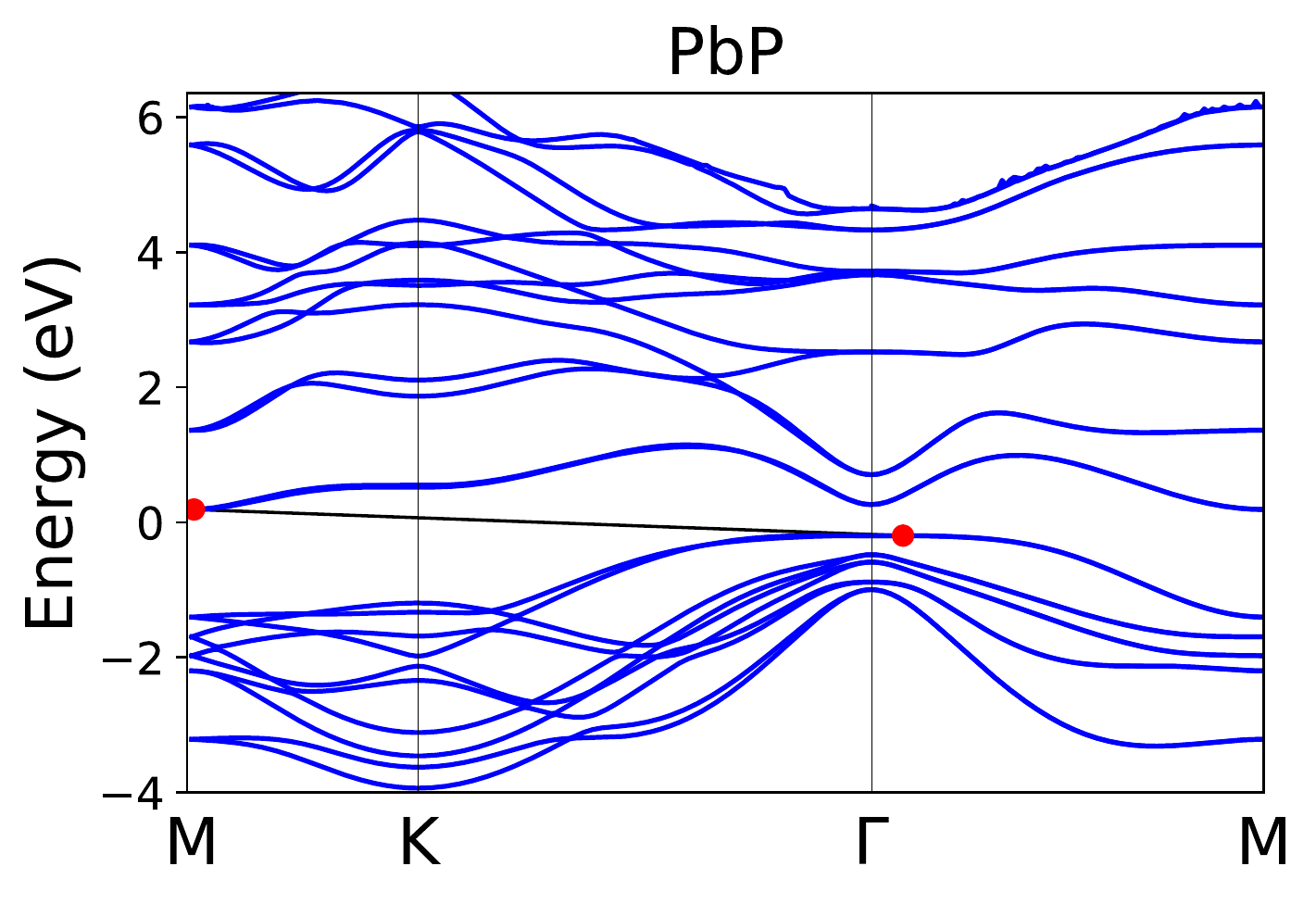}
\includegraphics[width=0.2\textwidth]{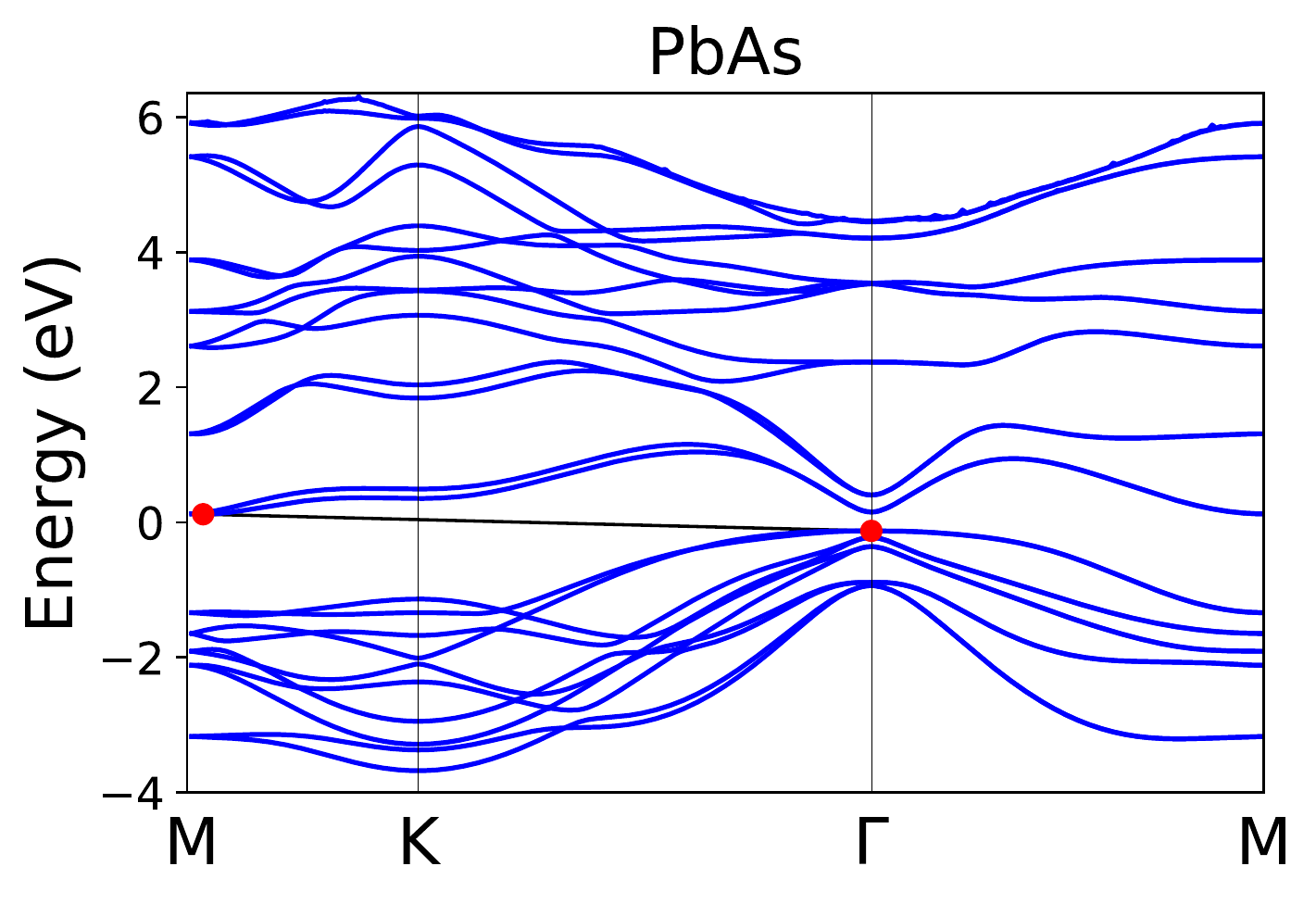}
\includegraphics[width=0.2\textwidth]{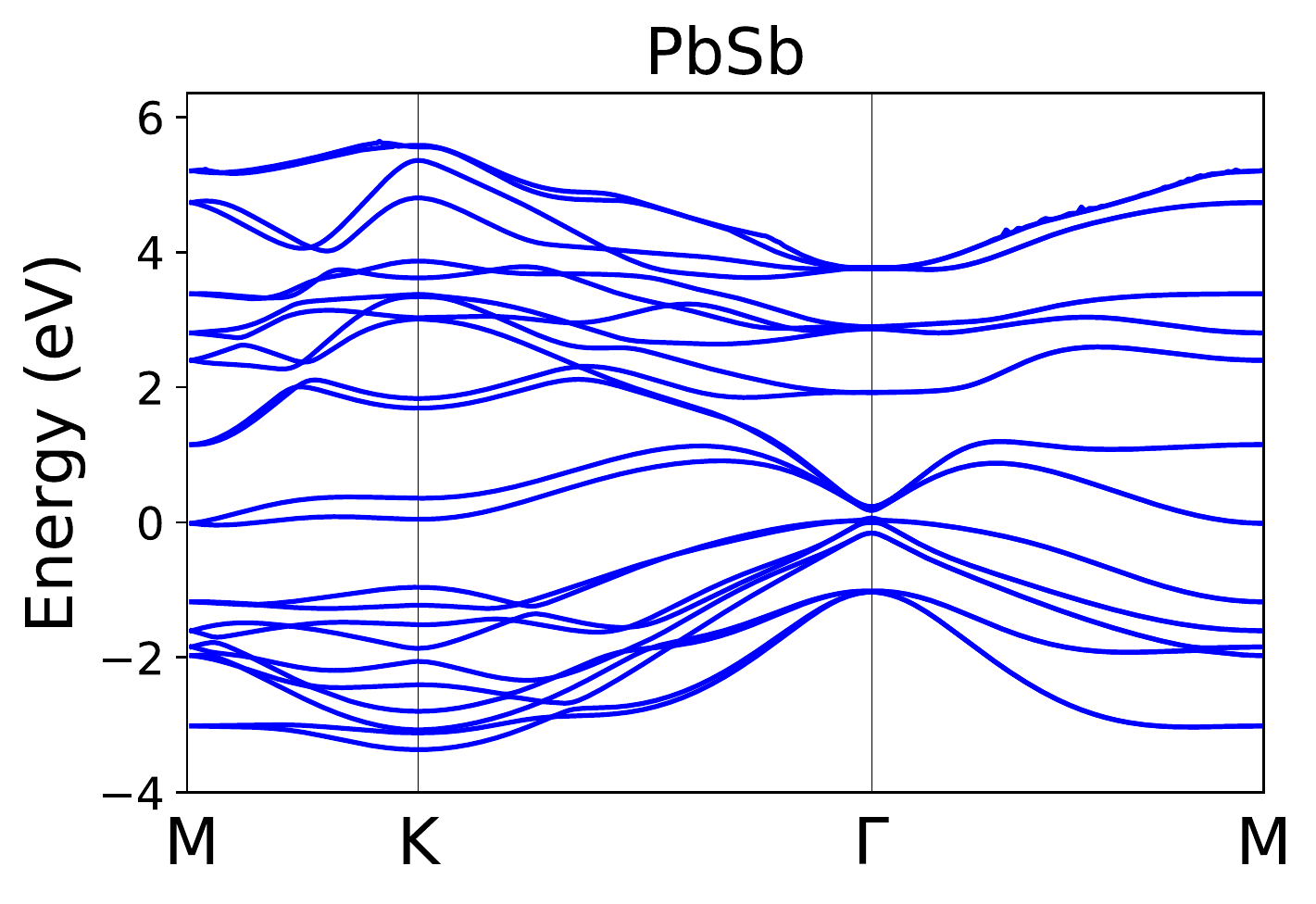}
\includegraphics[width=0.2\textwidth]{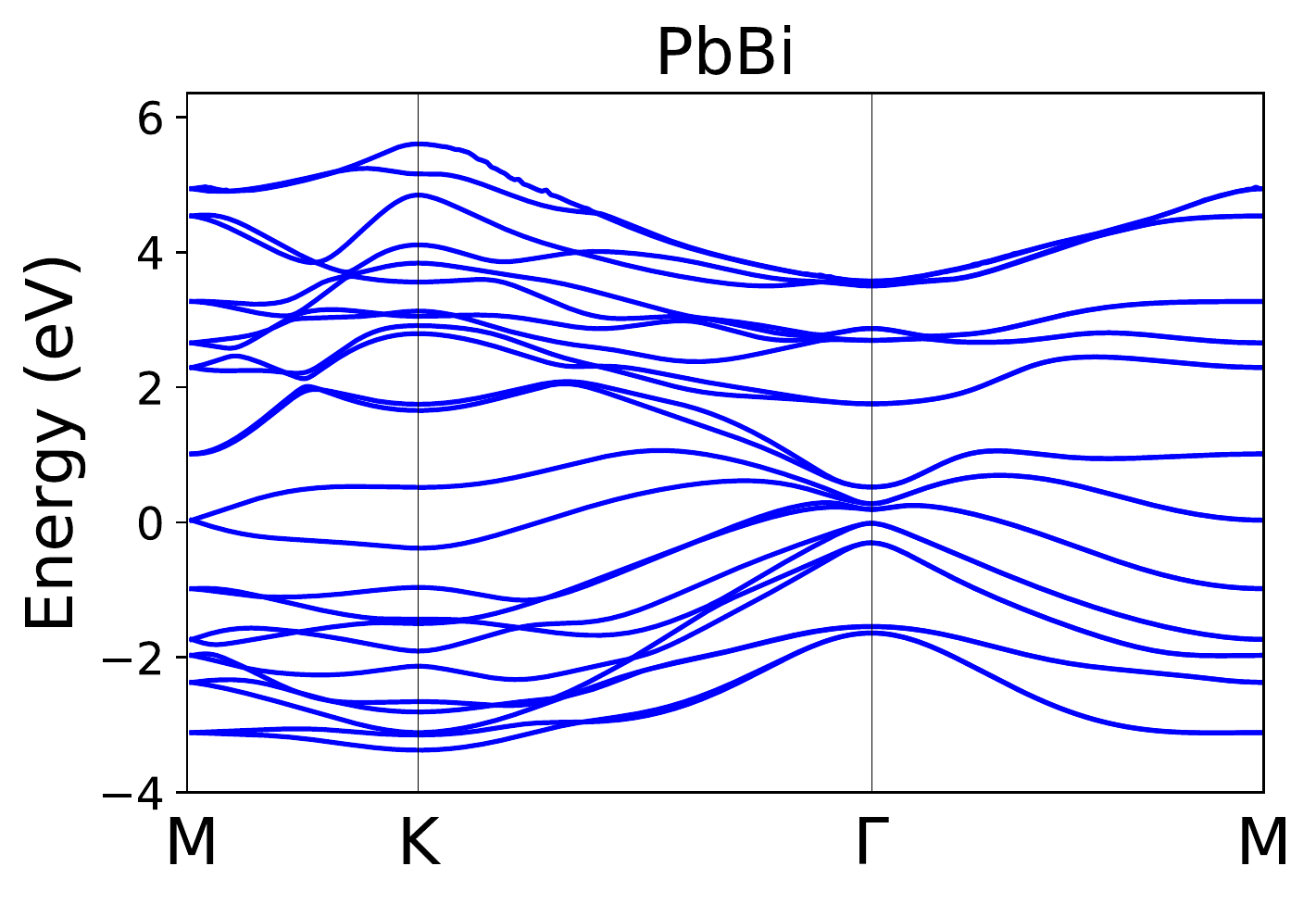}
\end{tabular}

\caption{Electronic band diagrams of $\alpha$-structures as obtained from PBE including SOC.}
    \label{fig:alpha-soc}
\end{figure}

\begin{figure}[t]
   \centering
\begin{tabular}{ccccc}
\\
\includegraphics[width=0.2\textwidth]{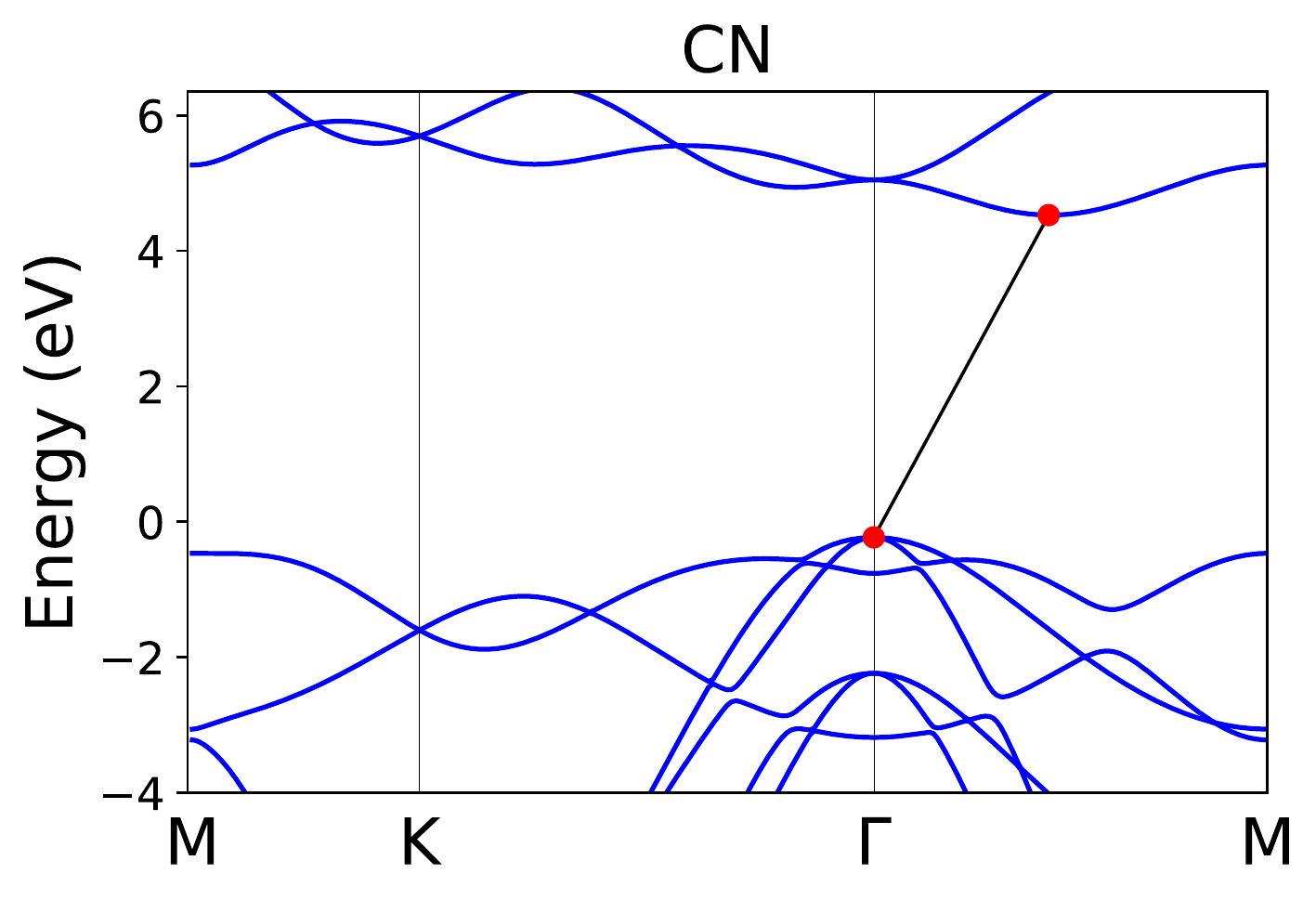}
\includegraphics[width=0.2\textwidth]{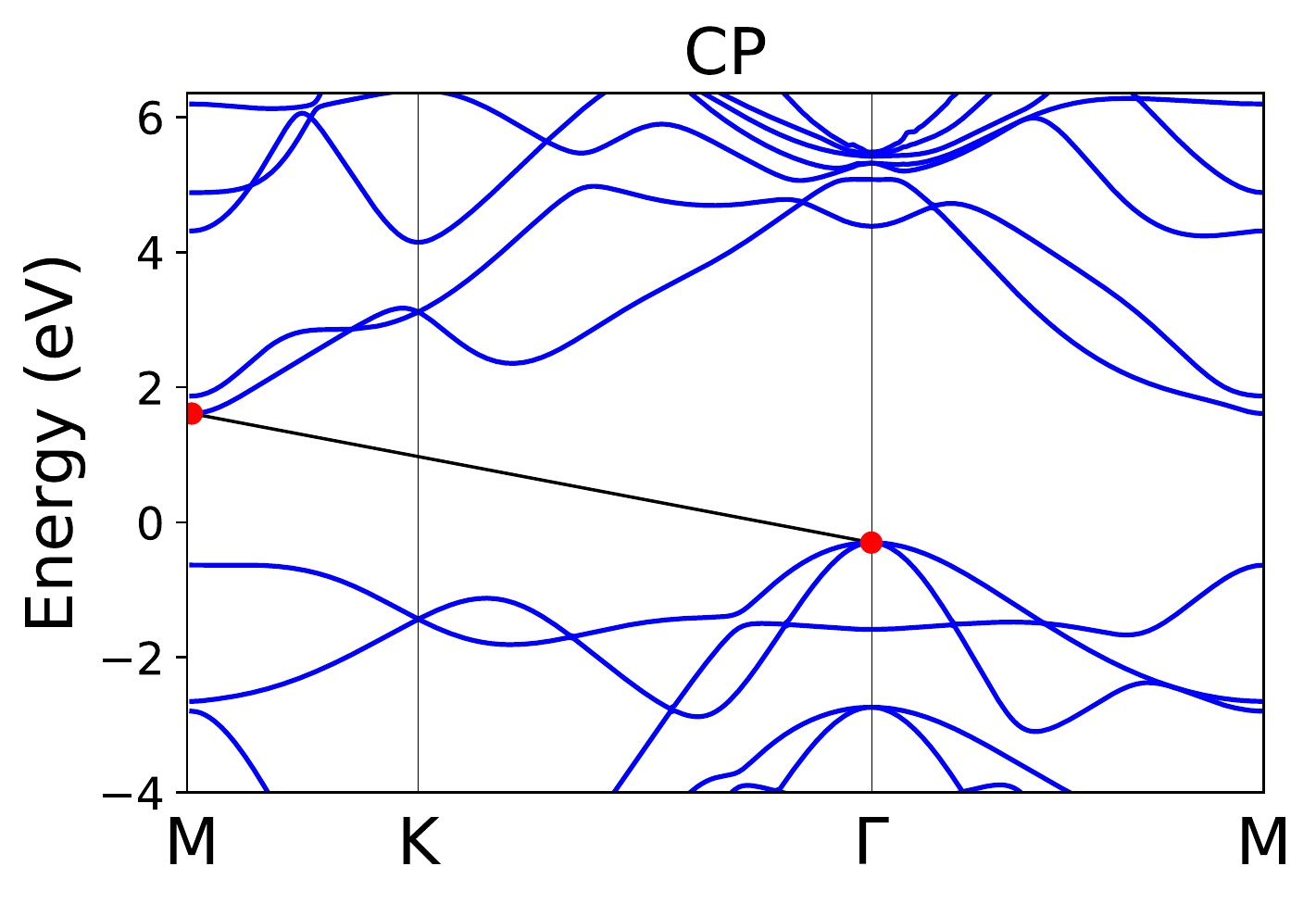}
\includegraphics[width=0.2\textwidth]{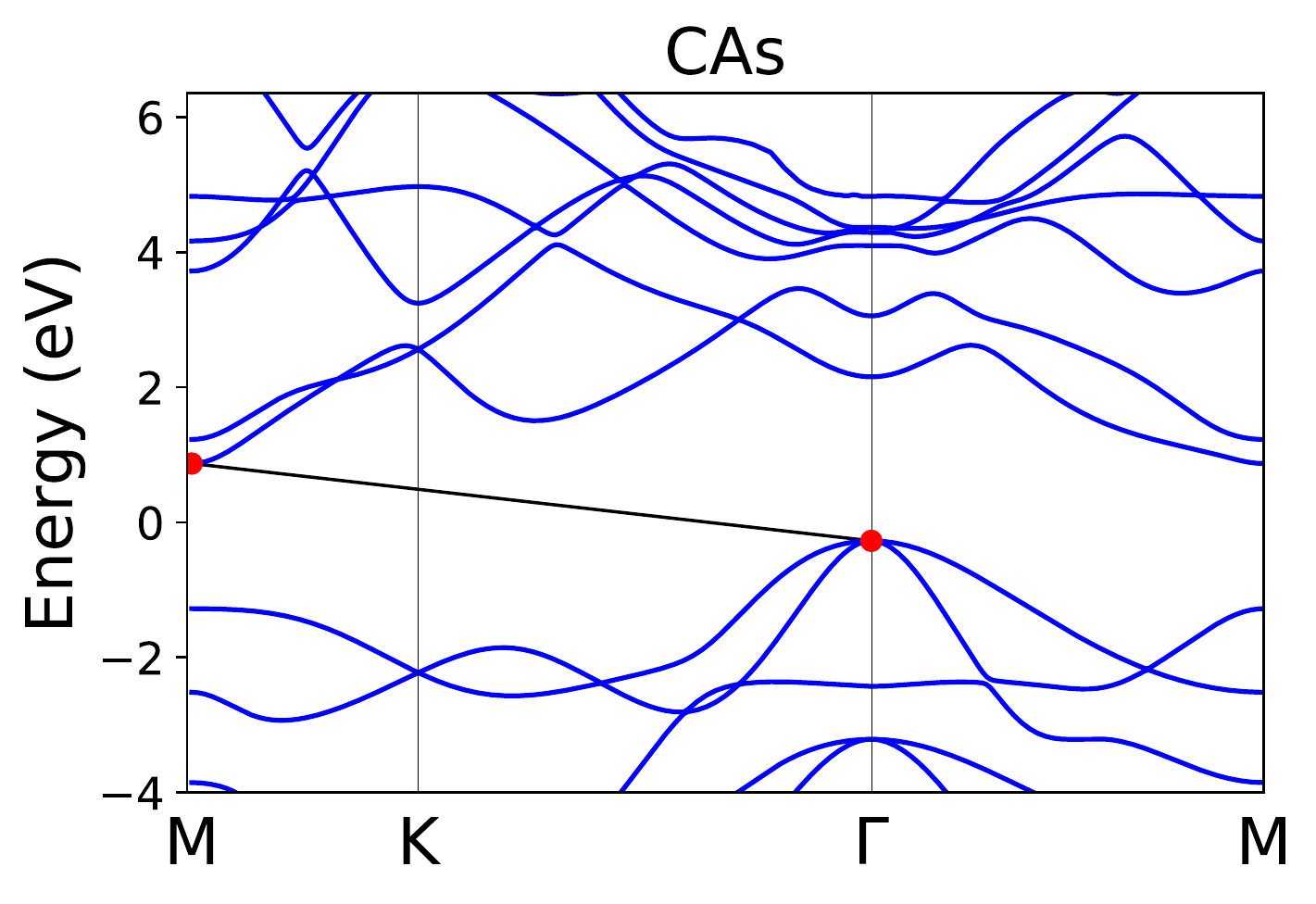}
\includegraphics[width=0.2\textwidth]{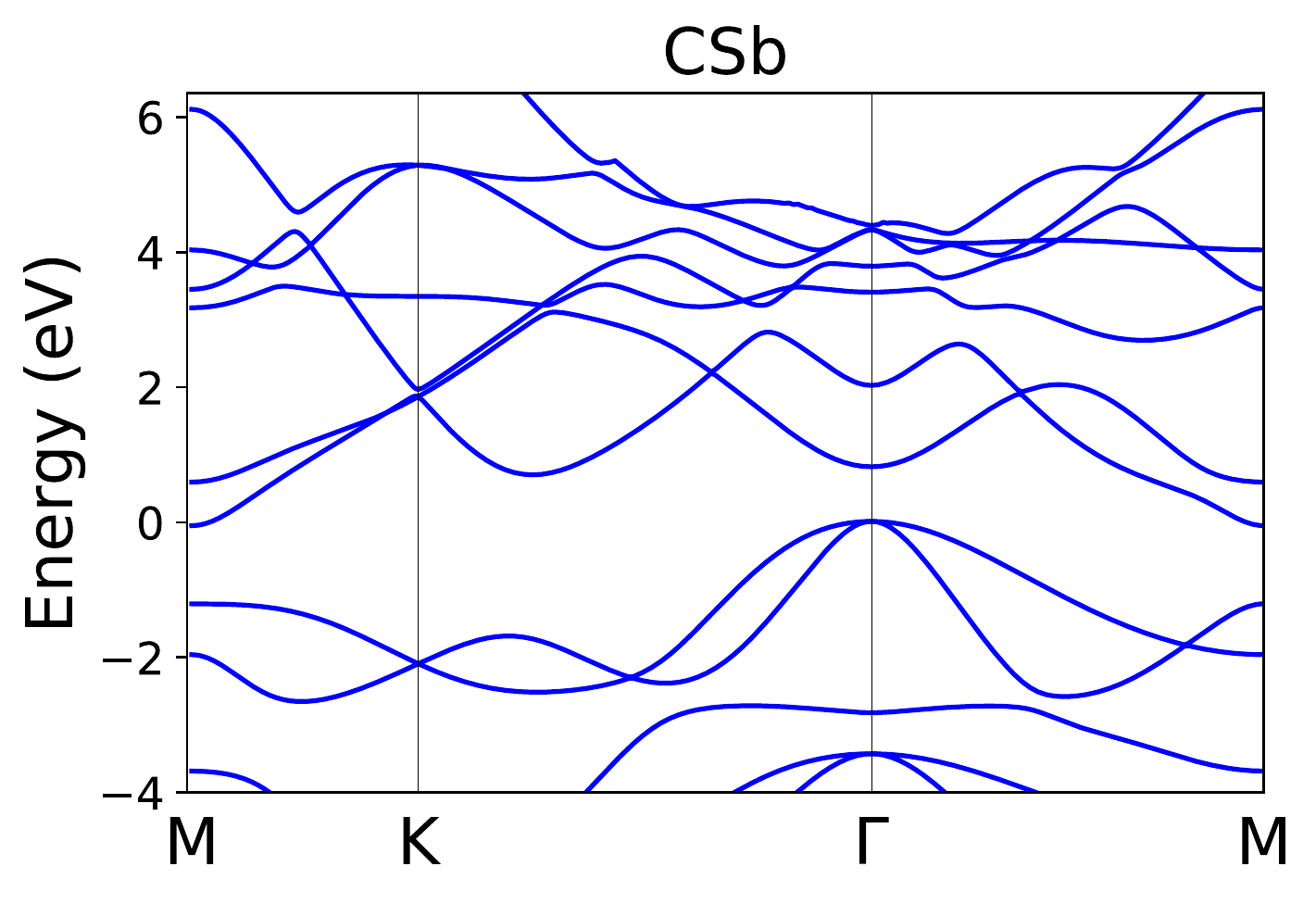}
\includegraphics[width=0.2\textwidth]{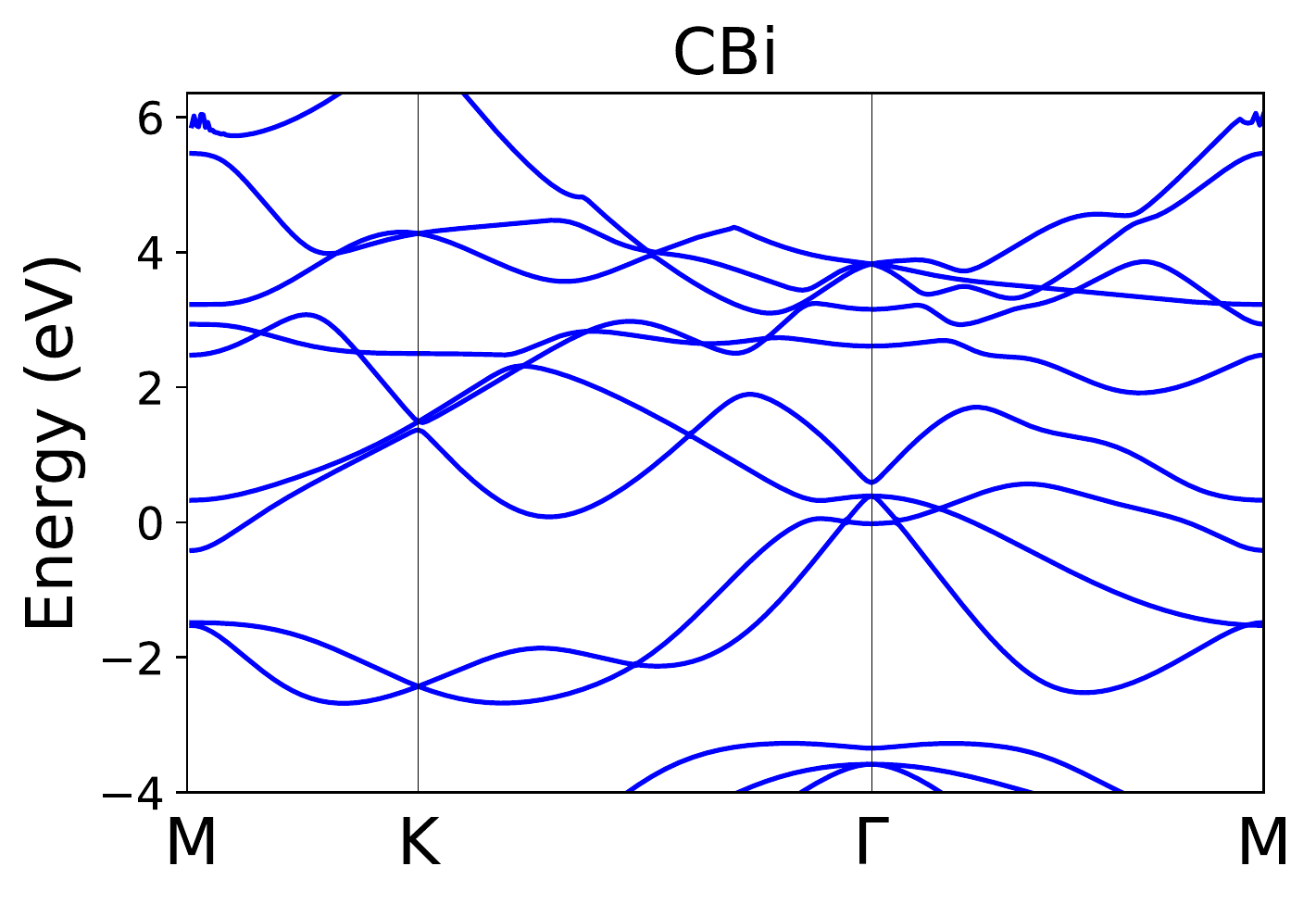}\\

\includegraphics[width=0.2\textwidth]{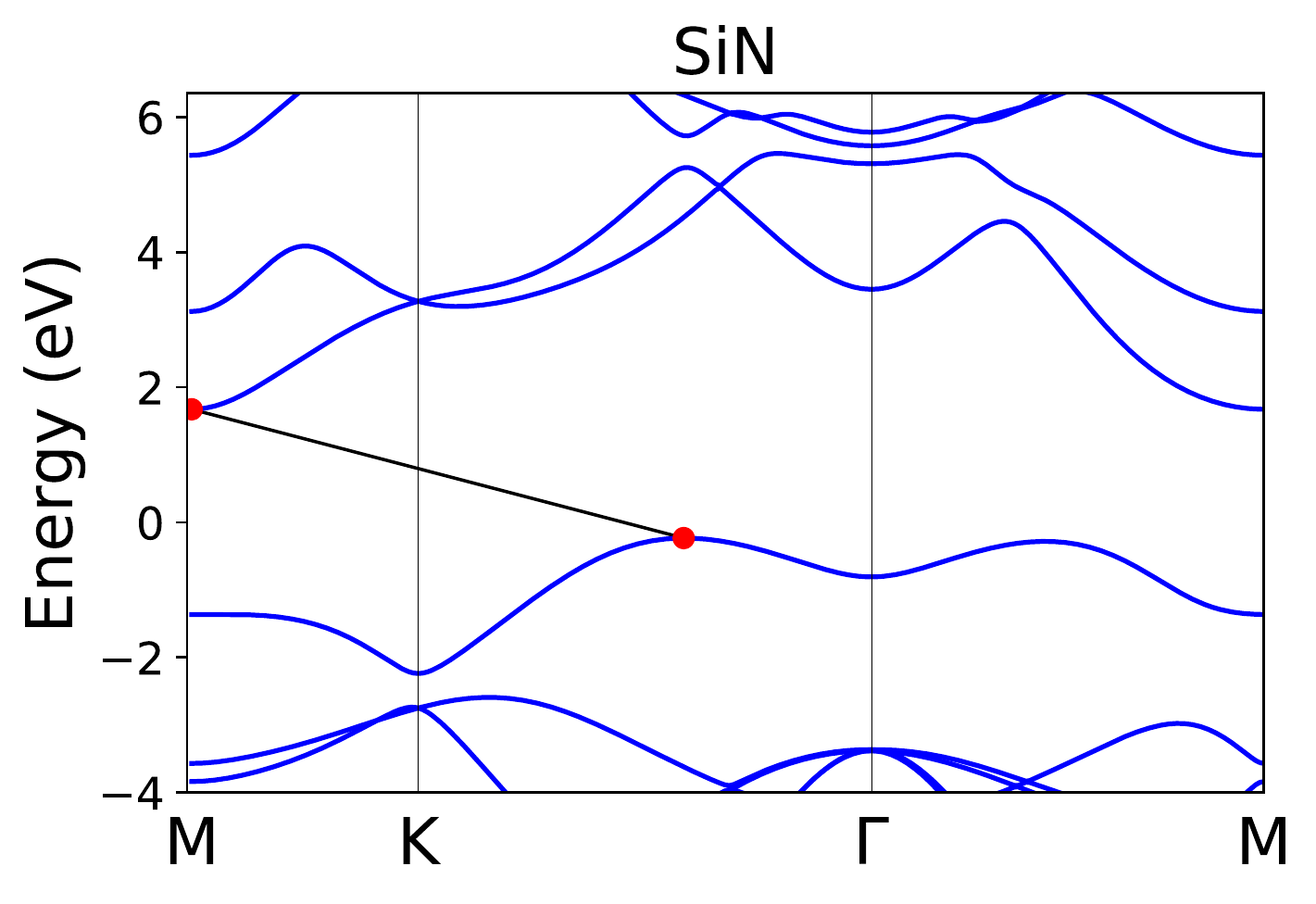}
\includegraphics[width=0.2\textwidth]{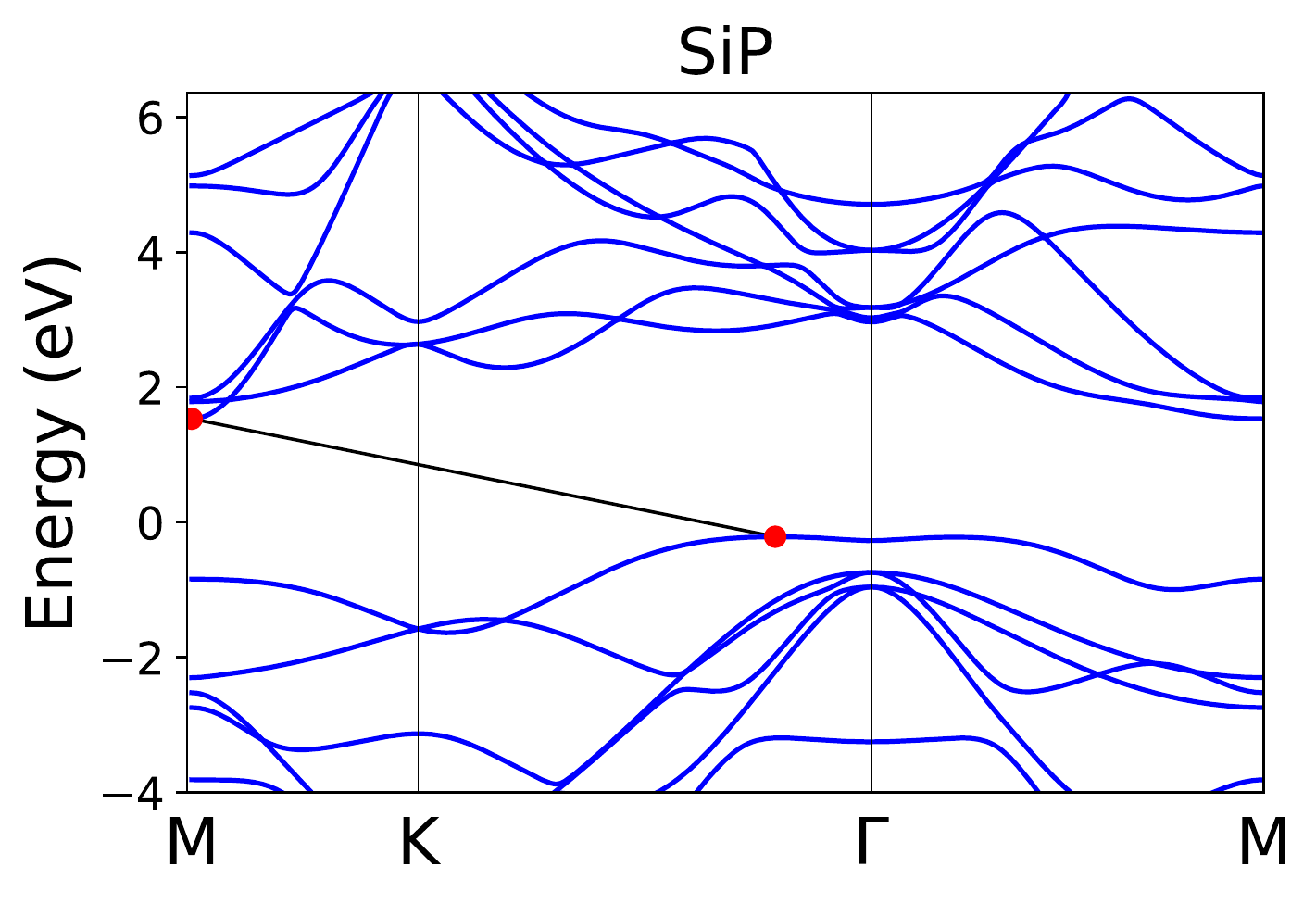}
\includegraphics[width=0.2\textwidth]{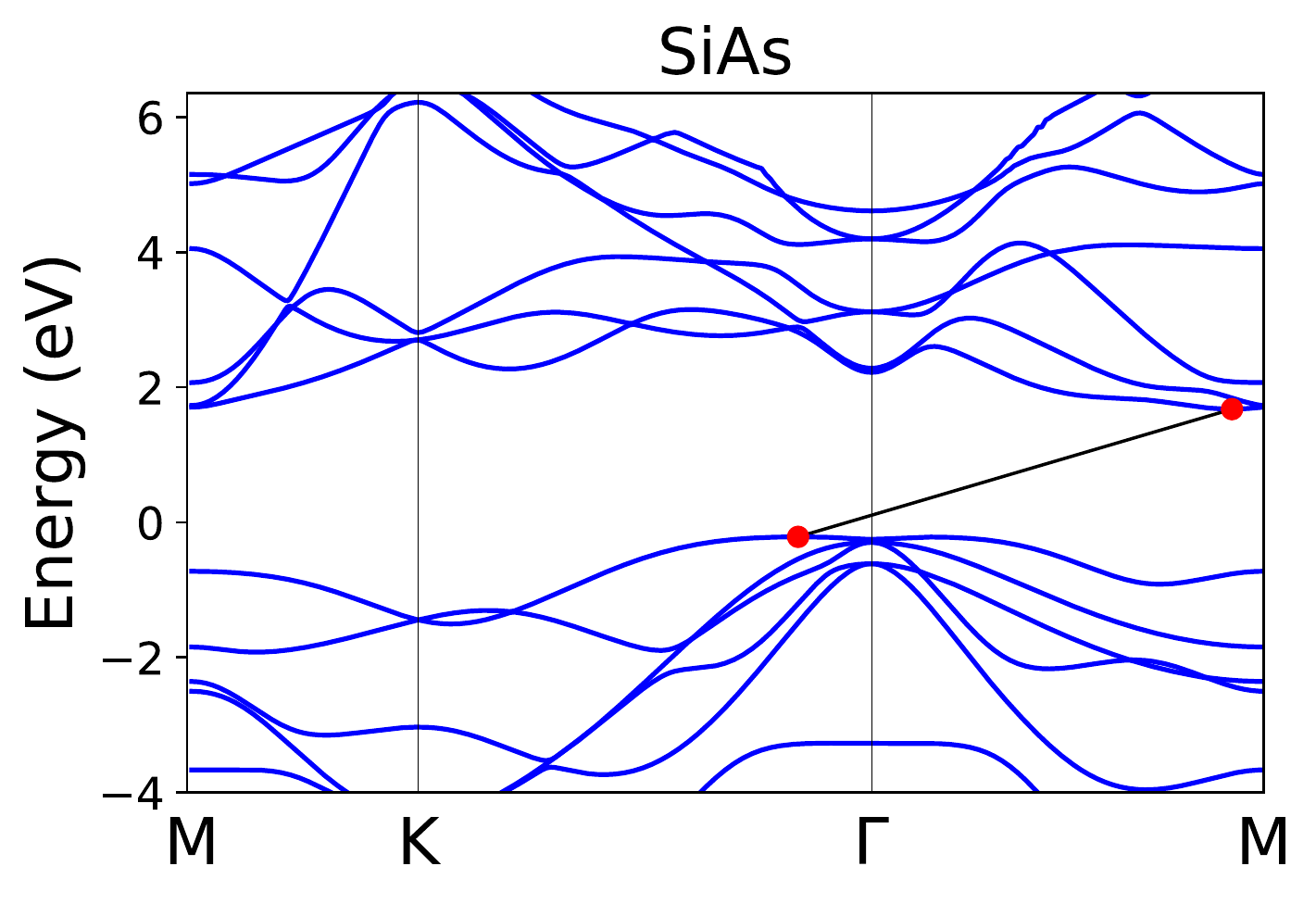}
\includegraphics[width=0.2\textwidth]{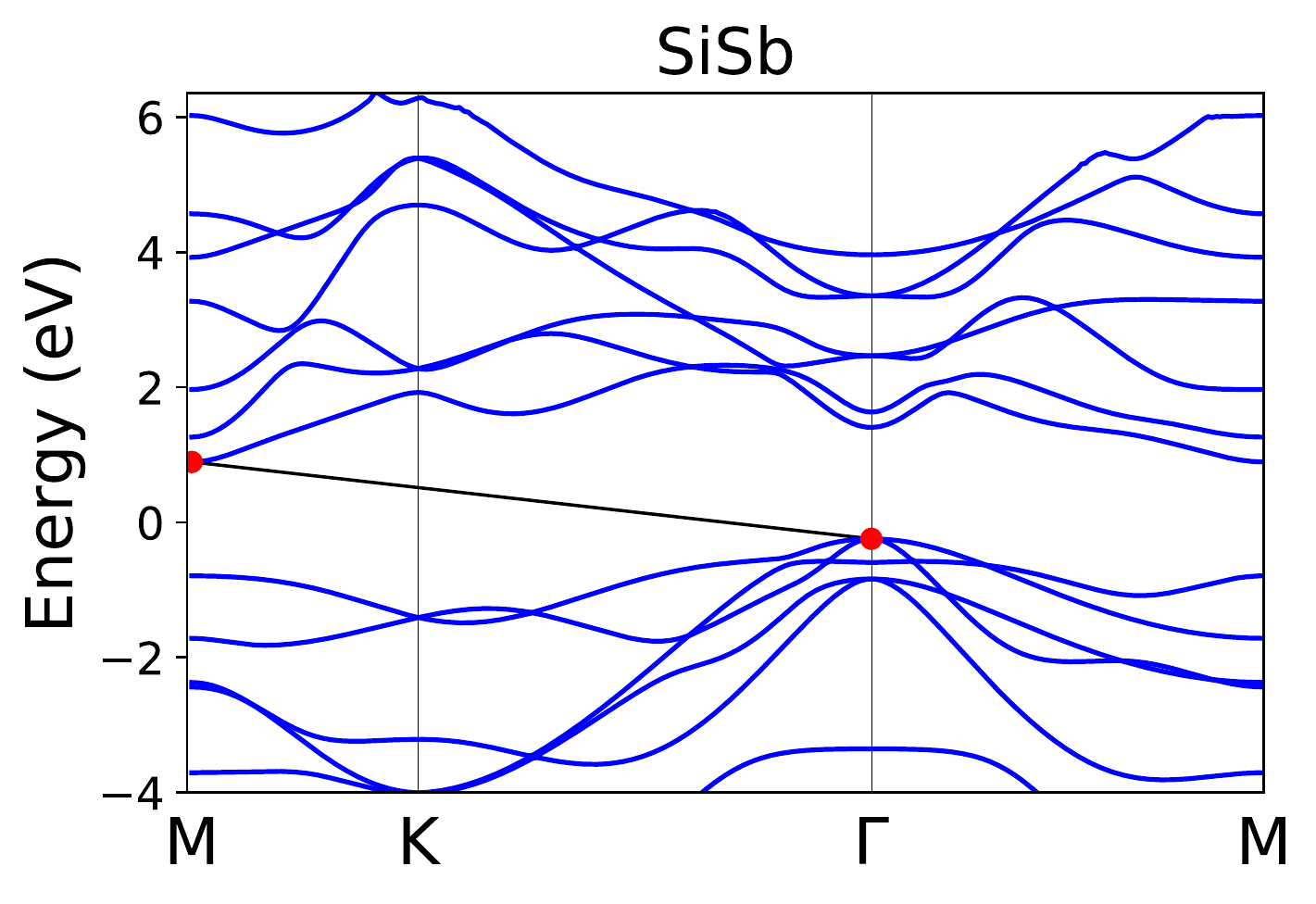}
\includegraphics[width=0.2\textwidth]{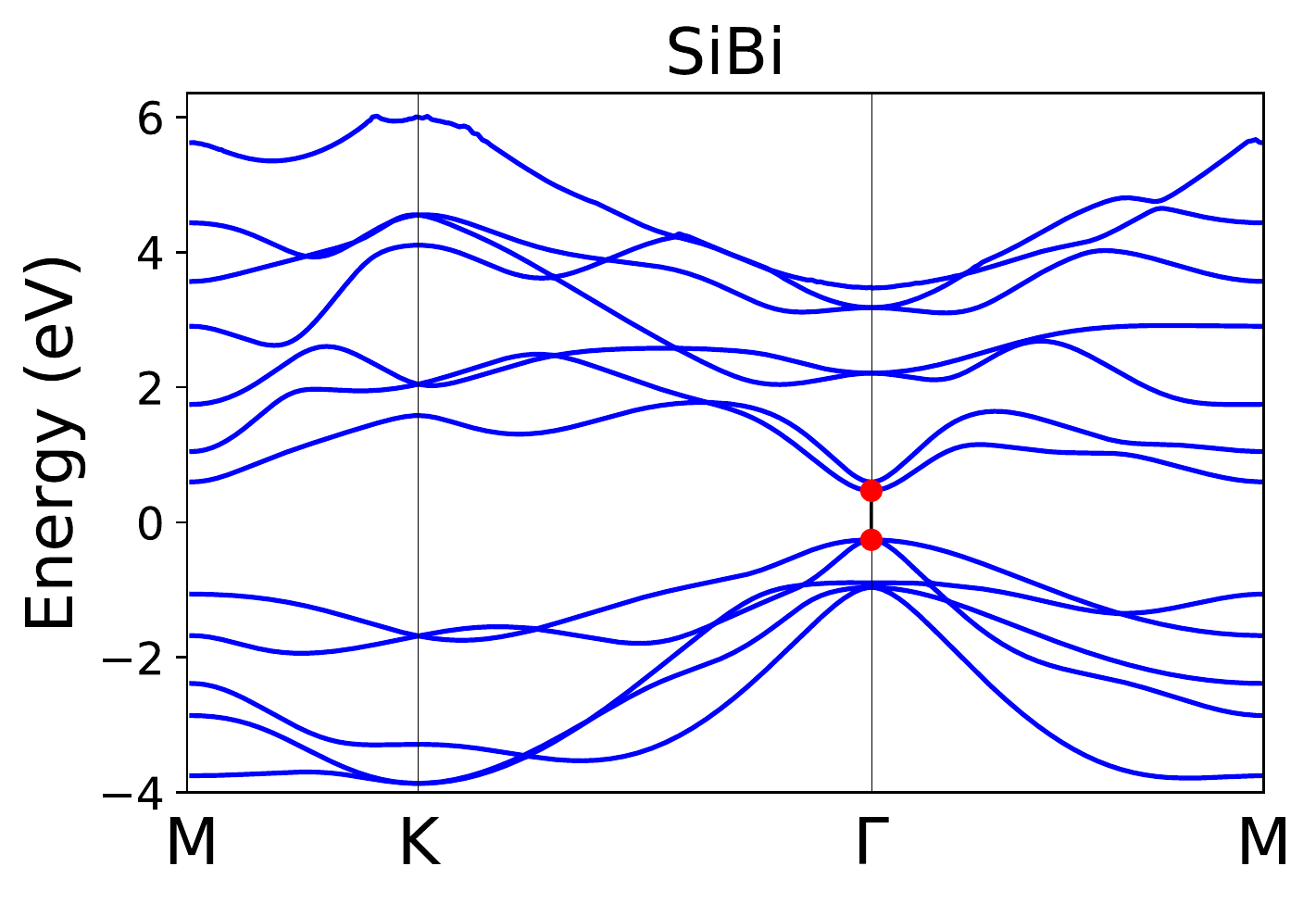}\\

\includegraphics[width=0.2\textwidth]{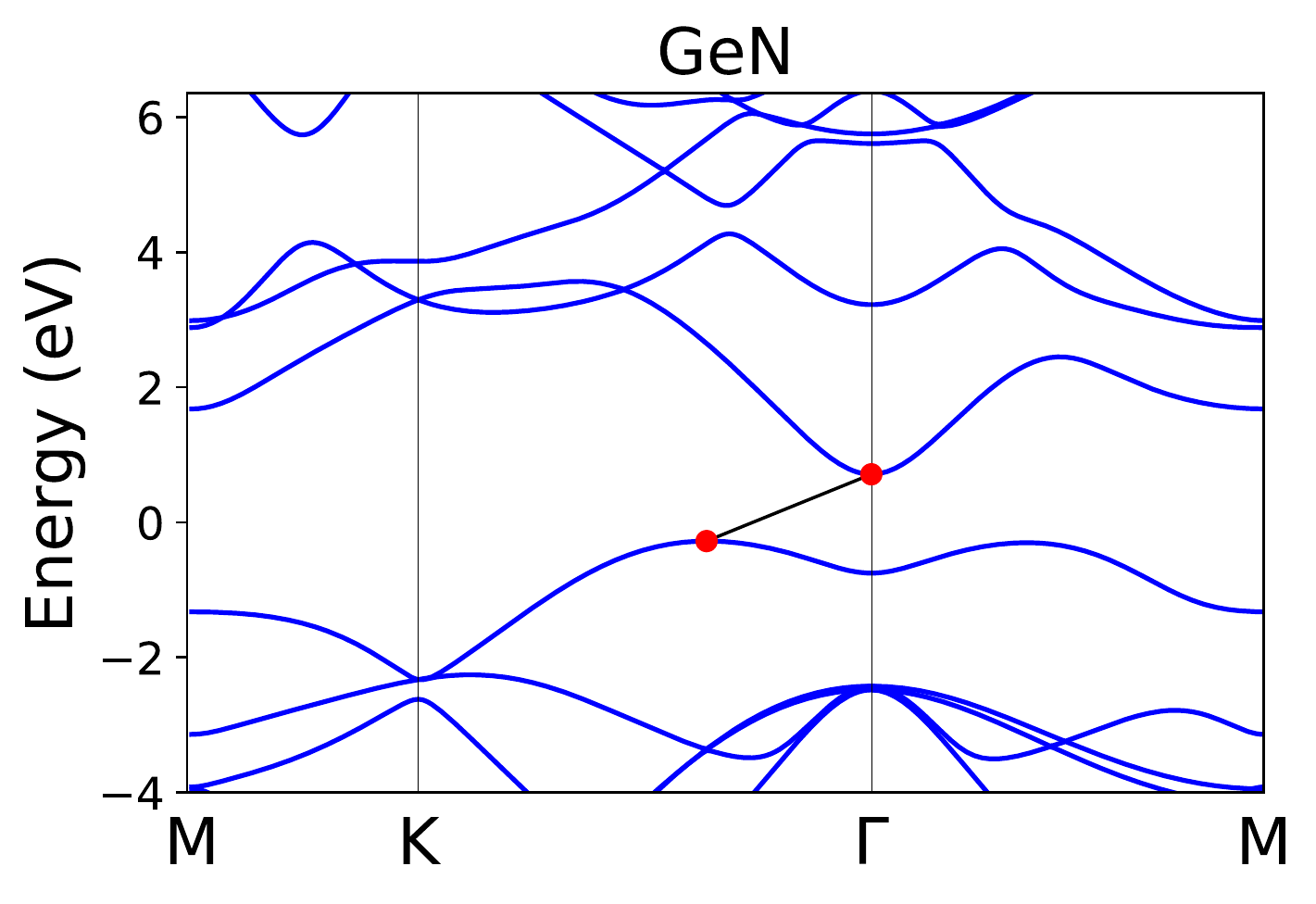}
\includegraphics[width=0.2\textwidth]{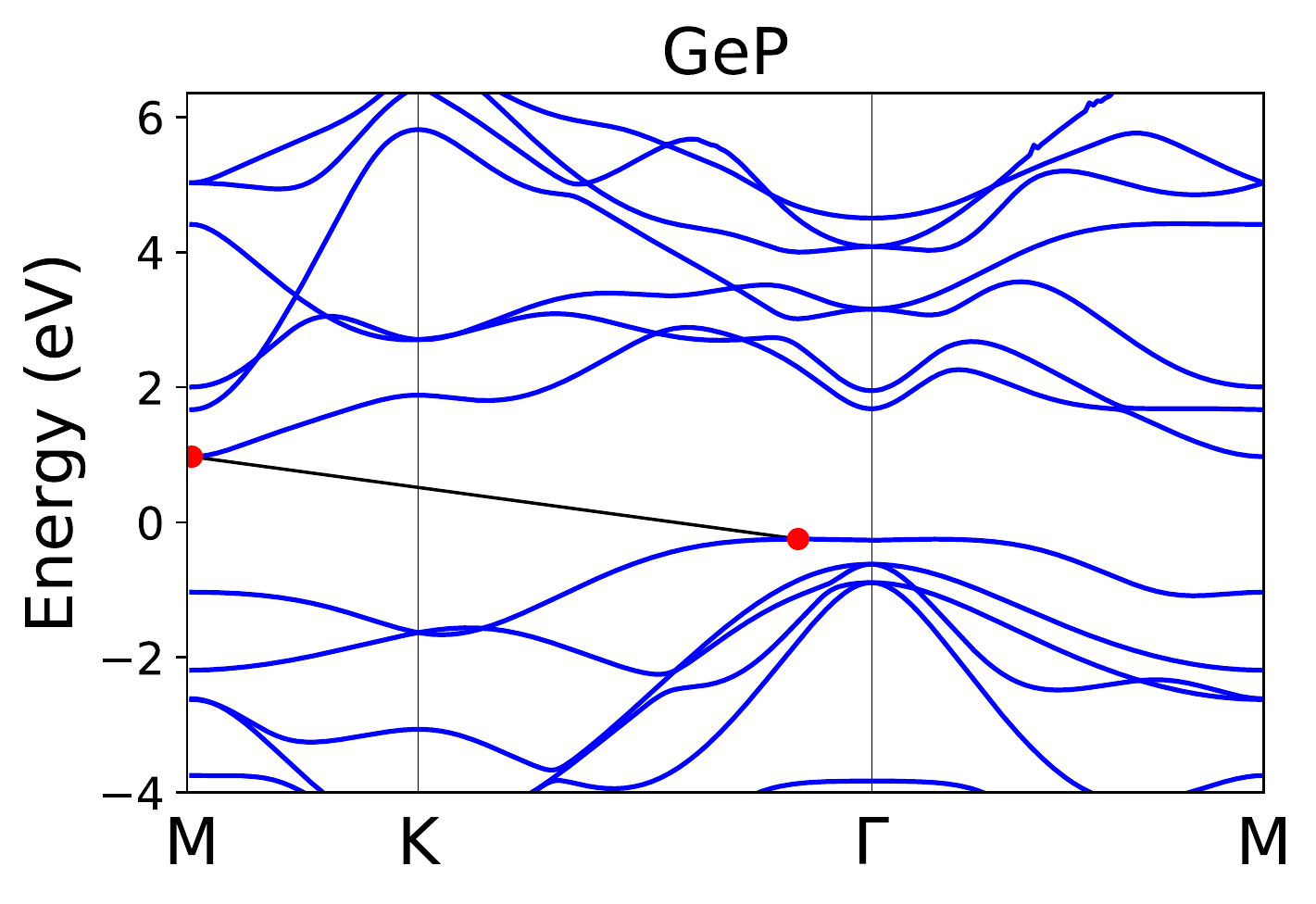}
\includegraphics[width=0.2\textwidth]{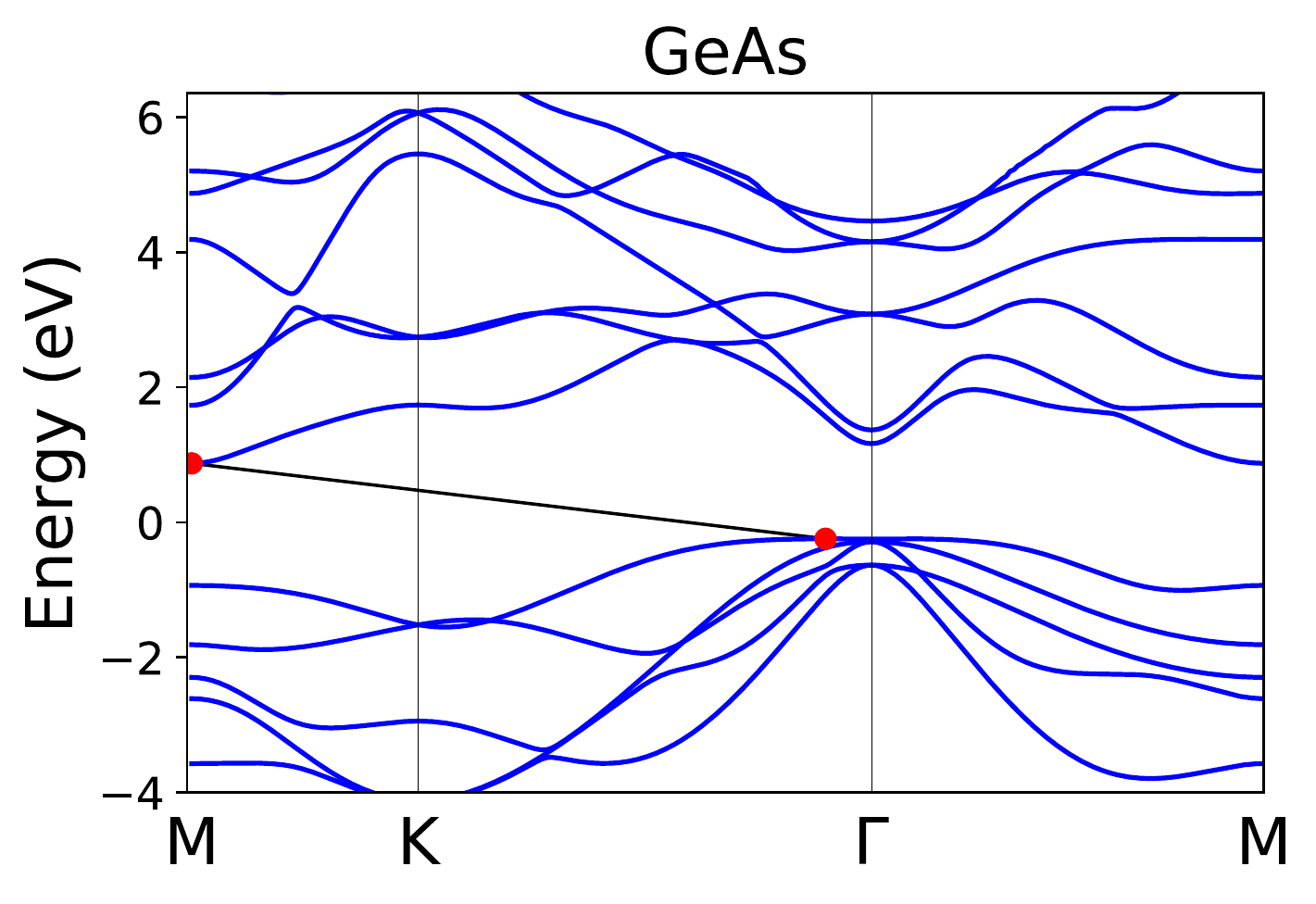}
\includegraphics[width=0.2\textwidth]{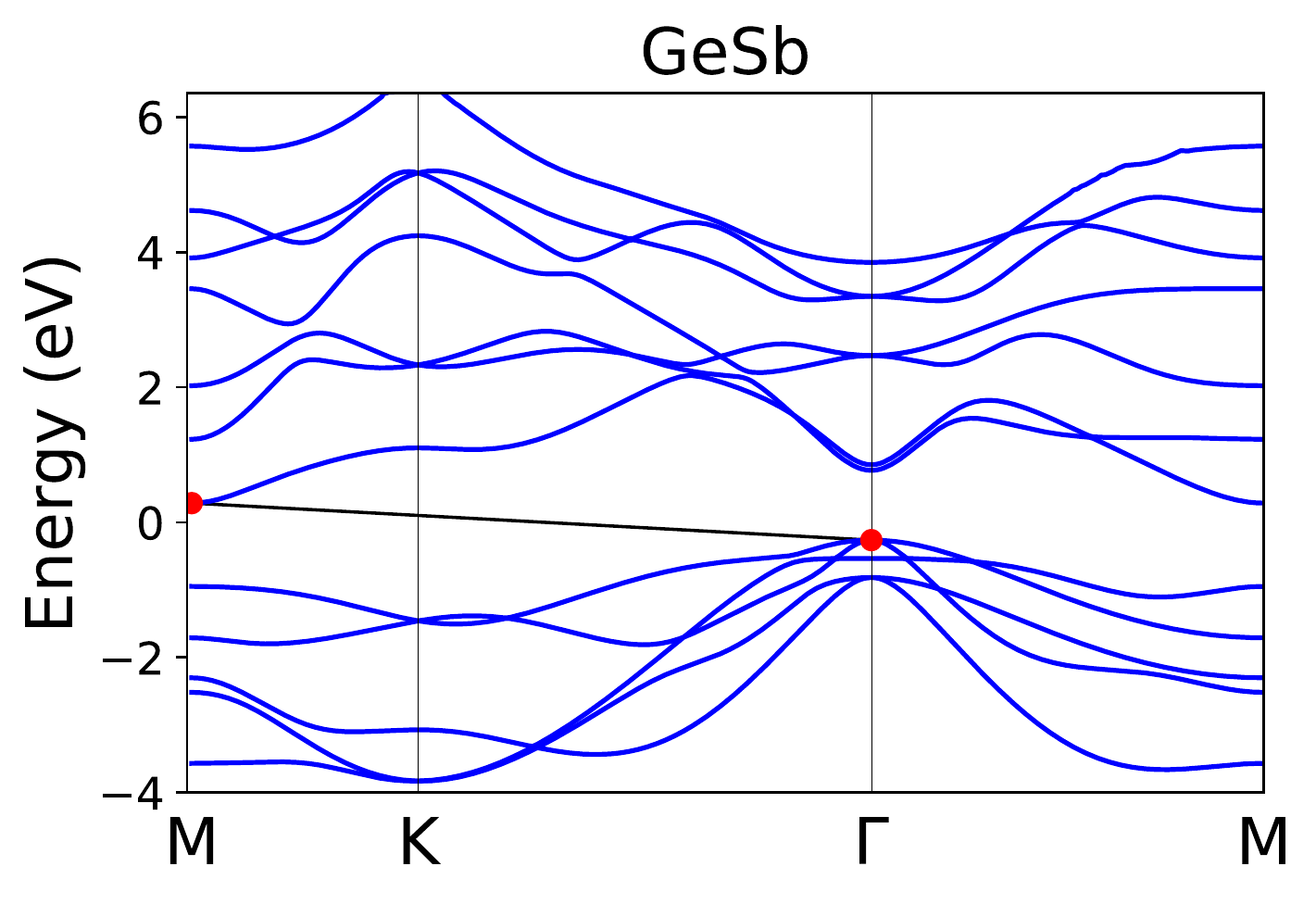}
\includegraphics[width=0.2\textwidth]{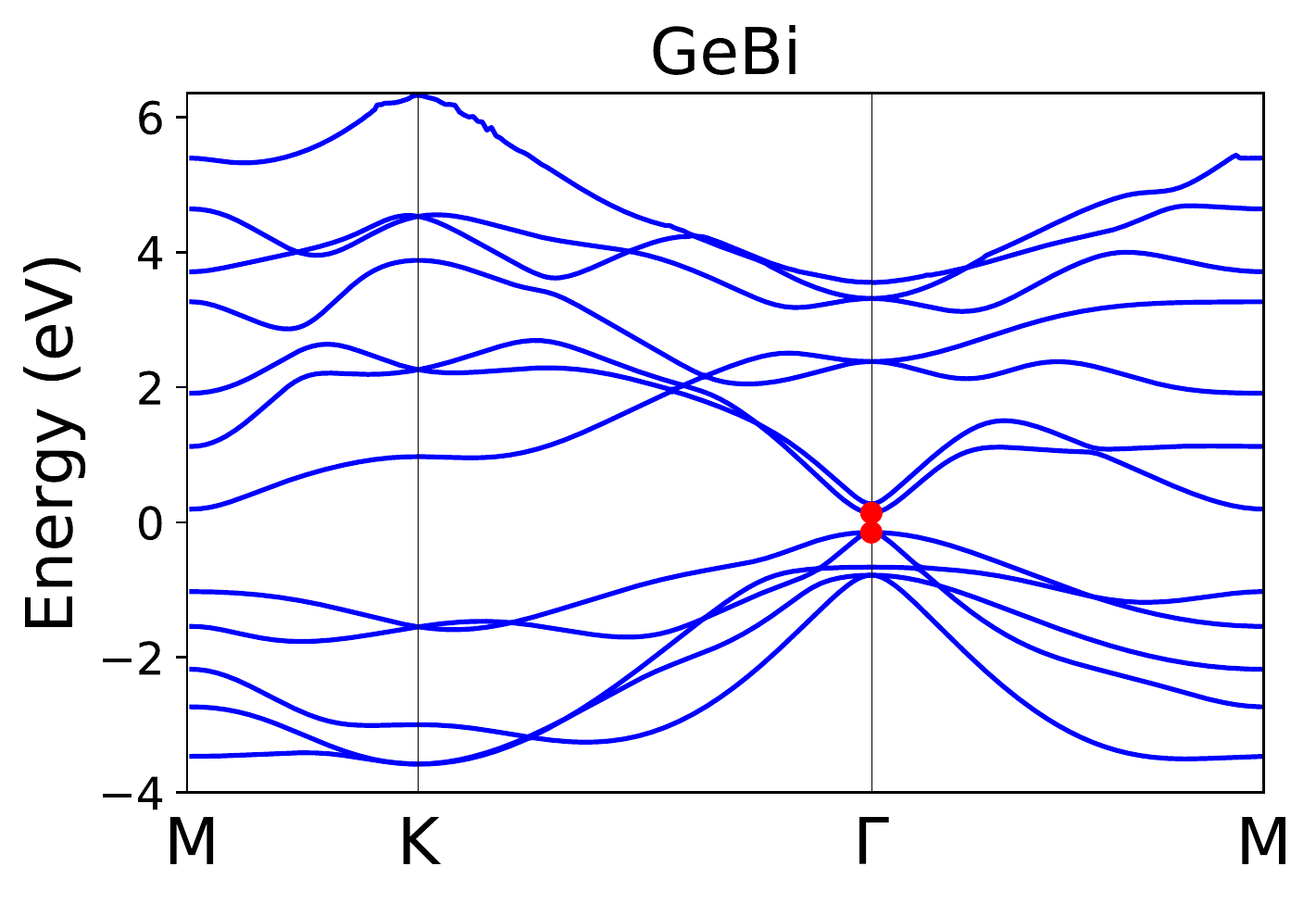}\\

\includegraphics[width=0.2\textwidth]{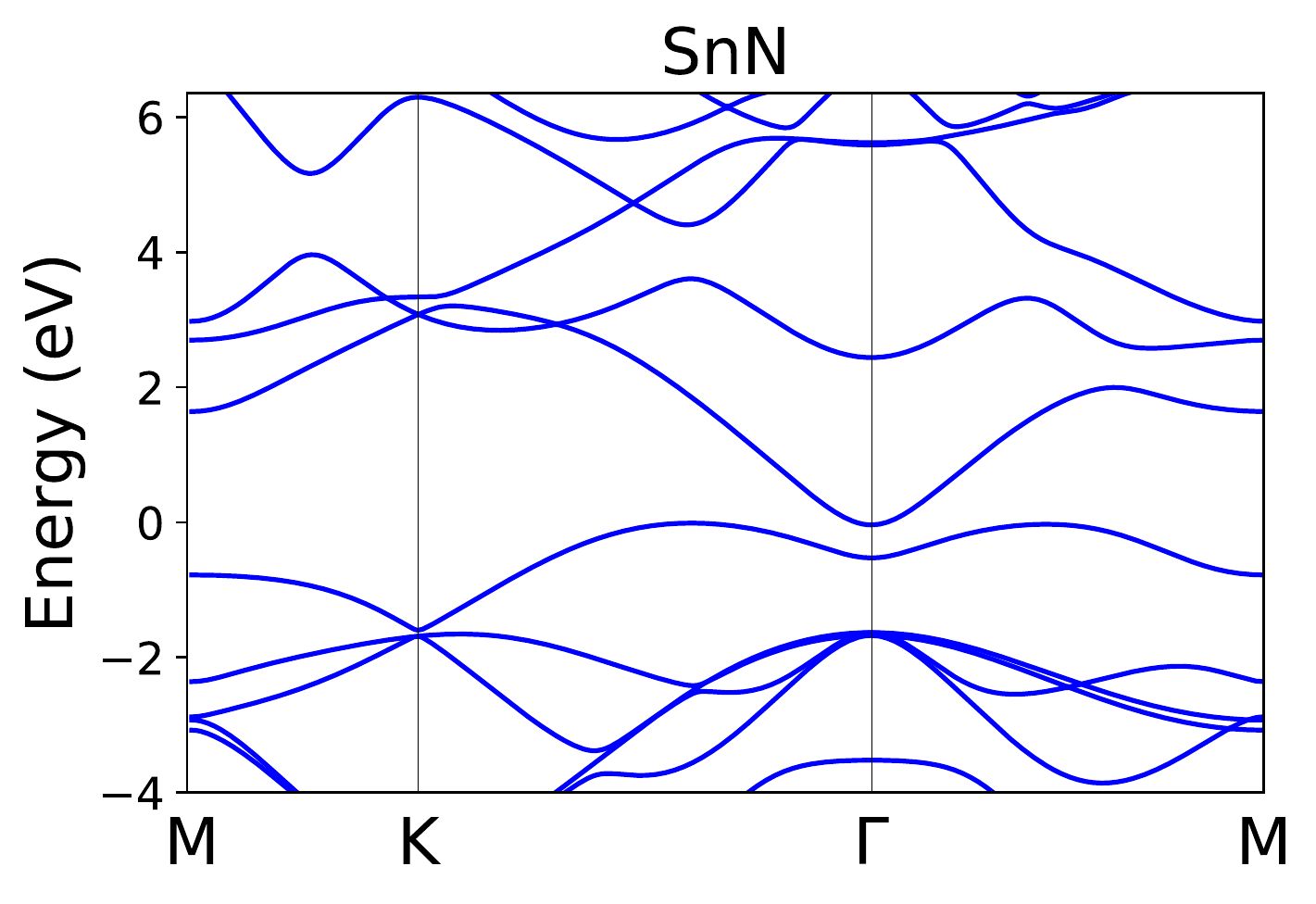}
\includegraphics[width=0.2\textwidth]{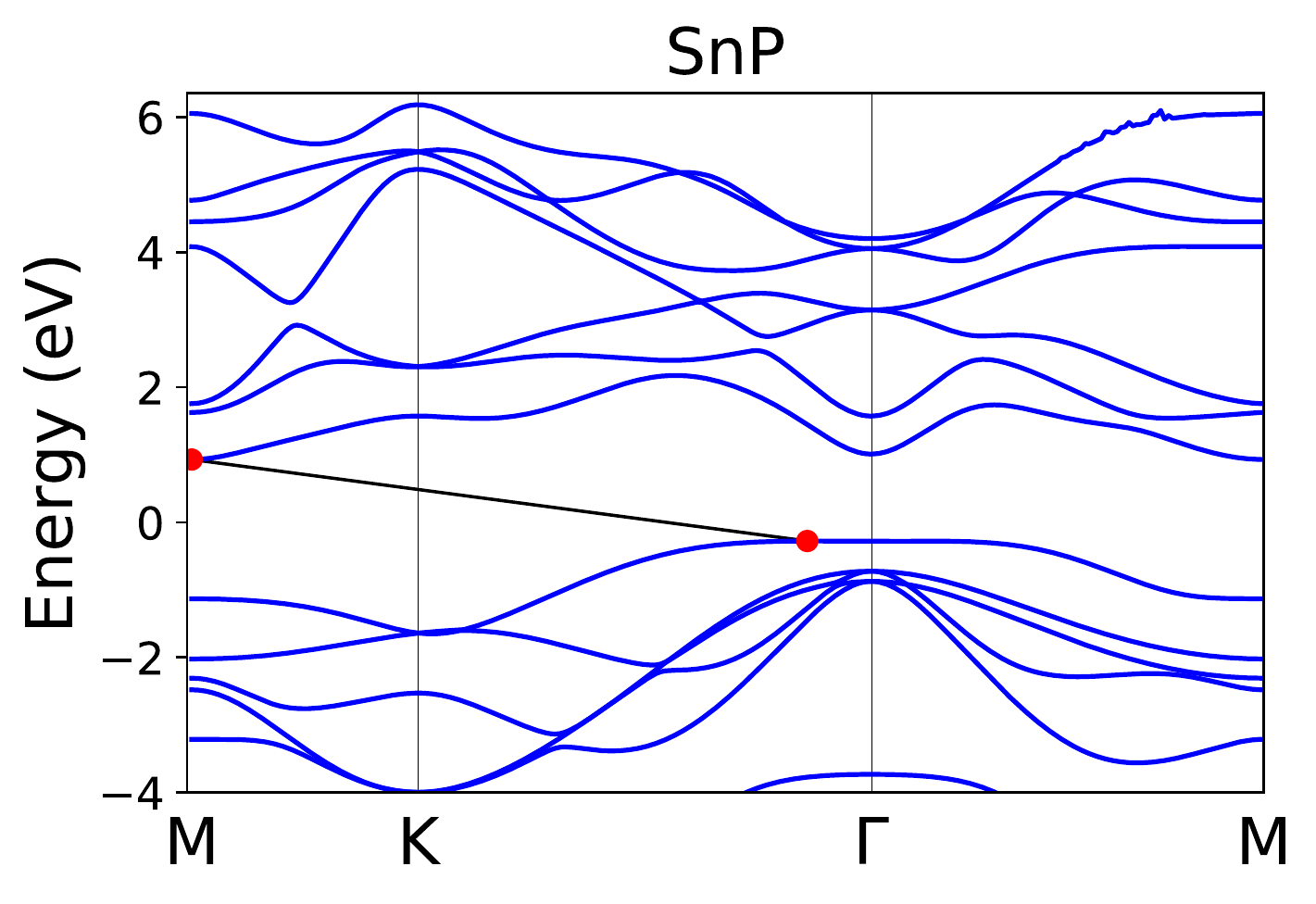}
\includegraphics[width=0.2\textwidth]{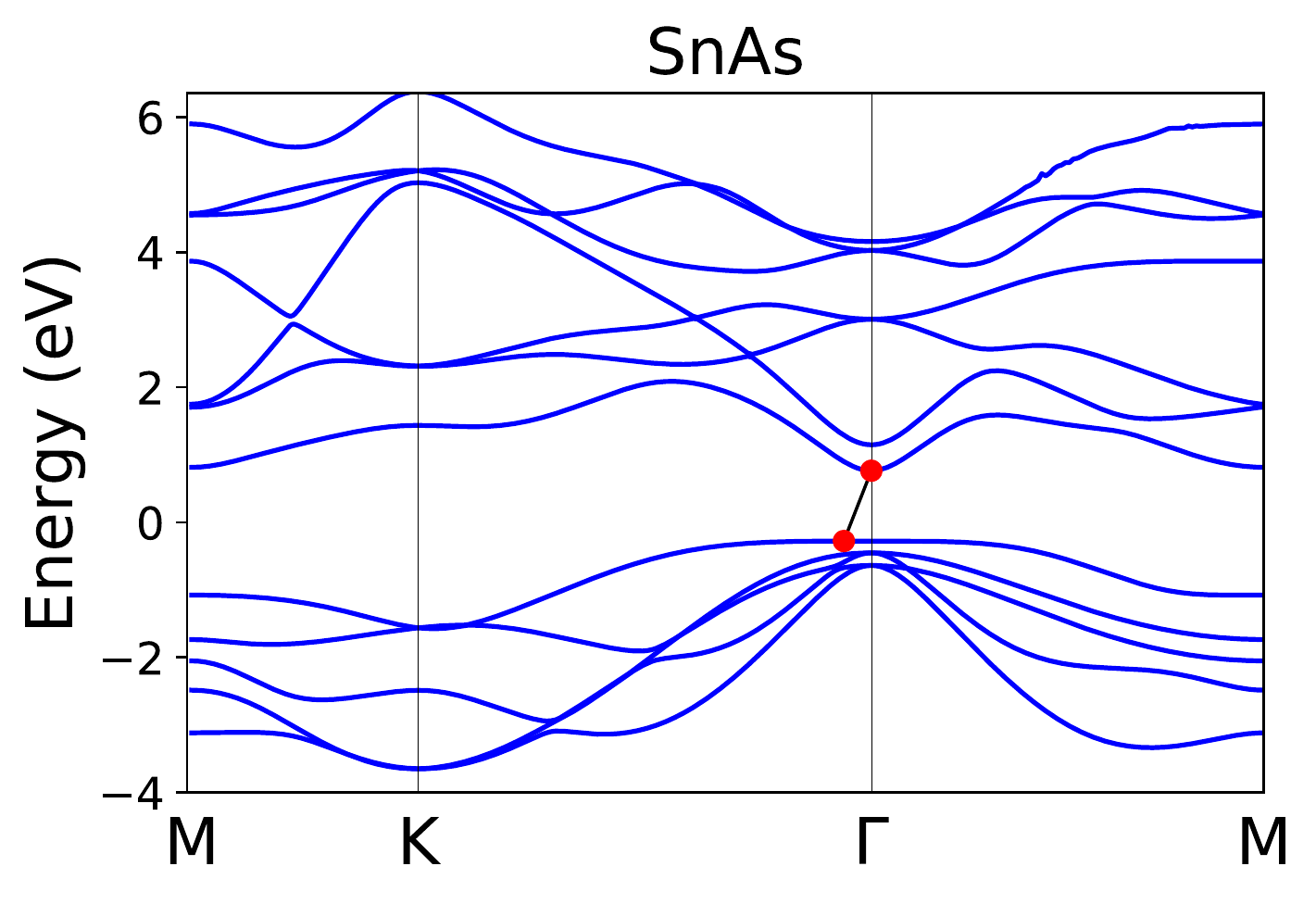}
\includegraphics[width=0.2\textwidth]{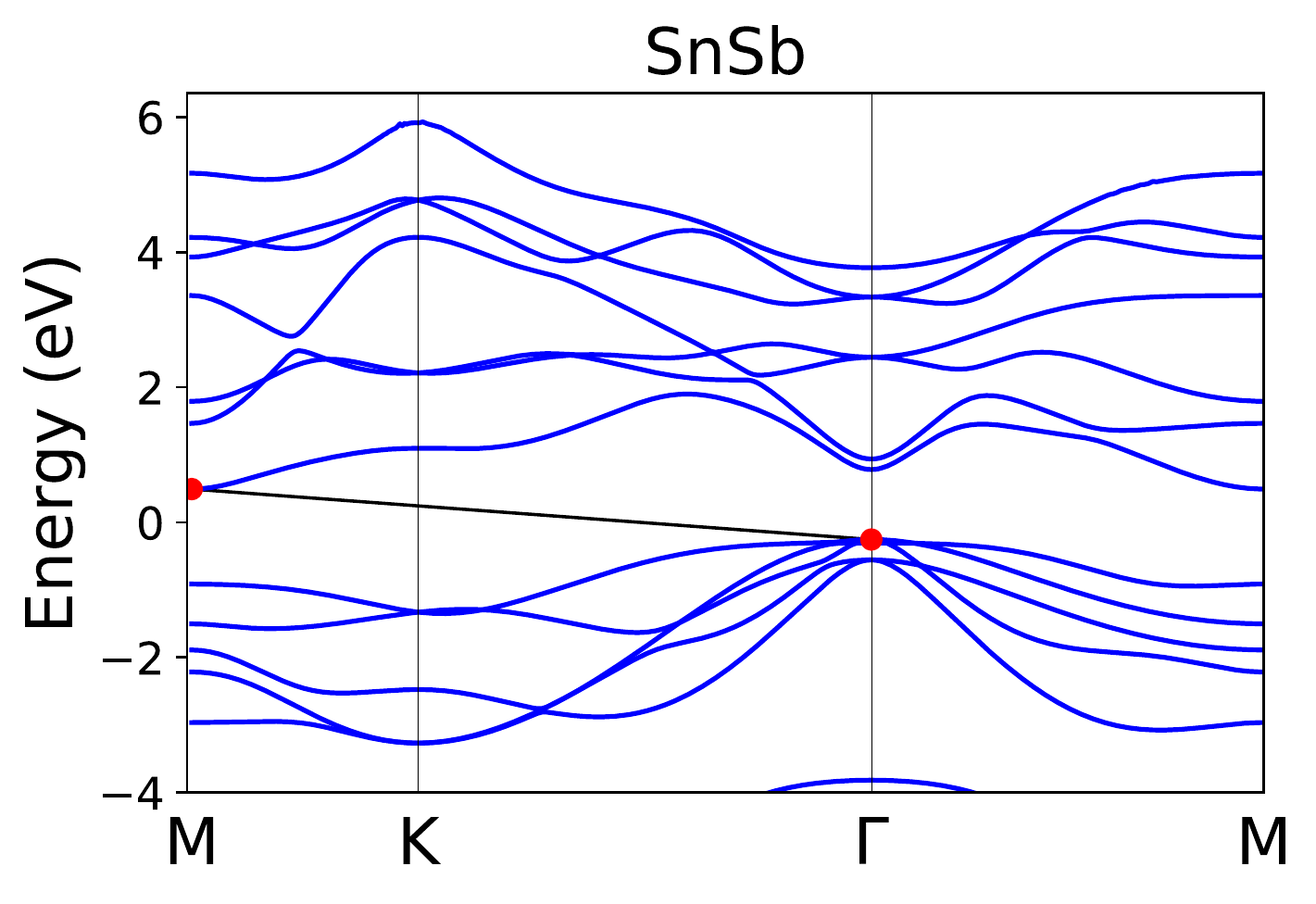}
\includegraphics[width=0.2\textwidth]{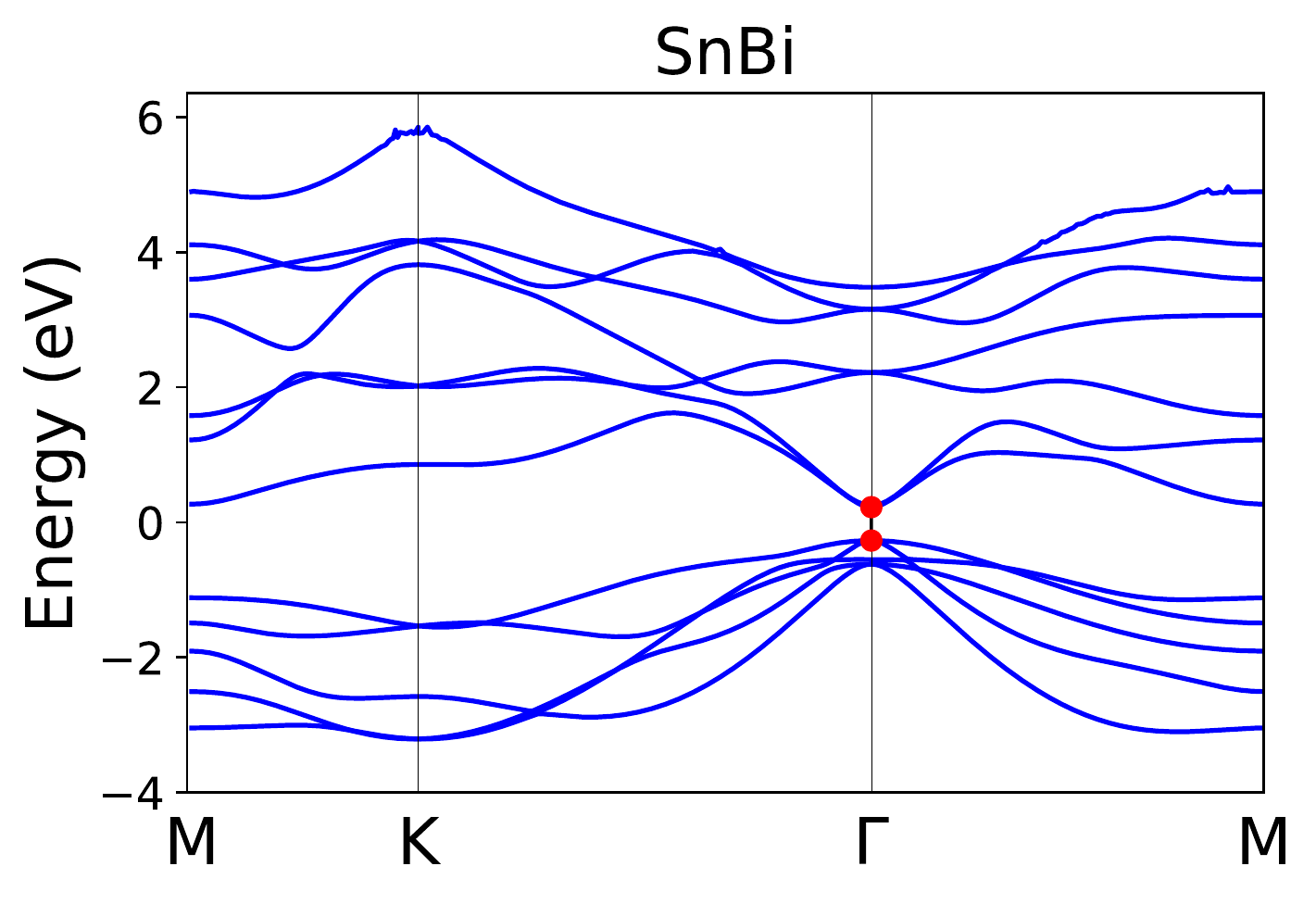}\\

\includegraphics[width=0.2\textwidth]{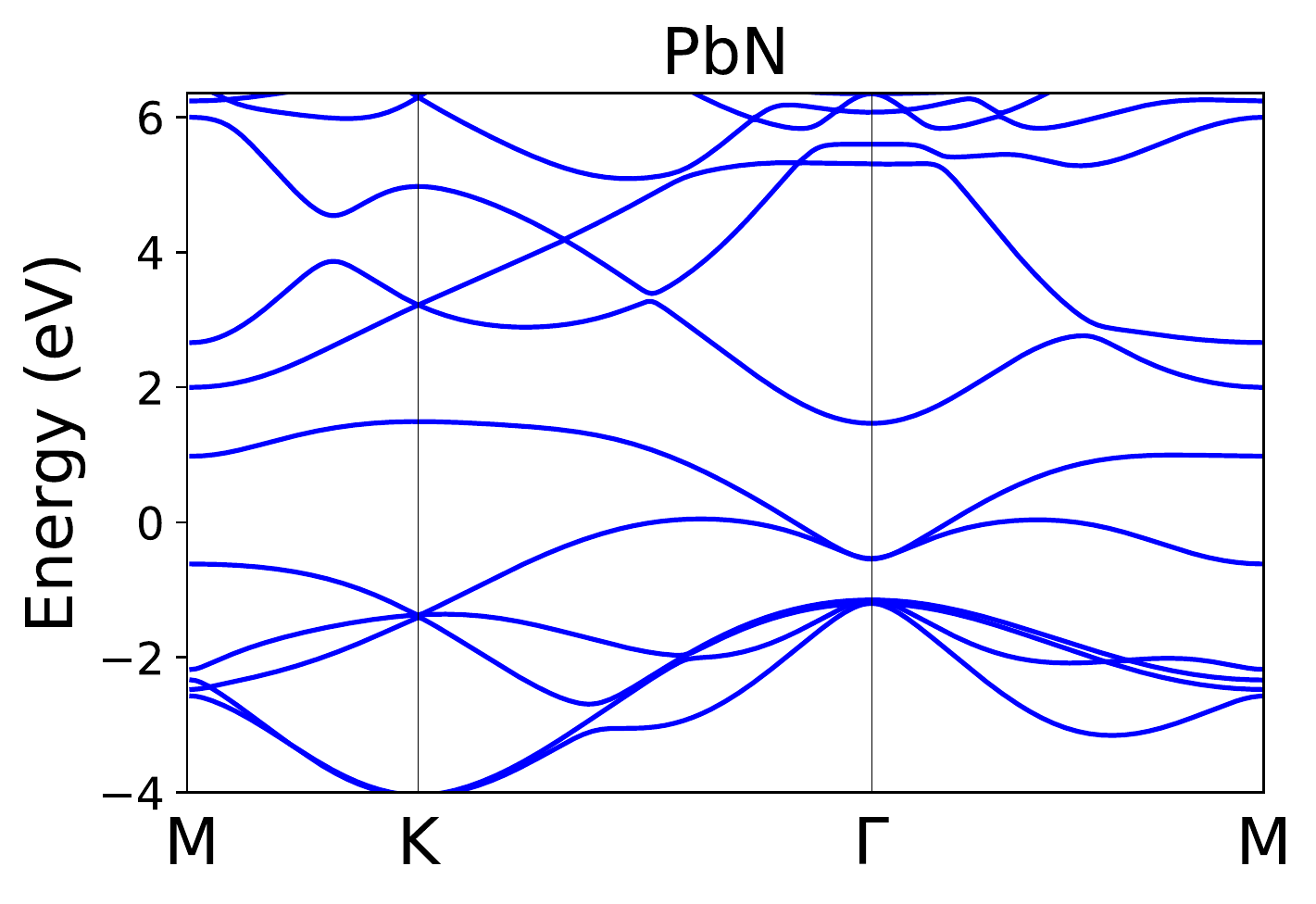}
\includegraphics[width=0.2\textwidth]{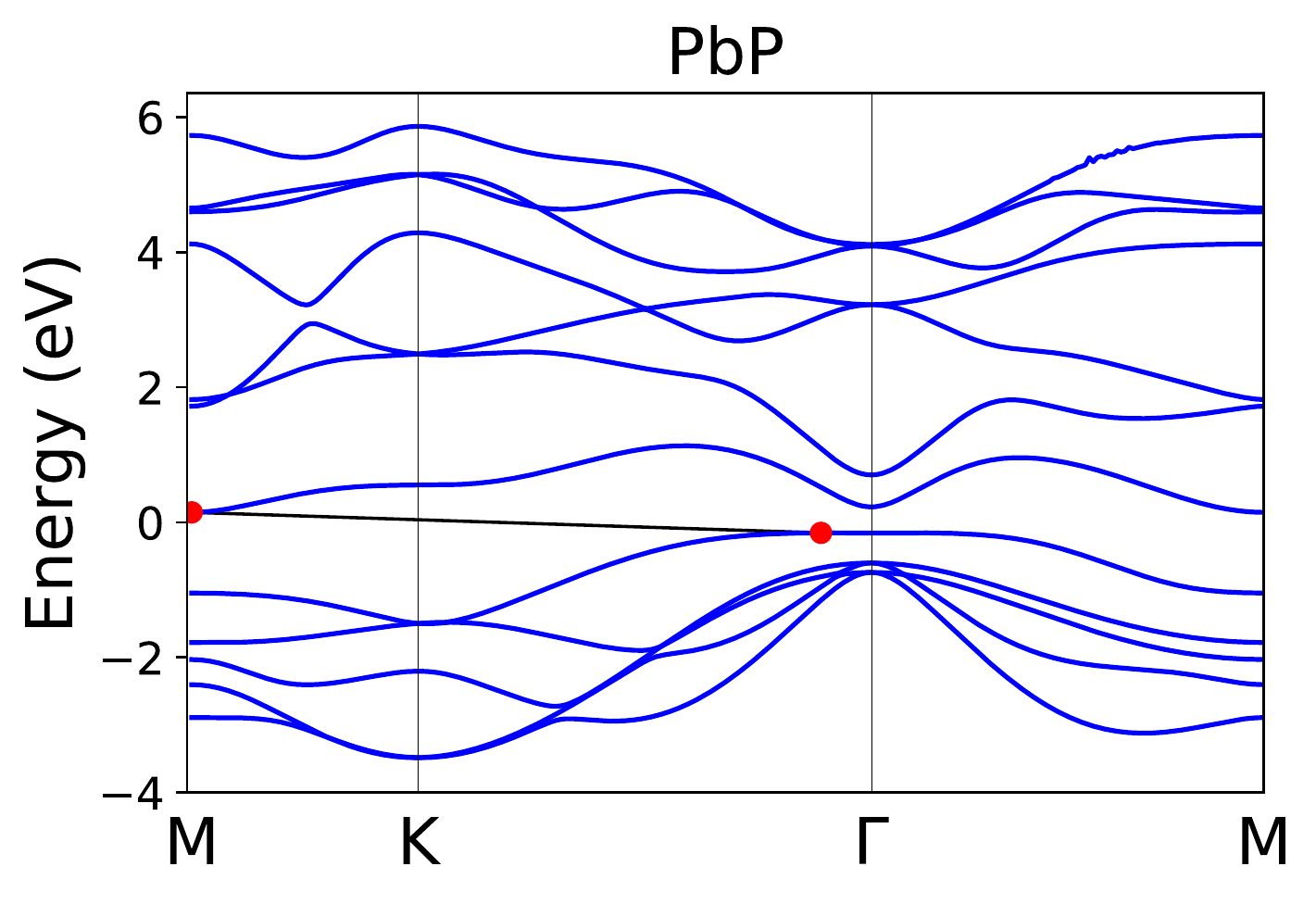}
\includegraphics[width=0.2\textwidth]{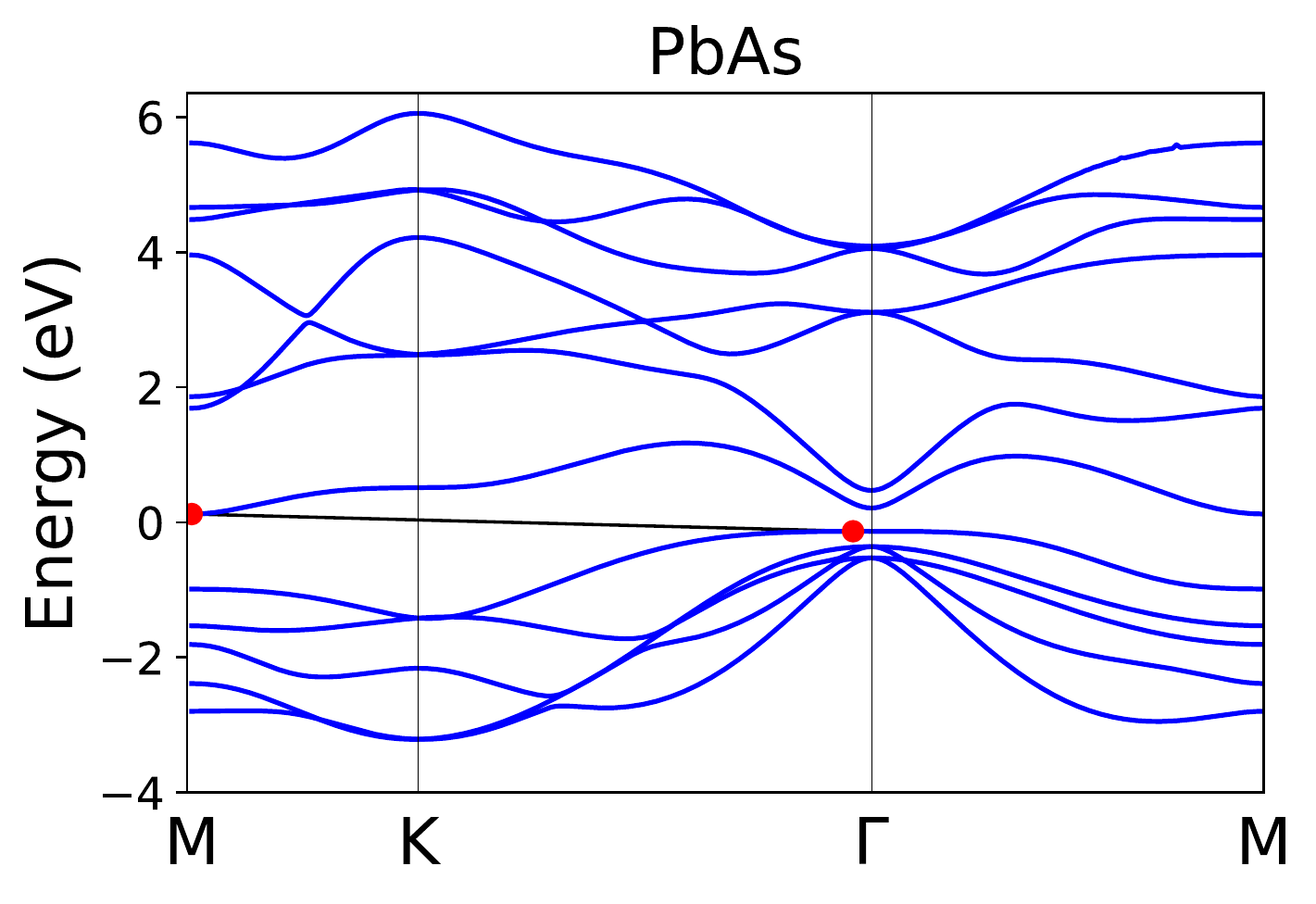}
\includegraphics[width=0.2\textwidth]{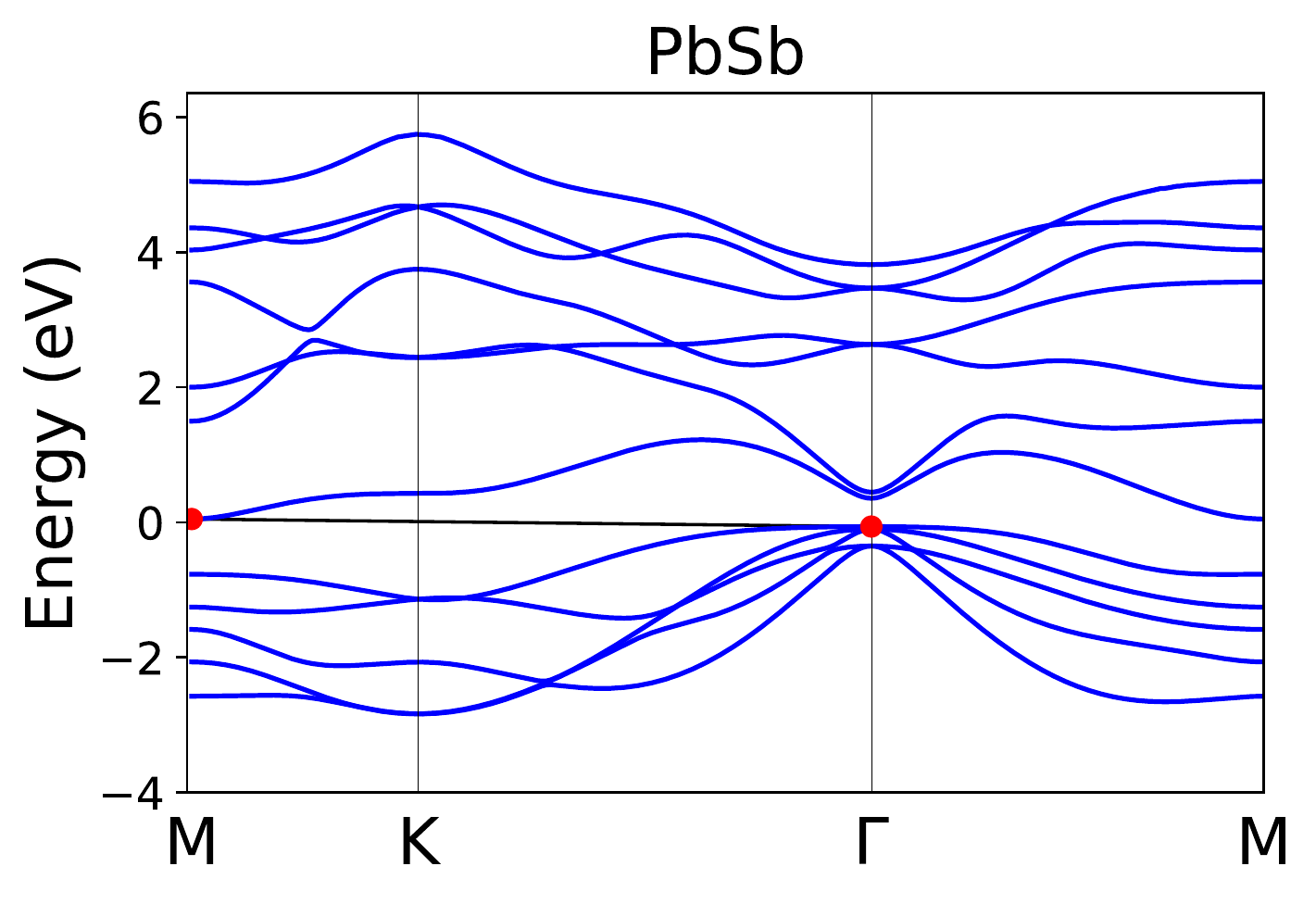}
\includegraphics[width=0.2\textwidth]{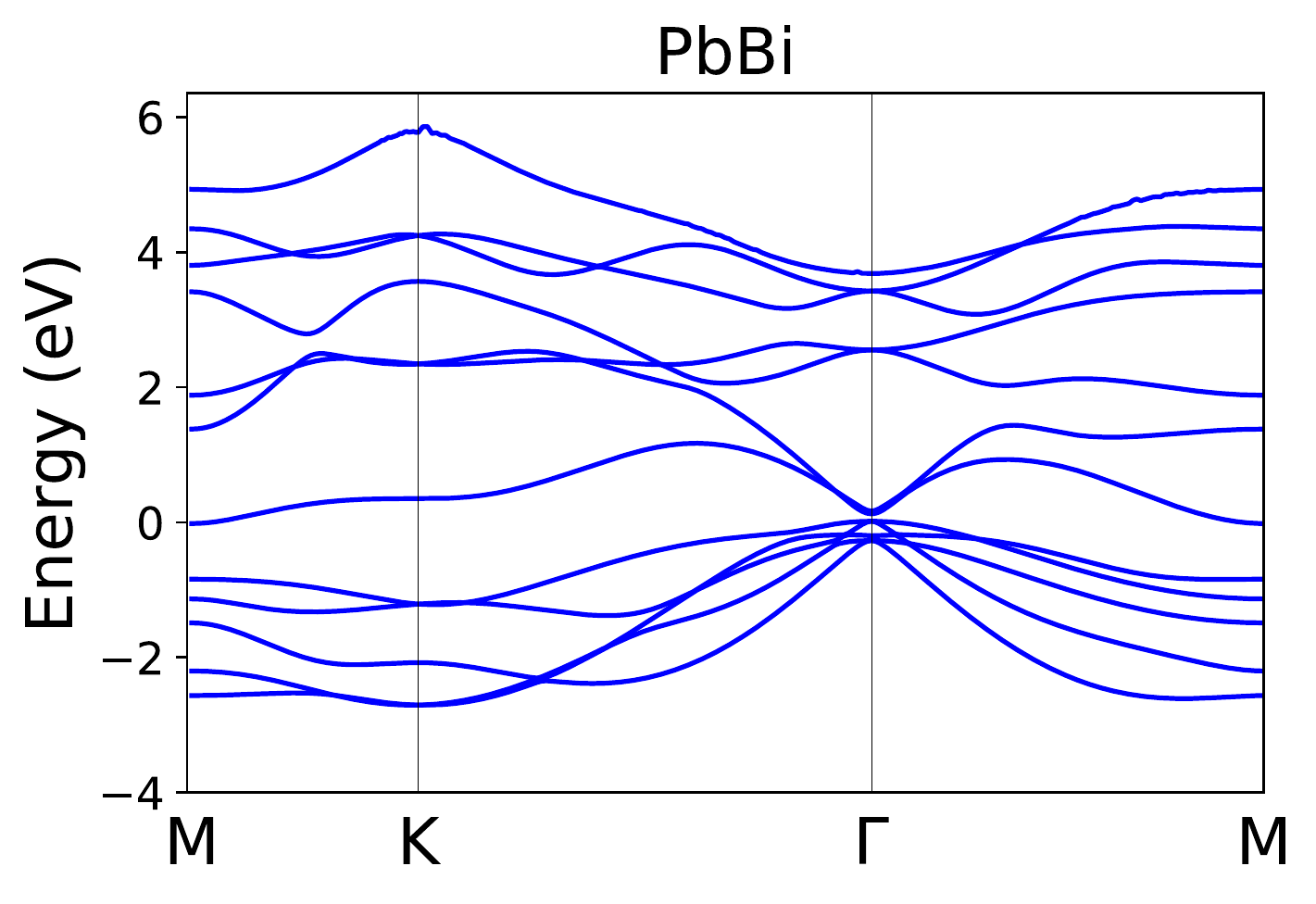}
\end{tabular}

\caption{Electronic band diagrams of $\beta$-structures as obtained from PBE.}
    \label{fig:beta-pbe}
\end{figure}

\begin{figure*}[h!]
   \centering
\begin{tabular}{ccccc}
\\
\includegraphics[width=0.2\textwidth]{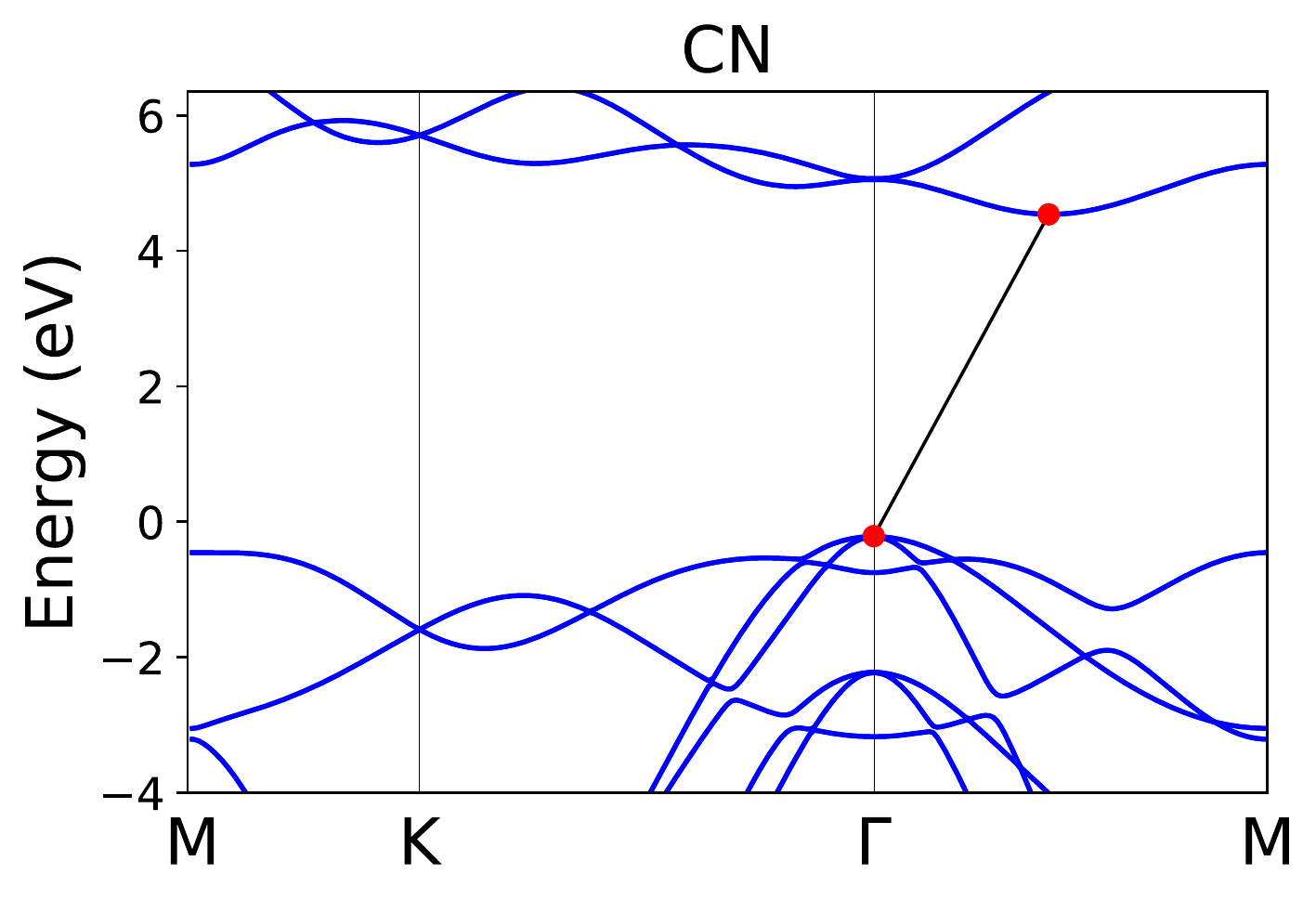}
\includegraphics[width=0.2\textwidth]{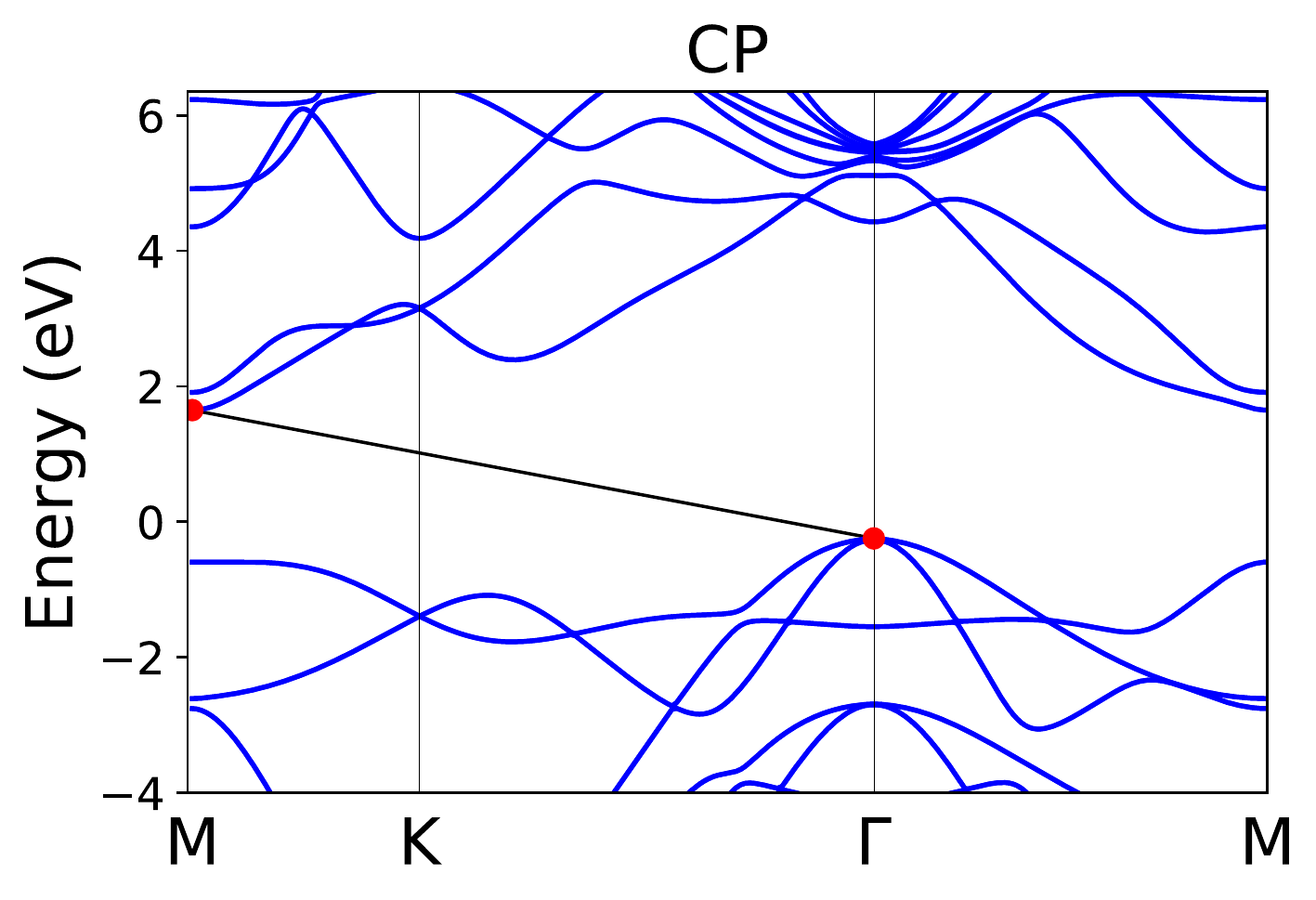}
\includegraphics[width=0.2\textwidth]{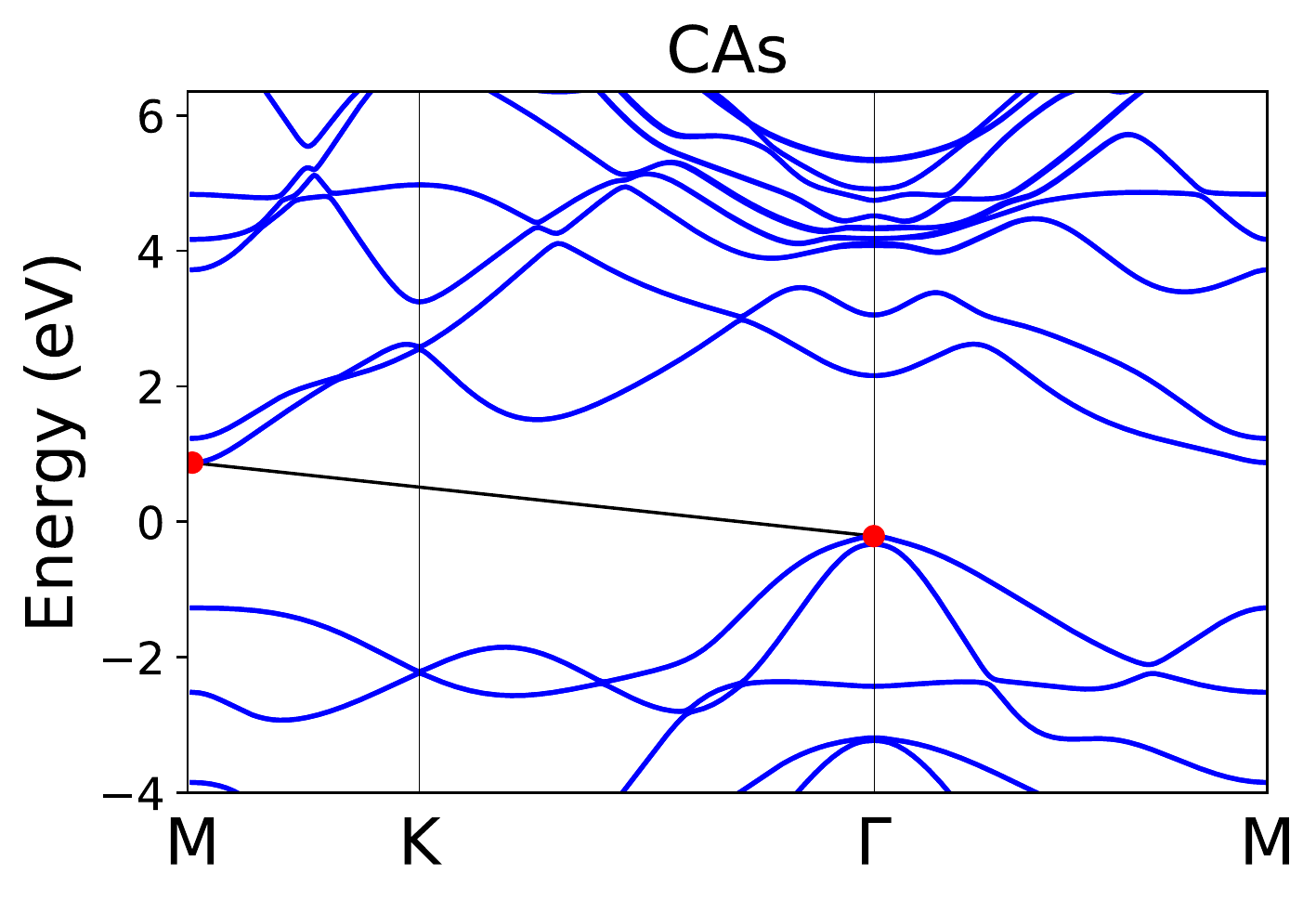}
\includegraphics[width=0.2\textwidth]{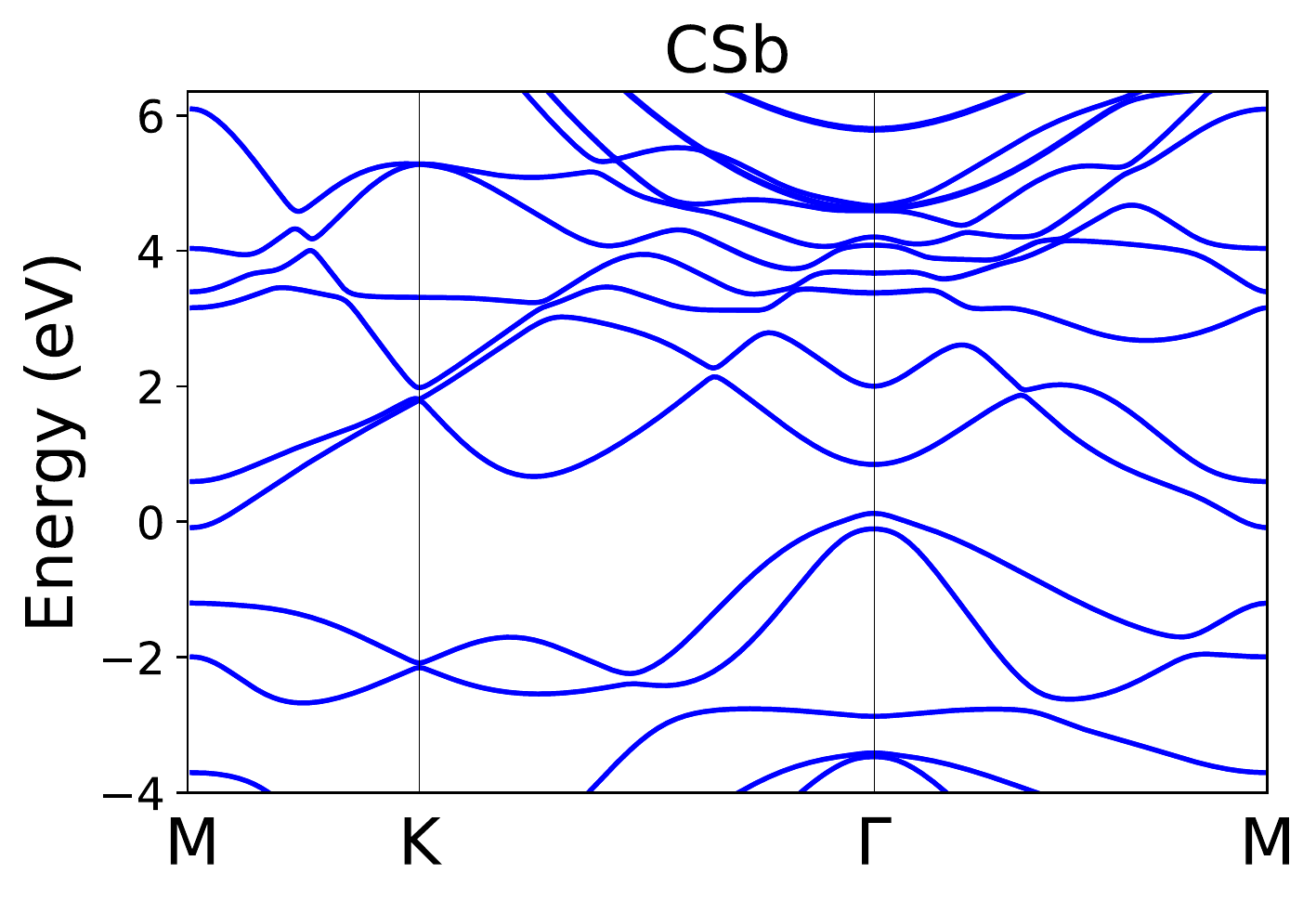}
\includegraphics[width=0.2\textwidth]{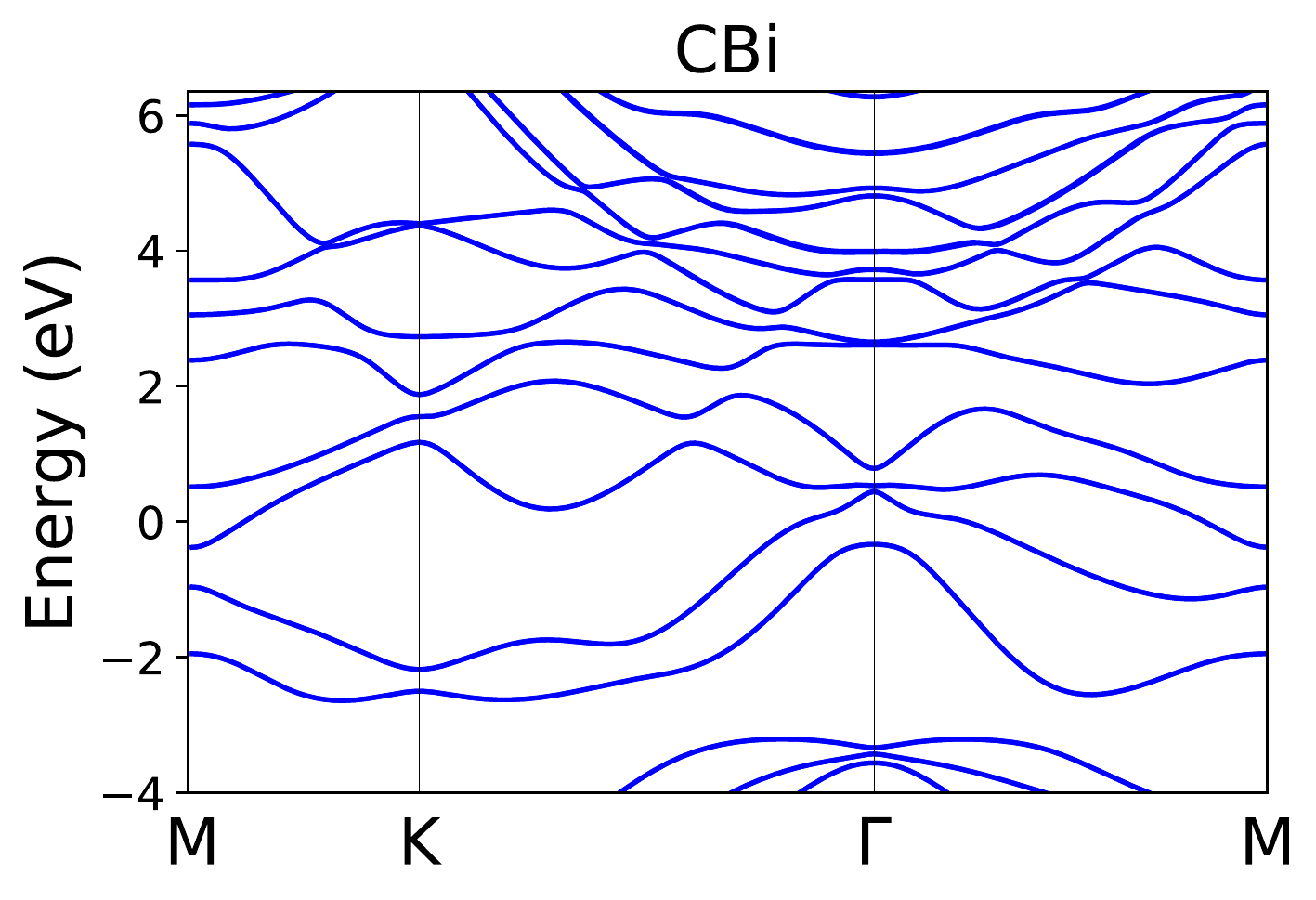}\\

\includegraphics[width=0.2\textwidth]{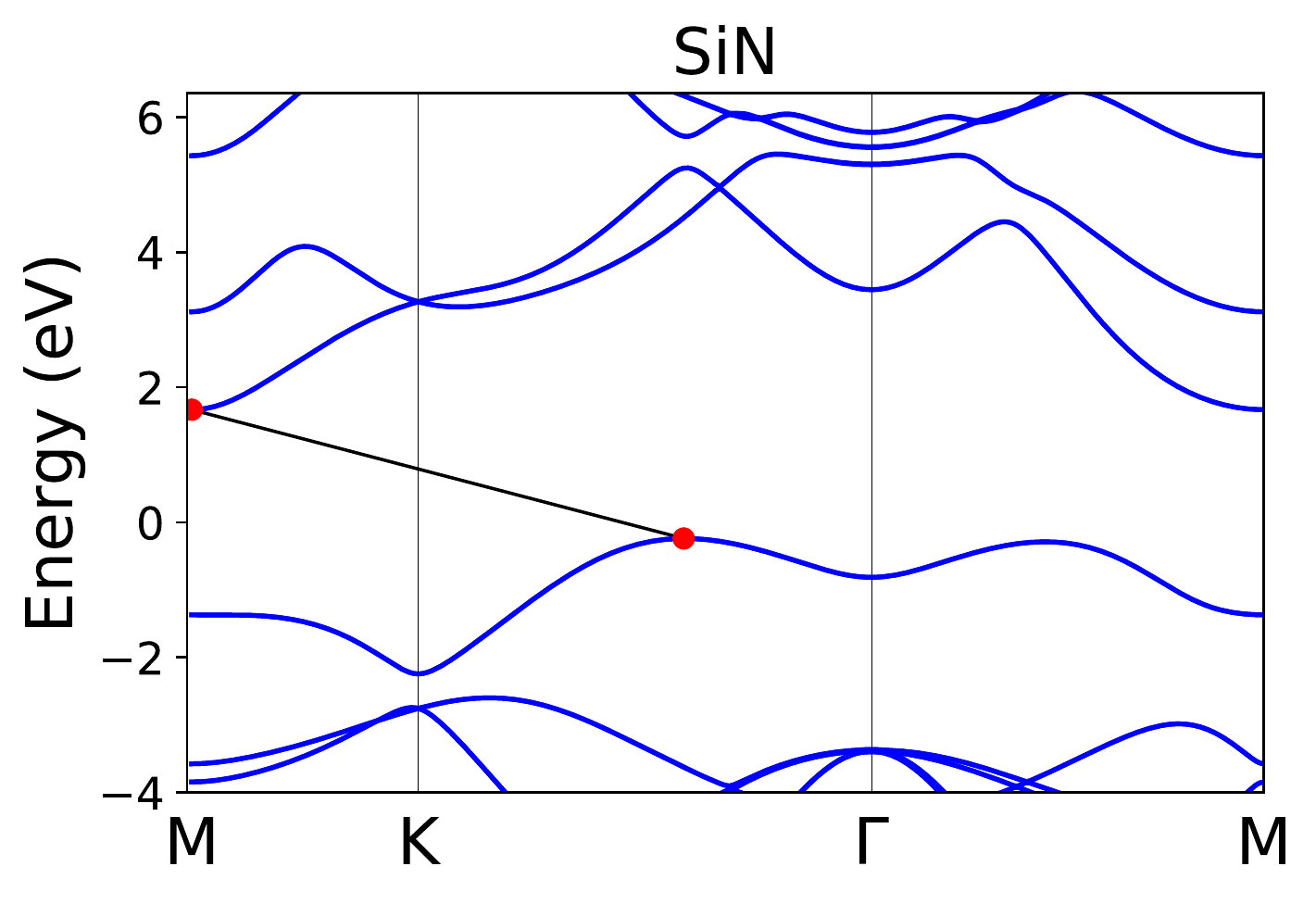}
\includegraphics[width=0.2\textwidth]{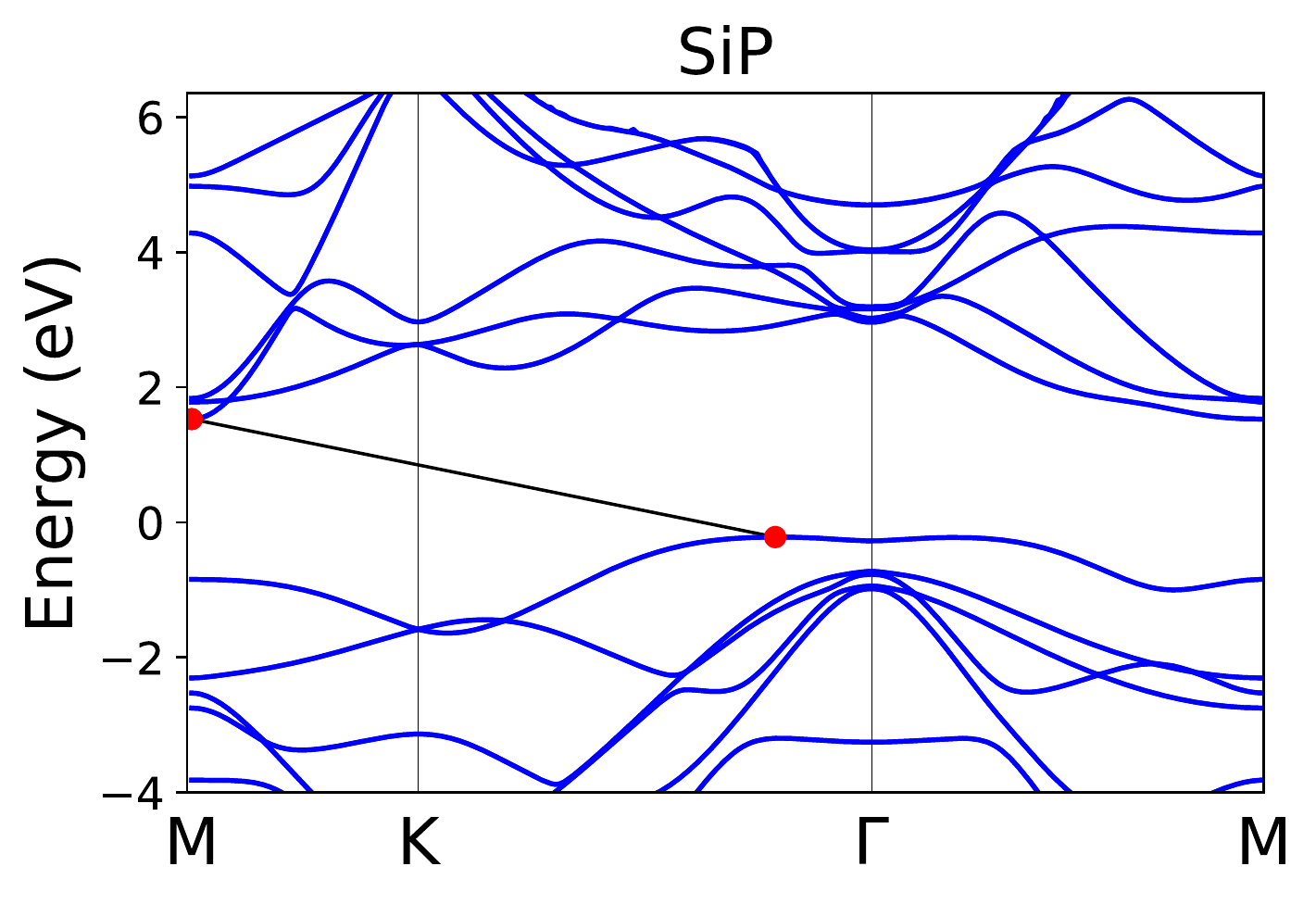}
\includegraphics[width=0.2\textwidth]{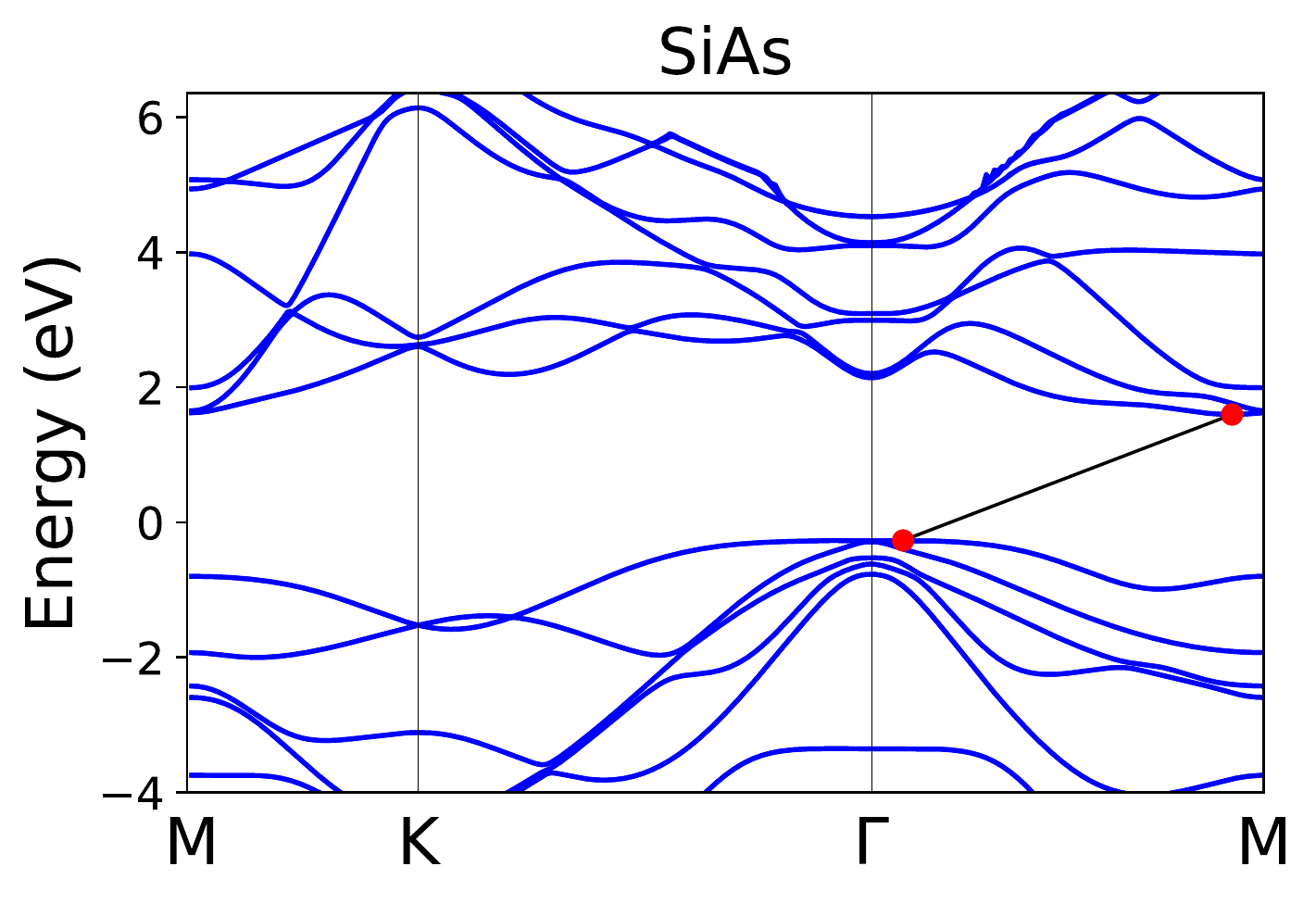}
\includegraphics[width=0.2\textwidth]{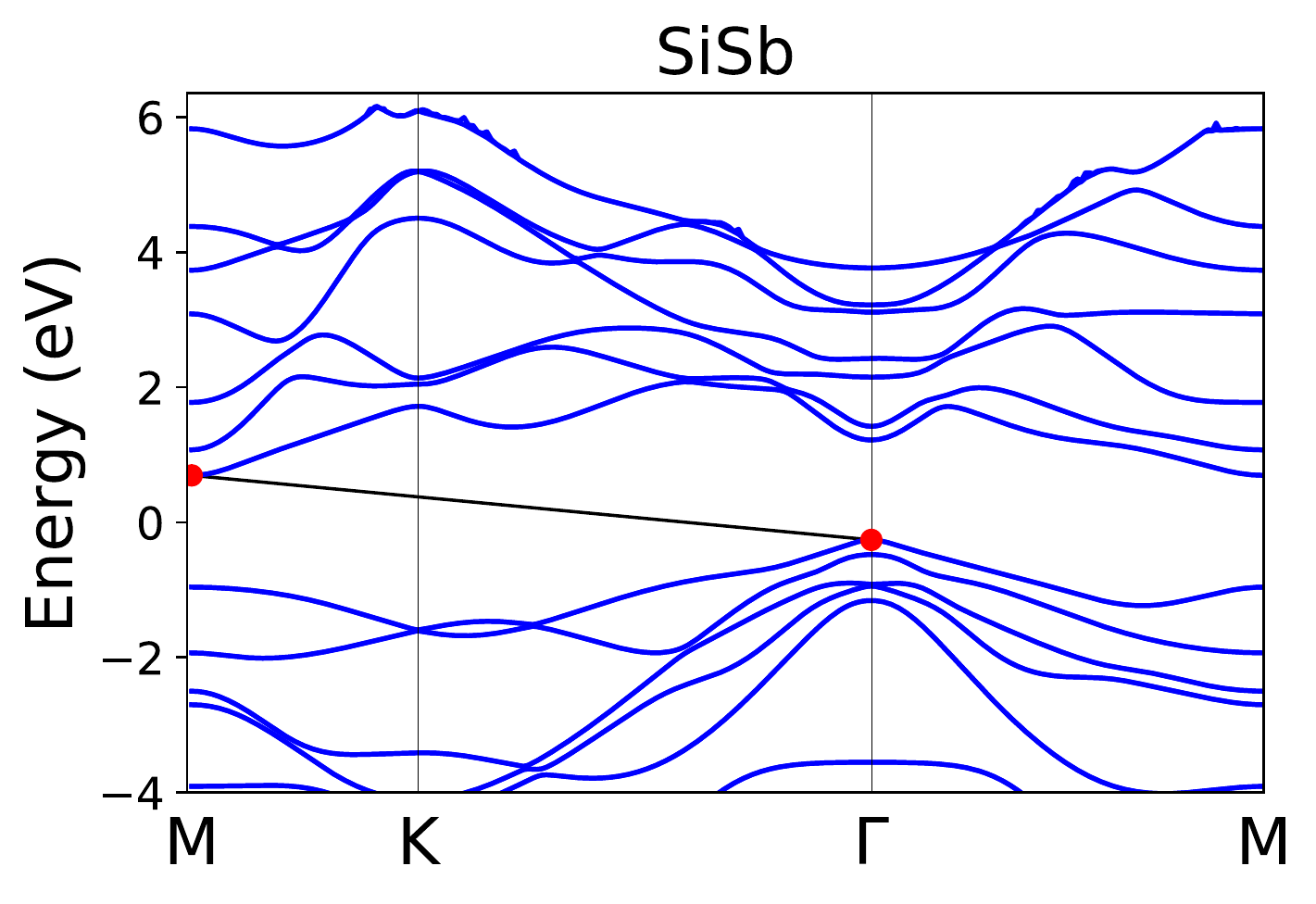}
\includegraphics[width=0.2\textwidth]{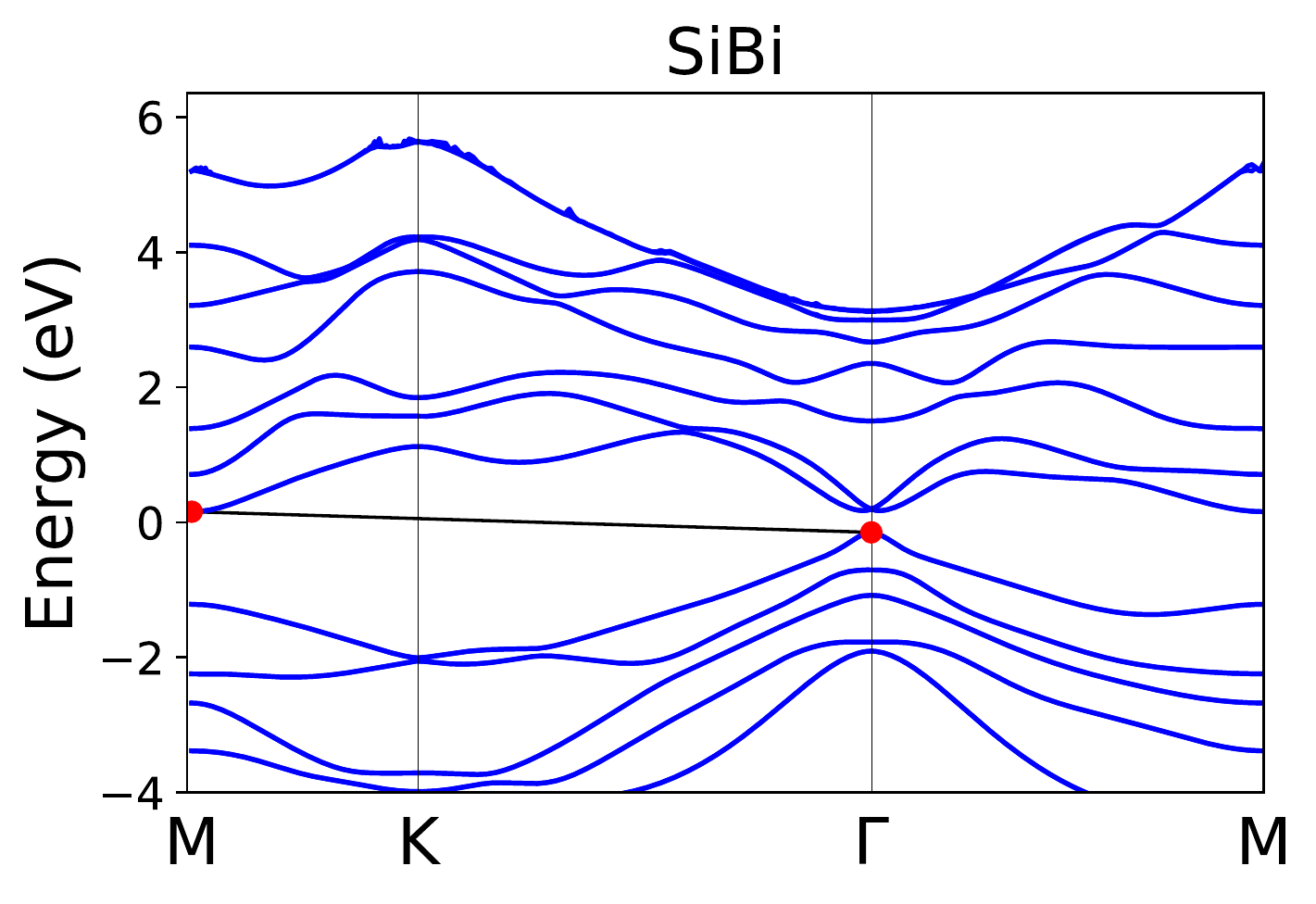}\\

\includegraphics[width=0.2\textwidth]{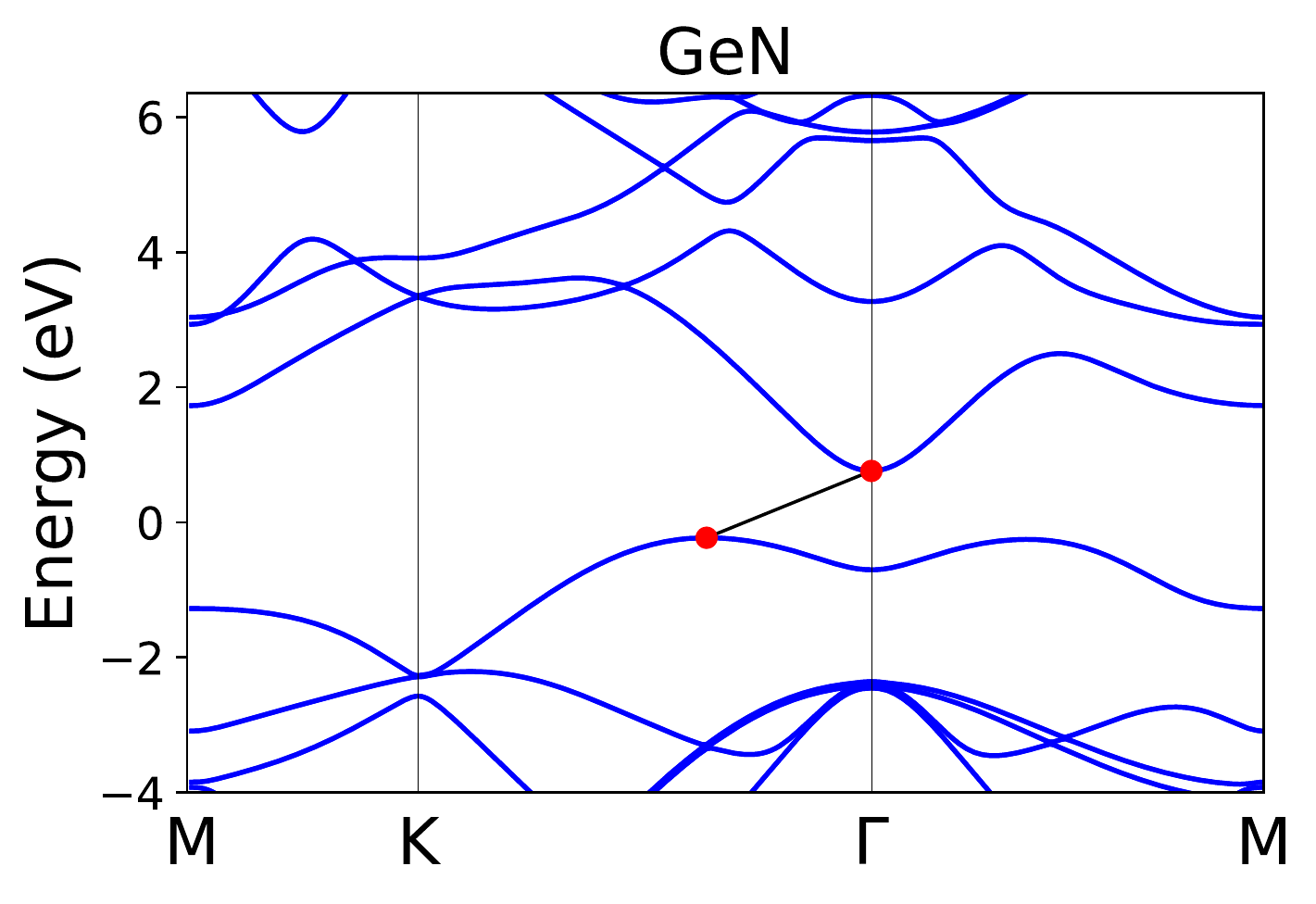}
\includegraphics[width=0.2\textwidth]{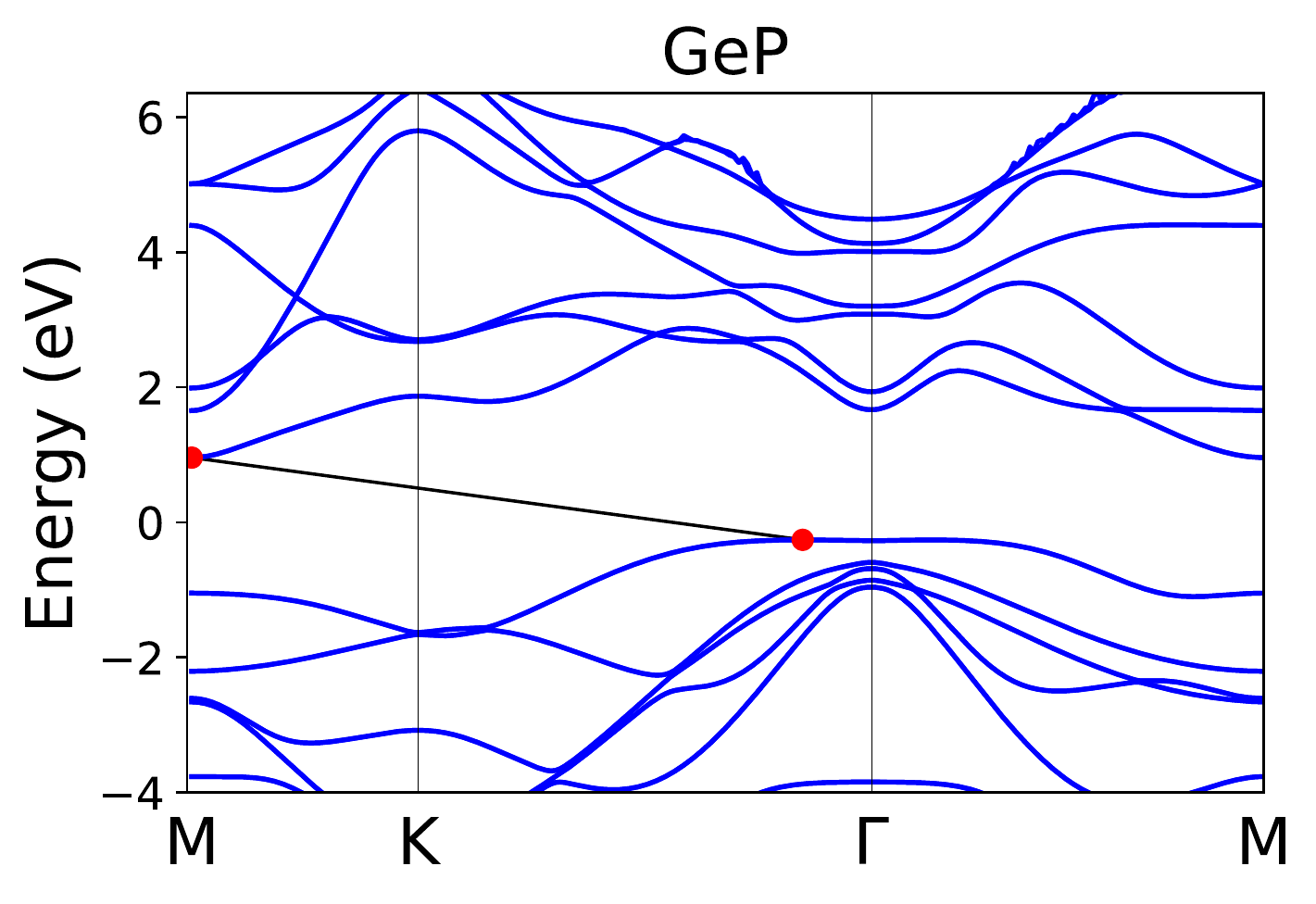}
\includegraphics[width=0.2\textwidth]{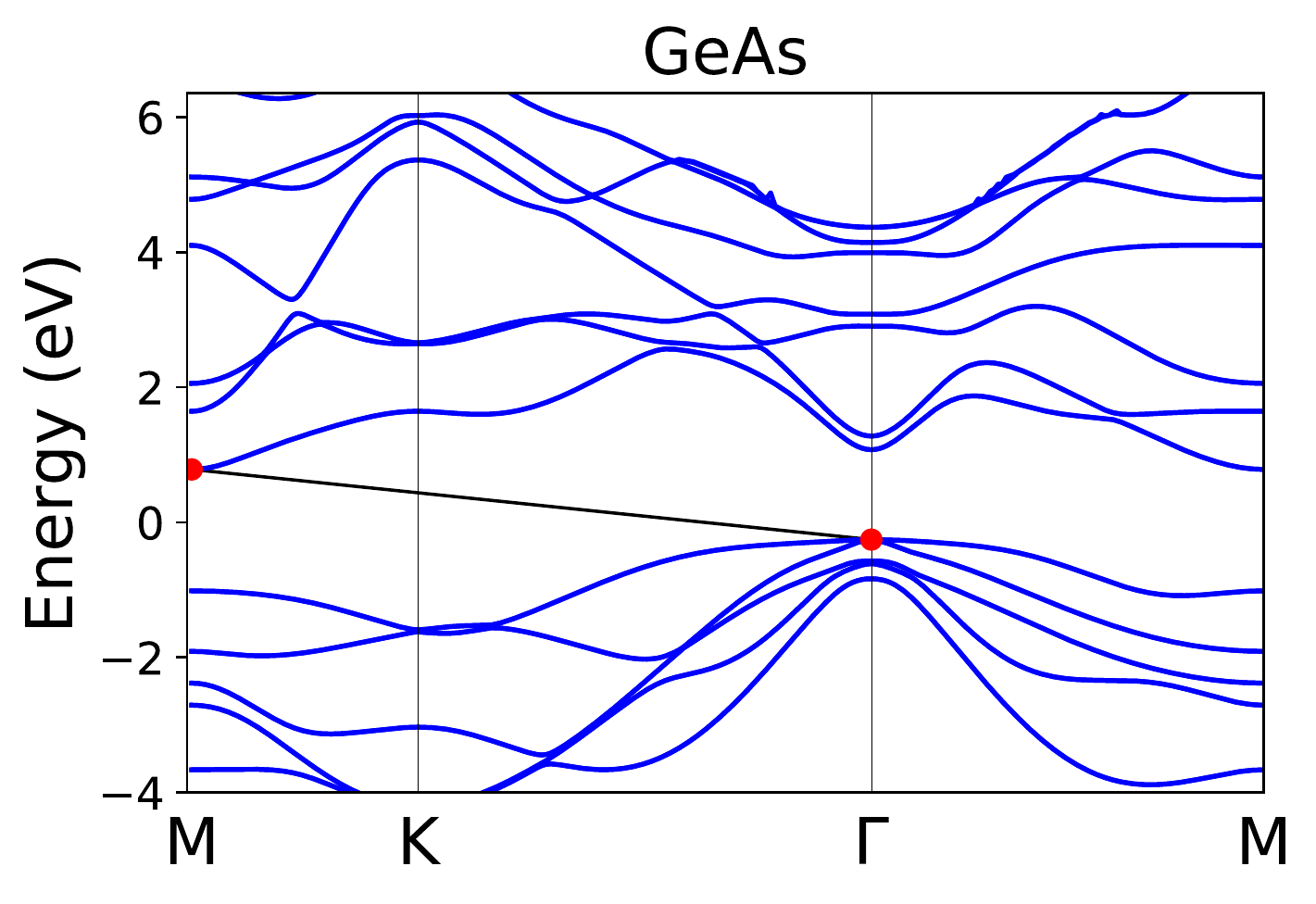}
\includegraphics[width=0.2\textwidth]{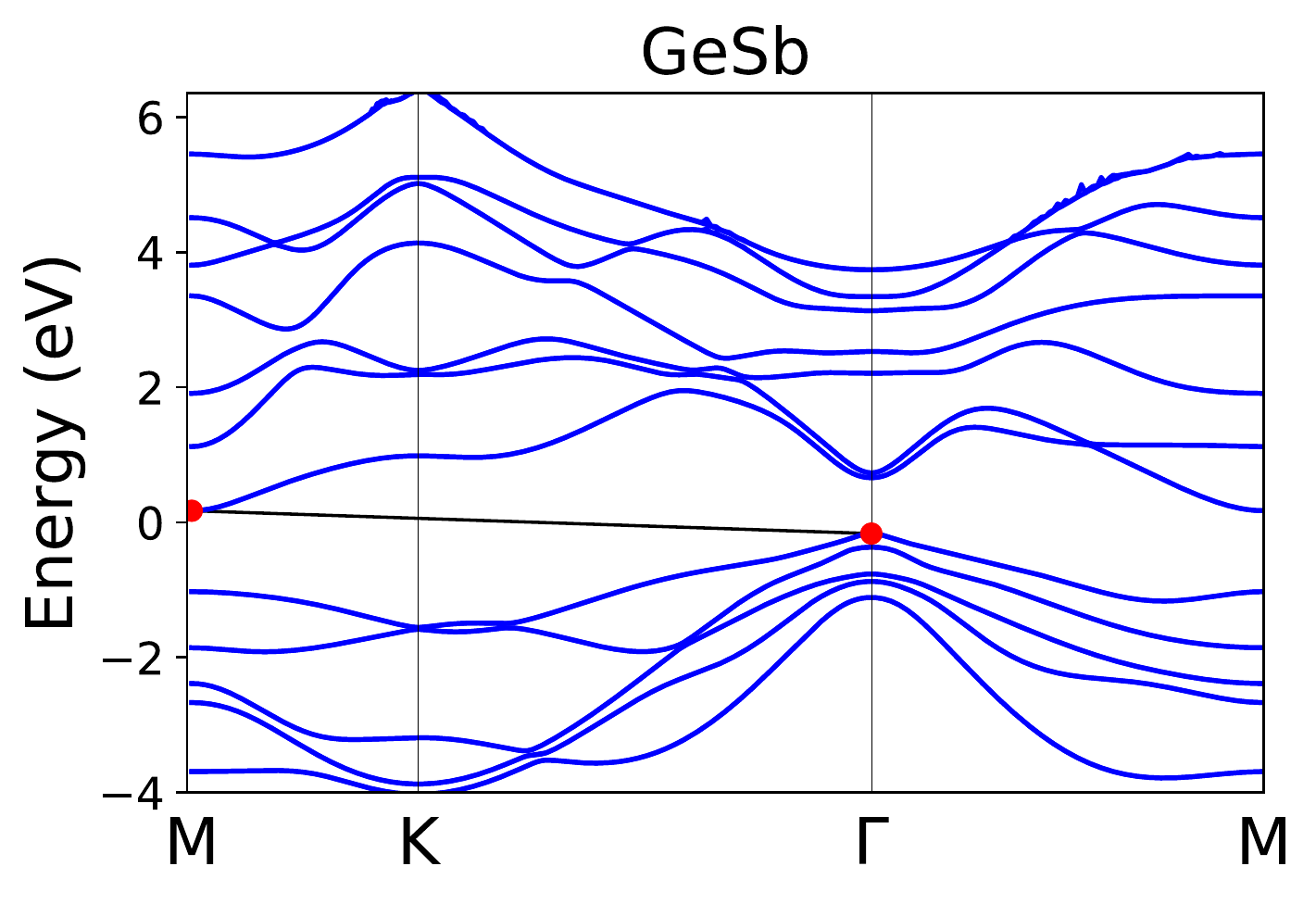}
\includegraphics[width=0.2\textwidth]{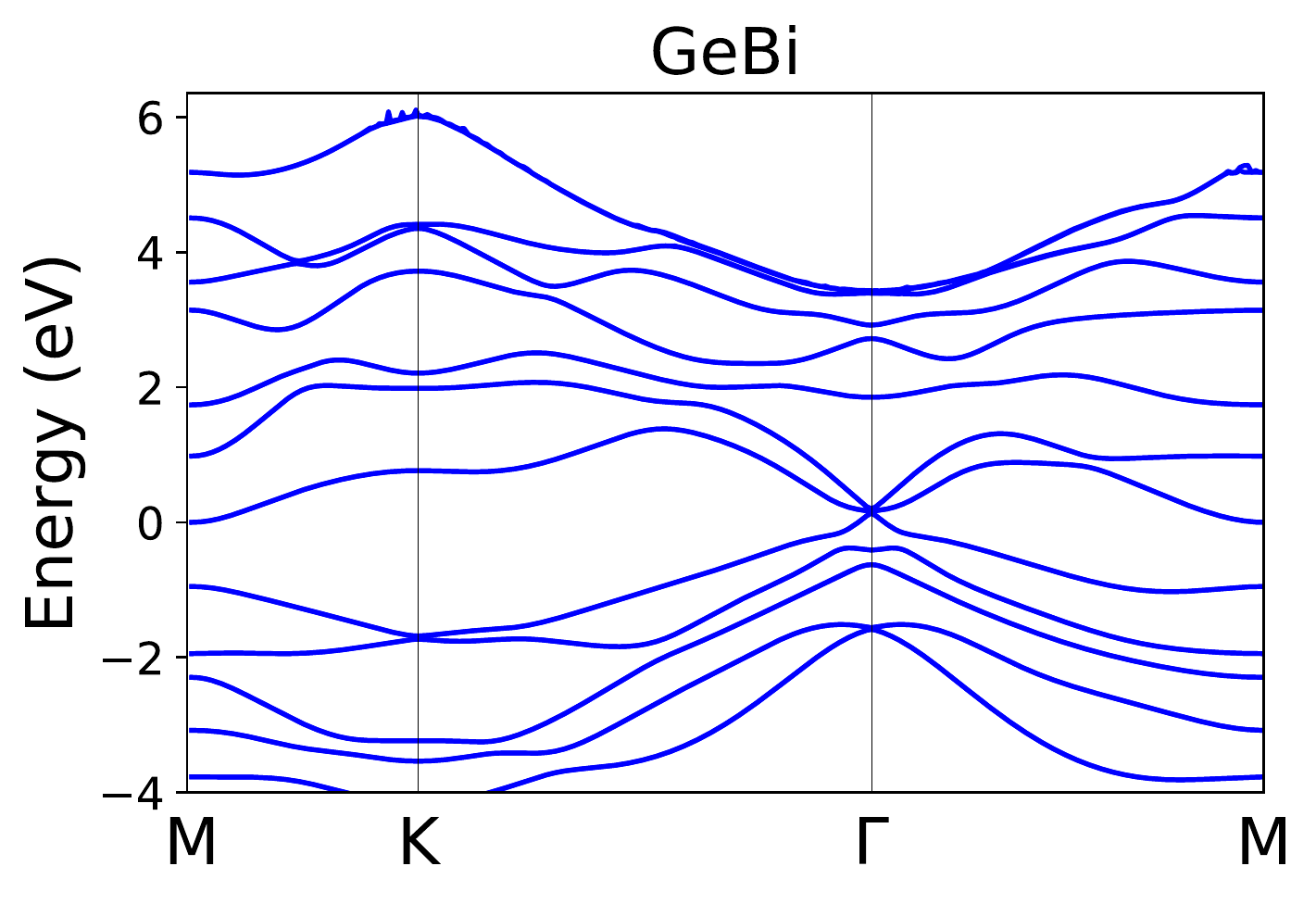}\\

\includegraphics[width=0.2\textwidth]{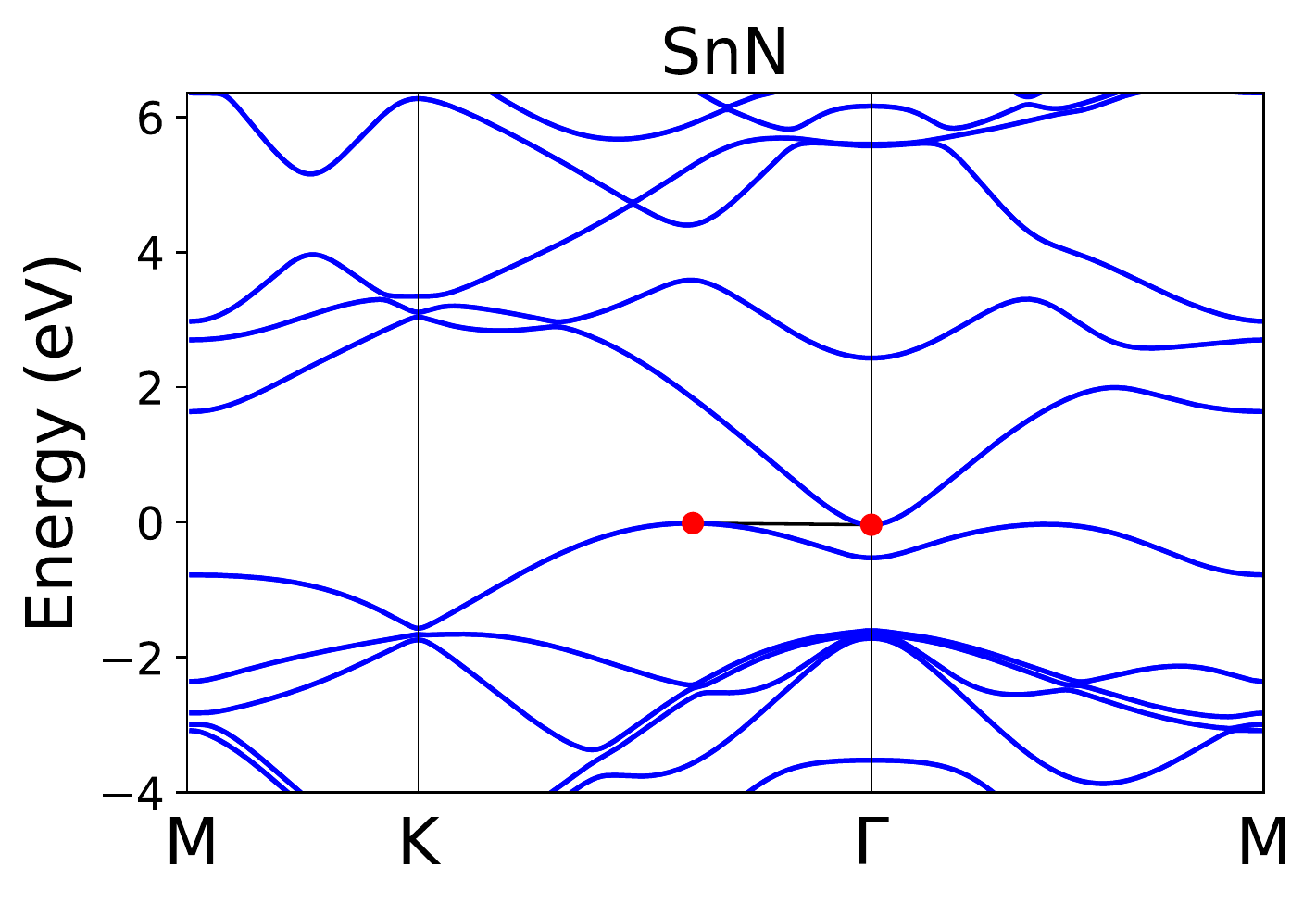}
\includegraphics[width=0.2\textwidth]{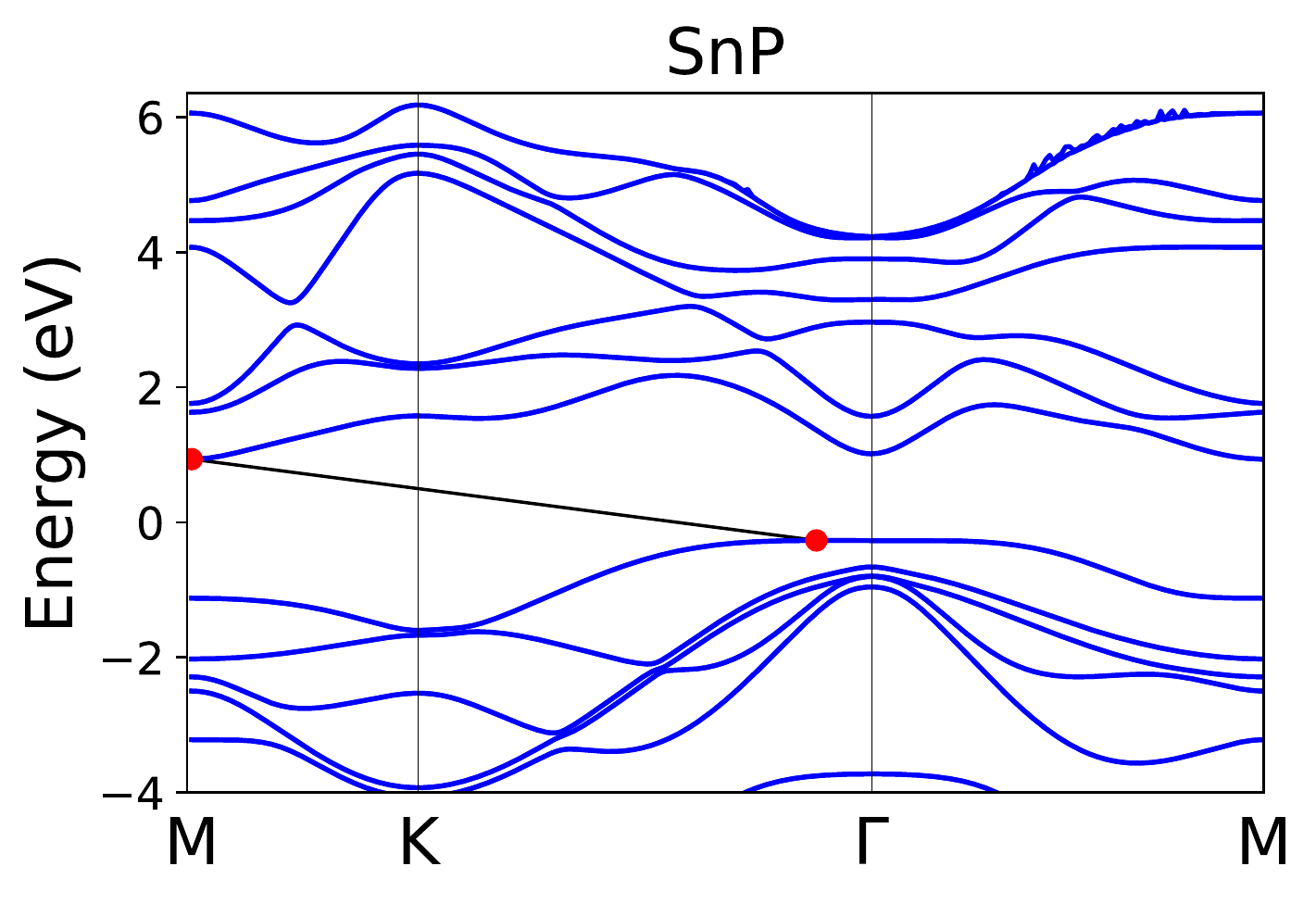}
\includegraphics[width=0.2\textwidth]{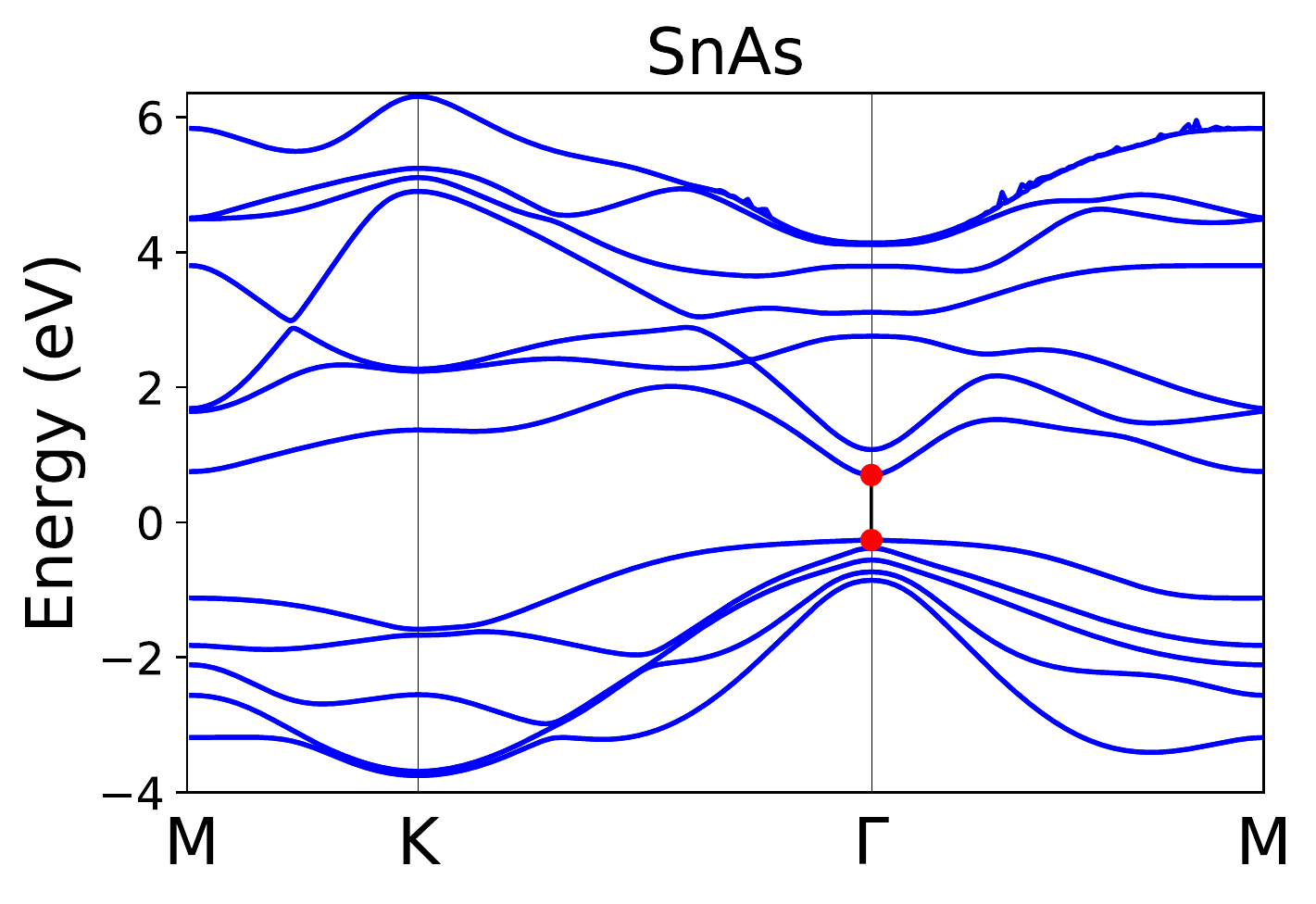}
\includegraphics[width=0.2\textwidth]{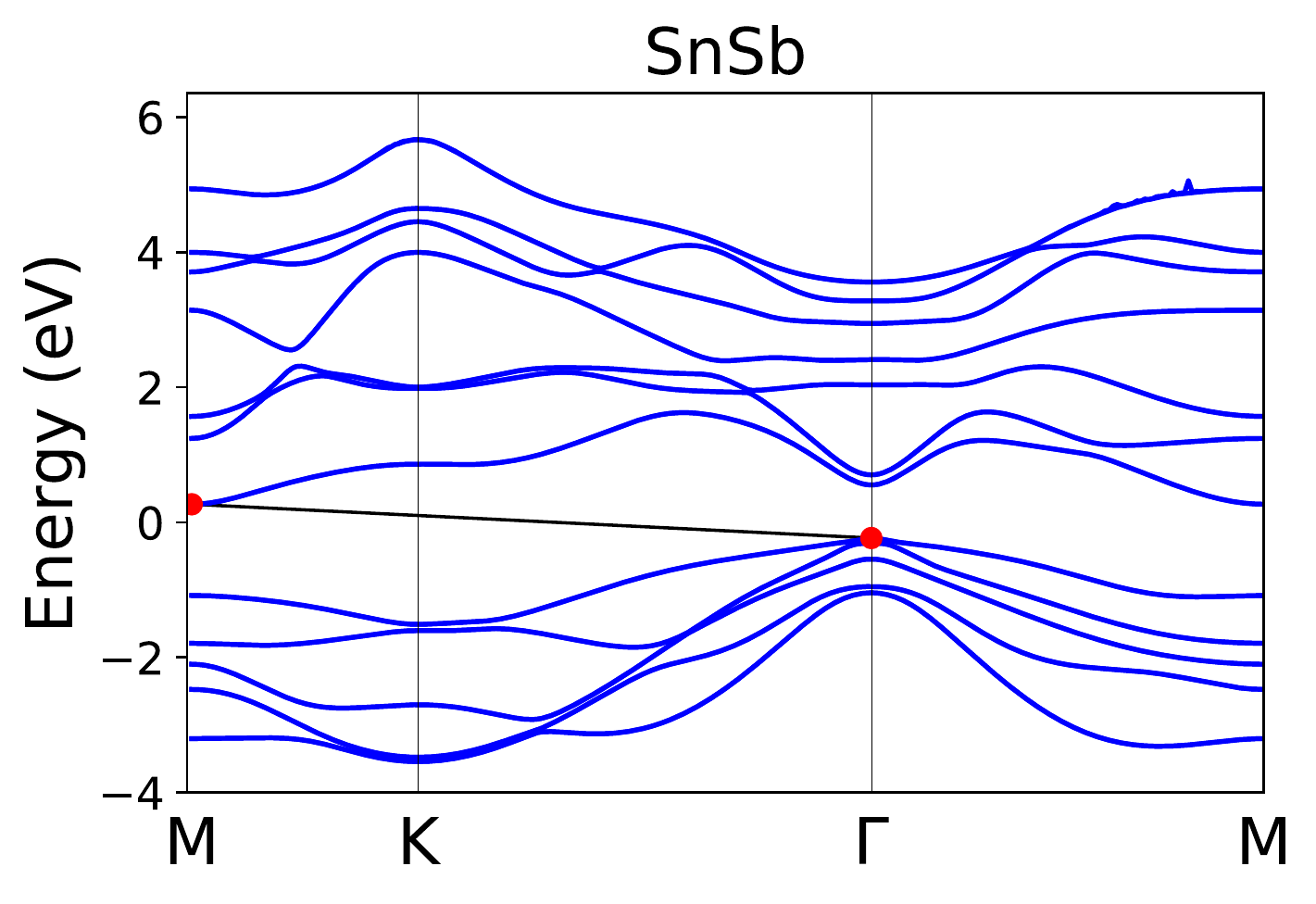}
\includegraphics[width=0.2\textwidth]{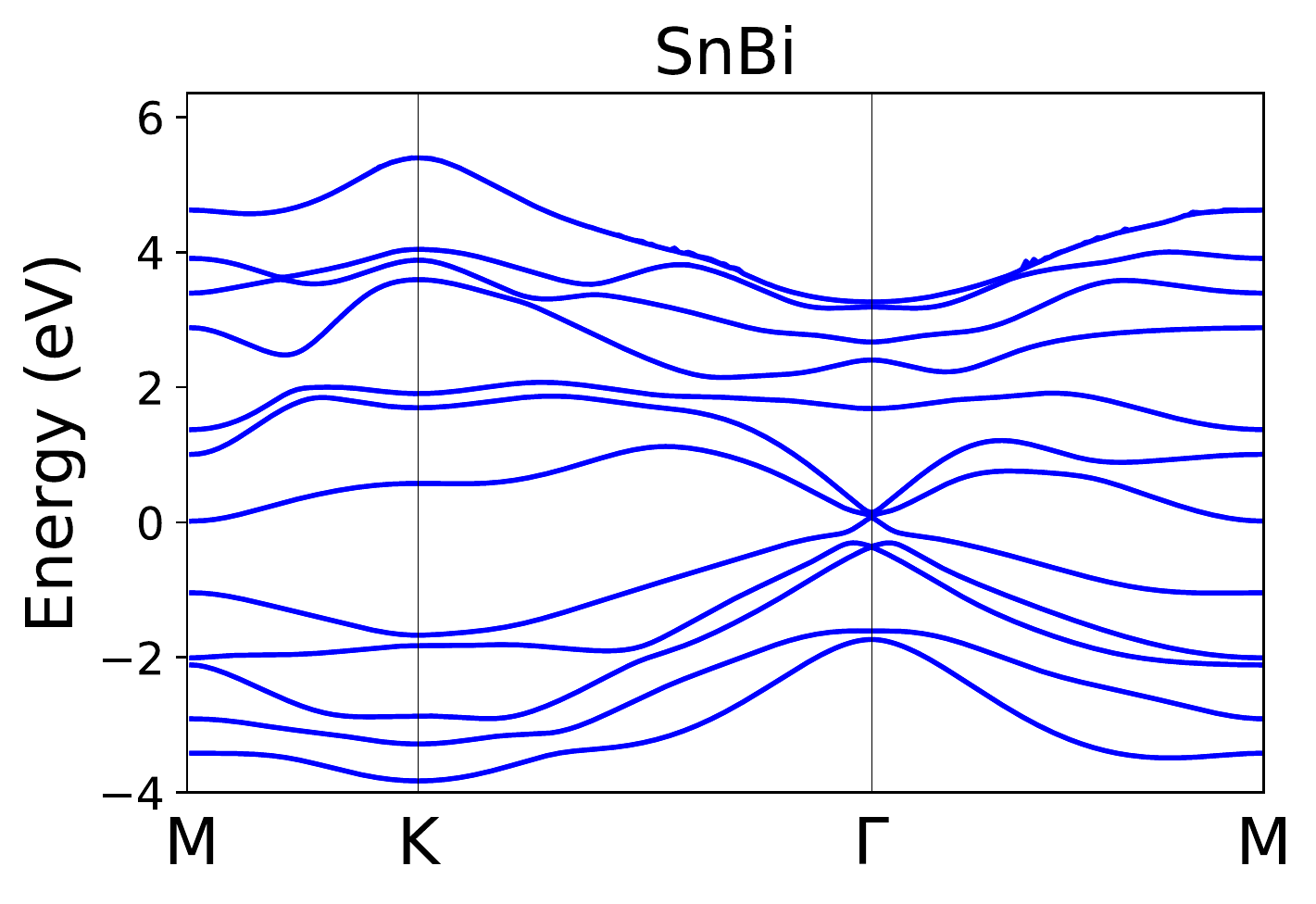}\\

\includegraphics[width=0.2\textwidth]{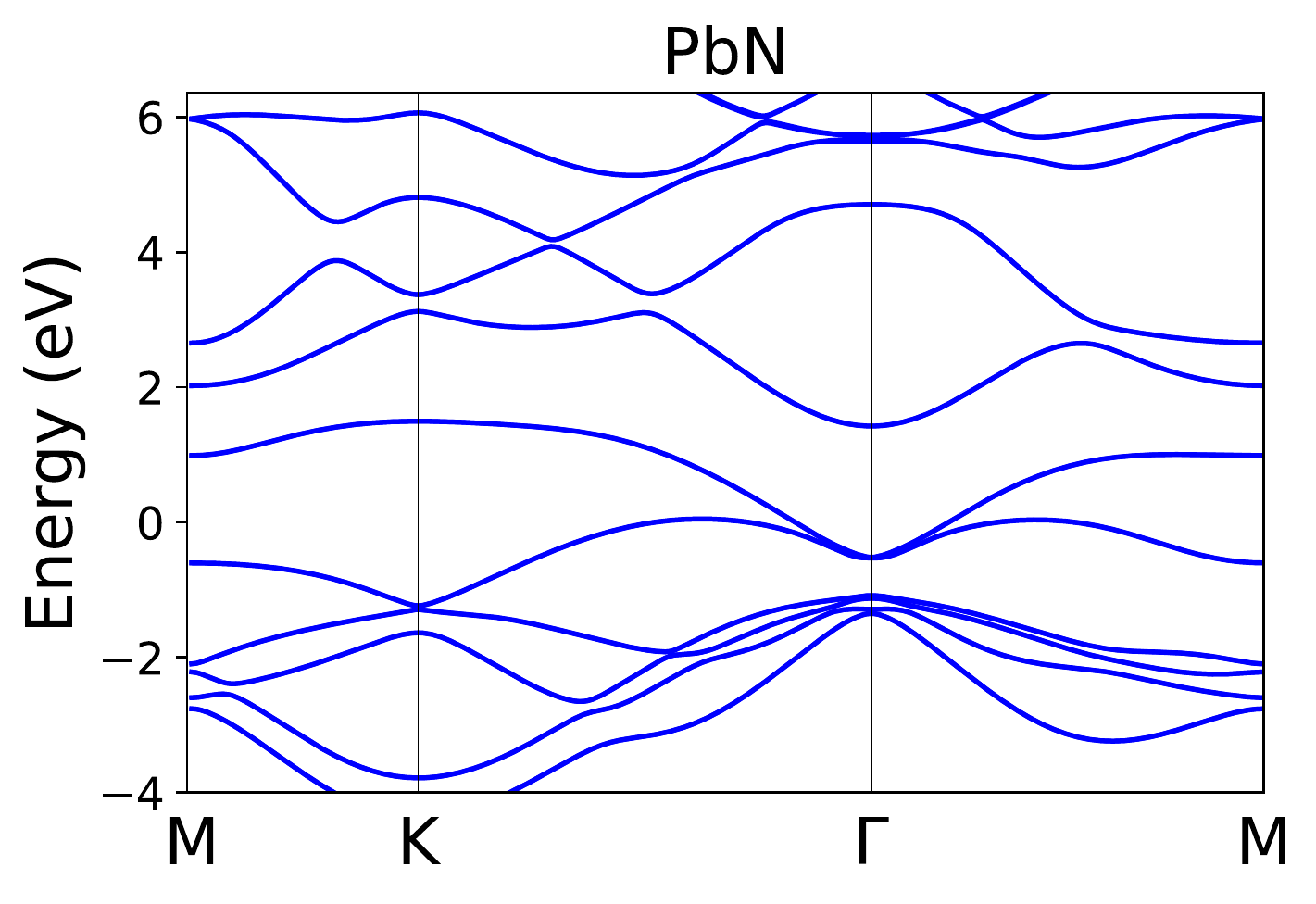}
\includegraphics[width=0.2\textwidth]{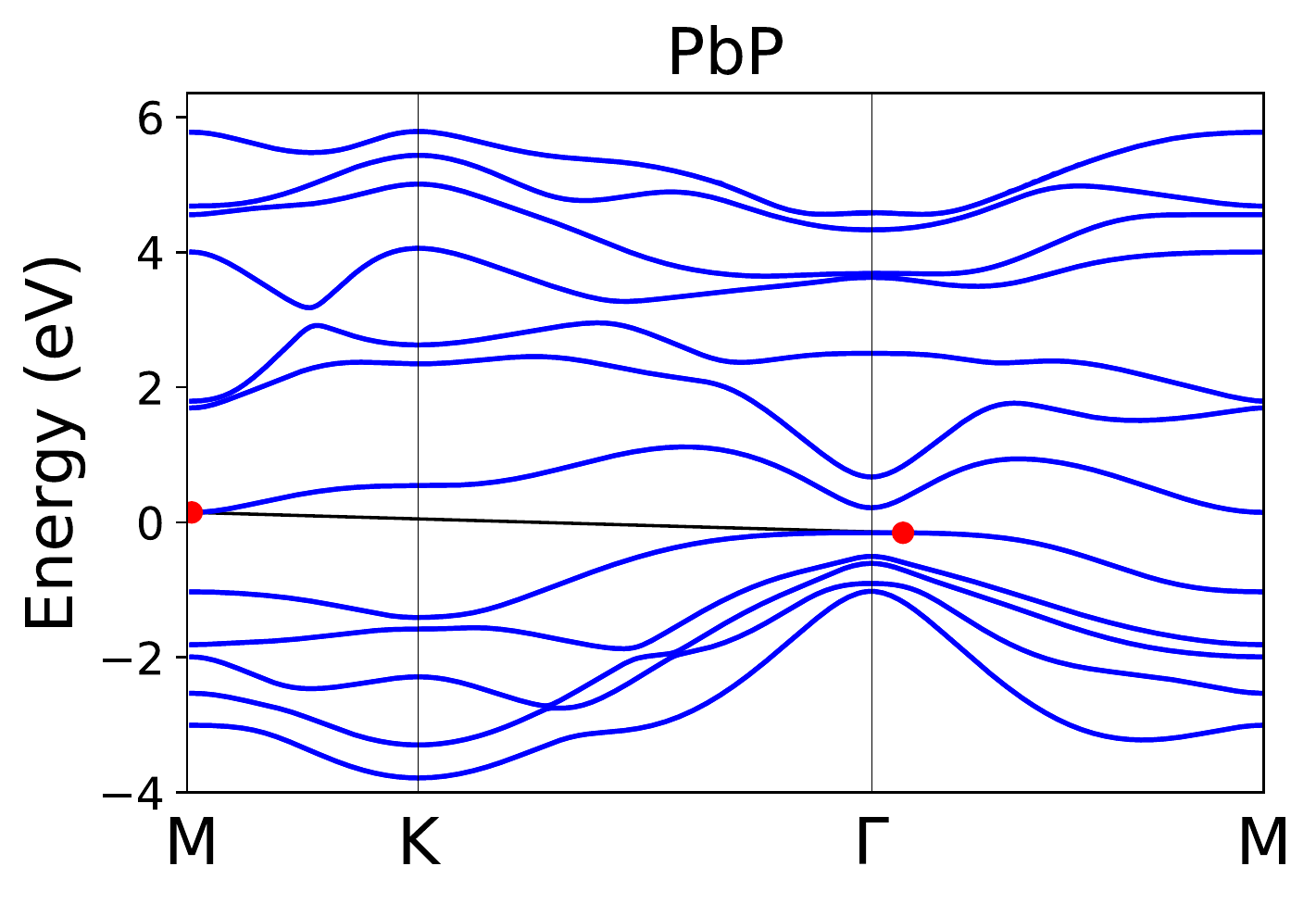}
\includegraphics[width=0.2\textwidth]{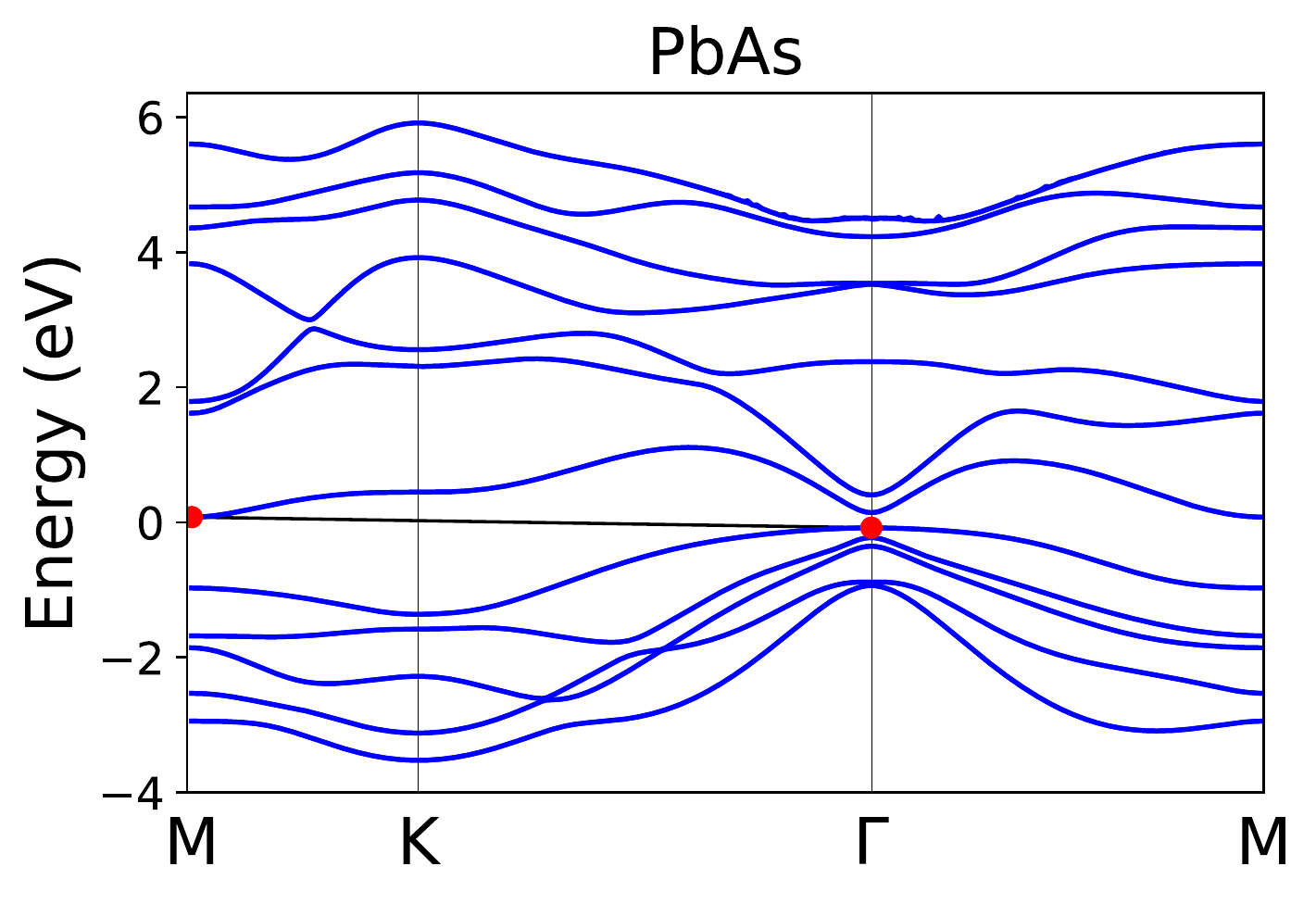}
\includegraphics[width=0.2\textwidth]{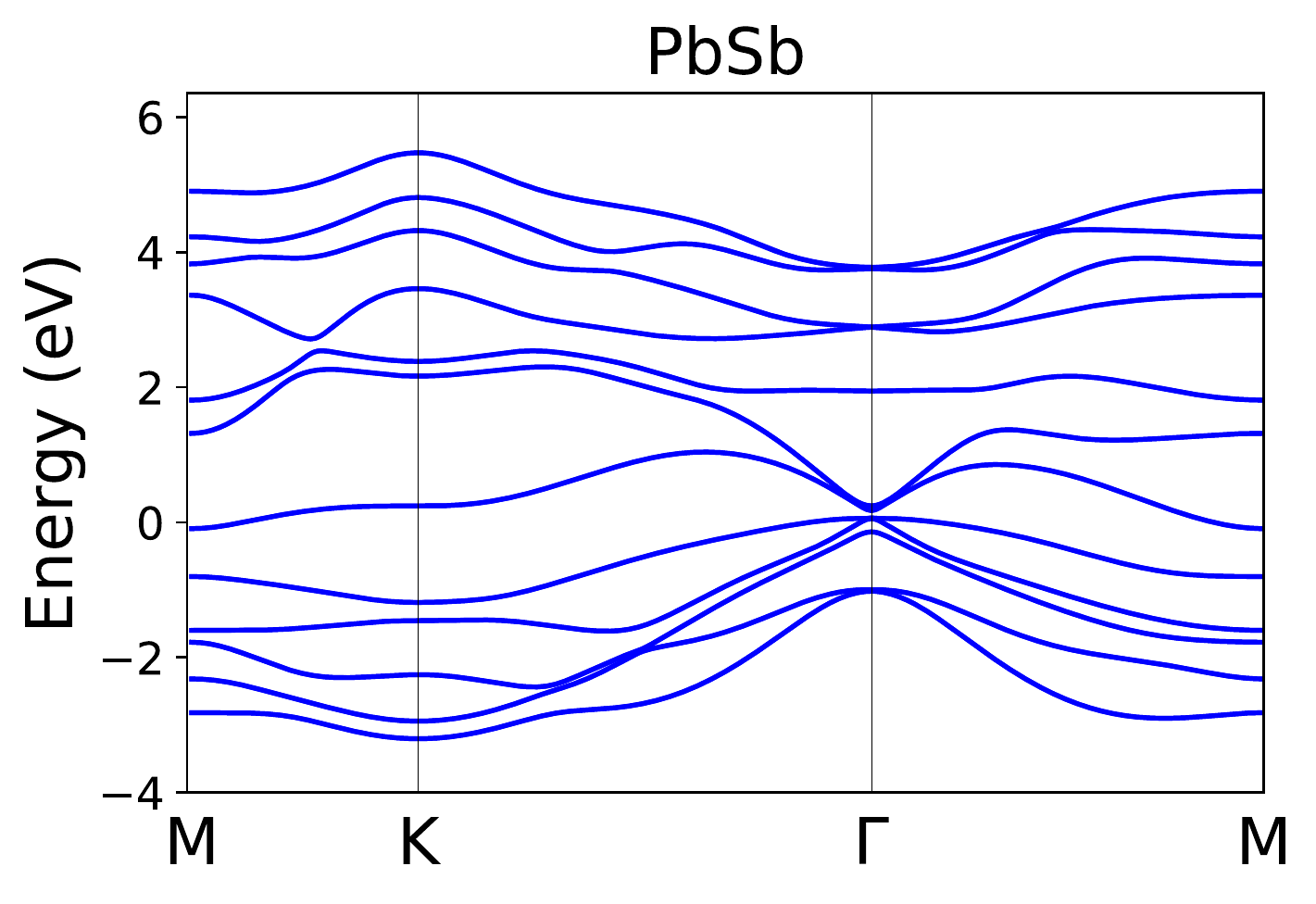}
\includegraphics[width=0.2\textwidth]{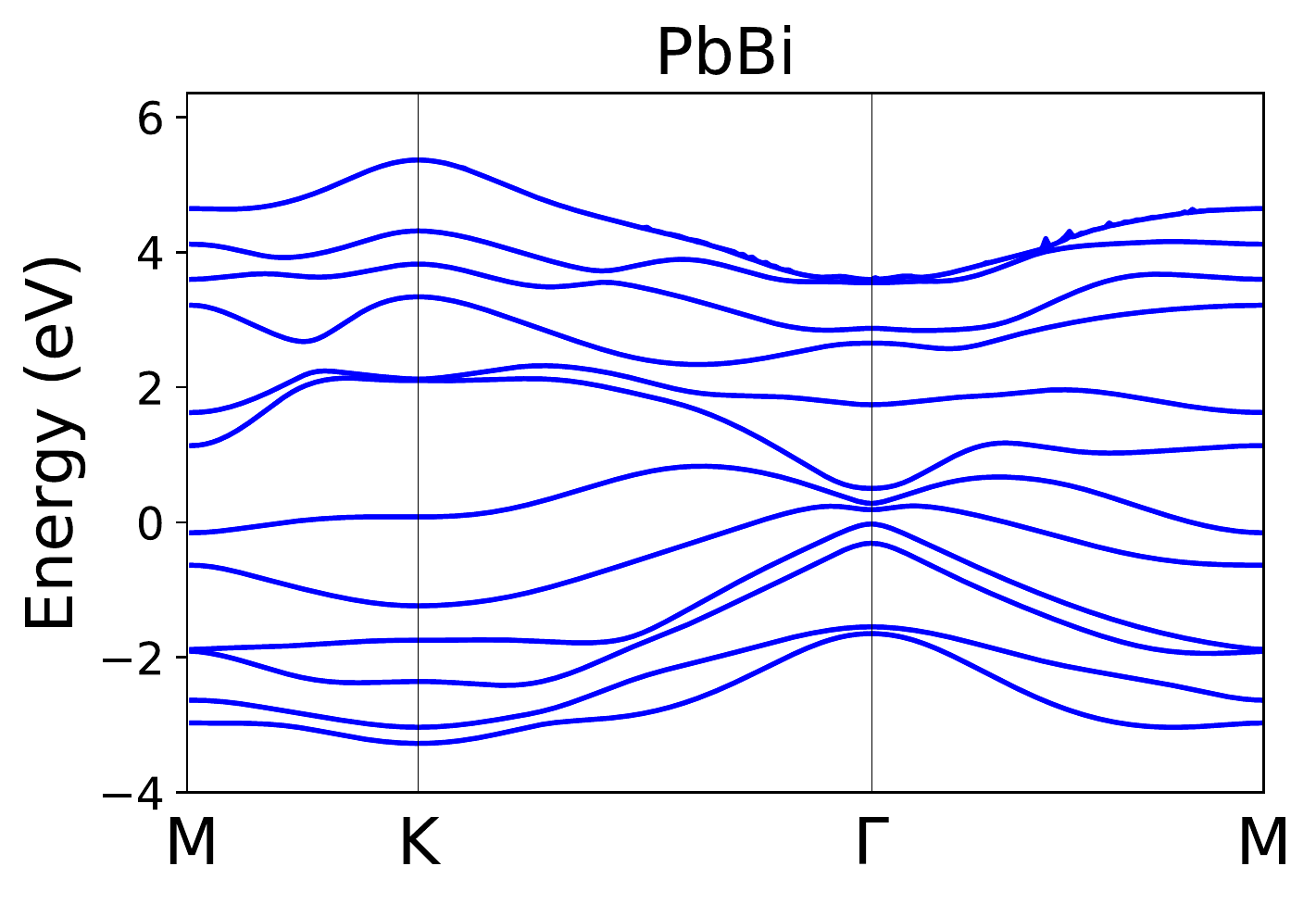}
\end{tabular}

\caption{Electronic band diagrams of $\beta$-structures as obtained from PBE including SOC.}
    \label{fig:beta-soc}
\end{figure*}

Here we present further details regarding the electronic structures of $\alpha$- and $\beta$-phases. In Figure~\ref{fig:alpha-soc} the band structures including the SOC are shown.
In Figure~\ref{fig:beta-pbe} the electronic band diagrams obtained from PBE are shown for $\beta$-phase compounds. The effect of SOC on the band structures of the $\beta$-phases are included in Figure~\ref{fig:beta-soc}. The features in the band structures are discussed in the main text.

\end{document}